\documentstyle[12pt,epsfig,aps]{revtex}

\begin{document}

\thispagestyle{empty}
\renewcommand{\thefootnote}{\noindent\fnsymbol{footnote}}
\setcounter{topnumber}{3}

\title{\baselineskip 16pt Measurements of the Proton and Deuteron Spin Structure Functions $g_1$ and $g_2$\footnote{\tenrm Work supported in part by Department of Energy contract DE-AC03-76SF00515.}}

\vskip -0.5cm
\author{
\baselineskip 14pt
K.~Abe,$^{15}$
T.~Akagi,$^{11,15}$
P.~L.~Anthony,$^{11}$
R.~Antonov,$^{10}$
R.~G.~Arnold,$^{1}$
T.~Averett,$^{16,\ddag\ddag}$
H.~R.~Band,$^{18}$
J.~M.~Bauer,$^{6,\S\S}$
H.~Borel,$^{4}$
P.~E.~Bosted,$^{1}$
V.~Breton,$^{3}$
J.~Button-Shafer,$^{6}$
J.~P.~Chen,$^{16,7}$
T.~E.~Chupp,$^{7}$
J.~Clendenin,$^{11}$
C.~Comptour,$^{3}$
K.~P.~Coulter,$^{7}$
G.~Court,$^{11,*}$
D.~Crabb,$^{16}$
M.~Daoudi,$^{11}$
D.~Day,$^{16}$
F.~S.~Dietrich,$^{5}$
J.~Dunne,$^{1,14}$
H.~Dutz,$^{11,**}$
R.~Erbacher,$^{11,12}$
J.~Fellbaum,$^{1}$
A.~Feltham,$^{2}$
H.~Fonvieille,$^{3}$
E.~Frlez,$^{16}$
D.~Garvey,$^{8}$
R.~Gearhart,$^{11}$
J.~Gomez,$^{14}$
P.~Grenier,$^{4}$
K.~A.~Griffioen,$^{10,17}$
S.~Hoibraten,$^{16,\S}$
E.~W.~Hughes,$^{11,\ddag\ddag}$
C.~Hyde-Wright,$^{9}$
J.~R.~Johnson,$^{18}$
D.~Kawall,$^{12}$
A.~Klein,$^{9}$
S.~E.~Kuhn,$^{9}$
M.~Kuriki,$^{15}$
R.~Lindgren,$^{16}$
T.~J.~Liu,$^{16}$
R.~M.~Lombard-Nelsen,$^{4}$
J.~Marroncle,$^{4}$
T.~Maruyama,$^{11}$
X.~K.~Maruyama,$^{8}$
J.~McCarthy,$^{16}$
W.~Meyer,$^{11,**}$
Z.-E.~Meziani,$^{12,13}$
R.~Minehart,$^{16}$
J.~Mitchell,$^{14}$
J.~Morgenstern,$^{4}$
G.~G.~Petratos,$^{11,\ddag}$
R.~Pitthan,$^{11}$
D.~Pocanic,$^{16}$
C.~Prescott,$^{11}$
R.~Prepost,$^{18}$
P.~Raines,$^{10}$
B.~Raue,$^{9,\dag}$
D.~Reyna,$^{1}$
A.~Rijllart,$^{11,\dag\dag}$
Y.~Roblin,$^{3}$
L.~S.~Rochester,$^{11}$
S.~E.~Rock,$^{1}$
O.~A.~Rondon,$^{16}$
I.~Sick,$^{2}$
L.~C.~Smith,$^{16}$
T.~B.~Smith,$^{7}$
M.~Spengos,$^{1,10}$
F.~Staley,$^{4}$
P.~Steiner,$^{2}$
S.~St.Lorant,$^{11}$
L.~M.~Stuart,$^{11}$
F.~Suekane,$^{15}$
Z.~M.~Szalata,$^{1}$
H.~Tang,$^{11}$
Y.~Terrien,$^{4}$
T.~Usher,$^{11}$
D.~Walz,$^{11}$
F.~Wesselmann,$^{9}$
J.~L.~White,$^{1,11}$
K.~Witte,$^{11}$
C.~C.~Young,$^{11}$
B.~Youngman,$^{11}$
H.~Yuta,$^{15}$
G.~Zapalac,$^{18}$
B.~Zihlmann,$^{2}$
D.~Zimmermann$^{16}$}

\address{
\baselineskip 14 pt
\vskip 0.1cm
{\rm The E143 Collaboration}\break
\vskip 0.1 cm
{$^{1}$The American University, Washington, D.C. 20016}  \break
{$^{2}$Institut f\" ur Physik der Universit\" at Basel,
  CH--4056 Basel, Switzerland} \break
{$^{3}$LPC IN2P3/CNRS,
  University Blaise Pascal, F--63170 Aubiere Cedex, France}  \break
{$^{4}$DAPNIA-Service de Physique Nucleaire
  Centre d'Etudes de Saclay, F--91191 Gif/Yvette, France} \break
{$^{5}$Lawrence Livermore National Laboratory, Livermore, California 94550}
\break
{$^{6}$University of Massachusetts,  Amherst, Massachusetts 01003}  \break
{$^{7}$University of Michigan, Ann Arbor, Michigan 48109} \break
{$^{8}$Naval Postgraduate School, Monterey, California 93943} \break
{$^{9}$Old Dominion University,  Norfolk, Virginia 23529} \break
{$^{10}$University of Pennsylvania,  Philadelphia, Pennsylvania
  19104} \break
{$^{11}$Stanford Linear Accelerator Center,
  Stanford, California 94309} \break
{$^{12}$Stanford University, Stanford, California 94305} \break
{$^{13}$Temple University, Philadelphia, Pennsylvania 19122}  \break
{$^{14}$Thomas Jefferson National Accelerator Facility, Newport News, Virginia 23606} \break
{$^{15}$Tohoku University, Sendai 980, Japan} \break
{$^{16}$University of Virginia, Charlottesville, Virginia 22901} \break
{$^{17}$The College of William and Mary, Williamsburg, Virginia 23187} \break
{$^{18}$University of Wisconsin, Madison, Wisconsin 53706}  \break 
\break \break
{\it Submitted to Physical Review D} \break
} 
 
\maketitle
\vskip -8.7 in
\vspace*{-.5in}
\begin{flushright}{\small
  SLAC-PUB-7753\\[-.15in]
  February 1998\\ \quad \\
}
\end{flushright}

\newpage

\begin{center}
\begin{minipage}{5.0in}  
\baselineskip 16pt 
\tenrm
Measurements are reported of the proton and deuteron spin structure 
functions $g_1^p$ and  $g_1^d$ at beam energies of 29.1, 16.2, and 9.7 GeV, 
and $g_2^p$ and  $g_2^d$ at a beam energy of 29.1 GeV. The integrals 
$\Gamma_p =
\int_0^1g_1^p(x,Q^2)dx$ and $\Gamma_d = \int_0^1g_1^d(x,Q^2)dx$ were 
evaluated at fixed $Q^2$ = 3~(GeV/c)$^2$ using the full data set to 
yield $\Gamma_p = 0.132 \pm 0.003{\rm (stat.)} \pm 0.009{\rm (syst.)}$ 
and   $\Gamma_d = 0.047 \pm 0.003 \pm 0.006$. The~$Q^2$ dependence of 
the ratio $g_1/F_1$ was  studied and  found to be small for 
$Q^2 > 1$~(GeV/c)$^2$. Within experimental precision the $g_2$ data 
are well-described by the twist-2 contribution, $g_2^{WW}$. Twist-3 
matrix elements were extracted and compared to theoretical predictions. 
The asymmetry $A_2$ was measured and found to be  significantly smaller 
than the positivity limit $\sqrt{R}$ for both
proton and deuteron  targets. $A_2^p$ is found to be positive and 
inconsistent with zero. Measurements of $g_1$ in the resonance region 
show strong variations with $x$ and $Q^2$, consistent with resonant 
amplitudes extracted from unpolarized data. These data allow us to study 
the $Q^2$ dependence of the integrals $\Gamma_p$ and $\Gamma_n$ below the 
scaling region.
\end{minipage}
\end{center}

\vskip 1cm
\renewcommand{\baselinestretch}{1.0}
\small
\normalsize

\tableofcontents

\newpage

\section{Introduction}
Inelastic lepton scattering from nucleons has been  used over 
the past thirty years to obtain an ever-increasing knowledge 
of the distribution of the partons that make up the nucleon, namely 
gluons and up, down, strange, and perhaps charmed quarks. It 
is one of the great successes of QCD that the same parton 
densities can be used to describe the unpolarized inelastic 
structure functions $F_1(x,Q^2)$ and $F_2(x,Q^2)$ as well as 
many other physical processes, such as the production of jets 
in pp collisions. The parton densities depend on the 
fractional momentum of the parton $x=Q^2/2M\nu$, where $-Q^2$ is
the four-momentum transfer squared, $M$ is the nucleon
mass, and $\nu$ is the lepton energy transfer. The measured 
$Q^2$ dependence at fixed $x$ of $F_1(x,Q^2)$ and $F_2(x,Q^2)$
has been shown to be in very good agreement with the 
QCD-based evolution equations \cite{dglap}.

The $F_1$ and $F_2$ structure functions are sensitive to the 
helicity-averaged parton densities. Recent improvements in polarized 
lepton beams 
and targets have made it possible to make increasingly accurate 
measurements of two additional structure functions, $g_1(x,Q^2)$ and 
$g_2(x,Q^2)$, which depend on the difference in parton densities
with helicity either aligned or anti-aligned with the spin of the
nucleon. Measurements of  $g_1^p$ have been made using electron beams 
at SLAC \cite{e80,e130,e143p,e143q} and muon beams at 
CERN \cite{emc,smcp}, while $g_1^n$ has been measured both
using polarized deuteron targets at SLAC \cite{e143q,e143d} and CERN 
\cite{smcd}, and a $^3$He target at SLAC \cite{e142,e154} and 
DESY \cite{hermes}. Measurements have
also been made of $g_2$ for both the proton and deuteron 
\cite{e142,e143t,smct,e154t}, although with limited statistical precision
compared to the $g_1$ measurements. This paper reports
final results for $g_1^p$, $g_1^d$, $g_2^p$, and $g_2^d$ 
from experiment E143 at SLAC, and includes more details of
the analysis procedure, as well as some auxiliary results not covered 
in the original short publications \cite{e143p,e143q,e143d,e143t,e143res}.

The earliest experiments \cite{e80,e130,emc} sparked 
considerable interest in the spin structure functions when it
was reported that, contrary to the quark model expectation,
the quarks contribute very little to the 
proton's spin (the so-called ``spin crisis'').
Subsequent precision measurements are consistent with the original
experimental results (with improved QCD corrections applied),
but the theoretical interpretation has
become more complex. It is now believed that in addition to the quarks, the
orbital angular momentum and gluons may contribute significantly to the 
proton's spin. There is still the unanswered question as 
to how much the gluons alone really contribute. 
The $g_1$ and $g_2$ structure functions are interesting 
not only in opening a new degree of freedom with which 
to explore the detailed structure of the nucleon, 
but also for making a precise test of QCD via the Bjorken sum 
rule which is a strict QCD prediction \cite{bj}.

In this paper we describe the theory 
and phenomenology of spin structure physics, and detail the
SLAC experiment E143, which  measured both $A_\|$ and $A_\bot$ 
for proton and deuteron targets over a wide range of kinematics.
The theory and experimental apparatus are described in Sections II 
and III. The analysis procedure is detailed in Section IV.
Results, their interpretation, and a discussion of systematic errors
are shown in Section V, and finally we present a summary and 
conclusions in Section VI.

\section{Interpretation and Theory}

\subsection{Formalism}
The structure functions $g_1(x,Q^2)$ and  
$g_2(x,Q^2)$ are typically extracted from asymmetry measurements.
Longitudinally  polarized leptons are scattered from a target 
that is polarized either  
longitudinally or transversely. The longitudinal ($A_\|$)  
and transverse ($A_\bot$) asymmetries are formed from  
combining data taken with opposite beam helicity:
\begin{equation}
A_\| = {{\sigma^{\downarrow\uparrow}-\sigma^{\uparrow\uparrow}}
\over {\sigma^{\downarrow\uparrow}+\sigma^{\uparrow\uparrow}}}, \hskip 1cm
A_\bot = {{\sigma^{\downarrow\rightarrow}-\sigma^{\uparrow\rightarrow}}
\over {\sigma^{\downarrow\rightarrow}+\sigma^{\uparrow\rightarrow}}}. 
\label{eq:rawasym}
\end{equation}
The polarized structure functions can be determined from these asymmetries:
\begin{eqnarray}\nonumber
   g_1(x,Q^2)&=&{F_1(x,Q^2) \over d'}\hskip .1in
   \big[A_\|+\tan(\theta/2) A_\bot\big]\ , \\[.1in] \label{eq:g12}
   g_2(x,Q^2)&=&{yF_1(x,Q^2) \over 2d'} \hskip .1in  \hskip  .1in 
   \bigg[{E+E^{\,\prime}\cos (\theta) \over E^{\,\prime} \sin  (\theta)} 
   A_\bot -A_\|\bigg]\ ,
\end{eqnarray}
where $E$ is the incident electron energy, $E'$ is the  
scattered electron energy, $\theta$ is the scattering  
angle, $y = (E-E^{\,\prime})/E$,  
$d' =[(1-\epsilon)(2-y)]/[y(1+\epsilon  R(x,Q^2))]$,
$\epsilon^{-1} =1+2[1+\gamma^{-2}]\tan^2(\theta/2)$,  
$\gamma =  2Mx/\sqrt{Q^2}$, 
$M$ is the nucleon mass and  
$R(x,Q^2)=\sigma_L /\sigma_T$ is the ratio of    
longitudinal  and  transverse virtual photon-absorption  
cross sections. $R(x,Q^2)$ is related to the spin-averaged
(or unpolarized) structure functions $F_1(x,Q^2)$ and 
$F_2(x,Q^2)$ by $R(x,Q^2)+1 = (1+\gamma^2)F_2(x,Q^2)/
[2xF_1(x,Q^2)]$.

The virtual  
photon-absorption asymmetries $A_1$ and $A_2$ are related to the 
measured asymmetries by
\begin{eqnarray}\nonumber
A_\|&=&D(A_1+\eta A_2), \\
A_\bot&=&d(A_2-\zeta A_1), \label{eq:asyms}
\end{eqnarray}
where the photon depolarization factor $D=(1-E^{\,\prime} \epsilon/E)/(1+\epsilon R)$, 
$\eta=\epsilon \sqrt{Q^2}/(E-E^{\,\prime} \epsilon)$, $d=D\sqrt{2\epsilon/(1+\epsilon)}$, 
and $\zeta=\eta(1+\epsilon)/2\epsilon$.
$A_1$ and $A_2$ can be expressed as:
\begin{eqnarray}\nonumber
A_1&=&{\sigma^T_{1/2}-\sigma^T_{3/2}\over \sigma^T_{1/2}+\sigma^T_{3/2}}
    = {\sigma_{TT}\over\sigma_T}
    = {A_\| \over D(1+\eta\zeta)} \hskip .03in - \hskip .03in {\eta
A_\bot \over d(1+\eta\zeta)}={g_1(x,Q^2) - \gamma^2 g_2(x,Q^2)\over F_1(x,Q^2)} \\ 
A_2&=&{2\sigma_{LT} \over \sigma^T_{1/2}+\sigma^T_{3/2}}  
    = {\sigma_{LT} \over \sigma_T}  
    ={\zeta A_\| \over D(1+\eta\zeta)} \hskip .03in + \hskip .03in
{ A_\bot \over d(1+\eta\zeta)} = {\gamma (g_1(x,Q^2) + g_2(x,Q^2))\over F_1(x,Q^2)},
\label{eq:aonetwo}
\end{eqnarray}
where $\sigma^T_{1/2}$ and $\sigma^T_{3/2}$ are the virtual  
photoabsorption transverse cross sections for total helicity  
between photon and nucleon of 1/2 and 3/2 respectively, 
$\sigma_{LT}$ is the interference term between the
transverse and longitudinal photon-nucleon amplitudes,
$\sigma_{T} = (\sigma^T_{1/2} + \sigma^T_{3/2})/2$, and 
$\sigma_{TT} = (\sigma^T_{1/2} - \sigma^T_{3/2})/2$.  
We see from Eq.~\ref{eq:aonetwo} that for low $x$ or high $Q^2$ (where $\gamma<<1$), 
$A_1 \approx g_1/F_1$. Positivity constrains $|A_1| \leq 1$ 
and $|A_2|\leq \sqrt{R(x,Q^2)}$. 
For the case where 
only the longitudinal asymmetry is measured, and a model is used for 
$g_2$, $A_1$ and $g_1$ can be expressed as
\begin{eqnarray}\nonumber
  A_1&=&{A_\| \over d'}\bigg[1 + {xMy \over E+E'{\rm cos}(\theta)}\bigg]
       -{g_2 \over F_1}\bigg[{4xME'{\rm cos^2}({\theta \over 2}) \over \nu 
(E+E'{\rm cos}(\theta))}\bigg], \\ \label{eq:a1g1}
  g_1&=& {A_\| F_1 \over d'}\bigg[{E+E' \over E+E'{\rm cos}(\theta)}\bigg]
       + g_2\bigg[{2Mx \over E+E'{\rm cos}(\theta)}\bigg]
\end{eqnarray}     

In the resonance region, $g_1$ and $g_2$ are well defined but are more 
properly interpreted in terms of the helicity structure of
the resonance transition amplitudes.  The $\gamma^*NN^*$ vertex for
electro-excitation of the resonance $N^*$ is generally given in terms
of three amplitudes, $A_{1/2}(Q^2)$, $A_{3/2}(Q^2)$ and 
$S_{1/2}(Q^2)$\cite{stoler,carlmuk}.
Here, $A$ denotes transverse photon polarization and $S$ indicates
longitudinal photons.  The index $1/2$ or $3/2$ refers again to the total
$\gamma^*N$ helicity. 
The virtual photon-nucleon cross sections for an isolated 
resonance can then be written in
terms of helicity amplitudes as
\begin{eqnarray}
\sigma^T_{1/2} &=& 
{{4\pi^2\alpha}\over {KM}} ( F_1 + g_1 - {{2Mx}\over\nu}g_2 ) = 
{2\pi {M\over W} b} |A_{1\over 2}|^2,\nonumber \\
\sigma^T_{3/2} &=& 
{{4\pi^2\alpha}\over {KM}} ( F_1 - g_1 + {{2Mx}\over\nu}g_2 ) =
{2\pi {M\over W} b} |A_{3\over 2}|^2,\nonumber \\ 
\sigma^L_{1/2} &=& \sigma_L = 
{{4\pi^2\alpha}\over K} \left[ {F_2\over\nu}(1+{\nu^2\over Q^2}) - {F_1\over M}\right]
= {2\pi {M\over W} b} {Q^2\over q^{*2}} |S_{1\over 2}|^2,\\
\sigma^{LT}_{1/2} &=& \sigma_{LT} = 
{{4\pi^2\alpha}\over K} {\sqrt{Q^2}\over M\nu} ( g_1 + g_2) =
\pi {M\over W} \sqrt{2}b{Q\over q^*}
S_{1\over 2}^*A_{1\over 2},\nonumber
\end{eqnarray}
in which $K$ is the incoming photon flux which is chosen
using the Hand convention such that the invariant mass squared of the
final state is $W^2 = M^2 + 2 M K$,
$b$ is the resonance line shape (unit area), and
$q^{*2} = Q^2+(W^2-M^2-Q^2)/4W^2$ is the squared magnitude of the
3-momentum transfer measured in the resonance rest frame. 
The electron scattering cross sections are then written
\begin{eqnarray}\nonumber
{{d\sigma}\over {dE^\prime d\Omega}} &=& \Gamma_V[\sigma_T + \epsilon\sigma_L], \\ 
\label{eqn:csdiff}
{{d\sigma^{\downarrow\uparrow}}\over {dE^\prime d\Omega}}
- {{d\sigma^{\uparrow\uparrow}}\over {dE^\prime d\Omega}}
&=& 2\Gamma_V D (1+\epsilon R)[\sigma_{TT} + \eta\sigma_{LT}],\\ \nonumber
{{d\sigma^{\downarrow\rightarrow}}\over {dE^\prime d\Omega}}
- {{d\sigma^{\uparrow\rightarrow}}\over {dE^\prime d\Omega}}
&=& 2\Gamma_V d (1+\epsilon R)[\sigma_{LT} - \zeta\sigma_{TT}],
\end{eqnarray}
where 
\begin{equation}
\Gamma_V = {\alpha\over{4\pi^2}}{K\over Q^2}{E^\prime\over E}{2\over{1-\epsilon}}.
\end{equation}

\subsection{The Deep-Inelastic Spin Structure Function $g_1(x,Q^2)$}
As will be shown below, the first moment of the spin structure function
$g_1(x,Q^2)$ is related to the net quark helicity $\Delta \Sigma$ which 
contributes to the proton spin. Angular momentum conservation requires
that
\begin{equation}
{1 \over 2}={\Delta \Sigma \over 2}+\Delta G +L_z,
\label{eq:angmomconv}
\end{equation}
where $\Delta G$ is the net gluon helicity, and $L_z$ is the 
orbital angular momentum.

\subsubsection{The Quark-Parton Model \label{sect:qpm}}

In the naive quark-parton model (QPM) the nucleon is composed of quarks 
which have no orbital angular momentum, and there are no polarized gluons present.
In this simple picture, the unpolarized structure function 
$F_1(x,Q^2)$ and the polarized structure function $g_1(x,Q^2)$ can be 
simply expressed as the charge-weighted sum and difference between 
momentum distributions for quark helicities 
aligned parallel ($q^\uparrow$) and antiparallel ($q^\downarrow$) to the 
longitudinally polarized nucleon:
\begin{eqnarray}\nonumber
   F_1(x)&=&{1\over 2}\sum_i e^2_i\big[q_i^{\uparrow}(x)+
   q_i^{\downarrow}(x)\big],\\
   g_1(x)&=&{1\over 2}\sum_i e^2_i\big[q_i^{\uparrow}(x)-
   q_i^{\downarrow}(x)\big]\equiv\sum_i e^2_i\Delta q_i(x)\ .\label{eq:eqqpm}
\end{eqnarray}
The charge of quark flavor $u, d,$ and $s$ is denoted by $e_i$, and 
$q_i^{\uparrow(\downarrow)}(x)$ are the quark plus antiquark 
momentum distributions.  The  quantity $\int_0^1  
\Delta q_i(x)dx=\Delta q_i$ refers to the helicity of quark 
flavor $i$ in the proton, and $\Delta \Sigma = \Delta u + 
\Delta d + \Delta s$ is the net helicity of quarks. Since $\Delta G = 0$ 
and $L_z = 0$, it follows from Eq.~\ref{eq:angmomconv}, that $\Delta \Sigma$ 
is expected to be unity in this model. In a relativistic quark-parton 
model \cite{jafman,beyer,schlumpf} (with no polarized gluons), the 
orbital angular momentum contribution is no longer zero and the quark
helicity contributions to the proton helicity are suppressed by a factor 
of about 0.75.

\subsubsection{Perturbative QCD and the Role of the Gluons}
The quark-parton model is useful for understanding some properties of the 
nucleon such as charge and isospin. However, it fails to adequately 
describe all properties, and it falls short in explaining the
dynamics of particle interactions. For this we need a more comprehensive
theory such as Quantum Chromodynamics (QCD) which can account for gluons and
their interactions with the quarks.

The operator product expansion (OPE) \cite{ale,jafji,shu} is a useful 
technique within QCD because it separates the physics into a perturbative 
part that is easily treatable and a non-perturbative part that is 
parameterized in terms of unknown matrix elements of Lorentz-covariant 
operators. At leading twist the first moment of $g_1(x,Q^2)$ can be 
expressed in terms of singlet ($a_0$) and nonsinglet ($a_3$ and $a_8$) 
proton matrix elements of the axial current:

\begin{eqnarray}\nonumber
\Gamma_1^p(Q^2) &=& \int_0^1 g_1^p(x,Q^2)dx = \bigg({1\over 12}a_3 +
                 {1\over 36}a_8\bigg)C_{ns} +
                 {1\over 9}a_0C_s,\\
\Gamma_1^d(Q^2) &=& \int_0^1 g_1^d(x,Q^2)dx = \bigg(1-{3\over 2}\omega_D\bigg)
               \bigg[{1\over 36}a_8C_{ns}
                +{1\over 9}a_0C_s\bigg],\label{eq:gamone}
\end{eqnarray}
where $\omega_D$ is the D-state probability in the deuteron, and
the factors $C_{ns}$ and $C_s$ are the $Q^2$-dependent non-singlet and 
singlet QCD corrections, which are discussed in more detail below.

Assuming that there are no polarized gluons contributing to the proton
spin, the singlet and nonsinglet proton matrix elements given in 
Eq.~\ref{eq:gamone} can be related to the quark helicities:
\begin{eqnarray}\nonumber 
a_0&=&\Delta u + \Delta d + \Delta s=\Delta \Sigma,\\ \label{eq:matax} 
a_3&=&\Delta u - \Delta d = F+D,\\ \nonumber
a_8&=&\Delta u + \Delta d -2\Delta s = 3F-D. 
\end{eqnarray}
Here, $F$ and $D$ are weak hyperon decay constants which can be
extracted from data assuming $SU(3)$ symmetry \cite{pdg,closea} 
\begin{eqnarray}\nonumber
F+D &=& g_A = 1.2601\pm 0.0025,\\
3F-D &=& 0.588\pm 0.033.
\end{eqnarray}
The error quoted above on $3F-D$ is the experimental error assuming 
SU(3) symmetry. It may be an underestimate because possible $SU(3)$ 
symmetry breaking effects could be significant. There have been
a number of attempts to estimate these effects 
\cite{jafman,rata,ratb,donoghue,ehrn,songa}.
According to Ratcliffe \cite{ratb}, symmetry breaking effects in the past
have always been found to be at most $10\%$. Assuming a generous $20\%$ 
systematic error from symmetry breaking combined with the above error 
in quadrature yields an error of $0.12$ on $3F-D$. This error is 
somewhat smaller than the range of possible values (0.40--0.84)
presented under various assumptions  \cite{rata,ratb,ehrn,songa}, some of 
which have come under criticism \cite{ratb}.

After combining Eq.~\ref{eq:gamone} and Eq.~\ref{eq:matax} 
\cite{closea} it is straightforward to extract the
singlet matrix element from the measured first moments of 
the proton:
\begin{equation}
a_0={9\over C_s}\Bigg[\Gamma_1^p(Q^2) - 
         {1\over 18}(3F+D)C_{ns}\Bigg], \label{eq:delqp}
\end{equation}
and the deuteron:
\begin{equation}
a_0={9\over C_s}\Bigg[{\Gamma_1^d(Q^2)\over 1- 
         {3\over 2}\omega_D} - 
         {1\over 36}(3F-D)C_{ns}\Bigg].\label{eq:delqd}
\end{equation}
\noindent
The nonsinglet QCD correction $C_{ns}$ \cite{larin} 
calculated in the $\overline{\rm MS}$ scheme 
to order three for three quark flavors is given by
\begin{equation}
C_{ns} = 1-{\alpha_s(Q^2)\over \pi}
-3.58\left({\alpha_s(Q^2)\over \pi}\right)^2
-20.22\left({\alpha_s(Q^2)\over \pi}\right)^3,
\label{eq:cns} 
\end{equation}
where $\alpha_s(Q^2)$ is the strong coupling constant. Fourth 
order QCD corrections have been estimated \cite{kataev} to be 
small at the kinematics of this experiment. 
The singlet QCD correction exists in two forms \cite{larin}, 
one which yields a $Q^2$-dependent $a_0(Q^2)$ in 
Eqs.~\ref{eq:delqp}-\ref{eq:delqd} and one which yields $a_0^{inv}$
which is the asymptotic high $Q^2$ limit of $a_0(Q^2)$.
These singlet QCD corrections have been calculated 
in the $\overline{\rm MS}$ scheme \cite{larin}:
\begin{eqnarray}\nonumber
C_s(Q^2) &=& 1 - {\alpha_s(Q^2)\over \pi}
            - 1.10\left({\alpha_s(Q^2)\over \pi}\right)^2,\\
C_{s}^{inv} &=& 1 - 0.3333\left({\alpha_s(Q^2)\over \pi}\right)
            - 0.5495\left({\alpha_s(Q^2)\over \pi}\right)^2.
            \label{eq:csinv}\label{eq:csq2}
\end{eqnarray}

The contribution of $\Delta G$ to $a_0$ (and thus to the first moment 
of $g_1$) is a factorization scheme-dependent quantity. In a 
gauge-invariant scheme such as the $\overline{\rm MS}$ scheme gluons do 
not contribute to the first moment of $g_1$ which means that 
$a_0 = \Delta \Sigma$. For chiral-invariant schemes such as
the Adler-Bardeen scheme \cite{ball}  the gluons do contribute to 
$\Gamma_1$. The physical quantity $g_1(x)$ is independent of the factorization 
scheme, however. In the Adler-Bardeen scheme \cite{ball}, the quantity 
$a_0\ne\Delta \Sigma$, and is instead written as
\begin{equation}
a_0 = \Delta \Sigma 
          - {3\over 2\pi}\alpha_s(Q^2)\Delta G(Q^2).\label{eq:gluon}
\end{equation}
This contribution of $\Delta G$ is called the gluon axial 
anomaly \cite{axial} or
the Adler-Bell-Jackiw anomaly (as applied to QCD from QED). 
The product $\alpha_s(Q^2)\Delta G(Q^2)$ is independent of 
$Q^2$ in leading order which implies that $\Delta G(Q^2)$
grows in $Q^2$ like $1/\alpha_s(Q^2)$, and $L_z$ compensates to
satisfy Eq.~\ref{eq:angmomconv}. Physically, this means that
as each quark radiates a gluon with some preferential helicity
the orbital angular momentum of the quark-gluon system must increase 
to conserve the total angular momentum. Thus, as more gluons 
are emitted, both $\Delta G$ and $L_z$ will grow, but with 
opposite signs.

Other quantities of interest are the helicity contributions from the
individual quarks. These quantities can be extracted from the measured
$a_0$, but may be subject to possible gluon contributions as in 
Eq.~\ref{eq:gluon}. Allowing for the possibility of gluon contributions, these quark helicities are calculated using:
\begin{eqnarray}\nonumber
\Delta u &=& {1\over 3}(a_0 + 3F +D)
       +{1\over 2\pi}\alpha_s(Q^2)\Delta G(Q^2), \\
\Delta d &=& {1\over 3}(a_0 - 2D)
       +{1\over 2\pi}\alpha_s(Q^2)\Delta G(Q^2),\\ \nonumber
\Delta s &=& {1\over 3}(a_0 - 3F +D)
       +{1\over 2\pi}\alpha_s(Q^2)\Delta G(Q^2).\label{eq:helicity}
\end{eqnarray}
If we include a contribution of $\Delta G$(1 (GeV/c)$^2)=1.6\pm 0.9$ 
\cite{abfr}, and use $\alpha_s(M_Z)=0.118\pm 0.003$ \cite{pdg}, we calculate
$\alpha_s \Delta G/2\pi = 0.13\pm 0.08$ and find good agreement with 
existing data. This model along with quark-parton model expectations are summarized in Table~\ref{tb:delqmod} and can be 
compared with data from this experiment in Table~\ref{tb:delta}. Note that the 
value used above for $\Delta G$ agrees well with a theoretical prediction based
on QCD sum rules \cite{mankps} which yields $\Delta G$(1 (GeV/c)$^2) = 2.1\pm 
1.0$ and on an earlier parameterization \cite{chiappetta} which yields $\Delta 
G$(1 (GeV/c)$^2) = 1.7$.

There are a number of other theoretical models which attempt to explain
how the quark helicity is distributed within the nucleon. Non-perturbative 
effects enhancing the role of intrinsic sea quarks have been proposed by 
several authors. Halperin and Zhitnitsky \cite{halperin} argue that a large 
portion of the nucleon spin comes from charm quarks by adding a term $2\Delta 
c$ to the $a_o$ term in Eq.~\ref{eq:gamone}. Brodsky and Ma \cite{brodma} contend that 
asymmetries in the light quark sea could generate the observed $\Delta\Sigma.$
The Skyrme model \cite{bek} predicts that 
$\Delta\Sigma=\Delta u +\Delta d + \Delta s$=0 
and $\Delta G=0$ and $L_z=1/2$, and should be accurate to O($1/N_c$) where 
$N_c=3$, the number of colors. Within its uncertainty this is consistent with the 
small observed value of $\Delta\Sigma.$ Other models include the chiral 
bag model \cite{hog}, the chiral quark model \cite{songa}, calculations 
based on QCD spectral sum rules\cite{narison}, or Pauli-exclusion 
principles \cite{bourrely}, and also lattice QCD predictions 
\cite{altmeyer,fukugita,dong,gockeler}.

\subsubsection{The Bjorken Sum Rule}

This sum rule was originated by Bjorken \cite{bj} 
using current algebra and isospin symmetry. It has since been 
re-derived in QCD and is a strict prediction made by this theory. 
It relates the integral over all $x$ at fixed $Q^2$ of the 
difference between $g_1^p(x,Q^2)$ and $g_1^n(x,Q^2)$
to the well-measured neutron beta decay coupling constant 
$g_A = 1.2601 \pm 0.0025$ \cite{pdg}, 
\begin{equation}
\Gamma_1^p(Q^2)-\Gamma_1^n(Q^2) = \int \left( g^p_1(x,Q^2) - 
g^n_1(x,Q^2) \right) dx\ =\ {1\over 6}\ {g_A}\ C_{ns}.
\label{eq:bj} 
\end{equation}
An experimental 
test of this sum rule provides a test of fundamental QCD 
assumptions. In addition, it is possible to use the measurement 
to extract a relatively accurate determination of 
$\alpha_s(Q^2)$ at low $Q^2$ (on the order of 2 to 10 
(GeV/c)$^2$) \cite{alphas}. A significant difference from other 
$\alpha_s(Q^2)$ determinations could indicate the presence
of interesting new physics.

\subsubsection{The Ellis-Jaffe Sum Rule}

The other sum rules of interest for $g_1$, although less rigorous 
than the Bjorken sum rule, are the Ellis-Jaffe sum 
rules\cite{ej} which were derived using SU(3)
symmetry and assuming the strange sea in the nucleons is 
unpolarized. These sum rules, including the necessary QCD
corrections, follow naturally from 
Eqs.~\ref{eq:gamone} and \ref{eq:matax} with $\Delta s =\Delta G=0$ such that
$a_0 = a_8 = 3F-D$:

\begin{eqnarray} \nonumber
  \Gamma_1^p(Q^2) &=& \int_0^1 g_1^p(x,Q^2)dx = {1\over 18}
  \left[C_{ns} (3F+D) +2C_s (3F-D)\right]\ , \\[.1in]                                  
  \Gamma_1^n(Q^2) &=& \int_0^1 g_1^n(x,Q^2)dx = {1\over 9}
   \left[ -DC_{ns} + C_s(3F-D)\right]\ .  \label{eq:ej}                                   
\end{eqnarray}

\subsubsection{$Q^2$ Dependence: Evolution and Higher Twist}
The quark-parton model does not inherently include gluons, and it is 
the interaction between the quarks and gluons which generates the 
observed $Q^2$ dependence of both the polarized and unpolarized
nucleon structure functions. 
The QCD theory which describes the quark-gluon dynamics gives 
predictions about how the parton distribution functions (and thus 
structure functions) evolve in $Q^2$ in the perturbative limit of 
small $\alpha_s$. The $Q^2$ evolution of the polarized parton 
densities is governed by the DGLAP \cite{dglap} equations which 
embody the emission of gluons by quarks. This gluon emission is 
responsible for the leading logarithmic $Q^2$ dependence. 
In addition, there are higher-twist contributions to the $Q^2$ 
dependence which are suppressed by powers of $1/\sqrt{Q^2}$. Higher 
twist corrections to $g_1$ have been estimated to be small 
\cite{balitsky,jiU,ehrnsperger,steing1} while higher twist corrections 
for $g_2$ have been estimated to be significant 
\cite{gockeler,balitsky,strat,song,steing2}. 
Fits to $\Delta u(x,Q^2)$, $\Delta d(x,Q^2)$, $\Delta s(x,Q^2)$, and 
$\Delta G(x,Q^2)$ have been made \cite{ball,grv} using
next-to-leading-order (NLO) DGLAP equations \cite{zijistra}. The 
results indicate that 
NLO fits are more sensitive to the strength of  the polarized gluon 
distribution function $\Delta G(x,Q^2)$ than leading order fits. 

\subsection{The Deep-Inelastic Spin Structure Function $g_2(x,Q^2)$}

\subsubsection{Physical Interpretation}

The interpretation of $g_2$ in the naive parton model is less straightforward 
than that of $g_1$. Feynman related the quantity 
$g_T(x) = g_1(x)+g_2(x) = A_2 F_1/\gamma$ 
to the distribution of quark polarizations aligned parallel ($k^\uparrow$) 
and antiparallel ($k^\downarrow$) to that of a transversely polarized 
proton \cite{jafji,feynman} by the expression
\begin{equation}
   g_T(x)=\sum_i e^2_i{m_q \over 2xM}\big[k_i^{\uparrow}(x)-
   k_i^{\downarrow}(x)\big], \label{eq:eqqpmg2}
\end{equation}
where $m_q$ is the quark mass. Leader and Anselmino \cite{leadans} subsequently 
derived the parton model expressions for $g_1$ and $g_2$ for a nucleon polarized 
at an arbitrary angle $\theta$ relative to the incident electron direction. 
Evaluated at $\theta=0$ their expression for $g_T(x)$ is
\begin{equation}
   g_T(x)=\sum_i e^2_i{m_q \over 2xM}\big[q_i^{\uparrow}(x)-
   q_i^{\downarrow}(x)\big],
\end{equation}
where $q_i$ are the same as in Eq.~\ref{eq:eqqpm}. Jaffe and Ji \cite{jafji} 
pointed out that claims \cite{ioffe,kane} that $g_T$ is small were generated 
by setting $m_q=0$ in Eq.~\ref{eq:eqqpmg2}. For consistency, if the quark 
momenta are taken to be along the longitudinal direction then $m_q = xM$ \cite{jafji,leadans}, 
which yields $g_2(x) = 0$. Because of Fermi motion, however, the quarks are 
off-shell and $m_q \ne xM$ in general. It is this large off-shell nature of 
the quark which produces large twist-3 effects in the MIT bag model 
calculation \cite{jafji}.

The naive parton model does not include transverse momentum or quark-gluon 
interactions. For this we can turn to a more advanced light-cone parton
model,\cite{mank,ralston} or an OPE analysis,\cite{jafji,jaffe} which 
indicates that there are 
three components (up to twist-3) contributing to $g_2$. These components 
include the leading twist-2 part $g^{WW}_2(x,Q^2)$ \cite{wand},
coming~from the same set of operators that contribute to $g_1$, 
another twist-2 part coming from the~quark 
transverse-polarization distribution $h_T(x,Q^2)$, and a 
twist-3 part coming from quark-gluon interactions $\xi(x,Q^2)$:
\begin{equation}
g_2(x,Q^2) = g^{WW}_2(x,Q^2) - \int_x^1{\partial\over \partial 
y}\biggl({m_q\over M}h_T(y,Q^2)+\xi(y,Q^2)\biggr)\ {dy\over y}\ .
\label{eq:eqtot}
\end{equation}
The term containing $h_T(y,Q^2)$ is usually neglected because it is 
suppressed by the quark mass $m_q$, and the $g_2^{WW}$ expression 
of Wandzura-Wilczek \cite{wand} is given by
\begin{equation}
g^{WW}_2(x,Q^2) =-g_1(x,Q^2)+\int_x^1{g_1(y,Q^2)\over y}\ dy\ . 
\label{eq:eqww} 
\end{equation}

\subsubsection{OPE Sum Rules and the Twist-3 Matrix Element}
Keeping terms up to twist-3, the OPE analysis of $g_1$ and $g_2$ yields an
infinite number of sum rules: 
\begin{eqnarray}
\Gamma_1^{(n)} = \int_0^1 x^n\ g_1(x,Q^2)\ dx\ =  &{\displaystyle a_n\over 
\displaystyle 2}\ ,\hfill \qquad\qquad\qquad\hfill &~~ n=0,2,4,\, \ldots\ ,
\nonumber\\ \Gamma_2^{(n)} = \int_0^1 x^n\ g_2(x,Q^2)\ dx\ 
= &\displaystyle{{1\over 2}\ {n\over  n+1}}\  (d_n
-a_n)\ , &~~ n=2,4,\, \ldots\ , \label{eq:ope}
\end{eqnarray}
where $a_n$ are the twist-2 and $d_n$ are the twist-3 matrix elements of the
renormalized operators. The OPE only gives information on the odd moments of the spin structure functions. Note that contributions involving $m_q/M$ (see Eq.~\ref{eq:eqtot}) have been left out of Eq.~\ref{eq:ope} as have target
mass effects discussed below. The twist-3 matrix elements follow from Eq.~\ref{eq:ope}:
\begin{eqnarray}\nonumber
d_n &=& 2\int_0^1 x^n\biggl[g_1(x,Q^2) + {n+1\over n}g_2(x,Q^2)\biggr]dx \hfill
\qquad\hfill n=2,4,\, \ldots\ , \\
&=& 2\int_0^1 x^n\biggl({n+1\over n}\biggr)\overline{g_2}(x,Q^2)dx 
\hfill\qquad\qquad\qquad\hfill ~~ n=2,4,\, \ldots\ ,\label{eq:dn}
\end{eqnarray}
where $\overline{g_2} = g_2-g_2^{WW}$. We see from Eq.~\ref{eq:dn} that 
if all $d_n = 0$ then $g_2$ is completely determined by $g_1$ because 
there are an infinite number of sum rules. This is how the quantity 
$g_2^{WW}$ was originally derived.

   There are a number of theoretical predictions for $d_2$ for both proton 
and neutron targets \cite{gockeler,balitsky,strat,song,steing2}. Some are 
based on bag models \cite{strat,song}, others on QCD sum rules \cite{balitsky,steing2}, 
and there is also a lattice QCD calculation \cite{gockeler}. Many of these models 
have predicted large values for $d_2$ which means there could be significant twist-3 
contributions to $g_2$. This makes the study of $g_2$ particularly  interesting.

\subsubsection{The Burkhardt-Cottingham Sum Rule}
The Burkhardt-Cottingham sum rule\cite{buco} for $g_2$ at large $Q^2$, namely
\begin{eqnarray}
\int_0^1 g_2(x)dx=0\ ,\label{eq:bc}
\end{eqnarray}
was  derived from virtual Compton scattering dispersion relations. This
sum rule does not follow from the OPE since the $n=0$ sum rule is not defined
for $g_2$ in  Eq.~\ref{eq:ope}. The validity of the Burkhardt-Cottingham sum 
rule relies on $g_2$ obeying Regge theory at low $x$,  which may not be a good assumption.
 A non-Regge divergence of $g_2$ at low~$x$ would invalidate this sum
rule \cite{ale,jafji}, although such a divergence could be very difficult to
detect experimentally.

\subsubsection{Target Mass Effects}
The OPE sum rules as given in Eq.~\ref{eq:ope} were derived in the limit 
$M^2 x^2/Q^2 \rightarrow 0$. These target mass effects can become significant 
when $M^2/Q^2$ is of order unity which is certainly the case for a subset of 
the data presented in this paper. These target mass effects for polarized 
electroproduction have been determined \cite{wandzura,matsuda}. The corrected 
Bjorken sum rule derived from these formulae is given by \cite{matsuda} 
\begin{eqnarray}\nonumber
{1\over 9}\int_0^1 &&dx {\xi^2 \over x^2}\biggl[5+4\sqrt{1+4M^2x^2/Q^2}\biggr]
            \biggl[g_1^p(x,Q^2) -g_1^n(x,Q^2)\biggr] \\
                   &&-{4\over 3}\int_0^1 dx {M^2\xi^2 \over Q^2} 
            \biggl[g_2^p(x,Q^2) -g_2^n(x,Q^2)\biggr] =\ {1\over 6}\ {g_A}\ C_{NS},
\label{eq:bjmass}
\end{eqnarray}
where the Nachtmann variable $\xi = 2x/(1+\sqrt{1+4M^2x^2/Q^2})$. This sum rule 
is now dependent on $g_2$. The size of the target mass effects \cite{kawa,piccione} 
to the uncorrected Bjorken sum rule formula are estimated to be of the same 
magnitude as higher-twist effects which are typically small. The target mass 
effects for the $g_2^{WW}$ calculation (missing in Eq.~\ref{eq:eqww}) have been investigated and are negligible for our kinematics \cite{todd}.

\subsection{Resonance Region Polarized Structure Functions}

The values of $A_1$ in the resonance region are a combination
of the asymmetries for individual resonances and for the nonresonant background. Resonance helicity amplitudes $A_{1/2}$ and $A_{3/2}$ 
are reasonably well measured at $Q^2=0$ for the prominent resonances 
\cite{pdg}.  Sparse data exist also for virtual photons\cite{BONNDESY}.
The excitation of the $\Delta(1232)$ resonance (spin-$\frac{3}{2}$)
includes both $\frac{1}{2}$ and $\frac{3}{2}$ spin projections. At 
low $Q^2$ the $\Delta(1232)$ excitation is expected to be primarily 
a magnetic dipole transition for which
$A_{3/2}/A_{1/2} = \sqrt{3}$ and $A_1 = (|A_{1/2}|^2-|A_{3/2}|^2)/(|A_{1/2}|^2+|A_{3/2}|^2)= -\frac{1}{2}$. 
For real photons $A_{3/2}/A_{1/2} = 1.064\sqrt{3}$\cite{pdg}.  Perturbative
QCD predicts that the ratio $A_{3/2}/A_{1/2}$ should go as $1/Q^2$ and
$A_1$ should approach unity as $Q^2\to\infty$. However, a recent analysis 
of pion electroproduction\cite{latifa} data shows that the magnetic 
dipole transition still dominates at $Q^2=3.2$~(GeV/c)$^2$. 
On the other hand, the $S_{11}(1535)$ resonance has no spin-$\frac{3}{2}$
projection, so $A_1$ should be unity at all $Q^2$.  Data \cite{BONNDESY} 
from Bonn, Daresbury and DESY have been used to extract $A_{1/2}$ and
$A_{3/2}$ up to $Q^2=3$~(GeV/c)$^2$ for the $S_{11}$, $D_{13}$, and 
$F_{15}$ resonances. Because of the large uncertainties of these extractions,
our knowledge of the $Q^2$ dependence of the helicity amplitudes is still
rudimentary.  The asymmetries $A_1$ for both
$D_{13}$ and $F_{15}$ make a transition from $A_1\approx -1$ to
$A_1\approx 1$ somewhere in the range $0<Q^2<3$ (GeV/c)$^2$\cite{BONNDESY}.

Less is known about $\sigma_{LT}$ and $A_2$. The positivity limit 
$A_2 < \sqrt{R(x,Q^2)}$ constrains $A_2$. The world average 
value\cite{keppel} for $R$ in the resonance region is $0.06\pm 0.02$
for $1 < Q^2 < 8$ (GeV/c)$^2$ and $W^2 < 3$ GeV$^2$.  Since this is
smaller by half than the deep-inelastic fit to $R(x,Q^2)$ \cite{whitlowr}
extrapolated into the resonance region, one might argue that $R(x,Q^2)$ for the 
resonances themselves is small.  However, little is known for 
$Q^2<1.3$~(GeV/c)$^2$.

A complete mapping of $g_1(x,Q^2)$ at low $Q^2$ where the resonances 
dominate is useful for two reasons.  First, these data provide 
important input for radiative corrections of the deep-inelastic data.  
Second, the evolution of the integral (defined to exclude elastic 
scattering) $\Gamma_1(Q^2) = \int_0^1 g_1(x,Q^2) dx$ for $Q^2\to 0$ 
should be determined by the Gerasimov-Drell-Hearn (GDH) sum rule \cite{GDH}
for real photons:
\begin{equation}
\int_{\nu_{th}}^\infty 2\sigma_{TT} d\nu / \nu = -{2\pi^2\alpha \kappa^2\over M^2}
\label{eqn:gdhnu}
\end{equation}
in which $\nu$ is the photon energy, $\nu_{th}$ is the threshold energy for
pion production, $\kappa$ is the nucleon anomalous magnetic moment, and $M$ 
is the nucleon mass.  A simple change of variables from $\nu$ to $ x$ in 
Eq.~\ref{eqn:gdhnu} and a reformulation of $\sigma_{TT}$ in terms of $g_1$ 
and $g_2$ yields
\begin{equation}
\lim_{Q^2\to 0}{\Gamma_1(Q^2)\over Q^2} = -{\kappa^2 \over 8 M^2}.
\label{eqn:gdhq}
\end{equation}
One important feature of Eq.~\ref{eqn:gdhq} is the sign.  $\Gamma_1^p$ for 
the proton is positive for $Q^2$ above 3~(GeV/c)$^2$ as measured in the 
deep-inelastic regime.  However, the GDH sum rule predicts that $\Gamma_1^p$
should become negative at small $Q^2$.  This implies that somewhere in the 
region $0 < Q^2 < 3$ (GeV/c)$^2$, $\Gamma_1^p$ must cross zero.  Exactly where 
this occurs depends crucially on the $Q^2$ evolution of the resonance helicity 
amplitudes which are presently not well known.  Predictions about how 
$\Gamma_1(Q^2)$ goes from the deep-inelastic values to the GDH limit have taken 
one of two paths:
1) theoretically motivated interpolation and 2) computations that include
all available knowledge of the resonance behavior.   Both need to be checked
with direct measurements of $\Gamma_1(Q^2)$ at low $Q^2$.

Recent theoretical work\cite{ji} indicates that at low $Q^2$, $\Gamma_1(Q^2)$ 
corrected to include the elastic contribution,
\begin{equation}
\Gamma_1^{el}(Q^2) = {1\over 2}F_1(Q^2)[F_1(Q^2)+F_2(Q^2)],\label{eq:gamelas}
\end{equation}
provides the twist-4 ($1/Q^2$) corrections to the Ellis-Jaffe sum rule.
$F_1$ and $F_2$ in Eq.~\ref{eq:gamelas} are the Dirac and Pauli elastic form
factors (not the deep-inelastic structure functions).

\section{The Experiment}
The goal of the E143 experiment was to determine the longitudinal 
and transverse cross-section asymmetries via deep-inelastic 
scattering of longitudinally polarized electrons from polarized 
protons and deuterons. Over a period of three calendar months data 
were taken at beam energies of 29.13 GeV (122 million events),  
16.18 GeV (56 million events) and 9.71 GeV (58 million events).  The 
longitudinal asymmetry $A_{\parallel}$ was obtained with the target 
polarization parallel to the beam momentum, whereas the 
transverse asymmetry $A_{\perp}$ (at $E=29.1$ GeV only) was obtained 
with the target polarization 
transverse to the beam momentum (right or left of the beam). 

The experimental apparatus employed consisted of five components: the 
polarized source, the accelerator and beam transport, the M\o ller 
polarimeter to measure the beam polarization in the End Station A (ESA), 
the polarized proton/deuteron target, and the two spectrometer arms to 
detect the scattered electrons. These components are discussed in the 
sections which follow.

\subsection{ The SLAC Polarized Electron Source }  

A polarized electron source for the SLAC Linear Accelerator was first 
developed in the early 1970's for experiments on the spin structure 
of the proton. Since 1978 the SLAC polarized electron source has 
been based on the principle of laser photoemission from a gallium 
arsenide (GaAs) photocathode. Strained GaAs photocathodes, which 
effectively doubled the polarization obtainable from an unstrained GaAs 
photocathode, were developed in 1991 \cite{source_1,source_2} and first 
used in a SLAC experiment in 1993. The design and operational characteristics 
of the SLAC polarized electron source are fully described in Ref.~\cite{sou}.  

\begin{figure}[h]
\centerline{\epsfig{file=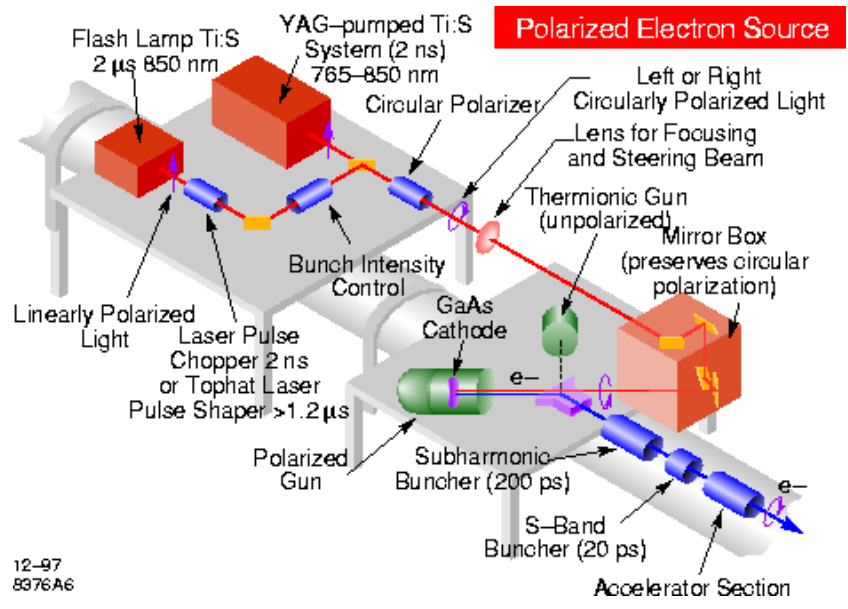,width=4.50 in}}\vskip 0.7cm
\caption{ The layout of the lasers and the polarized electron gun
at the accelerator injector is shown schematically.  Two types of
lasers are used, one for the SLC, which produces two {2 nsec}
pulses separated by 61 nsec, and one for the fixed target
experiments, which produces one pulse 2 $\mu$sec long.}
\label{fg:source}
\end{figure}

Figure.~\ref{fg:source} is a schematic of the layout of the laser and gun 
structure at the SLAC injector.  Electrons are photo-emitted from a
GaAs photocathode by illuminating the surface with a laser. The electrons 
are polarized with a helicity defined by the sign of the circular polarization 
of the incident laser light. Spin reversals are achieved at the source by 
reversing the circular polarization of the laser light with a Pockels cell.   
The pattern for the sign of the polarization is chosen to be a known 
pseudo-random sequence, permitting  validation of the sign of each pulse in 
the offline data stream. 

The polarization of the electrons is a consequence of the band structure 
of GaAs and the angular momentum selection rules that apply to this system.  
The presence of strain changes the lattice constant of the GaAs, shifts 
the energy levels, and breaks a spin state degeneracy in the valence band. 
Excitation by circularly polarized light near the band gap edge 
($\lambda \approx 850$ nm) will then result in only one set of spin 
states populating the conduction band provided the strain is sufficiently 
large. The strain in the active GaAs layer is achieved by growing a 100 
nm thick epitaxial layer of GaAs on a substrate of ${\rm GaAs_{(1-x)}P_x}$ 
(x=0.28). The lattice spacing for the GaAsP is about 1\% smaller than for 
GaAs, and the resulting lattice mismatch puts the GaAs epitaxial layer under 
a compressive strain sufficiently high to remove the spin state degeneracy. 
Under these conditions one expects that photo-emitted electrons will have close 
to 100\% polarization. In practice, the electron polarization averaged 85\% 
for the E143 experiment.

The laser system was designed and built at SLAC.  It consisted of a flash-lamp 
pumped titanium sapphire rod, producing light pulses which were optically 
chopped to a 2.3 $\mu$sec long pulse. The laser beam was transmitted through 
a lens system which allowed for steering and focusing on the cathode.  For 
the E143 experiment, the amount of laser power available was larger than 
needed, so the power was attenuated to about 10 watts peak, yielding 
approximately $4 \times 10^9$ electrons per pulse.  At this low electron 
intensity, the accelerator control system was unable to sense the presence 
of beam.  To allow the accelerator controls to operate, the beam was 
intentionally intensified to about $2 \times 10^{10}$ in one of the 
120 pulses generated per second. This ``witness pulse'' was then sent 
into a beam dump before reaching the target, and the experiment operated 
on the remaining 119 pulses per second. 

Possible systematic errors associated with reversal of the electron spin are 
important to this type of experiment. Correlations between beam current, 
beam energy, beam positions, and beam angles on the target were available
to the experimenters on a short time basis from beam monitors. For this 
laser-driven photoemission source, the reversal of the laser polarization 
is sufficiently free of unwanted effects such that all systematic errors from 
the source were negligible. 

\subsection{ The Electron Beam }

\subsubsection{Production and Transport}

The electrons produced by the polarized source were accelerated to energies
between 9 and 30 GeV in the linear accelerator. The electrons were then 
deflected through an angle of $24.5^\circ$ in the A-line beam transport and 
were directed onto the polarized target in the ESA.

Because of its anomalous magnetic moment, the spin of the electron precesses 
by an angle larger than that of the bend angle of the beam, according to the
formula:
\begin{equation}
\Delta\phi = \pi\biggl({24.5^\circ\over 180^\circ}\biggr)
\left( {g-2\over 2}\right)\left({ E\over m}\right)
 = \left({E\over 3.237}\right)\pi,
\end{equation}
where $g$ is the gyromagnetic ratio, $E$ is the energy (in GeV), $m$ is the 
mass of the electron, and $\Delta\phi$ is the angle between the electron spin 
and the momentum at the target. When $\Delta\phi$ is an integral multiple of
$\pi$, the electron spin is longitudinal at the target. The experiment was
run at energies of 9.71, 16.18 and 29.13 GeV, corresponding to 3$\pi$, 5$\pi$ 
and 9$\pi$ respectively.  By varying the energy around the nominal value
and measuring the longitudinal polarization in the M\o ller polarimeter,
we verified that the chosen energy produced the maximum polarization.

\subsubsection{Beam Monitoring}
The incident flux of electrons was measured independently in two identical
precision toroidal charge monitors in the ESA.  These were frequently
calibrated with a known charge and agreed to better than 1\%. The response of 
the toroids is independent of the polarity of the beam.

The position of the beam at the target was monitored in two devices: a 
traveling-wave radio-frequency beam position monitor which was non-interfering 
and was placed just in front of the target, and a pair of secondary emission 
foil arrays with 1 mm spacing located 10.8 m downstream from the target. The 
former provided a direct measurement of beam centroid position, and was used 
in an automatic feedback system to keep the beam on target; the latter allowed 
a measurement of both the position and the transverse dimensions of the beam
by comparing the charge collected on the individual foils.

\subsubsection{Beam Rastering}
To minimize effects such as target depolarization from local beam heating and 
radiation damage, the beam was moved or ``rastered'' across the face of the 
target. The beam position was changed between pulses by means of a pair of 
air-core magnet coils located 67 m upstream of the target. The lack of iron in 
these magnets allowed the fields to be quickly changed under computer control. 
The beam at the target was rastered on a grid with a spacing of 1.2 mm inside 
a circle of radius 10.8 mm for a total of 253 points. Thus, each point in the 
target was illuminated only once every 2.1 seconds.  The raster pattern skipped 
every other point and row, so that subsequent pulses did not overlap, and the 
entire raster pattern was completed in four passes. Because the dimensions of the beam
(Gaussian $\sigma$ of 2 mm horizontally and 1 mm vertically) were comparable to 
the raster spacing, the overall illumination of the target was quite uniform 
inside the circle of the raster.

\subsubsection{The Beam Chicane}
For the measurement of $g_2$, the target was rotated by $90^\circ$ so that the 
target nucleons were polarized transversely to the beam direction in the 
scattering plane. In this configuration, the electrons in the beam passed 
through $\int B dl = 1.52$ T-m as they traversed the target. This was enough to 
deflect them through an angle of $0.90^\circ$ at 29.1 GeV and to rotate the 
polarization vector through an angle of $60^\circ$. As a result, the beam after
the target would have no longer been parallel to the nominal beam-line, and in
fact would have been about 30 cm low at the exit of the ESA. More 
significantly, the deep-inelastic scattering would have taken place at a 
different average angle and longitudinal polarization than in the parallel 
case.

To compensate for the effects of this magnetic field, we inserted four 
identical dipole magnets (the chicane) into the beam-line, three upstream and 
one downstream of the target. The first magnet deflected the beam down by 
$0.45^\circ$, and the second pair bent the beam back up by twice this amount.  
This caused the beam to arrive at the center of the target with both the 
momentum and polarization vectors horizontal.  After exiting the target, the 
beam was tilted downward, and the fourth magnet returned the beam to the 
horizontal so that it left the ESA parallel to the nominal beam-line displaced 
vertically by only 3.5 cm at 29.1 GeV.

\subsection{Beam Polarimetry}

A M\o ller polarimeter was used to measure the beam polarization during the 
E143 experiment. This is a practical and reliable approach based on 
$\vec{e} + \vec{e} \rightarrow e + e$ scattering, a spin-dependent QED process 
with large a cross-section and analyzing power. The expected cross-section 
asymmetry can be calculated with high precision~\cite{Wagner} and is not 
significantly modified by radiative processes~\cite{alpha4}. 

For a beam with longitudinal polarization $P_{B}$ and target with longitudinal 
polarization $P_{T}$, the beam polarization is measured by comparing the 
relative cross-section asymmetry for beam and target spins aligned parallel 
($\uparrow \uparrow$) and anti-parallel ($\uparrow \downarrow$): 
\begin{equation}\label{asym}
\epsilon=\frac{d \sigma^{\uparrow \uparrow}/d \Omega -
d \sigma^{\uparrow \downarrow}/d \Omega}{d \sigma^{\uparrow 
\uparrow}/d \Omega + d \sigma^{\uparrow \downarrow}/d \Omega}=
A_{ZZ}(\theta)P_{B}P_{T}.
\end{equation}
The relative cross-sections are determined by detecting either of the scattered 
electrons or both in coincidence. 

\subsubsection{Layout}

\begin{figure}[h]
\centerline{\epsfig{file=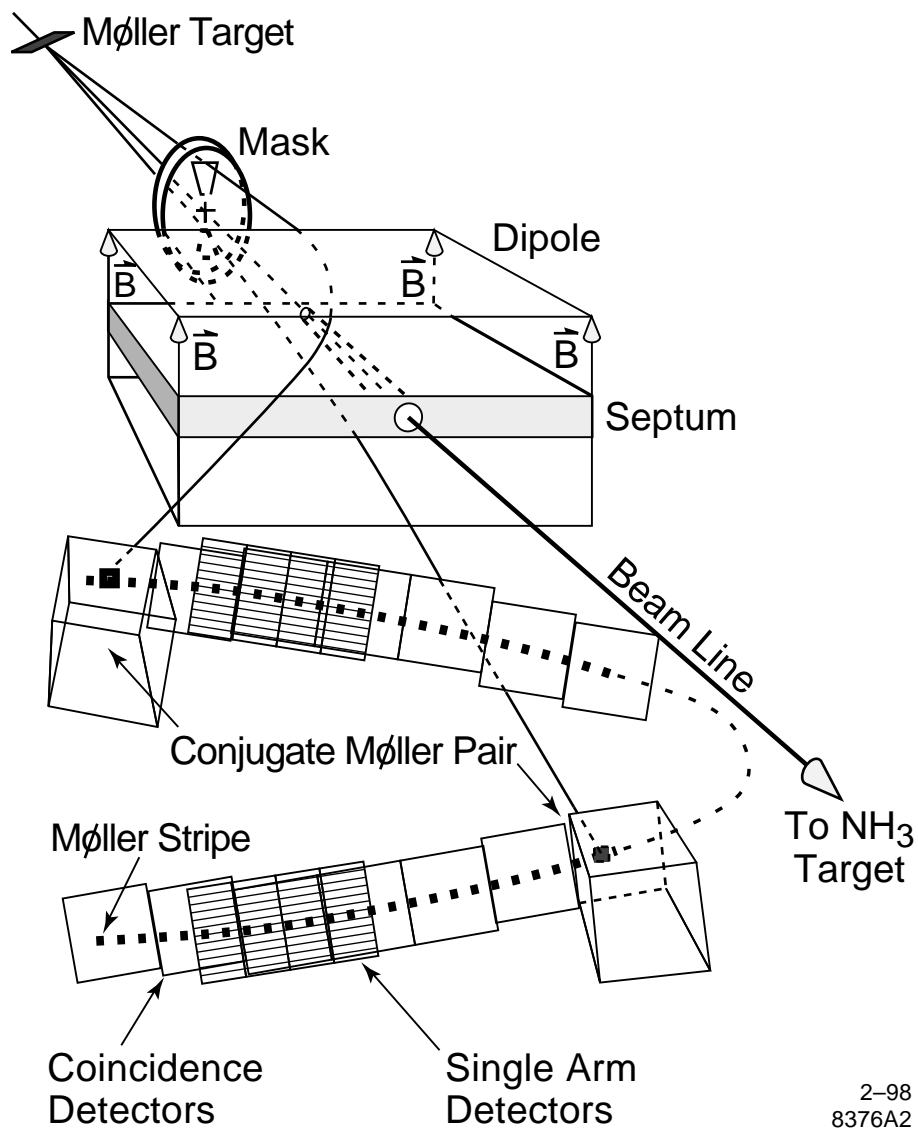,height=15cm}}\vskip 0.7cm
\caption{The layout of the M\o ller polarimeter systems used in the 
	E143 experiment (not to scale).}
\label{fg:moller-layout}
\end{figure}

A schematic of the polarimeter is shown in Fig.~\ref{fg:moller-layout}. The major components 
are a polarized electron target, an acceptance-defining collimator, a dipole 
magnet spectrometer, and two independent detector systems. One system detected 
the M\o ller electrons in coincidence, whereas the other integrated the single 
electrons over the duration of the beam spill.

The polarized electron target consisted of six magnetized ferromagnetic foils 
of different thickness which could be moved into the beam. The foils were 
magnetized to near saturation by Helmholtz coils providing nearly 100 gauss at 
the target center. The permendur (49\%~Fe, 49\%~Co, 2\%~V) foils were 3 cm wide 
and varied in thickness from 20~$\mu$m to 150~$\mu$m. The target electron
polarization (typically 0.082) was determined to a relative accuracy of 
1.7\%~\cite{E154moller} from foil magnetization measurements.

The tungsten collimator which was 20 radiation lengths thick (see 
Fig.~\ref{fg:moller-layout}) had a central opening to allow the main beam to 
pass and wedge shaped apertures of constant azimuthal acceptance (0.2 radian 
top, 0.22 radian bottom) above and below the beam-line to select  M\o ller 
electrons scattered transverse to the bend plane of the downstream dipole 
magnet. The vertical acceptance was 3.6--9~mrad in the lab (corresponding to 
70--116$^\circ$. in the center-of-mass). The 2.1~T-m dipole field separated the 
scattered electrons according to momenta. Since M\o ller scattering is elastic, 
the x and y position of the scattered electrons at the detector plane are 
correlated as shown in Fig.~\ref{fg:moller-layout}.

The detector hut was situated 27~m downstream from the M\o ller target. The 
single-arm detector package of three radiation lengths of lead and a single 
plane of position-sensitive silicon detectors was placed immediately in front 
of the coincidence detectors. The coincidence package consisted of two arrays 
of seven lead glass blocks (SF-6), each with a 10$\times$10~cm$^{2}$ entrance 
area and 25~cm of depth ($\sim$15 radiation lengths).

\subsubsection{The Measurements}

Measurements of the beam polarization were performed every one to two days. 
Each measurement period typically consisted of four runs using two target foils 
(thin and thick) and  
opposite target polarization directions. This made it possible to look for 
rate or helicity dependent effects in the data. For polarization measurements, 
the beam rastering was turned off and the beam focus was moved to the M\o ller 
target. Otherwise, the beam conditions were identical to that of the main experiment. Data were obtained from almost 200 runs over a range of 
luminosities (more than a factor of 8) through different combinations of foil 
thickness and beam current. The coincidence polarimeter obtained a typical 
statistical precision of 0.010 (absolute) per run whereas the single arm 
polarimeter achieved 0.019 per run. Both detector systems took data at 29.1 
and 16.2 GeV. The single arm collected data with the 9.7 GeV beam.

\subsubsection{The Coincidence Polarimeter}
  
The segmented lead glass arrays provided good energy and timing information 
and made it possible to accommodate the high instantaneous rates of several 10s 
of MHz characteristic of the low duty factor (10$^{-4}$) at SLAC. The 
combination of \v{C}erenkov light in the glass blocks, fast photomultiplier 
tubes, and a clipping circuit resulted in signal pulse widths as narrow as 
five nanoseconds. The signals were fed into an Ortec~935 constant fraction 
discriminator with the threshold set at 30--40\% of typical M\o ller signal 
amplitude. The discriminator output was then fed into a fast multiplexing 
circuit with a fanout to three or four TDC channels.  The time of each event 
was recorded by a LeCroy 2277 multi-hit TDC which has a least significant bit
time of one nanosecond.  The multiplexor was required to decrease the dead-time 
and increase the maximum hit capability of the individual TDC channels. A laser 
pulser, triggered randomly $\leq$1/spill, was fed into each detector block 
simultaneously through fibers to provide both time calibration and detector 
dead-time information for the analysis.

The data were recorded on tape on a spill-by-spill basis as a series of event 
times and corresponding TDC channels. The analysis identified coincidence 
events by the arrival of single event times within a predefined time window of 
$\pm$4~ns, determined by the resolution of the TDC's. Frequently, the 
analysis encountered ambiguous coincidence pair combinations where a single 
event of one detector could be combined with events in two or more other 
detectors. Such ambiguities arose as a result of cross-talk between adjacent 
detectors due to shower sharing or due to random coincidences, particularly at 
higher luminosities. In the case of ambiguities, the cluster of all possible 
coincidences was subjected to a decision making routine which selected the most 
probable combinations of events. In the case of cross-talk events the full 
weight of the single coincidence event was shared with the adjacent coincidence 
pairs. Background contributions consisted of random coincidences between 
M\o ller or Mott type electron events. Their contribution, typically $\le$1\% 
was estimated from the product of the singles rates in each conjugate detector.

The dead-time measurement was obtained using the laser pulser system which sent 
a known pulse to all 14 detectors  simultaneously. The efficiency at which both 
detectors of a pair saw the pulser event yielded the live-time for that pair. 
It was also necessary to correct for the possibility of two M\o ller events 
occurring in a given pair within the same coincidence window. In such a case 
the system is only capable of seeing one of the pairs, an inefficiency which 
would be unaccounted for in the dead-time correction. To correct for this 
effect an estimate of the number of M\o ller coincidence events occurring 
during the pulser event was added to the known number of pulser events for each 
pair. 

\begin{figure}[h]
\centerline{\epsfig{file=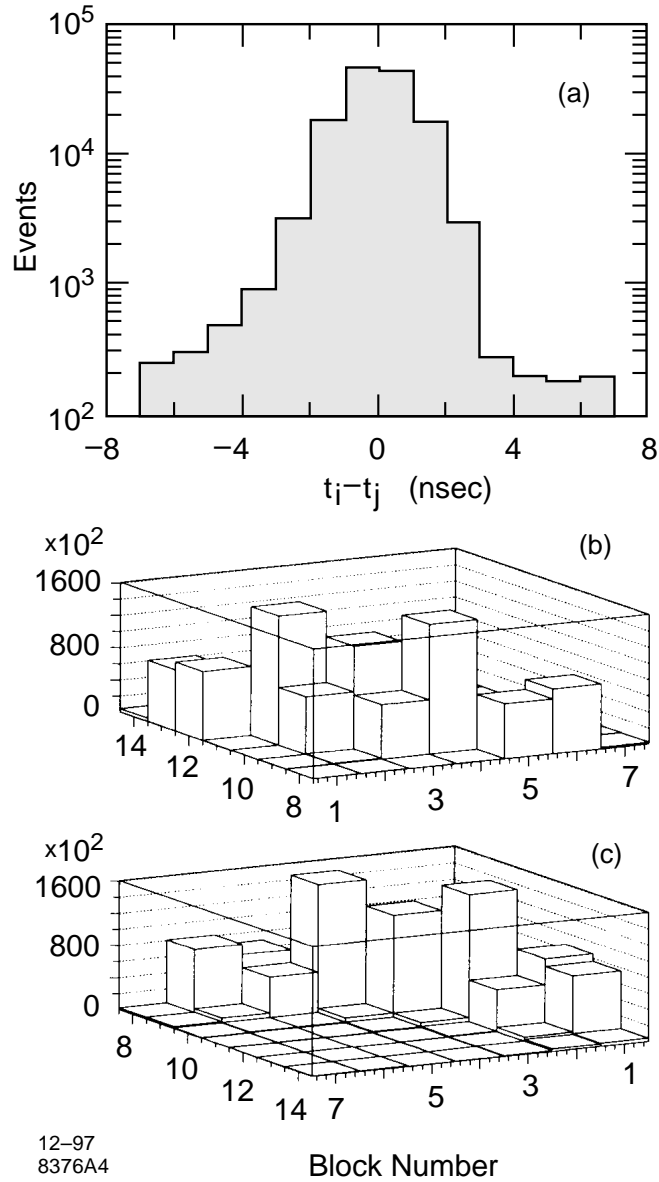,width=3.5 in}}\vskip 0.7cm
\caption{(a) A typical M\o ller coincidence event time difference spectrum. 
(b)-(c) Two views of the distribution of coincidence events in the 
7$\times$7=49 possible combinations. View (c) shows the hidden back-side 
of view (b). True M\o ller events were constrained to occur in the crest.}
\label{fg:moller-twoarm}
\end{figure}

A typical M\o ller coincidence time difference spectrum is shown in 
Fig.~\ref{fg:moller-twoarm}a. Two views of a typical distribution of 
coincidence events in the two detector arms are given in 
Figs.~\ref{fg:moller-twoarm}b-c for a run at 29~GeV.  True M\o ller events were 
kinematically restricted to occur only in 11 (9 for the 16~GeV data) of the 
possible (7$\times$7) 49 pairs. Radiative effects did allow some true M\o ller 
coincidences to occur on the low momentum side of the ridge 
seen in Fig.~\ref{fg:moller-twoarm}c, but these events were not considered 
in the total event yield due to poor signal-to-noise ratio and greater 
uncertainties in the analyzing powers.

The beam polarization for each coincidence pair was determined from the 
asymmetry in the yield corrected for background, dead-time, charge asymmetry, 
the effective analyzing power of each pair, the target polarization and the 
target angle. The polarization for a run was taken from the weighted average 
over all the pairs. 

A Monte Carlo analysis~\cite{Swartz:95} was used to determine the effective 
analyzing powers of each coincidence pair and  to evaluate the sensitivity of 
the analyzing powers to possible systematic influences such as the atomic 
motion of the target electrons~\cite{levchuk} and typical shifts in the beam 
position or focus. The analyzing powers were found to range from 0.776 to 0.690 
for the different pairs. The effect of the target electron motion was to
increase the average analyzing power by $<$0.5\%. Typical beam parameter shifts 
resulted in changes to the average analyzing power within $\pm$0.6\%. Since 
only one set of analyzing powers was used at each beam energy, the $\pm$0.6\% 
variation was included as a source of systematic error. The large acceptance of
the detectors reduced the sensitivity of the analyzing powers to these 
systematic influences. 

The possibility of rate dependence was investigated in two studies. One study 
compared low and high luminosity runs taken during a run set where little 
variation in polarization was expected during the set. In this study the internal 
agreement between all measurements of a run set was very good, resulting in an 
average $\chi^{2}$ per degree of freedom (df) of 1.1  for all the run sets. 
Another study tested the effectiveness of the analysis routine in dealing with 
ambiguities in the data which were most prevalent at high luminosity. 
Data taken from subsequent spills of a low luminosity run  were artificially superimposed to 
create a fictitious spill of high luminosity. After imposing the detector 
dead-time on the single events the data were analyzed as a normal run, and the 
yields could be compared with the original luminosity analysis. In both studies 
it was possible to rule out a rate dependence at better than $\pm$0.5\%.

Despite the excellent agreement of the polarization results within a run set, 
a large fluctuation in the average polarization values obtained from each 
coincidence pair was observed. Although the sources of these fluctuations 
likely cancel in the average, their origins are not clear. As a result a 
maximum error contribution of 1.3\% was included. This contribution reduced  
the $\chi^{2}$/df of the pair-dependent polarization distribution to unity. 
This uncertainty was combined in quadrature with the uncertainties estimated 
for the analyzing powers and the limit on a possible rate dependence to obtain 
a total systematic uncertainty of 1.5\%  for the coincidence polarimeter 
measurement. 

\subsubsection{The Single-Arm Polarimeter}

The single arm detectors had four silicon pad detectors above and below the 
beam height. A lead converter absorbed soft photon backgrounds and amplified 
the M\o ller signal. Each detector consisted of two 4 (x) by 6 (y) cm
silicon devices approximately 300 $\mu$m  thick. Each device was segmented
into 7 pads (channels) 8.70 mm wide and 40 mm long. Only 12 contiguous channels 
were instrumented in each detector. The detectors were tilted by 
$-10.5^\circ$ (top) and $+11.0^\circ$ (bottom) to align the channels along the 
M\o ller scattered electron stripe. Since each detector was formed from two 
silicon devices there was a 5.3 mm gap between channels seven and eight.

The silicon detector channels were connected to 96 charge sensitive
preamplifiers which integrated over the entire 2300 ns beam pulse. The 
preamplifier outputs were brought into ADCs to measure the peak of the 
preamplifier signal and were recorded together with the sign of the beam 
polarization for each beam pulse.

\begin{figure}[h]
\centerline{\epsfig{file=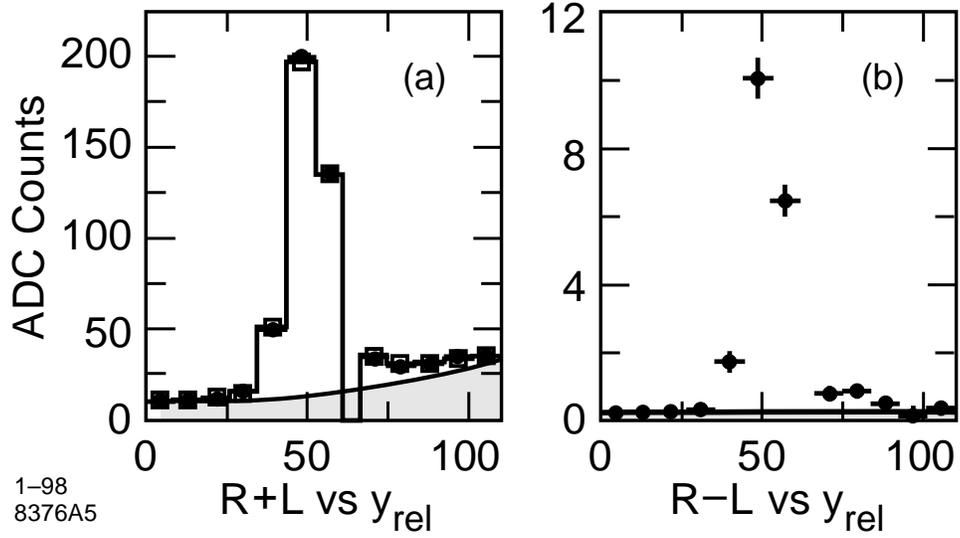,width=5in}}\vskip 0.7cm
\caption{Typical measured single-arm
M\o ller line-shapes in detector 7 at  29.1 GeV.
The (R+L) data (points), (R+L) fit (histogram), and (R+L) background 
(shaded region) are plotted in (a) versus the relative y 
position of the channel. The (R--L) data (points) and (R--L) 
background (shaded region) are plotted in (b).}
\label{fg:moller-sarm}
\end{figure}

The M\o ller analysis proceeded through two steps. The first-pass analysis 
calculated average pedestal subtracted pulse heights and errors for each 
channel from the pulse-by-pulse data. Separate averages were made
for pulses tagged by right (R) and left (L) handed polarization bits. 
Correlations between channels were calculated and recorded. A very loose beam 
current requirement was made before including the pulse in the overall 
averages. A summary file containing the ADC averages and errors as well as 
useful beam and polarimeter parameters was written for each run. A second-pass 
analysis read the summary file, applied channel by channel gain corrections, 
and formed sum (R+L) and difference (R--L) averages and errors for each 
channel. Typical (R+L) and (R--L) line-shapes are shown in 
Fig.~\ref{fg:moller-sarm} for data at 29.1 GeV.

The background under the unpolarized (R+L) M\o ller scatters was estimated by 
fitting the (R+L) line-shape to an arbitrary quadratic background plus the 
line-shape expected from unpolarized M\o ller scattering. The technique for 
estimating the unpolarized line-shape used the observed (R--L) line-shape and 
angular smearing functions \cite{E154moller} to generate a predicted (R+L) 
line-shape for M\o ller scatters. The observed (R+L) distribution was then 
fit by this predicted line-shape and a quadratic background. Since the observed 
(R--L) line-shape is already broadened by multiple scattering in the target 
material, beam windows, air, and helium, only corrections to the line-shape 
which are different~\cite{levchuk} for scatters from polarized and unpolarized 
target electrons are included in the smearing function.

An analyzing power for each detector was calculated from the target 
polarization and the expected M\o ller asymmetry determined by Monte Carlo 
simulations of the scattering process~\cite{Swartz:95} and detector response. 
The effect of the target electron momentum distribution ~\cite{levchuk} was to 
modify the expected asymmetries by 1.4\%. The measured asymmetry for each 
detector was calculated from the ADC averages by:
\begin{equation}
A_{meas.} =  {\sum_i (R-L)_i -\sum_i (BKG)_{R-L} \over 
{\sum_i (R+L)_i - \sum_i (BKG)_{R+L}}},
\end{equation}
where the sum is over the central five channels including the  M\o ller peak.
The (R--L) background was estimated by averaging the channels far from the 
M\o ller peak. The (R+L) background subtraction increased the measured asymmetry 
by 17--24\%. The full covariance matrix calculated from the pulse-by-pulse 
data was used to determine the statistical error of $A_{meas}$. The beam
polarization was calculated from the measured asymmetry divided by the 
analyzing power.

To check for possible systematic biases in the single arm analysis, both the 
number of channels included in the sum over the M\o ller peak and the shape of 
the background fit to  the unpolarized (R+L) line shapes were varied. From the 
observed spread in calculated polarizations, the sensitivity of the single 
arm analysis to the choice of fit parameters was estimated to be 1.3\%.
The polarization determined individually by each detector agreed with the 
overall mean within statistical errors. The total systematic error of the 
single arm analysis includes contributions from: the detector analyzing 
power, known to $\approx 0.5$\%,  possible nonlinearities in the preamplifier 
and ADC response which could change the computed  polarization by $\leq 1.0$\%,
and the sensitivity to analysis parameters, 1.3\% as discussed above. The total 
systematic error of the single arm analysis is estimated to be 1.7\%.

\subsubsection{Results \label{sect:beampol}}

\begin{figure}[h]
\centerline{\epsfig{file=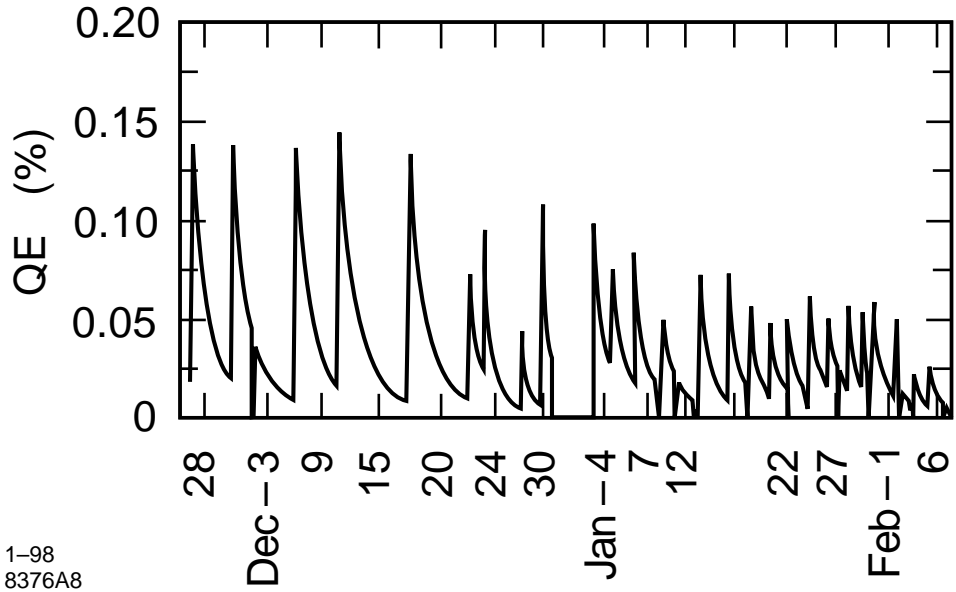,width=5in}}\vskip 0.7cm
\caption{Variation of the quantum efficiency (QE) of the polarized 
source over the course of the E143 experiment. }
\label{fg:beam-qe}
\end{figure}

The polarization values measured by the single and coincidence arm polarimeters 
were in good agreement, although the results from the coincidence system were 
on average 0.6\% lower than the single arm. Both systems measured the same 
polarization dependence on the quantum efficiency of the polarized source, 
resulting in a linear decrease of source polarization with increasing quantum 
efficiency. The polarized source quantum efficiency time history is shown in 
Fig.~\ref{fg:beam-qe}. The  variations in polarized source quantum efficiency 
were related to the frequent cesiation treatments which were applied to the 
source in order to maintain the source quantum efficiency at an acceptable
level. 

\begin{figure}[h]
\centerline{\epsfig{file=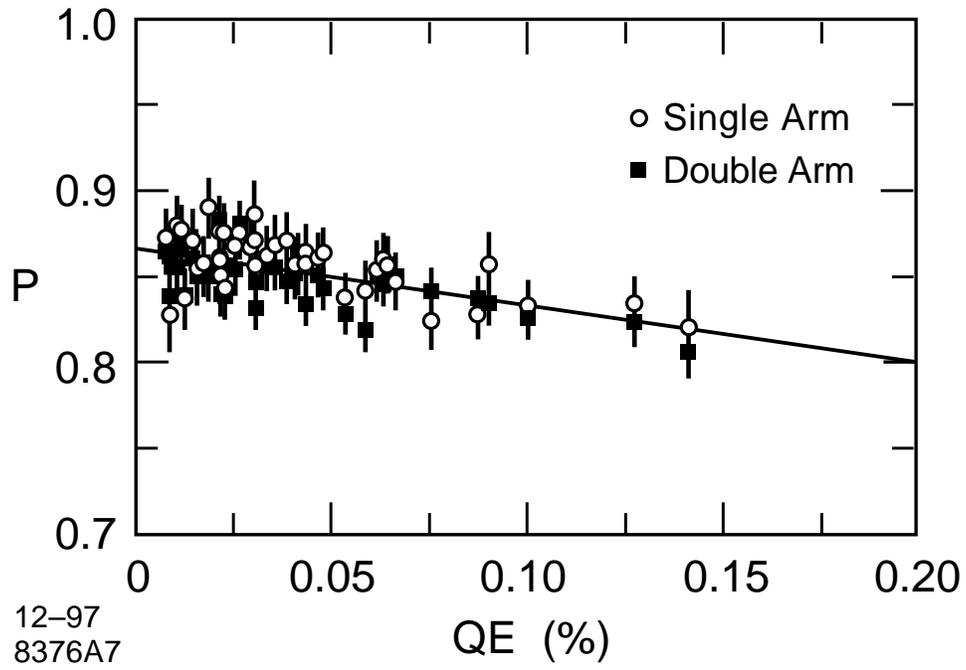,width=5in}}\vskip 0.7cm
\caption{Beam polarization versus QE
for the the single (circles) and double-arm (squares)
polarimeter systems.}
\label{fg:moller-final}
\end{figure}

To obtain the beam polarization for the main analysis, the average polarization 
value for each run set was computed separately for each polarimeter. The 
results from both polarimeters are shown in Fig.~\ref{fg:moller-final} plotted 
as a function of source quantum efficiency. The plotted errors are a 
combination of the computed statistical errors and an additional 0.8\% 
systematic error to account for non-statistical fluctuations in the data. 
 
A linear fit to the single arm and coincidence data as a function of polarized 
source quantum efficiency (QE) yields:
\begin{equation}
P_B = (0.866 - 0.34\times QE)\pm 0.003\pm 0.022,
\end{equation}
where the first error term is statistical while the second and dominant term is
systematic. The systematic error includes a contribution of $\pm 0.8\%$, as 
discussed above, a $\pm 1.6\%$ contribution from the average of the single arm 
and coincidence M\o ller systematic errors, and a $\pm 1.7\%$ contribution from 
the uncertainty assigned to the target foil polarization. The resultant 
systematic error is $\pm 2.5\%$.

\subsection{The Polarized Target}

\begin{figure}[h]
\centerline{\epsfig{file=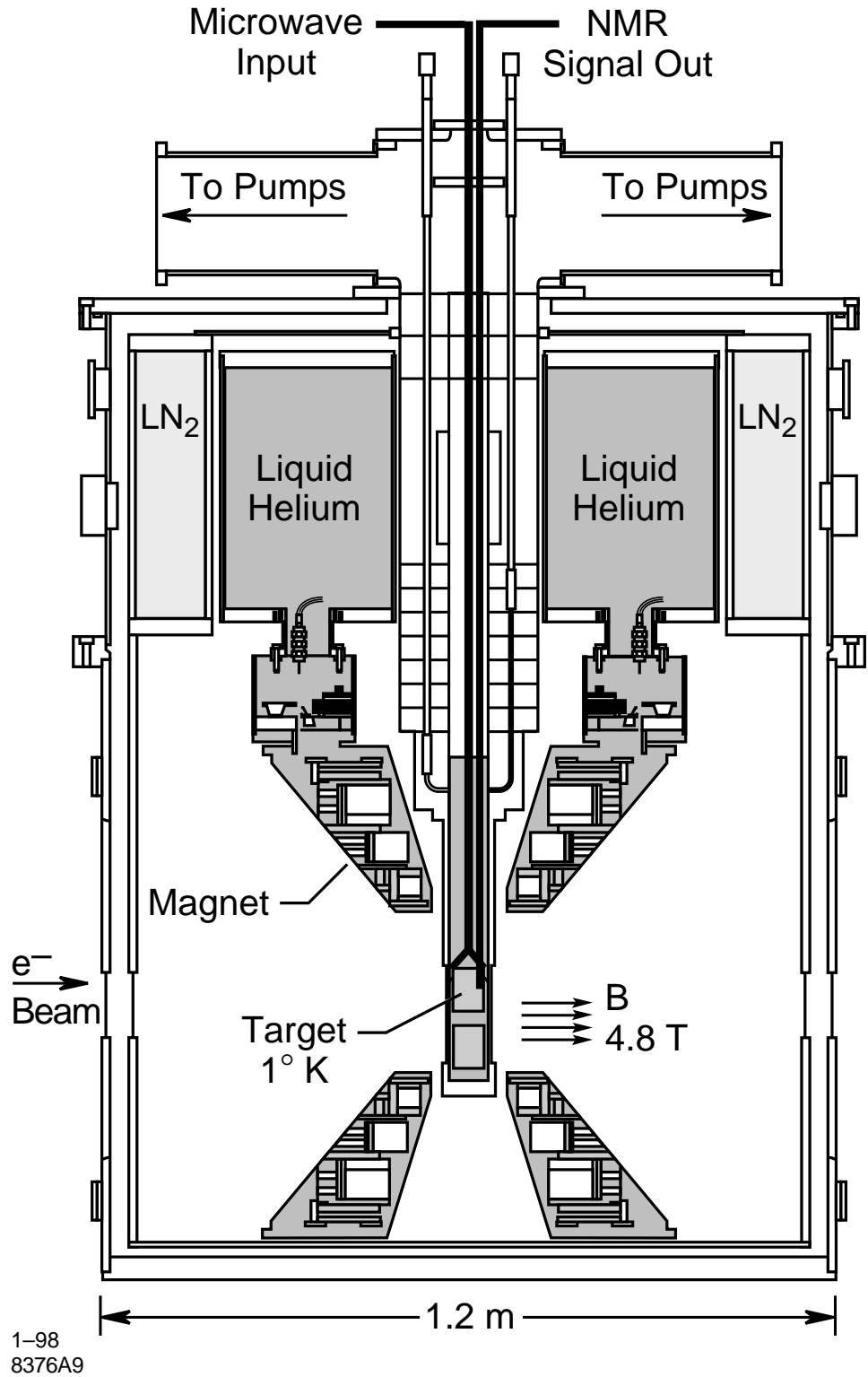,width=5 in}}\vskip 0.7cm
\caption{ E143 target schematic.}
\label{fg:targ}
\end{figure}

\begin{figure}[h]
\centerline{\epsfig{file=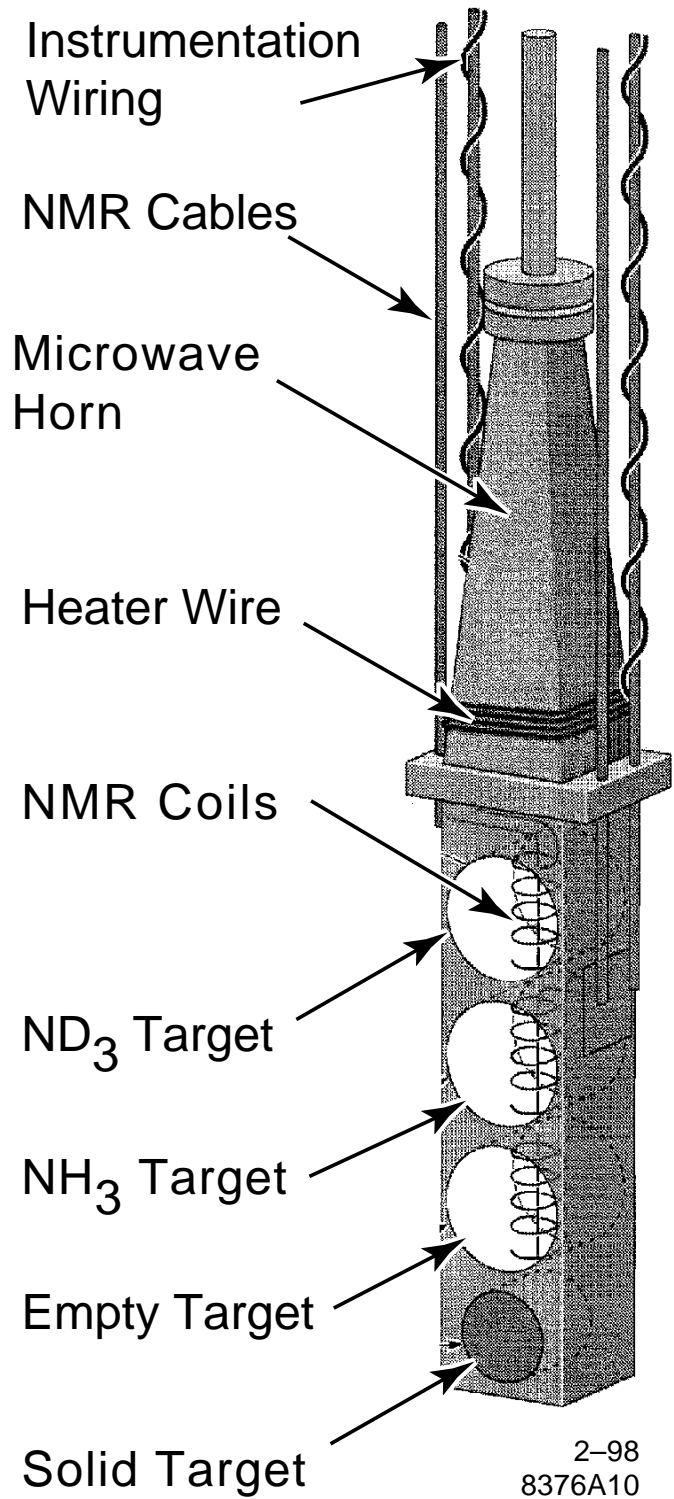,width=3.5 in}}\vskip 0.7cm
\caption{Target insert schematic.}
\label{fg:targinsert}
\end{figure}

The polarized target required a high-power $^4$He evaporation refrigerator 
operating near 1 K, and a 5 T superconducting split pair magnet.\footnote{%
Oxford Instruments, Eynsham,UK} The target material, frozen $^{15}$N ammonia,
was polarized using dynamic nuclear polarization (DNP). A schematic diagram of 
the target is shown in Fig.~\ref{fg:targ} \cite{craday}. The magnet is shown 
with its field direction along the beam momentum direction. The refrigerator is 
positioned vertically and along the axis of the magnet. It is connected to a 
large Roots blower pumping system. The target insert lies along the central 
axis of the refrigerator. This insert was slid up and down to position any one 
of four targets in the beam. The targets were (from the top position) 
$^{15}$ND$_3$, $^{15}$NH$_3$, an empty cell, and either carbon or aluminum. 
A fifth position having no target was also available. The target insert also 
carried coaxial cables for the NMR measurement, a waveguide to transmit 
microwaves to the target(s) for DNP, and various temperature sensors. A diagram 
of the target insert is shown in Fig.~\ref{fg:targinsert}.

\subsubsection{DNP and Ammonia}

The DNP process for polarizing protons, deuterons, or any nucleus possessing a 
magnetic moment, requires temperatures of $\sim$1 K or less and large
magnetic holding fields.  For thermal equilibrium at 1 K and 5 T, the proton 
polarization is only about 0.5\%. However, the polarization of the ``free'' 
electrons, associated with the paramagnetic radicals introduced into the target 
material, is greater than 99\%. The electron polarization can be transferred to 
the proton through a hyperfine transition by irradiating the target with 
microwaves at appropriate frequencies. The two polarization directions for the 
proton are reached by irradiation at frequencies slightly above or below the
electron Larmor frequency, $\approx $ 140 GHz at 5 T. Details of the DNP 
process can be found in the literature, e.g., Abragam and Goldman 
\cite{abragold} or Borghini \cite{borg}. In our case the magnetic field was 
held at 4.87 T to match the frequency range (136 - 137 GHz) of the microwave
tube\footnote{%
CPI, Georgetown, Ont., Canada} being used.

Ammonia was chosen as the target material because of its relatively large 
dilution factor compared to most other polarized target materials, its high 
polarizability, and its resistance to radiation damage. Furthermore, $^{15}$N 
ammonia (spin $\frac 12$) was chosen over $^{14}$N (spin 1) because in $^{15}$N 
the spin is carried by an unpaired proton, in contrast to  $^{14}$N where the 
spin is carried by a proton-neutron pair. Using $^{15}$NH$_3$ reduces the 
systematic errors on the proton spin structure functions by eliminating 
unwanted contributions from the neutron asymmetry. In addition, the $^{15}$N 
polarization is easier to measure.

The $^{15}$NH$_3$ and $^{15}$ND$_3$ targets were both prepared in the same way:
First, the ammonia gas was slowly frozen in a test tube; the resulting solid lump of ammonia ice was crushed while immersed in liquid nitrogen and sifted to  select granules of approximately 2 mm size. Smaller pieces were recycled in the 
same apparatus.

The paramagnetic radicals necessary for DNP were introduced by irradiation 
using various electron beams. Each sample was immersed in liquid argon and 
given a dose of about $3 - 5 \times 10^{16}$ electrons cm$^{-2}$. Targets for 
E143 were irradiated at Bates (at an electron energy of 350 MeV), at the Naval 
Postgraduate School, Monterey (65 MeV), and at the High Energy Physics 
Laboratory at Stanford (30 MeV). Samples of $^{14}$N ammonia were irradiated at
Saskatoon (250 MeV) for the initial tests. All irradiated samples were packed 
into thin-walled (0.0127 cm) torlon cylinders with 0.0025 cm aluminum end-cap windows. 
Each cylinder was 3 cm long and 2.5 cm in diameter, and contained two NMR coils 
made from 70/30 Cu/Ni tubing of 0.5 mm outer diameter and 0.0178 cm wall 
thickness. A straight piece of tubing was used to measure the proton
polarization in the NH$_3$ cell and the residual proton polarization in the
ND$_3$ cell. A coil of three to four turns with a 1 cm diameter measured the
deuteron polarization and $^{15}$N polarization in the ND$_3$ target, while
a similar one measured the $^{15}$N polarization in the NH$_3$ cell. During the 
course of E143 only the proton and deuteron polarizations were measured; the $^{15}$N and residual proton polarizations were checked after the experiment.

\subsubsection{Polarization Measurement and Performance\label{sect:targpol}}

The polarization was measured via NMR with a series-tuned Liverpool Q-meter
\cite{gcourt}. Each spin species in the targets was measured with its own
separately tuned Q-meter. Only one Q-meter could measure at a given time,
taking one polarization measurement per minute. The Q of the tuned
circuit is changed by the presence of the appropriate polarized nuclei, and
the integral of this response is proportional to the polarization. The response 
function was determined by subtracting the Q-curve measured when the magnetic 
field was moved off resonance from the Q-curve obtained when the magnetic field 
was moved on resonance. The integral is normalized by comparing to the signal 
area at thermal equilibrium (TE) where the polarization (P$_{TE}$) can be 
calculated.  For the proton, P$_{TE}$ = tanh$\left[ \frac{\mu B}{kT}\right] $,
where $\mu $ is the magnetic moment of the proton and $k$ is Boltzmann's 
constant. Therefore,  P$_{TE}$ = 0.0034 for $B = 5$ T and $T = 1.5$ K.

\begin{figure}[h]
\centerline{\epsfig{file=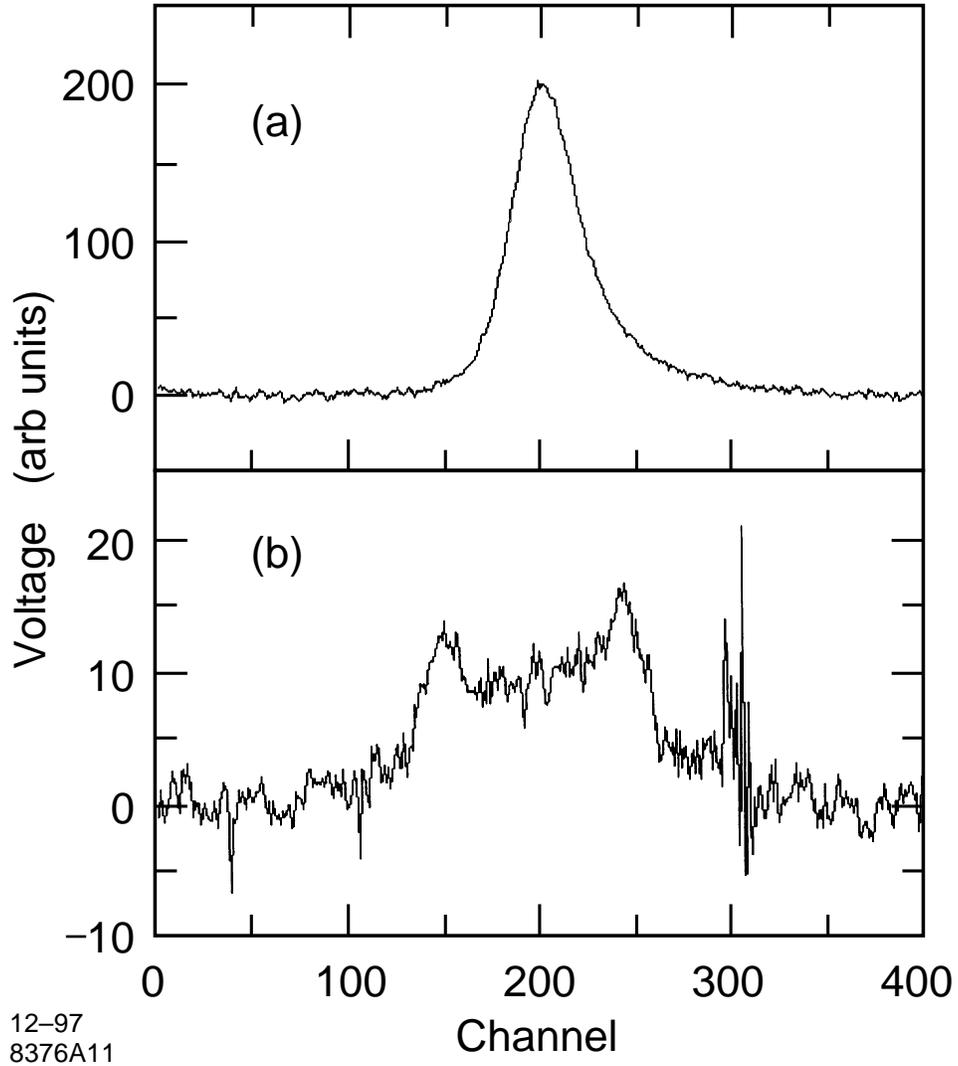,width=5 in}}\vskip 0.7cm
\caption{Typical TE signals measured from (a) polarized protons
and (b) deuterons. The spikes in the deuteron signal are an artifact
of the synthesized signal generator that was used.}
\label{fg:TE}
\end{figure}

The TE signal for the proton is relatively easy to observe and measure, but the 
deuteron TE signal is about 500 times smaller than this, and thus requires 
advanced techniques of noise and drift suppression and signal averaging for
a credible measurement\cite{dhawan}. Measurements were made of the proton TE 
signal area to a precision of about 0.2\%, but repeated measurements, over a 
period of many weeks, showed considerable fluctuations in the mean value. 
Including this scatter, the overall precision of measuring the TE polarization 
was $\pm$ 2.5\%. For the deuteron the precision of measuring the TE signal area
was $\pm$ 3\% and $\pm$ 4\% overall. The fluctuation in signal area was
attributed to small changes in the distribution of ammonia granules around
the NMR coils. Typical TE signals for polarized protons and deuterons are shown 
in Fig.~\ref{fg:TE}.

\begin{figure}[h]
\centerline{\epsfig{file=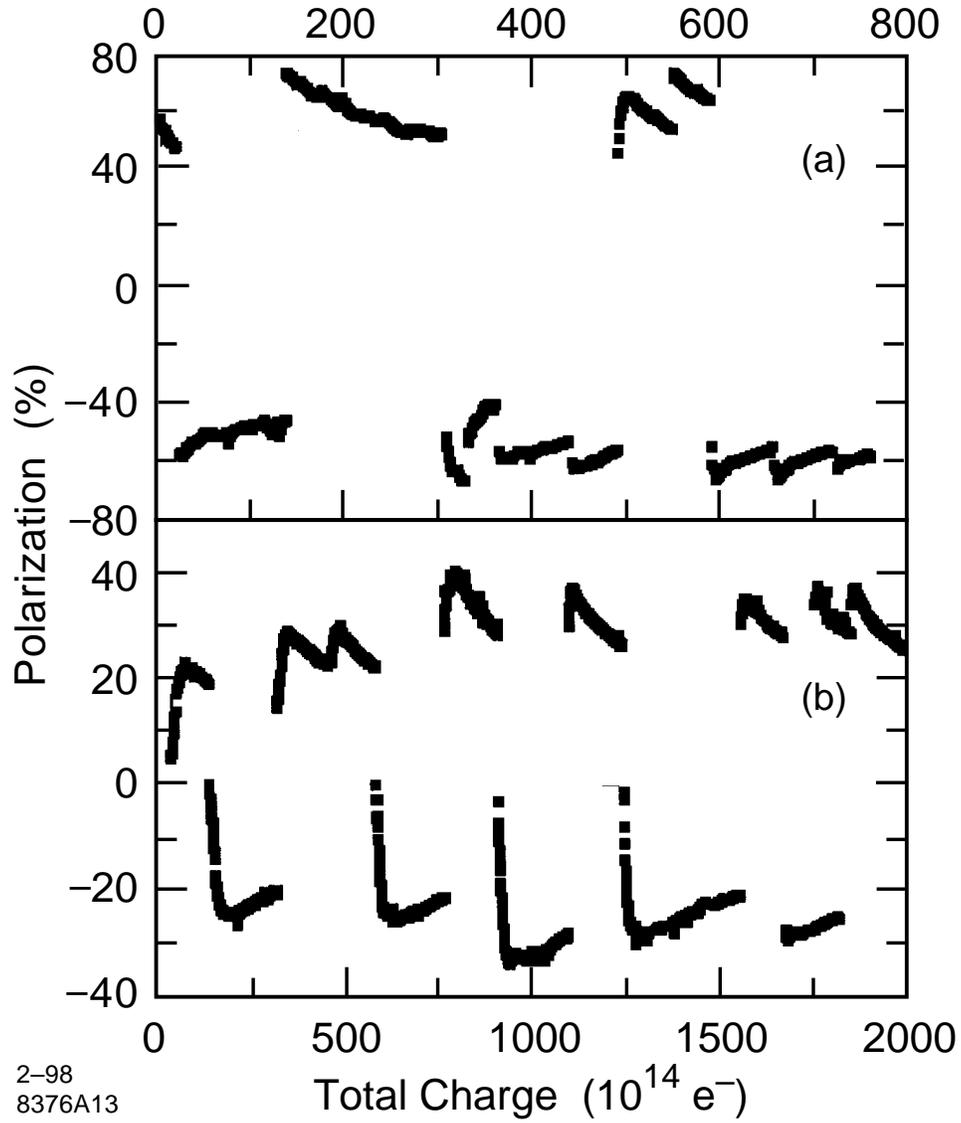,width=5.0 in}}\vskip 0.7cm
\caption{The polarization history over the course of a few days 
is shown as a function of received charge for (a) proton
and (b) deuteron targets.}
\label{fg:polhist}
\end{figure}

In our initial measurements of $^{15}$N ammonia, the proton polarization
performance was similar to that seen previously \cite{crabb} ($>$90\%), but the 
deuteron only reached 13\%.  The maximum deuteron polarization was expected to 
increase with {\em in situ} irradiation \cite{boden}. Figure~\ref{fg:polhist},
which verifies this expectation, shows how the proton and deuteron 
polarizations performed as a function of beam dose for 5 x 10$^{11}$
electrons/sec rastered over the face of a target. 

The polarization decayed with beam dose as the ammonia became 
radiation-damaged. Once the polarization fell below a predetermined value, the 
other target was put into the beam until its polarization dropped to a 
specified level. Then both targets were annealed by warming them up to a 
temperature between 80 and 90 K. The sequence of polarization, irradiation and annealing affected $^{15}$NH$_3$ and $^{15}$ND$_3$ differently. For the
proton, annealing brings the target back to its starting polarization, and
there was no evidence of change over the period of irradiation. On the other
hand, the deuteron polarization improved after each anneal, ultimately
reaching a polarization of 42\%. This value was obtained with frequency
modulation of the microwave source. Previously, at CERN, in the SMC
experiment\cite{smcfm}, frequency modulation had been found to improve the
polarization of deuterated butanol by almost a factor of two. In deuterated
ammonia the gain is more modest, with a factor of two improvement in the rate 
of polarization which  leads to a gain in absolute polarization of $3-5\%$.  
The level of proton polarization in Fig.~\ref{fg:polhist} is lower than the 
expected maximum of  more than 90\% seen in the early measurements. This was 
because the NH$_3$ target was situated below the ND$_3$ target which absorbed 
some fraction of the microwave power.

After the experiment the $^{15}$N polarizations were measured as a function of 
both proton and deuteron polarizations. Residual proton polarizations were 
measured in the deuteron case. In addition, the protons in the torlon target 
cups became polarized once the electron beam created paramagnetic centers in 
that material. This led to a 3\% correction of the proton polarization. The 
polarization values were also corrected for effects arising from 
inhomogeneities in target polarization due to local beam heating as discussed
below.

The average polarizations for the entire experiment were 0.70 with a relative 
precision of 2.5\% for the proton and 0.25 with a relative precision of 4\% for 
the deuteron.

\subsubsection{Beam Heating Corrections}

As the beam passes through the polarized target, the temperature of the ammonia
granules increases, and the polarization drops. By rastering the beam over the 
face of the target, this depolarization effect is greatly reduced. The average 
polarization measured by the standard NMR technique is generally not the same 
as what the beam sees locally. One reason for this is that ammonia granules 
outside the raster radius do not experience the same depolarization from beam 
heating as the granules inside the raster radius. The measured polarization, 
however, reflects a combined polarization of all the target granules. Another 
reason is that the polarization during the beam spill may be lower than during 
the time between spills when no beam heats the target. This latter
effect has been studied in detail\cite{tj} and has been shown to be very small.
Hence, it has been neglected in the present analysis.

If $z$ is the relative contribution of the rastered granules to the NMR signal, then 
\begin{equation} \label{eq:avepol}
P_{m}=zP_{T}+(1-z)P_{i},
\end{equation}
where $P_{i}$ is the initial polarization with no incident beam, $P_{m}$ is the 
measured polarization with incident beam, and $P_{T}$ is the true polarization 
of the rastered granules. We define a correction to the measured polarization 
$C_{heat}$ as
\begin{equation}\label{eq:ref0}
1-C_{heat} \equiv P_{T}/P_{m} = {{P_m - (1-z)P_i} \over {zP_m}}.
\end{equation}
The parameter $z$ depends on the geometry of the NMR coils, which is different 
for NH$_3$ and ND$_3$ targets, and on the direction of the target polarization 
(longitudinal or transverse). Values for $P_i$, $P_m$, $P_T$, $z$, and
$C_{heat}$ and the corresponding errors are given in Table~\ref{tb:beamheat} 
for a maximum beam intensity of $4\times  10^9$ electrons/pulse. The errors on 
$z$ include uncertainties for the target granule settling effect and for the 
rastering radius due to the finite size of the beam spot. For the ND$_3$ 
targets there is an additional uncertainty in the diameter of the 4 turns of 
the NMR coil. The corresponding corrections at other beam intensities can be 
extracted using the knowledge that the measured target depolarization is 
proportional to beam intensity.

\subsection{Spectrometers}\label{spect}

\begin{figure}[h]
\centerline{\epsfig{file=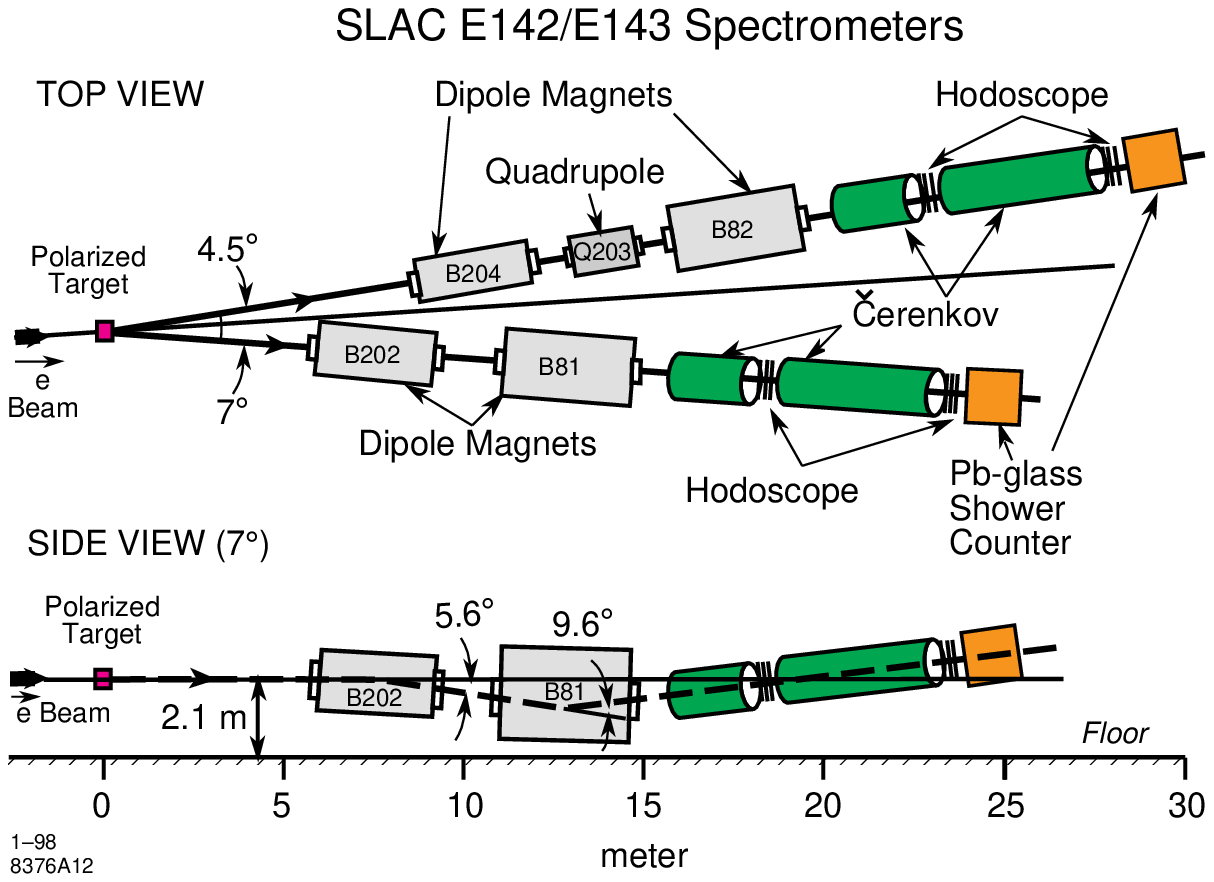,width=6.0 in}}\vskip 0.7cm
\caption{A schematic of the E142/E143 spectrometer layout.}
\label{fg:spect}
\end{figure}

Two large acceptance spectrometers, situated at $4.5^\circ$ and $7.0^\circ$,
were used to detect the electrons scattered from the polarized target. The momentum acceptance of each spectrometer arm ranged from 7 to 20 GeV/c. Each 
spectrometer contained two dipole magnets, bending in opposite directions in the vertical 
plane, two gas threshold \v{C}erenkov detectors, two scintillation hodoscope 
packages, each consisting of several planes, and an array of lead glass total 
absorption shower counters which were 24~radiation lengths in depth.  The 
$4.5^\circ$ spectrometer also contained a quadrupole magnet which was needed to spread the scattered electrons over a larger detector area. A schematic of the 
spectrometers is shown in Fig.~\ref{fg:spect}.

The two-bend design was chosen to have maximum acceptance over a wide momentum 
range, and to shield the detectors from the considerable photon background 
produced by the electron beam interacting in the thick target. The \v{C}erenkov 
detectors allowed discrimination against a large pion background. The 
hodoscopes were used to reconstruct the trajectory of each particle, which in 
turn could be used to determine the momentum and other kinematic variables. 
Finally the shower array provided the energy measurement as well as particle 
identification information. 

The spectrometers were almost identical to those used in the E142 experiment 
\cite{e142}. The magnets, however, were operated at somewhat higher fields to 
accommodate the larger momenta of scattered electrons due to the higher beam 
energy. Also, the hodoscopes were modified to handle a higher instantaneous 
rate.

The two scintillator hodoscope arrays provided the track information of the 
incident particles for each spectrometer, and consisted of horizontal ($y$), 
vertical ($x$), and slanting ($u$) planes of fingers. The upstream hodoscope 
array contained four planes: $u$, $x$, and two $y$; the downstream hodoscope 
array contained an $x$, $y$, and $u$ plane. The first $y$ plane in each hodoscope 
array consisted of scintillator elements of 3.0 cm width, while the elements of 
the second $y$ plane in the upstream hodoscopes were 4.76 cm wide. The elements 
in the first $x$ plane of the 4.5$^\circ$ spectrometer were 2.0 cm wide, and the 
remaining $x$ plane elements were 3.0 cm wide. The $u$ planes 
contained elements that were 4.5 cm and 7.5 cm wide for the front and back 
hodoscopes respectively. Within each plane the hodoscope fingers overlapped by 
1/3 of the width on both edges, resulting in a bin width of 1/3 of the element 
width. The moderately fine hodoscope segmentation ($\sim$210 scintillator
elements per spectrometer) was chosen to tolerate the large photon and
neutron backgrounds and to reconstruct with sufficient resolution the 
trajectory of the scattered particles. The signal from each finger was 
discriminated and fed into a multi-hit TDC which recorded all signals in a 
100 ns window around each trigger.

The separation of the two hodoscopes was 5.0~m in the $4.5^\circ$ spectrometer 
arm and 5.1~m in the $7.0^\circ$ arm. The scattering angle resolution at the 
target in the non-bend plane was ~0.3~mrad for both spectrometers, whereas for 
the bend plane, it was $\pm$0.9 mrad for the $4.5^\circ$ arm and $\pm$0.3 
mrad for the $7.0^\circ$ arm. The ideal momentum resolution was dependent on 
the absolute value of momentum and varied from $\pm$0.3\% to $\pm$3.2\% for the 
$4.5^\circ$ arm and from $\pm$0.6\% to $\pm$3.8\% for the $7.0^\circ$ arm.

The upstream \v{C}erenkov counters were 2.24 m long aluminum tanks filled with 
nitrogen gas at a pressure of 6.3 psi for a pion threshold of 9 GeV, and the 
downstream counters were 4.3 m tanks containing nitrogen at 3.0 psi for a 
13~GeV pion threshold.  Pions below these threshold momenta did not emit 
\v{C}erenkov light.  The shorter tanks had inner radii of 60 cm and effective 
radiator lengths of 2.0 m, while the larger tanks had inner radii of 80 cm to 
cover the large spectrometer acceptances, and had effective radiator lengths of 
4.0 m. To minimize $\delta$-ray production and multiple scattering effects, 
thin tank entrance/exit windows were made from 1 mm thick aluminum. 

Inside the tanks, spherical mirrors were positioned to reflect all of the 
emitted \v{C}erenkov light back onto a single Hamamatsu R1584-01 five-inch 
photomultiplier tube coated with a p-terphenyl wavelength shifter and 
maintained at a base voltage of -2600 V.  The mirrors had a radius of curvature 
of 1.63 m and 1.2 m for the large and small counters, respectively, and had 
reflectivity close to 90\%.  The large counters contained three mirrors 
vertically stacked and mounted on an adjustable frame for focusing purposes, 
and the small counters contained two mirrors mounted similarly. Signals from 
each photomultiplier tube were discriminated at four levels corresponding to 
0.6, 1.5, 3, and 4 photoelectrons and fed into four channels of multi-hit TDC's as well as an ADC.

An electromagnetic shower calorimeter was positioned at the downstream end of 
each spectrometer.  Each detector consisted of two hundred 
$6.2\times6.2\times75.0$ cm Schott type F2 lead glass blocks stacked 10 wide 
and 20 high in a fly's eye 
configuration.  The glass had a radiation length of 3.17 cm and refractive index 
of 1.58. The incident electrons created showers via bremsstrahlung and 
$e^+/e^-$ pair production in the lead glass. Electrons (and positrons) produced 
\v{C}erenkov light in an amount proportional to the incident energy. The light 
was collected by phototubes attached to the back of the glass.  To monitor the 
blocks, a high intensity Xe flash lamp system (Hamamatsu L2360) was installed 
in each calorimeter.  The lamp delivered a luminous signal to each block via 
plastic optical fibers.  The signal from one of the fibers, as well as that 
from a $^{241}$Am source, was read out by a monitoring photomultiplier tube and 
sent to ADCs to detect possible Xe lamp intensity fluctuations and to monitor ADC gain changes by looking at shifts in averaged signals.

\subsection{Trigger}

The trigger consisted of a triple coincidence between discriminated signals 
from the two \v{C}erenkov counters and the analog sum of the shower counter 
elements. The shower discriminator threshold was set to be greater than 99\% 
efficient for the lowest energy electrons and the \v{C}erenkov thresholds were 
set to be efficient for one photoelectron signals. Up to four triggers could be 
generated in each beam spill.  Each shower and \v{C}erenkov counter signal was 
fanned out to four separate ADCs, and each trigger gated a different set of 
these ADCs. The detector signals to the multi-hit TDC's were  filtered  by a 
sub-trigger to reduce noise hits. Additional triggers were used to record a 
small fraction of the pions and to measure detector efficiencies.

\hfill

\subsection{Data Acquisition}

The data acquisition (daq) was distributed over a number of computers linked 
together by an ethernet-based network which implemented DECnet for 
communications. The distributed nature of the daq allowed us to build a system 
which could service interrupts at 120 Hz, read typically 3 KB of data for each 
interrupt,  write data to tape at a sustained rate of nearly 300 KB/sec, 
control the electron beam position on target on a pulse-to-pulse basis, and 
analyze a substantial fraction of the event data online.

A VAX4000.200 qbus computer, referred to as real time front end (RTFE), was 
interrupted at 120 Hz, read data from three CAMAC branches, built an event, and 
then sent it via network to the data logger computer, a VAX4000.60 workstation. 
The RTFE ran an application which was developed with DEC's VAXeln development 
toolkit. The VAXeln application was able to access hardware resources more 
efficiently than usually possible under the DEC VMS operating system.  
Also, task scheduling was under 
programmer control. The data logger computer controlled two SCSI EXB8500 
Exabyte tape drives. Event data received from the RTFE were packed into 
record-size buffers (approximately 32 KB in size) and written to tape. The data 
logger computer also distributed a sample of the event data via network to two 
VAX4000.60 workstations, one for online data analysis in each spectrometer. The 
event data were analyzed and various histograms and tables were presented for 
viewing in X11/Motif windows. Special purpose analyses could be performed by 
other VAX workstations which connected to the network. The electron beam was 
monitored and controlled by a microVAX II computer which, like the RTFE, was 
loaded with a VAXeln application specially developed for this task. A VAX 
cluster boot node, VAX4000.300 computer, was used to control, monitor, and log 
information on the spectrometer magnets and their power supplies, detector high 
voltage power supplies, NIM and CAMAC crate voltages, scalers, target 
parameters, pedestals, etc.

\section{Data Analysis}

\subsection{Event Selection}

Events which produced a trigger were further analyzed to identify electrons 
amid a background of mostly pions, and to determine the energy, momentum, and 
scattering angle. Raw asymmetries were then formed from the number of scattered 
electrons coming from each of the two states of incident electron polarization 
directions. These asymmetries are a function of $x$, $Q^2$, and beam energy.

Particle tracking was performed using the spatial and timing information 
provided by the hodoscopes and shower counter. Once a track was found, the 
particle's momentum and scattering angle were reconstructed. The efficiency of 
the hodoscope package was found to be $91\%$ for the $4.5^\circ$ spectrometer 
and $96\%$ for the $7.0^\circ$ spectrometer. The tracking efficiency was about 
$98\%$ for the $4.5^\circ$ and $99\%$ for the $7^\circ$ spectrometer. The 
hodoscope and tracking efficiencies were worse for the small angle spectrometer 
due to the higher count rate.

The shower counter was used to measure the energy deposited by the incident 
particle and to provide electron identification.  In order to use the shower 
counter for energy measurements, it was necessary to calibrate each block for 
differences in phototube, lead glass, and ADC channel responses to the
electrons. This was achieved using clean electron events which were selected 
using knowledge from the other detectors. In an iterative process, a set of 
calibration constants for the glass blocks was determined by requiring that 
the total energy of the cluster be equal, on average, to the momentum of the 
event. 

\begin{figure}[t]
\centerline{\epsfig{file=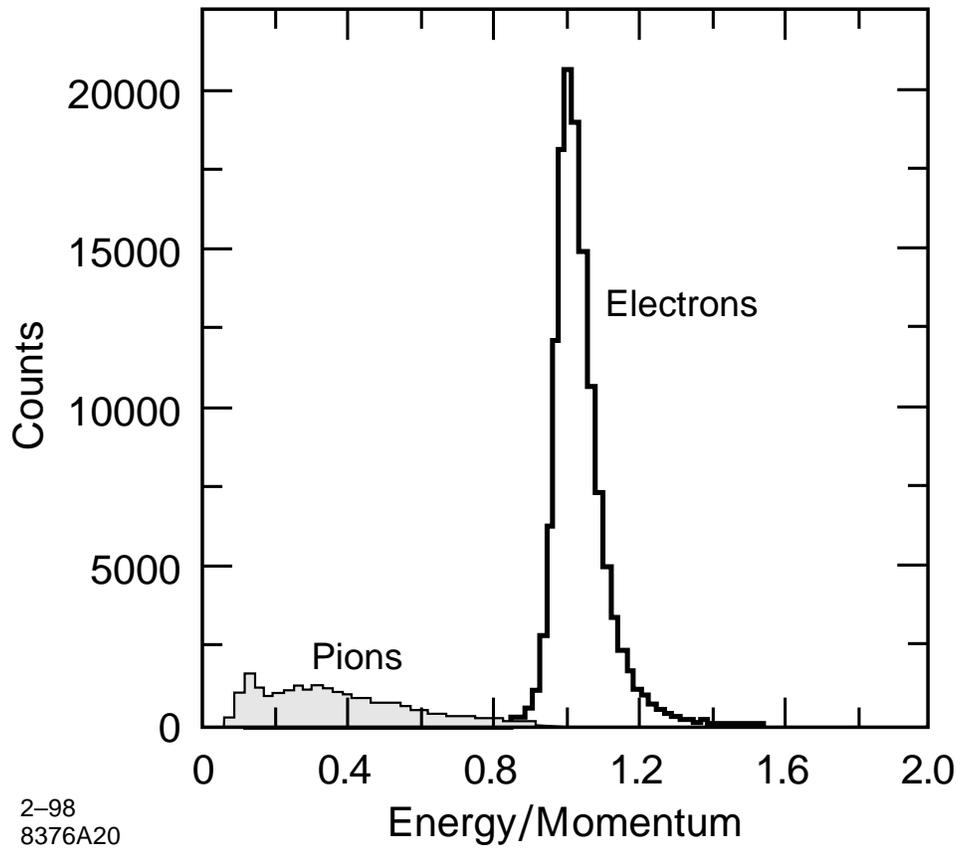,width=5.0 in}}\vskip 0.7cm
\caption{Sample plot of events versus the ratio of energy to momentum as
measured by the shower counter and tracking, respectively. The electron events
are peaked at unity. The lower energy background pion events are removed when all cuts are applied to the data.}
\label{fg:eoverp}
\end{figure}

Once calibrated, the shower counter was used to select electrons by comparing 
the energy of the particle as measured by the shower counter ($E'$) to the 
momentum of the particle as measured by the hodoscope tracking system ($P$). 
Rejection of pions was achieved since typically electrons deposit all of their 
energy in the shower counter while pions do not. Thus, the electron events had 
an $E'/P$ peak centered around unity, whereas the $E'/P$ values for pions were 
in general much less than one. By making a cut around the electron peak of 
$0.8 < E'/P < 1.25$, we were able to reject the majority of pion contaminants 
left in our data sample. The $E'/P$ requirement was approximately 96\% 
efficient for electrons, and left a pion contamination of less than 1\%. A 
sample plot of $E'/P$ for this experiment is shown in Fig.~\ref{fg:eoverp}.

Electrons could also be identified over pions using the spatial profile of the 
shower formed by the incident particle. These profiles vary significantly 
depending on the type of incident particle.  In particular, pion shower 
clusters are much smaller than electron clusters, and many of them are fully 
contained in one block.  Electron clusters are typically contained in nine 
glass blocks.  For electrons, the central block contained $50\%$ to $90\%$ of 
the energy and the eight neighbors contained the rest. We used a shower counter 
neural network algorithm \cite{vincent} which modeled a typical electron 
cluster profile to determine which events were electron events and which were 
background. The neural network was approximately 98\% efficient for identifying 
electrons and left a pion contamination of about $0.5\%$. 

To further ensure a clean electron sample, spatial, timing, and pulse height 
cuts were made. A cut requiring a minimum \v Cerenkov ADC pulse 
height of 40 (more than two photoelectrons) was made yielding an efficiency of 
95-99\% for the four counters. Next, we required that the track used for the 
momentum measurement was within 40 mm, horizontally and vertically, of the 
cluster from the shower counter that was used for the energy measurement. The 
track and the cluster were required to be within 10 ns of each other, and the 
track was required to point back to the target to within 13 mm to eliminate bad 
tracks. In addition, the few events with clusters on the outer edges of the 
shower counter were rejected due to the possibility of energy leakage from the 
sides of the counter.

\subsection{Asymmetries and Corrections}

Electrons passing the event selection cuts were binned in $x$ such that the 
resolution in $x$ was slightly finer than the binning.  The electrons were also 
tagged according to their relative target and beam helicity states,  
$N^{\uparrow\downarrow(\uparrow\uparrow)}$, and which spectrometer they entered. 
The asymmetries $A_\parallel$ and $A_\perp$ were formed:
\begin{equation}
 A_\parallel\,(\hbox{or } A_\perp) = C_1 \biggl({1 \over f P_b P_t } 
{ N_L - N_R \over N_L + N_R } - C_2 \biggr) + A_{rc}.\label{eq:asym}
\end{equation} 
Here $f$ is the dilution factor, $P_b$ and $P_t$ are the beam and target 
polarizations, $A_{rc}$ is the radiative correction to the asymmetry, and 
$C_{1(2)}$ are the corrections needed due to the presence of nitrogen in the 
targets, with $C_2$ disappearing for the proton target. These corrections are 
discussed in more detail below. Here $N_{L(R)}$ is the number of left or 
right-handed helicity events corrected as
\begin{equation}
N_{L(R)} = {N_{L(R)}^{(raw)} {d_{L(R)} \over Q_{L(R)}} }
\end{equation}
where $d_{L(R)}$ is the appropriate dead time correction and $Q_{L(R)}$ is the 
appropriate incident charge.

\subsubsection{Polarized Nitrogen and Residual Proton Corrections}

In measuring the proton and the deuteron asymmetries, it was necessary to 
correct for events which scattered from other polarizable nuclei in the target 
aside from the desired protons or deuterons \cite{johannes}.  
The targets were made of 
$^{15}$NH$_3$ and $^{15}$ND$_3$, and both the $^{15}$N and the $\approx$2\% 
contamination of $^{14}$N were polarizable.  In addition, the $^{15}$ND$_3$ 
target contained $\approx$1.5\% of unsubstituted or residual polarizable 
protons from $^{15}$NH$_3$. 

The polarization of $^{15}$N and the residual protons was measured after the 
experiment. The unpaired proton in $^{15}$N contributes to the measured proton 
asymmetry proportionally to the nitrogen polarization and with a negative sign 
because of the negative magnetic moment of $^{15}$N. For the target material 
$^{15}$NH$_3$, the following fit was used to express the $^{15}$N polarization 
$P_N$ in terms of the polarization of the protons $P_p$:
\begin{equation}
 P_N = 0.136 \,P_p - 0.183 \,P_p^2 + 0.335 \,P_p^3 \approx 0.12. 
\end{equation} 
The correction $C_1^p$ to the proton asymmetry ($C_2^p=0$) which is 
referred to in  Eq.~(\ref{eq:asym}) is given by:
\begin{equation}
C_1^p = 1 - {1\over3}{1\over3} {P_N\over P_p} g_{\rm EMC}(x) \approx 0.98. 
\label{eq:ncorrprot}
\end{equation}
Here $g_{\rm EMC}(x)$ is the correction for the EMC effect \cite{GOMEZ} 
taken at atomic mass 
number 15. The first factor $-{1\over3}$ comes from Clebsch-Gordan coefficients 
involving the nitrogen wave function.  The second factor $1\over3$ reflects the 
fact that ammonia has three hydrogen atoms for each nitrogen atom. The error on 
the second term in $C_1^p$ was estimated to be about 20\% relative which
yields a systematic error of 0.004 on $C_1^p$.  Here the contribution of $^{14}$N to the asymmetry was neglected. 

For the target material $^{15}$ND$_3$, the corrections were more complicated
because they account for both the residual protons and the unpaired proton in 
the $^{15}$N. For each case, the correction involved the measured proton 
asymmetry.

The $^{15}$N polarization $P_N$ is given by
\begin{equation}
 P_N = -0.40 P_d,  
\end{equation} 
where $P_d$ is the polarization of the deuterons. The residual proton 
polarization is expressed as:
\begin{eqnarray}\nonumber
P_p^{res} &=& \hbox to35mm { 0.191 + 0.683 $P_d$ \hfil} \hbox{for $P_d > 0.16$} 
\\ &=& \hbox to35mm { 1.875 $P_d$ \hfil} \hbox{for $P_d\le0.16$}   \end{eqnarray}
The polarization of $^{14}$N was obtained from the measured $^{15}$N 
polarization by assuming that the polarization was equal and opposite in sign 
to that of $^{15}$N\null. The corrections used in Eq.~(\ref{eq:asym}) for the 
$^{15}$ND$_3$ target are given by:
\begin{eqnarray}\nonumber
 C_1^d &=& {1\over1-\eta_p+D_n/(1-1.5\omega_D)} \approx 1.02\ , \\   
 C_2^d&=& {U_pF_2^p\over U_dF_2^d}(D_n-D_p)(A^p-A_{rc}) \approx -0.03(A^p-A_{rc}).    
\label{eq:ncorrdeut}
\end{eqnarray}
$A^p$ is the final proton (Born) asymmetry $A_\parallel$ or $A_\perp$, and by 
subtracting the appropriate proton radiative correction $A_{rc}$ we are left 
with the radiated asymmetry. $U_p$ and $U_d$ are the radiative corrections to 
the unpolarized cross-sections. The remaining factors are defined as
\begin{eqnarray}\nonumber
\eta_p &=& {\hbox{number of protons} \over 
   \hbox{number of deuterons  + number of protons}}\approx 0.015\ ,\\  
\nonumber
  D_n&=&\eta_N{P_N\over P_d}{g_{_{\rm EMC}}(x)\over9}\ ,\\
  D_p&=&\eta_p{P_p^{res}\over P_d}+(2\eta_N-1){P_N\over P_d}
       {g_{_{\rm EMC}}(x)\over9}\ ,\\ \nonumber
\eta_N   &=& {\hbox{number of $^{14}$N}
   \over{\hbox{number of $^{14}$N + number of $^{15}$N }}}\approx  0.02\ .
\end{eqnarray}
The error on $C_1^d$ was neglected since this value was very small and stable.  
The factor $C_2^d$ contains the proton asymmetry and was calculated for each
$x$-bin using the measured proton asymmetry and its error.

\subsubsection{Background Subtraction of Positrons and Pions}

The data collected in each of the spectrometers included background events 
coming from a small number of misidentified pions and from electrons produced 
in pair-symmetric processes (mostly $\pi^0 \rightarrow 2\gamma$, $\gamma 
\rightarrow e^- + e^+$). This background (mostly pair-symmetric) was 
responsible for up to 10\% of the events in the lowest $x$-bin, but close 
to zero events for $x > 0.3$. To measure the background, data were taken with 
the spectrometer magnets' polarity flipped to measure $\pi^+$ and $e^+$. The
same cuts 
were applied to eliminate the majority of pions as in the electron runs. A 
positive particle asymmetry $A_+$ was formed and was corrected just as in 
the case of the electron asymmetry $A_-$ for varying experimental conditions 
such as beam and target polarizations. This positive particle asymmetry was 
found to be consistent with zero. The background-corrected electron asymmetry 
was determined by
\begin{equation}
A = A_-{N_- \over {N_- - N_+}} - A_+{N_+ \over {N_- - N_+}},
\end{equation}
where $N_-$, $N_+$ are the number of events per incoming charge for electron
and positron runs.  The misidentified pion background was subtracted along with 
that of the positron background since $A_+$ also contained a measure of the 
misidentified pions and assuming $A_{\pi^+} \sim A_{\pi^-}$.

\subsubsection{False Asymmetries}

It is important to make sure that our experimental data are free from
significant false asymmetries which could systematically shift the data. During 
the experiment, data were taken (either longitudinal or transverse) with the 
target B-field pointing in either one of the two possible directions.  For each 
field direction, two different target polarization directions were used, 
parallel or antiparallel to the B-field.  We then had four different
configurations, and approximately the same amount of data were taken for each 
configuration, thus cancelling out the electroweak contributions to our 
measurement so that no correction to the data was necessary. 
The asymmetries in each configuration were compared by looking at 
the $\chi^2$ distributions of the asymmetry differences. For the proton, the 
$\chi^2$ distributions were all nicely centered at one, and the mean value of 
the asymmetry differences was approximately one standard deviation from zero. 
For the deuteron the results were slightly worse, yet still very reasonable. 
The $\chi^2$ distributions were centered around one with a few points greater 
than two, and the mean values of the asymmetry differences were within two 
standard deviations of zero.  We conclude that there were no significant 
systematic effects on the asymmetry due to changes in target B-field or target 
enhancement field directions. Also, no statistically significant variation of 
the asymmetry was found for either NH$_3$ or ND$_3$ targets as a function of 
raster position.

\subsubsection{Dilution Factor}

In general, incident electrons will scatter both from polarized target nucleons 
and unpolarized nuclei that are part of the target assembly. These 
unpolarized materials include liquid helium, $^{15}$N, NMR pick-up coils, and 
vacuum windows. Scattering from unpolarized materials will dilute the measured 
asymmetry, and a correction must be applied. The dilution factor $f$ is a 
function of $x$ and $Q^{2}$, and is defined as the ratio of the total event 
rate from polarizable nucleons, to the total event rate from all target 
materials. The measured asymmetry is then corrected for unpolarized events by 
dividing by the dilution factor.

For a material of density $\rho$ and thickness $z$, the event rate from Born 
processes was calculated as follows:
\begin{equation}
r(x,Q^{2})=\rho z
\left[ZF_{2}^{p}(x,Q^{2})+N
F_{2}^{n}(x,Q^{2})\right]g_{EMC}(x,Q^{2}),
\end{equation}
where $Z$ and $N$ are the number of protons and neutrons in the nucleus.  
$F_{2}^{p}$ and $F_{2}^{n}$ are unpolarized proton and neutron deep-inelastic 
structure functions. They were obtained from a parameterization of the NMC 
data~\cite{new_NMC}. The factor $g$ corrects for the ``EMC effect'' which
accounts for the difference in nucleon cross sections for free and bound 
nucleons.

With this model for rates, the dilution factor can then be calculated as 
follows:
\begin{equation}
f(x,Q^{2})=\left(\frac{r_{pol}(x,Q^{2})}{r_{pol}(x,Q^{2})+ 
\sum_{i}r_{i}(x,Q^{2})}\right)\times r_{c},\label{eq:dilution}
\end{equation}
where we are summing the rates from all unpolarized materials which contribute 
to the overall event rate. The factor $r_{c}$ corrects the dilution factor
for radiative effects and was typically less than a 5\% correction.

The target material was in the form of frozen granules which were tightly 
packed into a target cell.  The volume fraction  of the target cell which the 
target material occupied is known as the packing fraction, and was determined 
independently by three different methods. The first consisted of studying the 
difference in event rates for empty, carbon, and full target cells. The second method 
was a measurement of the attenuation of a mono-energetic X-ray beam as it 
passed through the target material. The attenuation of the incident beam is 
directly related to the thickness and attenuation coefficient of material it
passes through, and was therefore sensitive to the packing fraction. Finally, 
the target material was weighed, and the packing fraction was determined using 
the known volume of the target cell.  The measured packing fraction was 
different for each target used, and varied from $0.57$ to $0.64$.

Over the kinematic range of interest, the dilution factor typically varied from 
$0.15-0.19$ for the NH$_{3}$ target and from $0.23-0.25$ for the ND$_{3}$ 
target, with relative errors of $2\%$ and $1.5\%$, respectively. The error on 
the dilution factor comes from several sources. The packing fraction was known 
to 4\% for both targets. The relative error from the cross-section ratio 
$\sigma_n/\sigma_p$ was 1.0\% \cite{NMC-np}. This was one of the dominant errors for the NH$_3$ target and did not contribute to the ND$_3$ target. The 
ratio of nuclear to deuterium cross sections, the EMC effect, is known to 1.5\% 
relative and was another large source of error. This effect has a 1\% overall 
normalization and another 1\% uncorrelated error \cite{GOMEZ}. The small mass 
of the NMR coil ($\sim$0.1 gm) was known  to 20\% but did not contribute significantly to the 
overall error.

\subsubsection{Dead Time}

All the signals from various detectors went through discriminators before 
forming the various triggers. These discriminators have an output pulse width 
of 25 ns and a double pulse resolution of 8 ns. They were operated in an 
updating mode such that a second signal entering the discriminator after 8 ns 
and before 25 ns produced a single output pulse with an extended width. The 
effective dead time was 32 ns instead of 25 ns, due to slight mis-timing 
between various signals and signal jitter, especially from the shower counters. 
Rates measured with each beam helicity were corrected separately.

Using a Monte Carlo simulation, the probability matrix $M(i,j)$ for observing 
$i$ hits when there were really $j$ hits was generated using a typical spill 
length of 2200 ns and a dead time of 32 ns as inputs. The observed trigger 
frequency distribution $T_o(i)$ is related to the true distribution (without 
dead time) $T_t(i)$ by
\begin{equation}
T_o(i)=\sum_j M(i,j)\times T_t(j).
\end{equation}
Since there are practically no hits beyond 10, the sum is safely truncated at 
$j=16$. The matrix $M$ is inverted to solve for $T_t(i)$, and the dead time 
correction factor $d$ to the measured rates is defined as
\begin{equation}
d={\sum_{i=1}^{16} i T_t(i) \over 
{\sum_{i=1}^{4} i T_o(i) + 4\sum_{i=5}^{16} T_o(i)}}. 
\end{equation}
Here, the sum over $T_o$ is split into two parts because only four triggers 
could be recorded per beam spill.

The correction factor varies smoothly from 1 at very low rates to 1.07 at an 
average rate of 2 events/pulse. The systematic error on the corrections was 
calculated assuming upper and lower limits to the beam width of 2600 ns and 
1800 ns. The dead time was found to be accurate to a few parts in 1000, and the 
error for the corrected asymmetry by applying these factors is found to be less 
than $2\times 10^{-5}$, which is completely negligible.  No uncertainty in the 
dead time itself was considered because only the ratio of the beam spill length 
to the dead time is important.

\subsubsection{Radiative Corrections\label{sect:rc}}

Our experimental goal was to measure a single photon exchange process (Born) at 
specific kinematics. In reality there are higher order contributing processes
(internal), and the actual scattering kinematics can change due to energy 
losses in materials along the electrons' paths (external). The radiative 
corrections account for these unwanted effects. 

The radiative correction calculation is different for the unpolarized 
($\sigma^u$) and polarized ($\sigma^p$) components of the helicity-dependent 
cross sections which are given by $\sigma^{\uparrow \downarrow}$ and 
$\sigma^{\uparrow \uparrow}$ for a longitudinally polarized target, and 
$\sigma^{\rightarrow \downarrow}$ and $\sigma^{\rightarrow \uparrow}$ for a 
transversely polarized target. The longitudinal and transverse asymmetries can 
be written as
\begin{eqnarray} \nonumber
A_\parallel &=& {\sigma^{\uparrow \downarrow} - 
                 \sigma^{\uparrow \uparrow} \over 
                 \sigma^{\uparrow \downarrow} + 
                 \sigma^{\uparrow \uparrow} } = 
               {(\sigma^u + \sigma^{p\uparrow}) - 
                (\sigma^u - \sigma^{p\uparrow}) \over
                (\sigma^u + \sigma^{p\uparrow}) + 
                (\sigma^u - \sigma^{p\uparrow})} =
                {\sigma^{p\uparrow} \over \sigma^u}, \\ 
A_\perp &=& {\sigma^{\rightarrow \downarrow} - 
             \sigma^{\rightarrow \uparrow} \over 
             \sigma^{\rightarrow \downarrow} + 
             \sigma^{\rightarrow \uparrow} } = 
           {(\sigma^u + \sigma^{p\rightarrow}) - 
            (\sigma^u - \sigma^{p\rightarrow}) \over
            (\sigma^u + \sigma^{p\rightarrow}) + 
            (\sigma^u - \sigma^{p\rightarrow})} =
            {\sigma^{p\rightarrow} \over \sigma^u},
\end{eqnarray}
which is equally valid for Born, internally radiated, or fully radiated cross 
sections and asymmetries. For the remainder of the radiative correction 
discussion, quantities which are Born, internally radiated, or fully radiated 
are subscripted with 0, $r$, and $R$ respectively. Also, for simplicity, 
references to a particular target polarization are dropped such that $A$ 
could be either $A_\parallel$ or $A_\perp$, and $\sigma^p$ could be either 
$\sigma^{p\uparrow}$ or $\sigma^{p\rightarrow}$.

Calculation of the polarization-dependent internal corrections was done using 
code based on the work of Kuchto, Shumeiko, and Akusevich \cite{radcor}, who 
are also responsible for the development of their own code POLRAD. The two 
codes were carefully compared and found to be completely equivalent when the 
same input models were used. The calculation of the internally radiated cross 
sections can be decomposed into components:
\begin{eqnarray}\nonumber
\sigma^p_r = \sigma^p_0(1+\delta_v) + \sigma^p_{el}
           + \sigma^p_q + \sigma^p_{in}, \\
\sigma^u_r = \sigma^u_0(1+\delta_v) + \sigma^u_{el}
           + \sigma^u_q + \sigma^u_{in},
\end{eqnarray}
where $\delta_v$ includes corrections for the electron vertex and vacuum 
polarization contributions, as well as a term that is left after the 
infrared-divergent contributions are cancelled out. The vacuum polarization 
correction includes contributions from both leptons and light quarks. The terms 
$\sigma_{el}$, $\sigma_q$, and $\sigma_{in}$ are the radiative tails due to
internal bremsstrahlung (occurring within the field of the scattering nucleon) 
for elastic, quasielastic, and inelastic scattering processes.

The nuclear elastic tail for the deuteron was evaluated using fits to data 
\cite{d2ff} over a large range in $Q^2$. The elastic tail for the proton (and 
quasielastic for the deuteron) were evaluated using various form factor 
models~\cite{elff} which agree well with existing data over the kinematic 
region. Note that some of these models agree well with data for some of the four elastic form factors and not others, so different models were combined for 
the best representation of all 
four nucleon elastic form factors. Quasielastic cross sections were 
Fermi-smeared only for corrections to our resonance data since this smearing 
has a negligible impact on the radiative correction in the deep-inelastic 
region. Unpolarized cross sections were modeled using fits to structure 
function data in the deep-inelastic region \cite{new_NMC,whitlowr}, and fits to 
cross sections in the resonance region \cite{bodek,stuart}. The polarized 
component to the deep-inelastic cross sections was modeled using 
$Q^2$-dependent fits to $A_1$ as given in this paper. The polarized resonance 
region model was based on parameterizations of previous data and data presented
here. For the transverse contributions, we used $g_2 = g_2^{WW}$ \cite{wand} or 
$g_2 = 0$ which are both consistent with our data.

The external corrections account for bremsstrahlung radiative effects which 
occur as the electrons pass through material in their path. Ionization effects
were completely negligible at our kinematics. At any given interaction point 
within the target the radiative correction depends on the amount of material 
(in radiation lengths) the electron sees before ($t_b$) and after 
($t_a$) scattering. Because the radiation lengths before and after scattering 
did not vary significantly over the beam raster area, it was not necessary 
to integrate the external corrections over the raster area. Also, it was an 
excellent approximation to replace the target integration over the length of 
the target with the evaluation of the external radiative corrections at one 
point, namely the center of the target. At this point, $t_b=0.026$, $t_a=0.047$ 
for the $4.5^\circ$ spectrometer, and $t_a=0.040$ for the $7^\circ$ 
spectrometer. These radiation lengths, which are valid for both our proton and 
deuteron polarized target, are dominated by the target material, but also 
include contributions from various windows which are not part of the polarized 
target itself. The external corrections were thus calculated using \cite {tsai}
\begin{eqnarray}\nonumber
\sigma_R^p(E_0,E'_F,\theta) = 
\int_{E_{min}}^{E_o} \int_{E'_F}^{E'_{max}} &&I_b(E_0,E,t_b)
           \sigma_r^p(E,E',\theta)I_b(E',E'_F,t_a)[1-D(E_0,E,Z)]dEdE' \\
\sigma_R^u(E_0,E'_F,\theta) = 
\int_{E_{min}}^{E_o} \int_{E'_F}^{E'_{max}} &&I_b(E_0,E,t_b)
           \sigma_r^u(E,E',\theta)I_b(E',E'_F,t_a)dEdE'
\end{eqnarray}
where $E_0$ is the electron initial energy, $E'_F$ is the final scattered 
electron momentum, and $E_{min}$ and $E'_{max}$ are the minimum incident energy 
and maximum scattered energy as defined by elastic scattering. $I_b(E_1,E_2,t)$ 
is the probability \cite{tsai} that a particle with initial energy $E_1$ ends 
up with energy $E_2$ after passing through a radiator of thickness $t$, and 
$D(E_0,E,Z)$ is the electron depolarization correction \cite{olsen} which 
corrects for the depolarization of the electron beam due to the bremsstrahlung 
emission of polarized photons. This correction depends weakly on the $Z$ of 
the target material.

An additive correction $A_{rc}$ to the data was formed by taking the difference
between the fully radiated and Born model asymmetries
\begin{equation}
A_{rc} = A_0 - A_R = {\sigma_0^p \over \sigma_0^u} - {\sigma_R^p \over 
\sigma_R^u}.\label{eq:arc}
\end{equation}
Our fits to $A_1$ and the radiative corrections were iterated until they 
converged. For the purposes of statistical error propagation on our measured 
asymmetries, a ``radiative correction dilution factor'' $f_{rc}$ was evaluated. 
This dilution factor is simply a ratio of events coming from deep-inelastic 
processes to all events and multiplies the usual dilution factor in 
Eq.~\ref{eq:asym}. We only used $f_{rc}$ for the error propagation and not for 
correcting the data directly. Systematic uncertainties were estimated by varying 
input models within reasonable limits and measuring how much the radiative 
correction changed.  These uncertainties for the various models were then 
combined in quadrature for each $x$ bin. Results for $A_{rc}$ are listed with data in Tables~\ref{tb:APpar29}-\ref{tb:ADpar9},  \ref{tb:syserg10}-\ref{tb:sysergf23}, and 
\ref{tbl:A1}-\ref{tbl:A1err}.

  \subsection{Analysis of Resonance Region Data}

The resonance data \cite{raines} were taken with a 9.7 GeV  beam. The 
spectrometer angles of
4.5$^\circ$ and 7$^\circ$ corresponded to $Q^2 \simeq 0.5$ and $1.2$~(GeV/c)$^2$ in the resonance region ($W^2<5$~GeV$^2$), respectively. We have extracted $g_1$ from 
the measured asymmetries $A_\parallel$, and from the absolute cross-section 
differences given in Eq.~\ref{eqn:csdiff}. Each method has its own set of 
systematic errors. The difference method requires good knowledge of 
spectrometer acceptances, the number density of polarizable protons or 
deuterons in the target, and detector efficiencies. The asymmetry method 
requires knowledge of the dilution factor for the resonance region, which means  
an accurate model of the rapidly varying unpolarized cross sections is needed.  
We found that the two methods agreed to within a fraction of the statistical 
errors on each point (typically better than 3\%).  In our previous
report\cite{e143res} we have used the difference method.  The current 
reanalysis uses the asymmetry method, since we now believe that the systematic 
errors are slightly better in this case.  Other improvements on
Ref.~\cite{e143res} include better modeling of the resonance region for 
radiative and resolution effects.

The resonance asymmetries were calculated as specified in Section IV.B for the 
deep-inelastic analysis. In the present case, we have determined the dilution 
factor $f$ using a Monte Carlo routine as described below.  The term $A_{rc}$ 
also included a resolution correction in the resonance region.

We developed a Monte Carlo code which simulated all relevant aspects of the
experiment.  It was used to predict total count rates and count rate 
differences from a set of tables of cross sections and asymmetries generated by
the radiative corrections routine.  The unpolarized cross sections came from 
parameterizations for the resonance region \cite{bodek} and the deep-inelastic 
region \cite{new_NMC,whitlowr}. The asymmetries contain both resonant and 
non-resonant contributions. The resonance contribution was calculated using the 
code AO\cite{AO}, which includes parameterizations of all of the existing
resonance data; however, the helicity amplitudes $A_{1\over 2}$ and 
$A_{3\over 2}$ for $S_{11}$ and $D_{13}$ were tuned to agree with our data. The 
non-resonant part came from a parameterization of all existing deep-inelastic 
data (Fit III of Ref.~\cite{e143q}), which was extrapolated into the resonance 
region. Specifically, for $W^2<2.5$ GeV$^2$, $A_1$ was given by the tuned AO
result alone;  above $W^2=3.0$ GeV$^2$, $A_1$ was taken as the sum of the AO 
resonant contribution and the Fit III inelastic background; and in the region 
$2.5 < W^2 < 3.0$ GeV$^2$ the two extremes were linearly interpolated.  $A_2$ 
was calculated using $g_2^{WW}$, which yields values close to zero.  The 
model-dependence of this choice for $A_2$ was determined by alternately 
considering $g_2 = 0$ and $A_2= 0$.

The observed raw parallel asymmetry $A^{raw}_\parallel$ is proportional to the 
combination of photon asymmetries $A_1+\eta A_2$.  Therefore, we first
extracted $A_1+\eta A_2$ from the data, and then deduced $g_1$ from this using 
various assumptions about $g_2$. The following steps were required to produce 
$A_1+\eta A_2$ and $g_1$:

\noindent 1)
The radiative corrections code was run with the options as specified above to 
create the Born cross sections, the Born asymmetries, and the predicted values 
of  $A_1$, $A_2$, $g_1$ and $g_2$ for both NH$_3$ and ND$_3$ targets at E143 
kinematics.

\noindent 2) 
The radiative corrections code was run to create tables of cross sections and 
asymmetries over a wide range of kinematics, fully internally radiated, to use
as input to the Monte Carlo generator.

\noindent 3) 
The Monte Carlo routine was run for both polarized protons and deuterons alone, 
and for full ND$_3$ and NH$_3$ targets. This simulation included external 
radiation, spectrometer acceptance, resolution, multiple scattering and Fermi 
motion, as well as the reconstructed kinematic variables and raw asymmetries.

\noindent 4) 
The raw data was corrected for efficiencies, polarization, polarized nitrogen 
and polarized protons in ND$_3$ using the standard E143 procedure. Then, the data was corrected by the dilution factor (the ratio of Monte Carlo events from
polarizable protons or deuterons to those from all target components), and the 
additive radiative correction term $A_{rc}$ (obtained from the difference 
between fully radiated Monte Carlo results and the model Born asymmetry) was 
applied in order to generate the fully corrected values of $A_\parallel$.

\noindent 5) 
The ratios $( A_1+\eta A_2)/A_\parallel$ and $g_1/ A_\parallel$ were used as 
calculated in the Born version of the radiative correction routine to find 
$g_1$ and $A_1 + \eta A_2$ for our data.
 
This extraction method required that the Monte Carlo routine provide a detailed 
and realistic simulation of the data, including resolution effects which are 
very important in the resonance region. Therefore, we performed a series of 
tests to insure that the Monte Carlo simulation described the data well, and 
provided radiative and resolution corrections with sufficient precision
compared to the statistical accuracy of our data. Without any normalization 
factors, the generated unpolarized counts versus $W^2$ agree with the data to 
better than 2.2\% (3.4\%) in the $4.5^\circ$ ($7^\circ$) spectrometer.  The
shape as a function of $W^2$ is in even better agreement in the region of the 
resonances. This provides confidence that the acceptance and resolution of the 
spectrometer are properly modeled. In addition, we have found that the measured 
and simulated count-rate differences agree well with each other. These rates
were integrated over the (quasi-)elastic region ($W^2<1$ GeV$^2$) where model 
uncertainties are minimal because of reasonably good knowledge of elastic form 
factors and radiative corrections. The overall $\chi^2$ for the four degrees of 
freedom corresponding to  p($4.5^\circ$), p($7^\circ$), d($4.5^\circ$) and 
d($7^\circ$) is 3.85.

\section{Results and Comparison to Theory}

Table~\ref{tb:APpar29} gives the results for $A_\parallel^p$ and $A_\perp^p$ 
with the proton target for the beam energy of 29.1 GeV and for the $4.5^\circ$ 
and $7^\circ$  spectrometers, respectively, along with the total radiative 
corrections to each point. Tables~\ref{tb:APpar16}-\ref{tb:APpar9} give the 
results for $A_\parallel^p$ and radiative corrections for the beam energies of 
16.2 and 9.7 GeV and for the two spectrometers. 
Tables~\ref{tb:ADpar29}-\ref{tb:ADpar9} give the corresponding results for the deuteron target. Since the 29 GeV data include both $A_\parallel$ and $A_\perp$, 
Eqs.~\ref{eq:aonetwo} are used determine the asymmetries $A_1$ and $A_2$, and  Eqs.~\ref{eq:g12} are used to determine the structure functions $g_1$ and $g_2$
and the ratio of structure functions $g_1/F_1.$ The NMC fit \cite{new_NMC} was
used for $F_2.$ The SLAC global analysis \cite{whitlowr} was used for $R.$ 
While the fit to $R$ was made to data with a limited $Q^2$ range and 
$x\geq 0.1,$ it is consistent with recent measurements at lower $x$ 
\cite{CCFR,NMC_R} and different $Q^2$\cite{E140x}. Estimated errors on these 
unpolarized structure functions are given in Section \ref{sec:syserr}. The 
neutron spin structure function  can be extracted from the deuteron and proton 
results in a manner similar to that used for the unpolarized structure 
functions. For both $g_1$ and $g_2$ we use the relation:
\begin{equation}
g^n(x,Q^2) = 2g^d(x,Q^2)/(1-1.5\omega_D) - g^p(x,Q^2), \label{eq:gn}
\end{equation}
where $\omega_D$ is the probability that the deuteron will be in a D-state.  We 
use $\omega_D=0.05\pm0.01$ \cite{WD} given by N-N potential calculations. No 
other nuclear contributions to $\omega_D$ are included. The neutron asymmetries
can then be calculated using Eqs.~\ref{eq:aonetwo}.

\subsection{$A_1$ and $g_1$ }\label{sec:a1g1}

For beam energies of 16.2 and 9.7 GeV there are no $A_\perp$ data available. We 
have assumed that  $g_2(x,Q^2)$ is given by either $g_2^{WW}$ or $g_2$=0, both of 
which are consistent with our $g_2$ data at 29 GeV.  These different assumptions 
lead to very similar results. We have then determined  $A_1$ and $g_1$ using 
Eq.~\ref{eq:a1g1}.

\begin{figure}[t]
\centerline{\epsfig{file=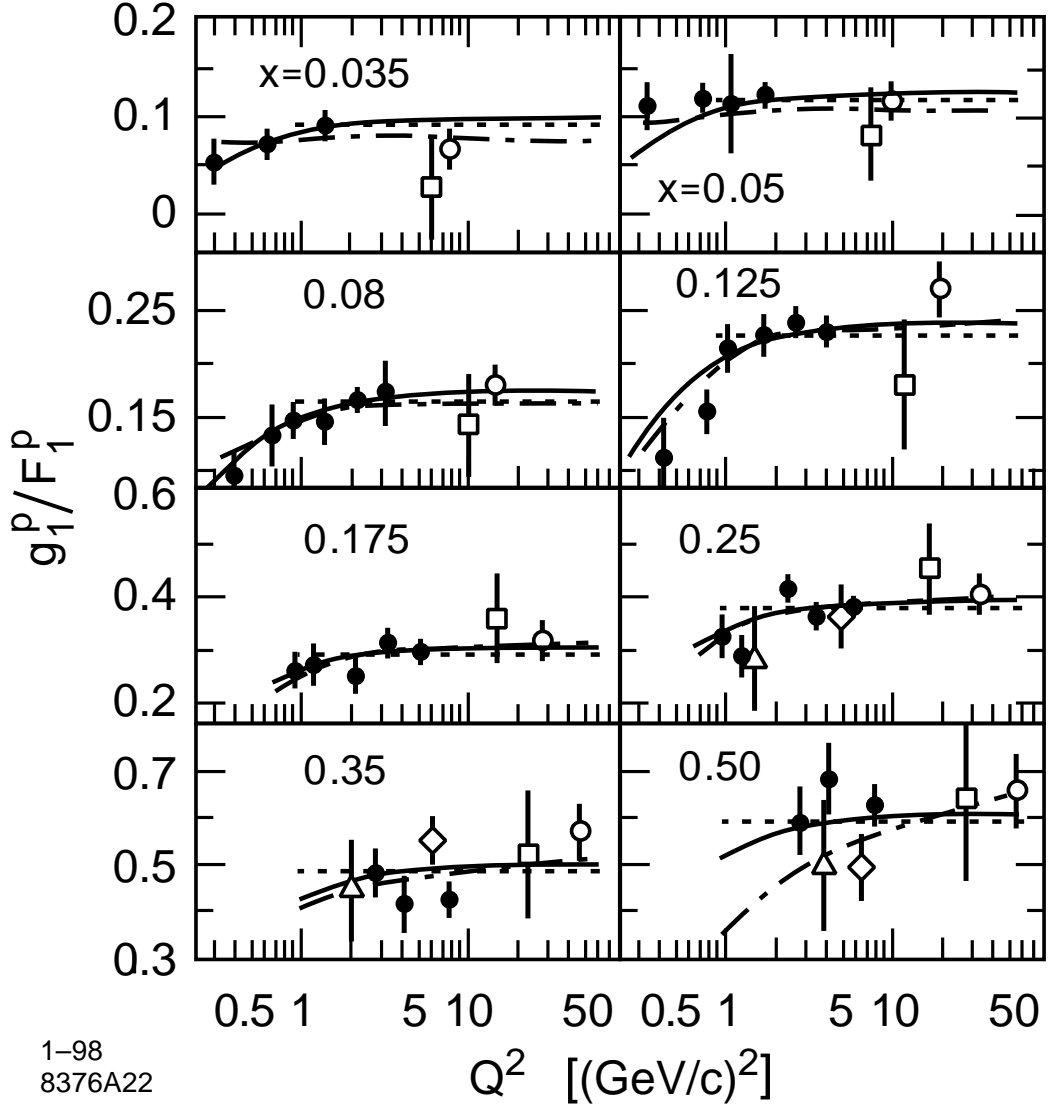,width=5.5 in}}\vskip 0.7cm
\caption{$g_1^p/F_1^p$ as a function of $Q^2$ for 8 different $x$ bins.
    The data are from this experiment (solid circle), SMC\protect\cite{smcp}
    (open circle), EMC \protect\cite{emc} (squares), SLAC E80
    \protect\cite{e80} (triangle), and SLAC E130
    \protect\cite{e130} (diamond).
    The dashed and solid curves correspond to global fits II
    ($g_1/F_1$ $Q^2$-independent) and III  ($g_1/F_1$ $Q^2$-dependent) in
    Table \protect\ref{tb:g1f1_fit}, respectively.  The E154  NLO 
    pQCD fit\protect\cite{e154nlo}
    is shown as the dot-dashed curve.}
\label{fg:g1F1p_Q2}
\end{figure}

\begin{figure}[t]
\centerline{\epsfig{file=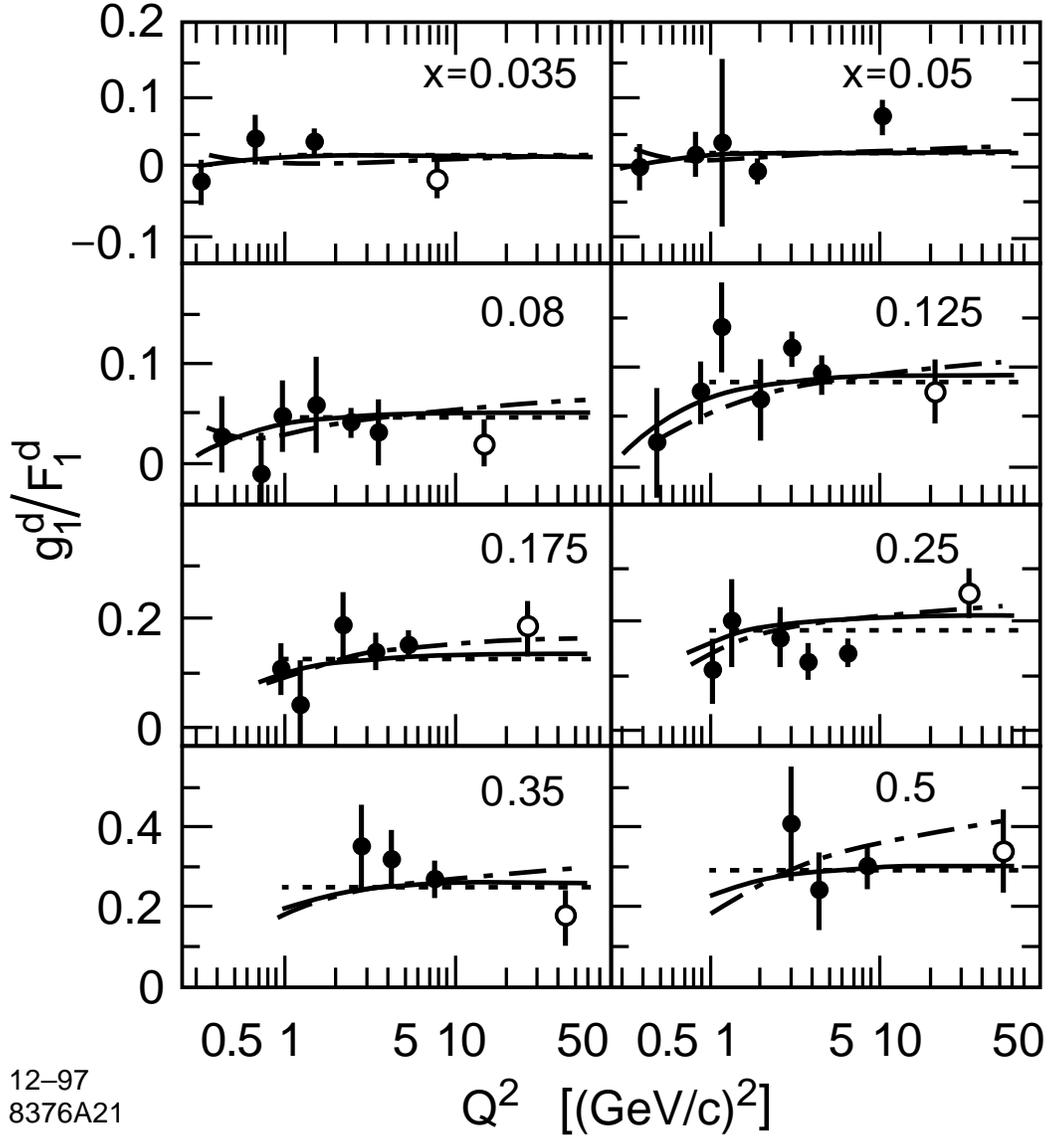,width=5.50 in}}\vskip 0.7cm
\caption{$g_1^d/F_1^d$ as a function of $Q^2$ for 8 different $x$ bins.
    The data are from this experiment (solid circles) and SMC\protect\cite{smcp}
    (open circles).
    The dashed and solid curves correspond to global fits II
    ($g_1/F_1$ $Q^2$-independent) and III  ($g_1/F_1$ $Q^2$-dependent) in
    Table \protect\ref{tb:g1f1_fit}, respectively.    The E154  NLO 
    pQCD fit\protect\cite{e154nlo}
    is shown as the dot-dashed curve.}
\label{fg:g1F1d_Q2}
\end{figure}

Tables~\ref{tb:g1f1} and  \ref{tb:a1} show the values of  $g_1/F_1$ and $A_1$ for 
deep-inelastic scattering ($W^2 \geq 4$~GeV$^2$) for all three beam
energies and both  spectrometers using $g_2$ = $g_2^{WW}$ ~for the 16.2 and 9.7 GeV 
data and $A_\parallel$ and $A_\perp$ ~for the 29.1 GeV data. Figures~\ref{fg:g1F1p_Q2} 
and \ref{fg:g1F1d_Q2} show $g_1/F_1$ for proton and deuteron as functions of 
$Q^2$ averaged into 8 $x$ bins. Data from EMC \cite{emc}, SMC 
\cite{smcp,smcd,smct},  SLAC E80\cite{e80}, and SLAC E130 \cite{e130} are also 
included. The results are consistent with  $g_1/F_1$ and $A_1$ being 
independent of $Q^2$ for $Q^2\geq $1~(GeV/c)$^2$. We  fit all the deuteron 
and proton  data, (including the SMC data at $Q^2\leq 1$) with the empirical 
parameterization $ax^\alpha(1+bx+cx^2)[1+Cf(Q^2)].$ The coefficients of the fit 
are shown in Table~\ref{tb:g1f1_fit}, and the fits are shown in 
Figs.~\ref{fg:g1F1p_Q2} and \ref{fg:g1F1d_Q2}. We chose three forms for 
$f(Q^2)$ shown in Column 3 of the table. They are
\begin{itemize}
\item
  $f(Q^2)=0$: no $Q^2$ dependence.
\item
  $f(Q^2)=1/Q^2$:  higher twist behavior.
\item
  $f(Q^2)={\rm ln}(1/Q^2)$: pQCD behavior.
\end{itemize}

The minimum $Q^2$ of the fits is shown in Column 2 and is either 0.3 or 1.0
(GeV/c)$^2.$ For the proton data,  the $\chi^2$/df is less than unity for all the fits 
except fit I, indicating that there is $Q^2$ dependence for $Q^2\leq 1$~(GeV/c)$^2$. 
Fit II indicates that there is no need for any $Q^2$-dependent term for $Q^2\geq 1$~(GeV/c)$^2$, which is our cut-off for deep-inelastic scattering. For the deuteron data, 
the fits are not as good, but still have a confidence level of about 10\%.
Fit~V, which has  $Q^2_{min}=0.3$~(GeV/c)$^2$, is used  to evaluate $g_2^{WW}$ and  to iterate the radiative corrections described above.

\begin{figure}[t]
\centerline{\epsfig{file=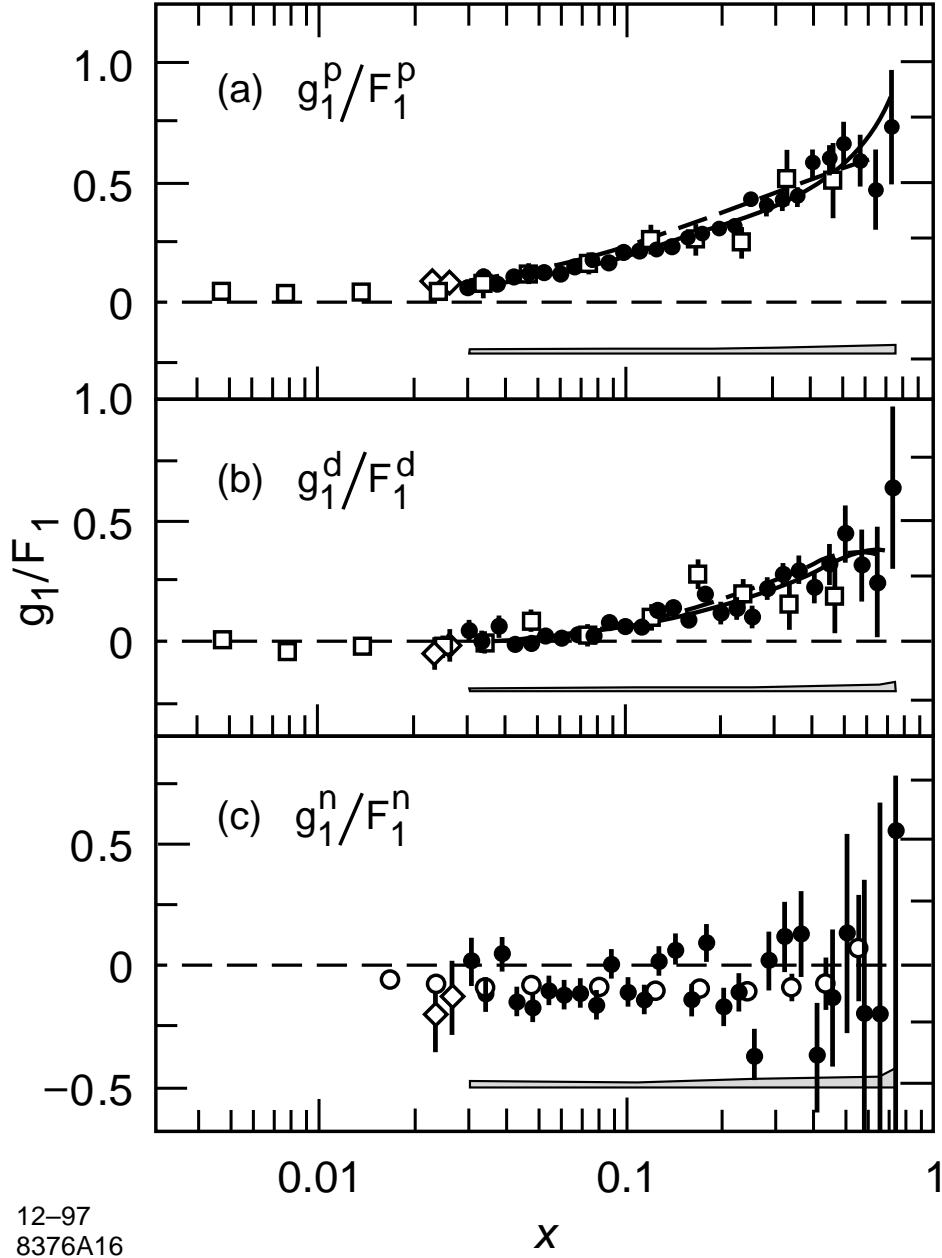,width=5.00 in}}\vskip 0.7cm
\caption{$g_1/F_1$ averaged over all beam energies and spectrometer 
angles as a function of $x$. The solid circles are at 
$Q^2\geq 1$ (GeV/c)$^2$, while the diamonds are at
$Q^2\leq 1$ (GeV/c)$^2$ for $x\leq 0.03.$ The data from SMC\protect\cite{smcp} (squares) are shown for proton and deuteron, and the data from E154\protect\cite{e154} 
(open circles) are shown for neutron. The curves are NLO pQCD fits by 
Altarelli, Ball, Forte, and Ridolfi \protect\cite{abfr} (solid) and 
Gl\" uck, Reya, Stratmann, and Vogelsang \protect\cite{grv} (dashed). }
\label{fg:g1F1AV}
\end{figure}

Also shown in Figs.~\ref{fg:g1F1p_Q2} and \ref{fg:g1F1d_Q2} are the results 
from the E154\cite{e154nlo} leading order pQCD evolution fit to world data 
including preliminary results from this experiment. It is in good agreement 
with the data, including the data for $Q^2\leq 1$~(GeV/c)$^2$ which was not 
used in the fit. However, for the proton it does have an exaggerated $Q^2$ 
dependence at the highest $x.$ Since $g_1/F_1$ and $A_1$ are both consistent with 
being independent of $Q^2$ for $Q^2 \geq 1$~(GeV/c)$^2$, we choose to combine 
our data at fixed $x$ by averaging them over all measured values for $Q^2\geq 
1$~(GeV/c)$^2$. Tables~\ref{tb:g1f1av} and ~\ref{tb:a1av} and 
Fig.~\ref{fg:g1F1AV} show these averaged values as a function of $x.$  The band 
at the bottom of Fig.~\ref{fg:g1F1AV} represents the size of the systematic 
errors. Also shown are results from other experiments \cite{emc,smcp,smcd,smct} 
averaged in a similar way. The various experiments are in agreement with each 
other. Results for $g_1/F_1$ and $A_1$ are similar at low $x$ and diverge slightly at 
high $x.$  For the proton, both $g_1/F_1$ and $A_1$ are small and positive at low $x$ 
and rise steeply toward unity as $x\rightarrow 1$. For the deuteron, both $g_1/F_1$ 
and $A_1$ are close to zero at low  $x$ and increase slowly with increasing $x.$ 
For the neutron, both $g_1/F_1$ and $A_1$ are negative over most of the $x$ region, 
showing almost no indication of becoming positive at high $x$ as expected from
earlier predictions \cite{closea}.

\begin{figure}[t]
\centerline{\epsfig{file=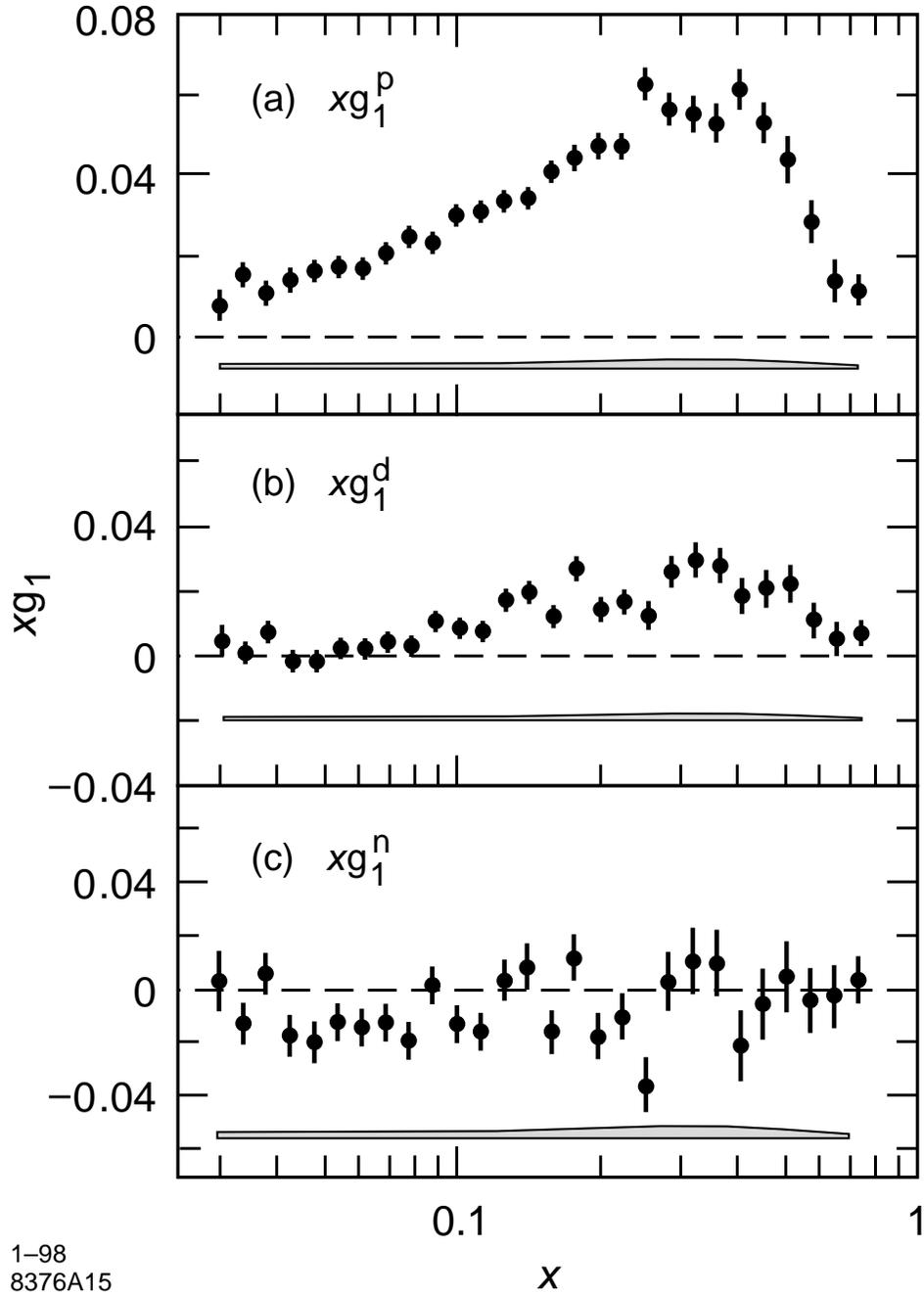,width=5.00 in}}\vskip 0.7cm
\caption{
$xg_1(x,Q^2)$ evaluated at the average measured $Q^2$ at each 
$x$.}
\label{fg:xg1}
\end{figure}

Table~\ref{tb:g1}  shows $g_1$ at the averaged measured value of $Q^2$ obtained 
from the average value of $g_1/F_1$. The quantity $g_1^n$ was obtained using
Eq.~\ref{eq:gn}. Figure~\ref{fg:xg1} shows $xg_1$ as a function of $\log x.$  The 
area between the data and zero is the integral forming the sum rules
$\int_0^1 g(x)dx$ = $\int_{-\infty}^0 xg(x) d\ln x.$

\subsection{Systematic Errors}\label{sec:syserr}

The systematic errors were calculated for $g_1/F_1$, $A_1$, and $g_1.$  Only the 
systematic error due to $A^\parallel$ was considered since the systematic 
errors due to $A_\perp$ were negligible compared to the statistical errors. 
Some of the errors were multiplicative and independent of $x$ while others
were $x$-dependent. The errors due to multiplicative factors (beam and
target polarization) are shown in Table~\ref{tb:norm_err}. The errors on $g_1$ 
and $g_1/F_1$ from these normalizations were obtained using a smoothed fit to 
$g_1/F_1$. The breakdown of the major sources of error for a sample of our $x$ bins 
is shown in Tables~\ref{tb:syserg10}-\ref{tb:sysergf23} for deuteron and 
proton targets for both $g_1$ and $g_1/F_1$. The radiative correction error dominated 
at low $x.$ The errors due to multiplicative factors were only significant when
either $g_1$ or $g_1/F_1$ were large at middle and high $x,$ respectively.

\noindent\underline{Multiplicative systematic errors:} 
\begin{itemize} 
\item 
The error of the beam polarization $P_B$ was estimated to 
be 0.024 (relative). See Section~\ref{sect:beampol}.
\item
The error of the target polarization $P_T$ had a relative systematic error of 
0.025 for protons and 0.04 for deuterons. The error was assumed to be~100\% 
correlated between runs, since the systematic error was obtained from the 
spread of the thermal equilibrium measurement results, each of which provided 
the calibration constants for large groups of runs. See 
Section~\ref{sect:targpol}.
\item
The proton in nitrogen correction (see Eq.~\ref{eq:ncorrprot}) contributed with 
a~0.004 relative systematic error since the correction $C_1$ was always around 
0.02, while the relative error on $C_1$ was estimated to be~0.2.  
\end{itemize}
\noindent\underline{Systematic errors dependent on $x$:} 
\begin{itemize}
\item
The error of the dilution factor $f$ came from several sources. The component 
dependent on our experimental setup (the amount of ammonia in the target cell)  
was known to 4\% for both targets. The relative error from the cross-section 
ratio $\sigma_n/\sigma_p$ was 1\% \cite{NMC-np}. It was one of the dominant 
errors for the NH$_3$ target and did not contribute for the ND$_3$ target. The 
ratio of nuclear to deuterium cross sections, the EMC effect, is known to 1.5\% 
relative and was another large error source. It has a 1\% overall normalization 
and another 1\% of uncorrelated error \cite{GOMEZ}. The small mass of the NMR 
coil was known to 20\% but did not contribute significantly to the overall 
error. This leads to an average error on $f$ of 2\% for  NH$_3$  and 1.5\% for ND$_3.$ 
\item
The nitrogen correction was applied via two factors, $C_1$ and $C_2$ 
(see Eq.~\ref{eq:ncorrdeut}). The error on $C_1$ was neglected since this value 
was very small and stable. The factor $C_2$ (ND$_3$ only) contained the proton 
asymmetry and was calculated for each $x$ bin using our measured proton 
asymmetry and its error. 
\item
The systematic error on the radiative corrections was calculated for each $x$ 
bin by varying several classes of input models. See Section~\ref{sect:rc} for 
details. It is  shown in  Tables~\ref{tb:syserg10}-\ref{tb:sysergf23} for 
typical values of $x$ at 29 GeV beam energy for $g_1$ and $g_1/F_1$.  
\item 
The error due to the structure function $R(x,Q^2)$ contributed to $g_1/F_1$ and $g_1$ quite differently due to the relationship between $F_1$ and $F_2.$ For $g_1$ the 
effect of $R$ is negligible whereas for $g_1/F_1$ it is one of the significant 
errors. Its systematic error was taken from the SLAC global analysis
\cite{whitlowr} and ranged from 3\% to about~7.5\%. While this fit to $R$ was 
made to data with a limited $Q^2$ range and $x\geq 0.1,$ it is consistent with 
recent measurements at lower $x$ \cite{CCFR,NMC_R} and different 
$Q^2$\cite{E140x}.
\item
The error in the structure function $F_2$ was obtained from the NMC fit
\cite{new_NMC}.  The error returned from the fit was taken as completely
correlated point-to-point.
\item
Pion and charge-symmetric backgrounds were treated as statistical errors from 
the measurement with spectrometers set at opposite polarity.  No systematic 
error was assigned to the model of charge symmetry.
\end{itemize}

When averaging $g_1/F_1$ over spectrometers and beam energies, the weight of each 
data point included statistical and point-to-point uncorrelated systematic 
errors. For the systematic error of the neutron structure function $g_1^n$ as 
well as of the difference $g_1^p-g_1^n$, the beam polarization error and the 
dilution factor errors due to the unpolarized cross sections were assumed
to be 100\% correlated, while the other errors were assumed to be uncorrelated.  

The systematic error on the integral was calculated using the separated
correlated and uncorrelated systematic errors. The systematic errors of the 
low-$x$ and high-$x$ extrapolations were added together with the systematic 
error for the data region. The sum was then quadratically combined with the fit 
errors for the low- and high-$x$ extrapolations to yield the total systematic 
error on the integral.  

\subsection{Integrals }

The Ellis-Jaffe (Eq.~\ref{eq:ej}) and Bjorken (Eq.~\ref{eq:bj}) sum rules 
involve integrals over all values of $x$ at a fixed $Q^2$.  The experimental 
results do not cover all $x$ at any single $Q^2$. In the measured region of $x$  
we must either interpolate or extrapolate our results from the measured $Q^2$ to
some fixed $Q^2_o.$  In the regions of $x$ above and below the measured
region, we use model-dependent extrapolations.
 
\subsubsection{Measured Region }

Several methods have been used to determine $g_1$ at fixed $Q^2_0.$
\begin{itemize}
\item{1)} Assume $g_1/F_1$ is independent of $Q^2$ and determine $g_1$ from
 $g_1(x,Q^2_o) = g_1/F_1\times F_1(x,Q^2_o).$
 \item{2)} Assume $A_1$ is independent of $Q^2$ and determine $g_1$ from
$[g_1(x,Q^2_o) = A_1(x)\times F_1(x,Q^2_0) + \gamma_o^2g_2^{WW}(x,Q^2)]$.
\item{3)} Fit the data to a functional form which has semi-empirical
dependencies on $x$ and $Q^2$ such as the fits described above.
\item{4)} Do a pQCD fit to determine the quark and gluon distributions and then 
calculate the change in $g_1$ going from the measured to the desired kinematics.
\end{itemize}

In this paper we will pursue the first two options with emphasis on the first. 
We note that the pQCD fits indicated in Fig.~\ref{fg:g1F1p_Q2} and 
\ref{fg:g1F1d_Q2} show little $Q^2$ dependence (compared to the errors of the 
experiments) for $g_1/F_1$ at $x\leq 0.5$ in the relevant $Q^2$ range. For 
$x\geq 0.6$, theoretical papers often use approximations in defining the 
relationship between $F_2$ and $F_1$ and sometimes use pQCD fits to $F_2$ 
instead of empirical fits to the data.  At $x=0.75$ typical pQCD 
fits \cite{grv,e154nlo,gs} show $g_1^p/F_1^p$ differing by 30-50\% between the 
measured $Q^2\sim 9.5$~(GeV/c)$^2$ and  $Q^2 =3$~(GeV/c)$^2$, but these predictions are questionable due to the assumptions used.

\begin{figure}[t]
\centerline{\epsfig{file=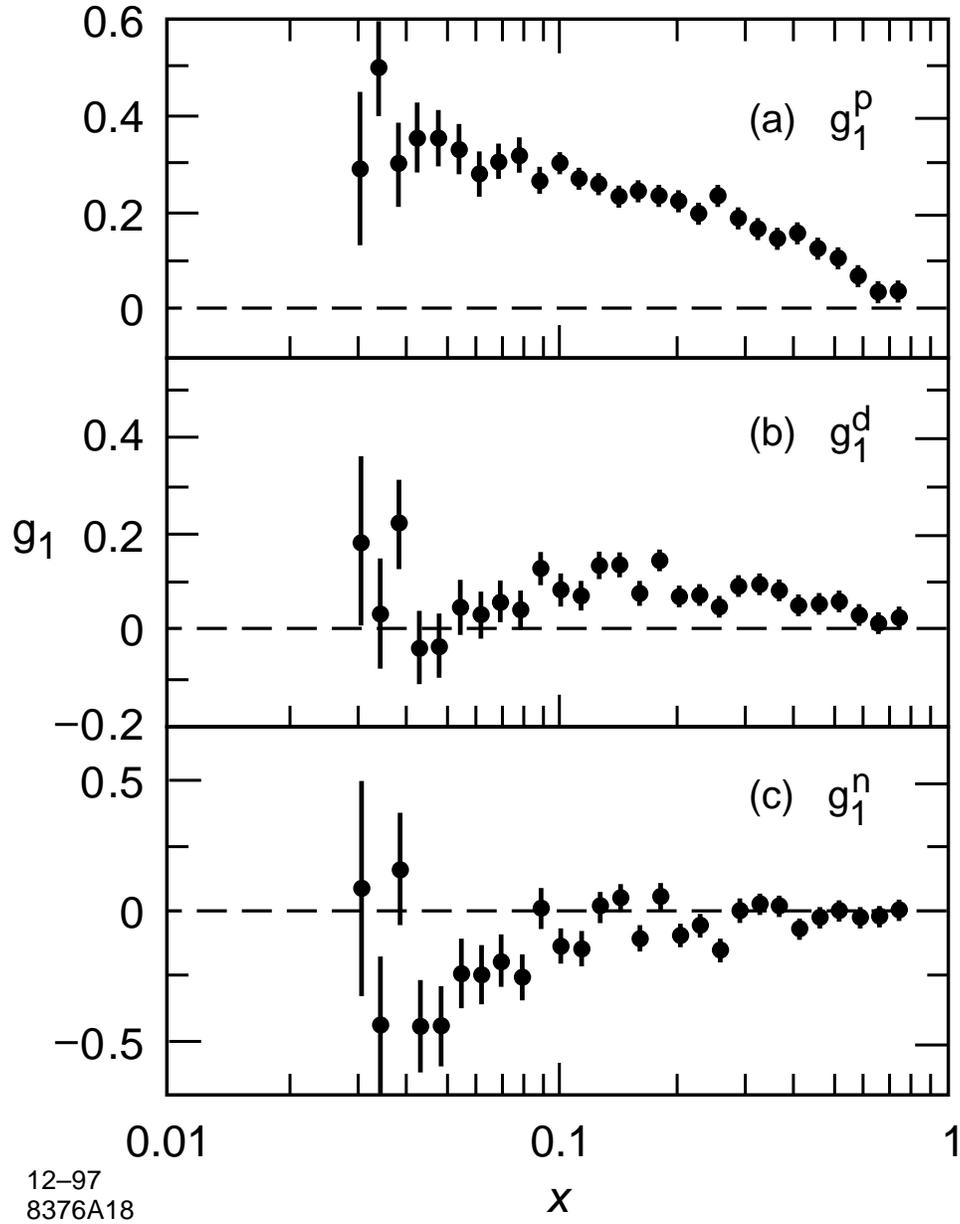,width=5.00 in}}\vskip 0.7cm
\caption{
$g_1(x,Q^2)$ evaluated at $Q^2=3$ (GeV/c)$^2$ (assuming $g_1/F_1$ is
independent of $Q^2$) as a function of $x.$}
\label{fg:g1Q0}
\end{figure}

Tables~\ref{tb:g1q0_1}-\ref{tb:g1q0_3} list $g_1$ as a function of $x$ at fixed 
$Q^2$ values of 2, 3, and 5~(GeV/c)$^2$ for proton, deuteron, and neutron. 
These results were evaluated by method 1 ($g_1/F_1$ independent of $Q^2$).
Figure~\ref{fg:g1Q0} shows the corresponding method 1 results for $g_1$ at 
$Q^2$=3~(GeV/c)$^2$. Results for  $\int_{0.03}^{0.8}{g_1(x)dx}$ using methods 1 
and 2 ($A_1$ independent of $Q^2$, $g_2^{WW}$) at the same three values of $Q^2$ are given in table~\ref{tb:gam1m}. Method 1 yields slightly larger 
results in magnitude than method 2, but the difference is smaller than the total error for all targets at fixed $Q^2\geq 3$~(GeV/c)$^2$. The components of 
the systematic error on the integral are shown in Table~\ref{tb:gam1_se}. The 
correlated systematic errors due to beam and target polarization and $F_2$ 
dominate. The radiative correction errors tend to be anti-correlated between 
low and high $x$ and thus partly cancel in the integral.

\subsubsection{Low $x$}

The evaluation of $\int_0^{0.03}g_1(x,Q^2_o)dx$ can be done by several methods.
\begin{itemize}
\item{1)} Using Regge trajectory-type behavior, $g_1 =x^{\alpha}g_1^0$ at fixed 
$Q^2$ and low $x$. The difference $g_1^p$-$g_1^n$ has isospin 1 and only one 
Regge trajectory contributes. The value of $\alpha$ is in the general range 
$0.5\geq \alpha \geq 0$ \cite{HEIMANN,KUAR,BASS}. For the individual proton and  
deuteron targets there may be more than one pole contributing \cite{HEIMANN}. 
There also may be even more complex behavior of the singlet term.

\item {2)}  Using SMC data from $0.003\leq x\leq 0.03$ and Regge extrapolations
below $x=0.003$.

\item{3)} Using the form $g_1 \propto \ln(1/x) \cite{BASS,closeb}.$

\item {4)}  Using the parameterization II from Table~\ref{tb:g1f1_fit}
(the form is Regge inspired at low $x$).

\item{5)} Using pQCD fits.
\end{itemize}

The Regge method requires a choice of $x$ range to determine the pole parameter 
and a choice of other possible Regge trajectories. In addition, if $g_1$ has 
Regge behavior at a given $Q^2$, it will not have Regge behavior at other 
$Q^2$ since $g_1$ evolves with $Q^2$ differently at different values of $x$.
Table~\ref{tb:lowx} shows the results of various options, 
including using the Regge form at $Q^2$=1 and 3~(GeV/c)$^2$. Constraining $\alpha=0$ gives good fits at both $Q^2$=1 and 3~(GeV/c)$^2$. However, 
requiring $\alpha=0.5$ gives a rather poor fit  $(\chi^2/df \approx 2)$ for the
proton. We take the average of the four fits in Table~\ref{tb:lowx} with 
$\alpha=0$ as the central value of the low $x$ extrapolation. The error 
encompasses all the other models indicated. These averages are shown in 
Table~\ref{tb:integral}. The values of the integral for proton, deuteron, 
neutron, and proton--neutron (p--n) may not add up exactly due to the 
non-linearity of the fits. 

Recent results from SLAC \cite{e154} indicate that $g_1^n$ may be behaving as 
$\sim x^{-0.8}$ at low $x.$ If the proton behaves in a similar way, then the 
above extrapolations would be open to question.

\subsubsection{High $x$}

The extrapolation to high $x$ was done by two methods: 1) assuming $g_1 \propto 
(1-x)^3$ \cite{brodsky} and fitting to the four highest $x$ bins; 2) assuming 
$A_1^p = 0.75(19-16F_2^n/F_2^p)/15$ and 
$A_1^n = 0.75(2-3F_2^n/F_2^p)/(5F_2^n/F_2^p)$ \cite{closea}. For both $g_1^d$ and 
$g_1^p$, both methods gave almost identical values of $\int_{0.8}^{1}{g_1(x)dx}.$ 
For the deuteron the value is 0.000 and for the proton 0.001. We assign an 
error of $\pm 0.001$ in both cases. For the neutron the value of the integral 
is $0.001 \pm 0.001$ where rounding errors account for inexact match with 
proton and deuteron results. The small value of the integral is mostly due to 
the small value of $F_1$ and not the properties of the individual models. The 
average values are shown in Table~\ref{tb:integral}.

\subsubsection{Total Integral}

Table~\ref{tb:integral} shows $\Gamma_1$, the total integral from $x=0$ to 1, 
in the last column for proton, deuteron, neutron, and the difference 
proton--neutron.  The experimentally measured portion of the integral makes the
largest contribution to $\delta\Gamma_1$ with the low $x$ extrapolation error a 
close second.  The correlation between the measurement errors at low $x$ and 
the extrapolation errors is small compared to the model dependence of the 
extrapolation.

The integrals  from this experiment, E142 \cite{e142}, and SMC\cite{smcp,smcd,smct} are compared in Table~\ref{tb:int_comp} at $Q^2$ 
values reported by the other experiments. Comparisons are made for the full range in $x$ of 0 to 1, as well as for the common measured $x$ range between 
experiments.  For each comparison, we evaluated the integral using the same assumptions about $Q^2$ dependence that the other experimenters used. (SMC 
results were calculated by us from their tables). In the experimental range  there is good agreement between this experiment and the other  results.  
$\Gamma_1$ for each experiment, as shown in Table~\ref{tb:int_comp}, are also 
in excellent agreement with the caveat that different $x$ ranges were measured 
and different extrapolations used in the unmeasured region.

\subsubsection{Ellis-Jaffe Sum Rule}

Ellis-Jaffe sum rule predictions for the integrals (Eq.~\ref{eq:ej}) are shown 
at three values of $Q^2$ in Table~\ref{tb:ej}. We use 
$\alpha_s(M_Z) =0.118\pm0.003$ \cite{pdg}, three active flavors, and
$3F-D =0.58$ \cite{closea} with uncertainties of either 0.03 (small) \cite{closea} or 0.12 (large) \cite{jafman}. This larger error, as discussed in 
Section~\ref{sect:qpm}, is likely to be an overestimate. For the Ellis-Jaffe 
sum rule, the values for both the ``invariant'' and $Q^2$-dependent singlet pQCD corrections (see Eq.~\ref{eq:csinv}) are given. They 
differ by an amount which is larger than the theoretical error due to  $\alpha_s.$ In the case of the deuteron, the experimental errors are
comparable to the theoretical difference. The measured values of $\Gamma_1^p$ 
and $\Gamma_1^d$ are shown in the table along with the derived value of 
$\Gamma_1^n.$ Using the small errors on $3F-D$ and the ``invariant'' singlet 
term the Ellis-Jaffe sum rules are violated by $0.023 \pm 0.007$ (deuteron) 
and $0.032\pm0.012$ (proton). A violation implies that there could be a significant SU(3) symmetry breaking effect or that there is strange and/or gluon spin contributing to the proton spin. If we consider the large $3F-D$ errors (larger error due to possible symmetry breaking) combined 
with the $Q^2$-dependent singlet term, the deviations from the sum rule reduce to $0.018 \pm 0.015$ (deuteron) and $0.027\pm0.019$ (proton).

\subsubsection{Bjorken Sum Rule}

The Bjorken sum rule integral (Eq.~\ref{eq:bj}) is given in Table~\ref{tb:ej}
for three different values of $Q^2$. The theoretical value involves only 
non-singlet pQCD corrections and is thus independent of ambiguities associated 
with the singlet corrections (invariant or $Q^2$-dependent). Theoretical errors 
depend only on the uncertainty in $\alpha_s.$  The measured values are from 
this experiment only and the errors include all correlations. The experimental 
errors are considerably larger than the theoretical errors. Experiment and
theory agree within one standard deviation.

A more precise result can be obtained by combining all the experiments 
\cite{smcp,smcd,emc,e154} which published a value of the integrals. SLAC 
experiment E154 \cite{e154} on the neutron was not included because they did 
not publish an integral of their data alone. We consistently used the method 
with $g_1/F_1$ independent of $Q^2$ to evolve the results to constant $Q^2$. In all 
experiments, the low-$x$ extrapolation errors were limited to $g_1$ being 
constant or approaching zero as $x\rightarrow 0.$ At $Q^2$=5~(GeV/c)$^2$, the 
combined results are: $\Gamma_1^{Bj}=0.170\pm0.012$,
$\Gamma_1^{p}=0.130\pm0.006$, and  $\Gamma_1^{n}=-0.040\pm0.008$
with a very small $\chi^2$/df.  $\Gamma_1^{Bj}$ is one standard deviation from 
the theoretical value of 0.182 determined from Eq.~\ref{eq:bj} with 
$\alpha_s(M_Z)=0.118.$ The addition of higher twist and other effects described 
below make the agreement even better. If we assume the Bjorken sum rule is true 
and solve  Eq.~\ref{eq:bj} for $\alpha_s$ we obtain 
$\alpha_s(M_z)=0.123^{+0.010}_{-0.006}.$

\subsection{Quark Polarization}

\begin{figure}[t]
\centerline{\epsfig{file=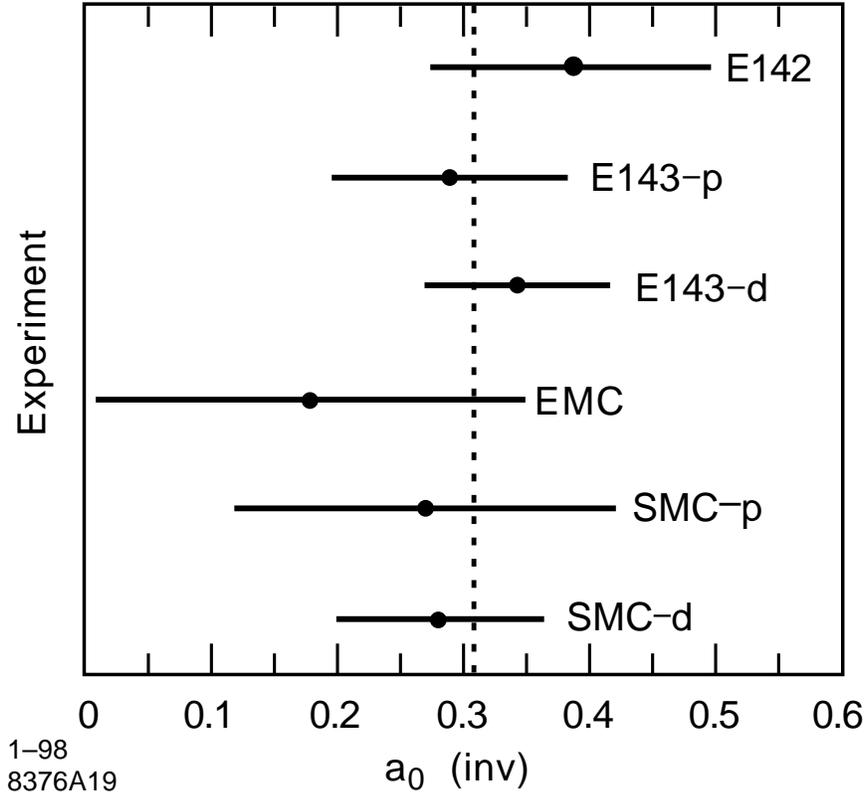,width=4.5 in}}\vskip 0.7cm
\caption{
Measured singlet matrix elements $a_0^{inv}$ from this experiment, 
E142 \protect\cite{e142}, SMC \protect\cite{smcp,smcd}, and EMC/E80/E130 
combined \protect\cite{emc}. These results were calculated from the published
first moments of $g_1$ using up-to-date ``invariant'' singlet pQCD corrections
and $\delta (3F-D)= 0.032.$ The dashed curve indicates the world 
average of $0.31\pm 0.04$.}
\label{fg:delta}
\end{figure}

We used  Eqs.~\ref{eq:delqp}-\ref{eq:delqd} to extract our measured value
of $a_0$ from the proton and deuteron first moments. Then using Eq.~\ref{eq:helicity} (with $\Delta G=0$), we extracted the individual polarizations of the quarks. 
It is important to remember that these polarizations have meaning within the 
quark-parton model where $a_0=\Delta\Sigma$. In pQCD the interpretation
becomes scheme-dependent and depends on whether $\Delta G(x)$ contributes to 
$\Gamma_1.$ The results are shown in Table~\ref{tb:delta} and 
Fig.~\ref{fg:delta}.  Results for both the ``invariant'' and $Q^2$-dependent pQCD 
singlet coefficients are shown in the table while the figure shows 
``invariant'' results.  The quantities $a_0$, $\Delta u$, and $\Delta d$ are 
relatively insensitive to the values of $F$ and $D$, but $\Delta s$ is very 
dependent on them. The last two columns of Table~\ref{tb:delta} show the errors 
on $\Delta s$ with two different estimates on the errors on $3F-D.$ As seen in 
Fig.~\ref{fg:delta} the results from the deuteron and proton targets are 
consistent with each other (there is only a small correlation between the 
errors). The differences between the ``invariant'' and $Q^2$-dependent results 
are smaller than the present experimental errors. The negative polarization of the strange 
quark sea of about $-0.08$ is very significant only if the smaller estimates of 
$F$ and $D$ are used. Our averaged proton and deuteron results for 
$a_0^{inv}=0.33\pm0.06$, while the world average yields $a_0 \sim 0.31\pm0.04$.
The results for $a_0$ are significantly smaller than the naive parton model 
prediction of $\Delta\Sigma = 1$, the relativistic parton model prediction of 
0.75, the Ellis-Jaffe sum rule prediction of 0.58, and a Quenched  Lattice 
Calculation of $0.60\pm0.05$\cite{lat96}. With such a low value of $a_0$, 
angular momentum conservation (see Eq.~\ref{eq:angmomconv}) requires that the 
nucleon spin be dominated by a combination of gluon polarization and orbital 
angular momentum, or a large charm polarization not included in the formalism 
above (Eqs.~\ref{eq:delqp}-\ref{eq:helicity}). There have been several 
approaches to understanding the low value of $a_0$. These are described in 
Section II.

\subsection{$A_2$ and $g_2$ }
\begin{figure}[t]
\centerline{\epsfig{file=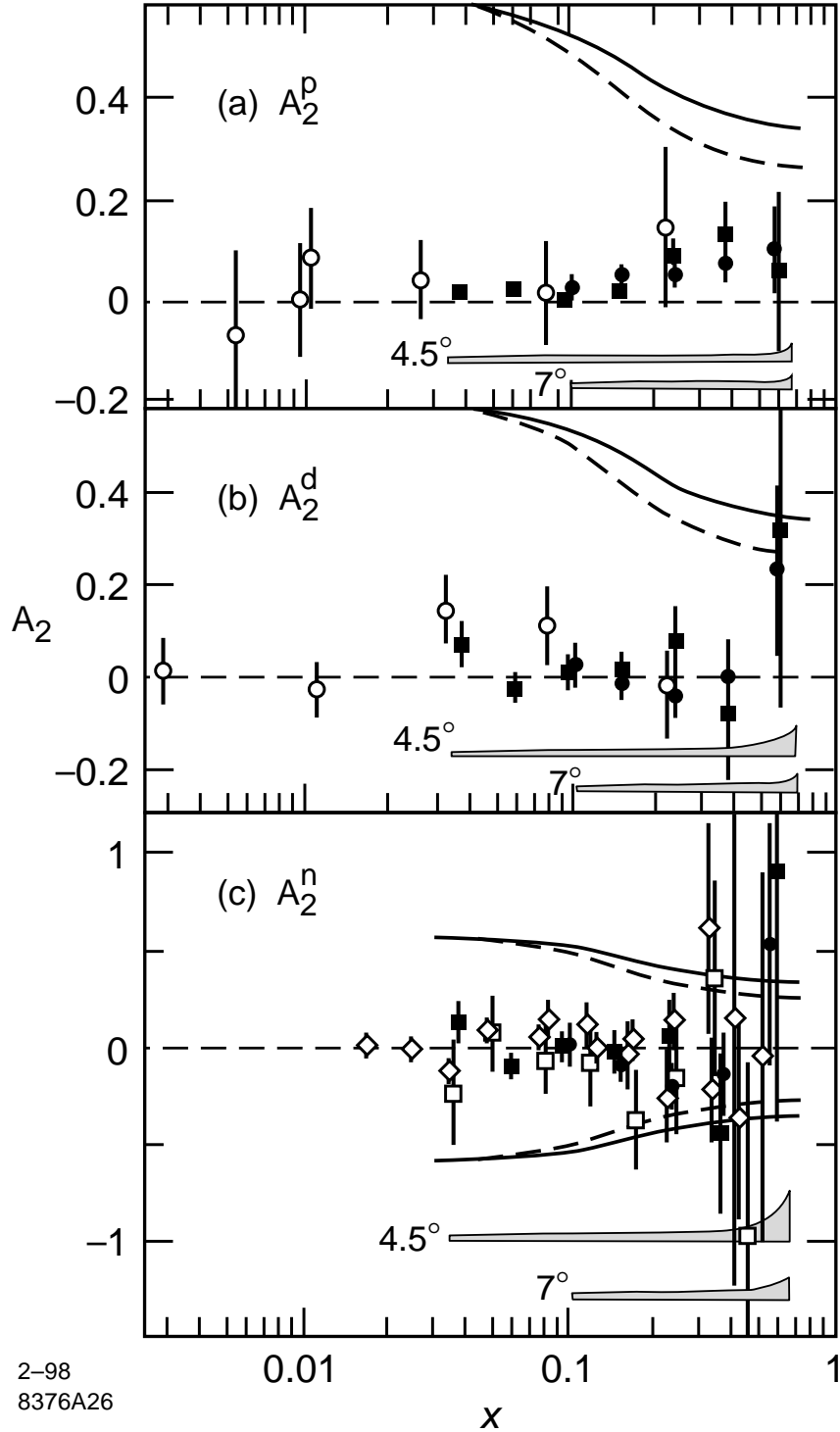,width=4.50 in}}\vskip 0.7cm
\caption{The results for $A_2(x,Q^2)$~ as a function of $x$ for 
this experiment (solid squares for $4.5^\circ$, solid circles 
for $7^\circ$). Also shown are SMC (open circles) \protect\cite{smct},
E142 (open square) \protect\cite{e142}, and 
E154 (diamond) \protect\cite{e154t}. The solid and dashed curves 
correspond to the positivity constraint at $4.5^\circ$ and 
$7^\circ$ kinematics, respectively. Overlapping data have been 
shifted slightly in $x$ to make errors clearly visible. The 
bands indicate systematic errors for the two E143 data sets.}
\label{fg:A2}
\end{figure}

\begin{figure}[t]
\centerline{\epsfig{file=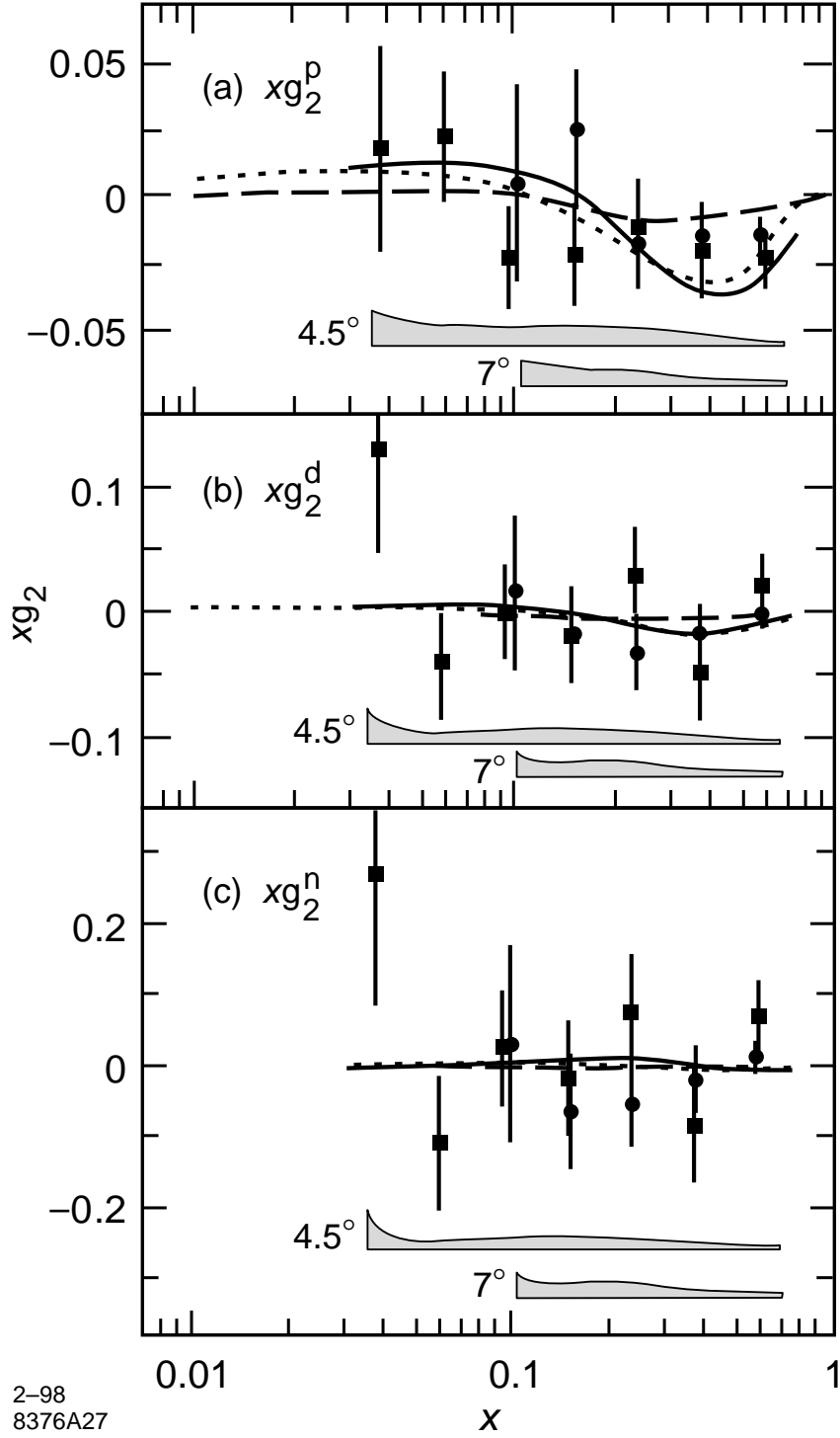,width=4.50 in}}\vskip 0.7cm
\caption{The results for $xg_2(x,Q^2)$ as a function of $x$ for 
this experiment (squares for $4.5^\circ$, circles for $7^\circ$).
Systematic errors are indicated by bands. Overlapping data have 
been shifted slightly in $x$ to make errors clearly visible. 
The solid curve shows the twist-2 $g_2^{WW}$ calculation for 
the kinematics of the $4.5^\circ$ spectrometer. The same curve 
for $7^\circ$ is nearly indistinguishable. The bag model 
calculations at $Q^2 = 5.0$ (GeV/c)$^2$ by 
Stratmann \protect\cite{strat} (short dash) and Song and
McCarthy \protect\cite{song} (long dash) are indicated. The
curves on the neutron plot are difficult to distinguish from zero.}
\label{fg:g2}
\end{figure}

Tables \ref{tb:g2A2p}-\ref{tb:g2A2n} show $A_2$ and $g_2$ for each target and 
spectrometer (and the derived neutron) with beam energy of 29.1 GeV. 
Figures~\ref{fg:A2}-\ref{fg:g2} show $A_2$ and $xg_2$ for each target and
spectrometer. The systematic errors are indicated by the bands. Also shown in 
Fig.~\ref{fg:A2} are the results from SMC \cite{smcd,smct} for the deuteron and 
proton and from E142 \cite{e142} and E154 \cite{e154t} for the neutron. All 
results are shown at their measured $Q^2$. There is good agreement between the 
various experiments. As seen in Fig.~\ref{fg:A2}, the values of $A_2^d$ are
consistent with zero, while the values of $A_2^p$ deviate significantly from 
zero for $x\geq 0.1.$  For this experiment the average value for all $x$ of 
$A_2$ is $0.031\pm0.007$ for the proton, $0.003\pm0.013$ for the deuteron, and 
$-0.03\pm0.03$ for the neutron. The measured $A_2$ obeys the $\sqrt{R}$ bound within errors, and at almost all kinematics the absolute value of the measured 
values are significantly lower than the bound. The dashed curve is a 
calculation of $g_2^{WW}$ from Eq.~\ref{eq:eqww} using $g_1$ evaluated from a 
fit to world data discussed in Sec.~\ref{sec:a1g1} of this paper. The 
$g_2^{WW}$ curves for the $7^\circ$ and the $4.5^\circ$ kinematics are 
indistinguishable on the figure. The other theoretical curves are bag model 
predictions \cite{strat,song} which include twist-2 and twist-3 contributions 
for $Q^2=5$~(GeV/c)$^2$. At high $x$ the E143 results for $g_2^p$ indicate a 
negative trend consistent with the expectations for $g_2^{WW}$ with a $\chi^2$ 
of 43 for 48 degrees of freedom.  However, the results are also consistent with 
$g_2^p$=0 with a  $\chi^2$ of 52. The deuteron and neutron results are less 
conclusive because of the larger errors and are also consistent with both 
$g_2^{WW}$ and $g_2=0$. The moments of $g_2$ will be discussed below along with 
the moments of $g_1$.

\subsection{Higher Moments of $g_1$ and $g_2$}

Using our results for both $g_1$ and $g_2$, we have computed the third moment 
of the OPE sum rules (Eq.~\ref{eq:ope}), and solved for the twist-3 matrix 
element $d_2$ and the twist-2 matrix element $a_2$. For the measured region 
$0.03<x<0.8$, we evaluated $g_1,$ corrected the twist-2 part of $g_2$ to fixed 
$Q^2$=5~(GeV/c)$^2$ assuming $g_1/F_1$ is independent of $Q^2,$ and have 
averaged the two spectrometer results. Possible $Q^2$ dependence of 
$\overline{g_2}$ was neglected. We neglect the contribution from the region 
$0\leq x<0.03$ because of the $x^2$ suppression factor. For $0.8<x\leq 1$, we 
assume that both $g_1$ and $g_2$ behave as $(1-x)^3$ since at high $x$, $g_2 
\approx -g_1$ from Eq.~\ref{eq:aonetwo} and $F_1\rightarrow 0$, and we fit the 
data for $x>0.56$. The uncertainty in the extrapolated contribution is taken to 
be the same as the contribution itself. The results are shown in
Table~\ref{tb:moments}. Our extracted values for $d_2$ are consistent with 
zero, but the errors are large. For comparison, in Table~\ref{tb:moments-th} we 
quote theoretical predictions\cite{balitsky,steing1,strat,song,steing2} for
$d_2^p$ and $d_2^d$. For $d_2^d$ the proton and neutron results were averaged 
and a deuteron D-state correction  was applied. We note that the results for 
$d_2^p$ and $d_2^d$ differ in sign from the theoretical QCD sum rule 
calculations \cite{balitsky,steing1,steing2}. The bag model predictions 
\cite{strat,song}, however, are of the same sign as the data. Ali, Braun and 
Hiller \cite{abh} showed that $g_2$ obeys an evolution equation in the limit 
that $N_c\rightarrow \infty.$ However, this program of calculation has not been 
carried out yet.

To test the Burkhardt-Cottingham sum rule, Eq.~\ref{eq:bc}, we have evaluated 
the integrals $\int_{0.03}^1 g_2^p(x) dx=-0.014\pm0.028$ and $\int_{0.03}^1 
g_2^d(x) dx=-0.034\pm0.082$ using the same high-$x$ extrapolation as discussed 
above. These results are consistent with zero. To evaluate the integral for 
$x\leq 0.03$ is theoretically challenging. A double logarithmic approximation 
has been used \cite{dla} to calculate both $g_1$ and $g_2$ in the low-$x$ 
region yielding $g_2\propto x^{-0.75}$ at a fixed $Q^2$ of a few (GeV/c)$^2$. 
Then  $\int_0^{0.03} g_2(x)dx$ is negligible and the sum rule is confirmed.

\subsection{Higher Twist Effects}
 We have compared  our experimental integrals with theoretical predictions
using pQCD for the finite $Q^2$ corrections to various sum rules, which were
originally derived at infinite  $Q^2$.  At low $Q^2$ it is possible that higher 
twist effects could also influence the evolution of $g_1.$  These terms 
generate a multiplicative term of the form $\{1+ C/[Q^2(1-x)]\}$ \cite{htBrod}. 
When going to very high order in pQCD there is a confusion between resummation 
effects generating $1/Q^2$ terms and the higher twist terms. There have been 
several calculations of the corrections to $\Gamma_1$ using QCD sum rules and 
the bag model (see the reviews \cite{hinchliffe}). These take the form of an 
additive correction to the sum rule of the form $\mu^t/Q^2$ where t=p, n, or d
for proton, neutron, or deuteron. From QCD sum rules \cite{shu,ehrnsperger} the 
higher twist contribution to $\Gamma_1$ is:
\begin{equation}
\Gamma_1^{HT} = {M_N^2 \over Q^2}(a_{2} + 4d_{2} +4f_{2}) + O({M_N^4\over Q^4})
              = {\mu\over Q^2} + O({M_N^4\over Q^4})
 \label{eq:gamht}
\end{equation}
\noindent
where $a_{2}$ (twist-2) and $d_{2}$ (twist-3) have been calculated from data 
above and $f_{2}$ is twist-4. The contribution to $\mu$ from  $a_{2}$ and 
$d_{2}$ is 0.004 for the proton and 0.002 for the deuteron, which are quite
negligible at our average $Q^2=3$~(GeV/c)$^2$ and small compared to the 
estimated contributions from $f_{2}.$ Table~\ref{tb:moments-th} shows 
calculated values of $\mu_2^p$ and $\mu_2^d$ using bag models and QCD sum 
rules. The sum rule calculations average about $-0.02$ for the proton and 
$-0.013$ for the deuteron and thus would have an effect on the calculation of 
$\Gamma_1$ at our average $Q^2$ of 3~(GeV/c)$^2$ comparable to our experimental 
error. The bag model calculations are similar in magnitude but opposite in sign 
to the sum rule calculations. A different type of calculation, using a diquark 
model\cite{anselmino} gives a higher twist contribution of a different form 
than Eq.~\ref{eq:gamht}, which numerically is  1\% or less of both
$\Gamma_1^{Bj}$ and  $\Gamma_1^{p}$ for $Q^2\geq 2$~(GeV/c)$^2$. Using data 
from this experiment, Ji and Melnitchouk \cite{ji} have extracted values for 
the twist-4 matrix element $f_{2}$. Combining this with results for $a_2$ and 
$d_2$ they find for the proton $\mu_{2}^p = 0.04\pm 0.02$, and for the neutron 
$\mu_{2}^n = 0.03\pm 0.04$.

The QCD sum rule higher-twist correction and a Pade summation of the 
perturbative terms have been applied to the Bjorken sum rule (Eq.~\ref{eq:bj}) 
by Ellis {\it et al.} \cite{pade}. They then use world data, including the 
preliminary results from this experiment, and find excellent agreement between
experiment and theory.  Working backwards, they determine the best value of
$\alpha_s $ is $0.117^{+0.004}_{-0.007}\pm0.002$ where the first set of errors
is experimental and the second theoretical, in excellent agreement with the 
world average of $0.118\pm0.003.$

\subsection{Pion Asymmetry}

\begin{figure}[t]
\centerline{\epsfig{file=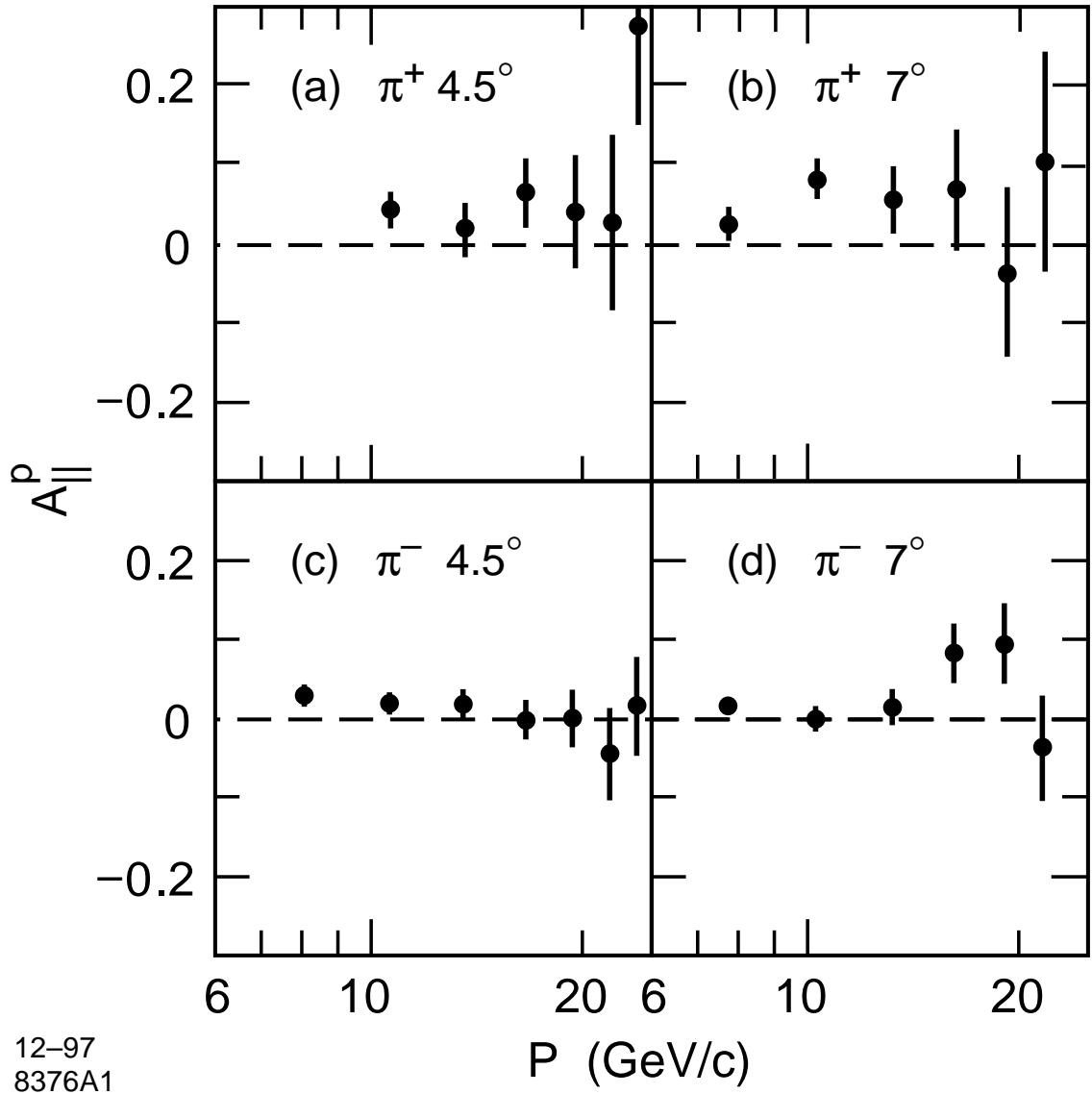,width=6.0 in}}\vskip 0.7cm
\caption{$A_\parallel$ versus momentum measured from polarized 
protons for $\pi^-$ and $\pi^+$, at $E=29$ GeV in both spectrometers.}
\label{fg:pi29p}
\end{figure}

\begin{figure}[t]
\centerline{\epsfig{file=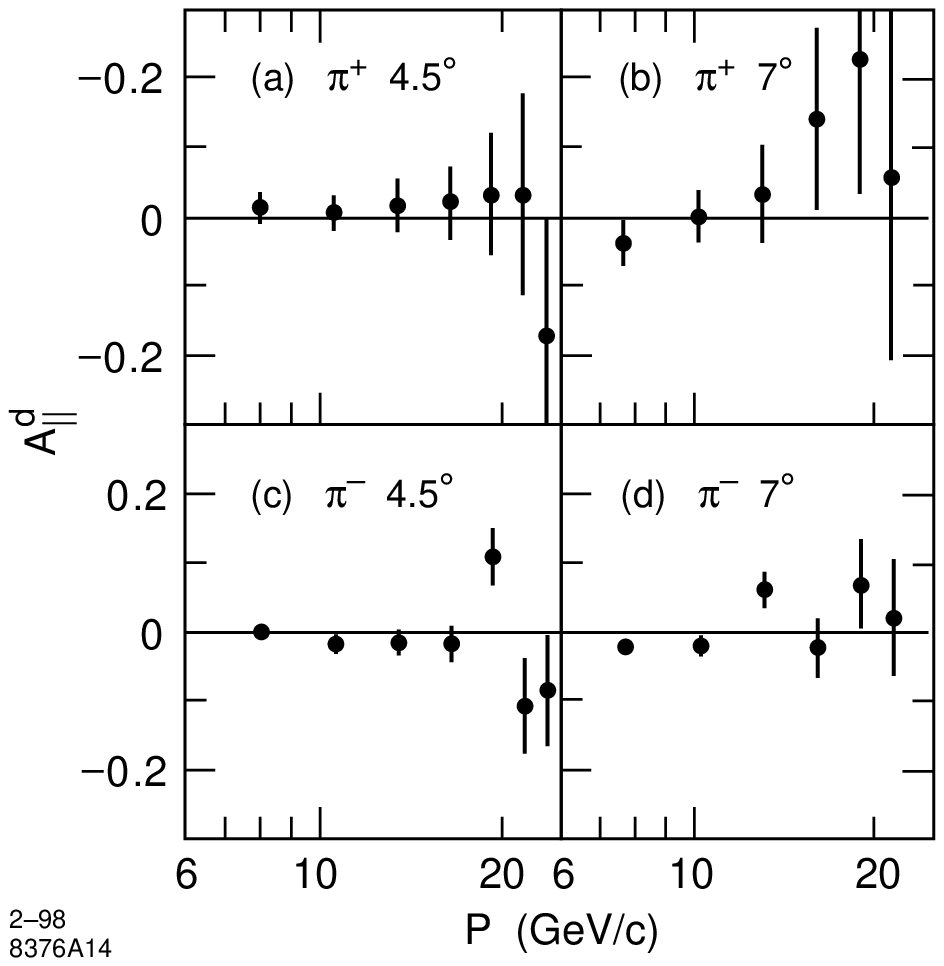,width=6.0 in}}\vskip 0.7cm
\caption{$A_\parallel$ versus momentum measured from polarized 
deuterons for $\pi^-$ and $\pi^+$, at $E=29$ GeV in both spectrometers.}
\label{fg:pi29d}
\end{figure}

The asymmetries for $\pi^+$ and $\pi^-$ for our primary energy of 29 GeV 
corresponding to target polarization parallel and anti-parallel to the beam 
direction ($A^\pi_\parallel$) are shown in Figs.~\ref{fg:pi29p}-\ref{fg:pi29d}.
These data were measured using our polarized NH$_3$ and ND$_3$ targets, and 
were corrected for beam and target polarizations as well as dilution to
obtain the asymmetries from polarized protons and deuterons. The asymmetries 
are small, but for the proton  may be slightly positive for both signs of 
pions at small momentum. Table~\ref{tb:ppion} gives the 
pion asymmetry at the beam energy of 29.1 GeV for proton and deuteron targets. 
Data for other beam energies and for the target spins oriented perpendicular to 
the beam direction have much larger statistical errors and are consistent with zero.

  \subsection{Resonance Region} 
\begin{figure}[h]
\centerline{\epsfig{file=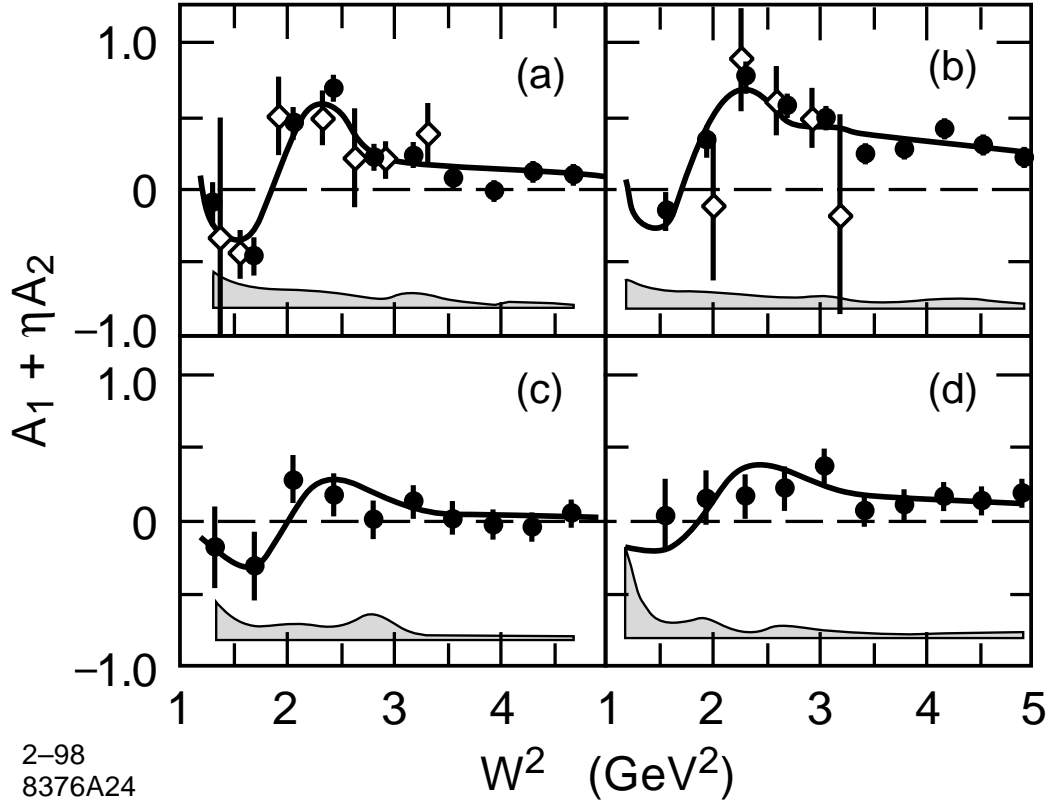,width=5.50 in}}
\vskip 0.7cm
\caption{Extracted values for $A_1 + \eta A_2$ (circles) in the
resonance region for the proton 
at (a) $4.5^\circ$  and (b) $7.0^\circ$; and for the deuteron at
(c) $4.5^\circ$ and (d) $7.0^\circ$. Also shown are the Monte 
Carlo predictions (solid line) and the data of Baum {\it et al.} 
\protect\cite{baum} (diamonds).
Error bars correspond to statistical errors only, whereas the
bands below the data correspond to the systematic errors.}
\label{fig:res4}
\end{figure}

Results for $A_1+\eta A_2$ extracted via the asymmetry method for the resonance region ($W^2 < 5$~GeV$^2$) are shown in Table~\ref{tbl:A1} and in Fig.~\ref{fig:res4}. Also shown in Fig.~\ref{fig:res4} is the data of Baum {\it at al.} \cite{baum}. The two data sets 
agree within the errors of both measurements. The asymmetry is negative and 
close to the expected value $A_1=-{1\over 2}$ for the $\Delta$ resonance. In 
the region of the $D_{13}$ and the $S_{11}$ resonances ($W^2 \sim 2.34$~GeV$^2$) $A_1+\eta A_2$ is large and 
positive.  Although $\eta$ is large  for our kinematics, a small value of $A_2$ 
would imply that $A_1+\eta A_2$ is $\approx A_1$.

Table~\ref{tbl:A1err} lists the systematic errors on $A_1 + \eta A_2$ by
category.  The procedure for estimating these is as follows:
   a) radiative corrections ($A_{rc}$): the maximum deviation in the radiative
	correction resulting from a 50\% change in the input asymmetries;
   b) model dependence (model): the worst-case change due to
	using various  cross-section models in the extraction of
	$A_1 + \eta A_2$ from $A_\parallel$;
   c) central angle ($\theta$): uncertainty due to the location of the central
	angle of the spectrometers;
   d) energy calibration ($E^\prime$): uncertainty due to the spectrometer
	energy calibration;
   e) spectrometer resolution (Resol): the maximum difference obtained by
	varying the width of the hodoscope fingers by 20\% and re-running
	the Monte Carlo routine;
   f) polarization ($P_b P_t$): combined uncertainty in the beam and
	target polarizations;
   g) dilution factor ($f$): uncertainty based on the variations in the
	calculated dilution factor with various cross-section models and the
	stated uncertainty in the target composition;
   h) $R(x,Q^2)$ ($R$): uncertainty arising from lack of knowledge
	of $R = \sigma_L/\sigma_T$;
   i) no transverse data ($g_2$): uncertainty in the extraction of $g_1$ from
	$A_1 + \eta A_2$ due to the lack of knowledge about $A_2$.  This was 
	estimated using the maximum deviations in $g_1$ assuming $A_2=0$ and 
	$g_2=0$. Even if $A_2$ were as large as 0.3, the extracted values of 
	$g_1$ would shift by less than 0.014, which is small compared to the 
	statistical errors on each point.
By far the largest error comes from radiative and resolution  corrections.

\begin{figure}[h]
\centerline{\epsfig{file=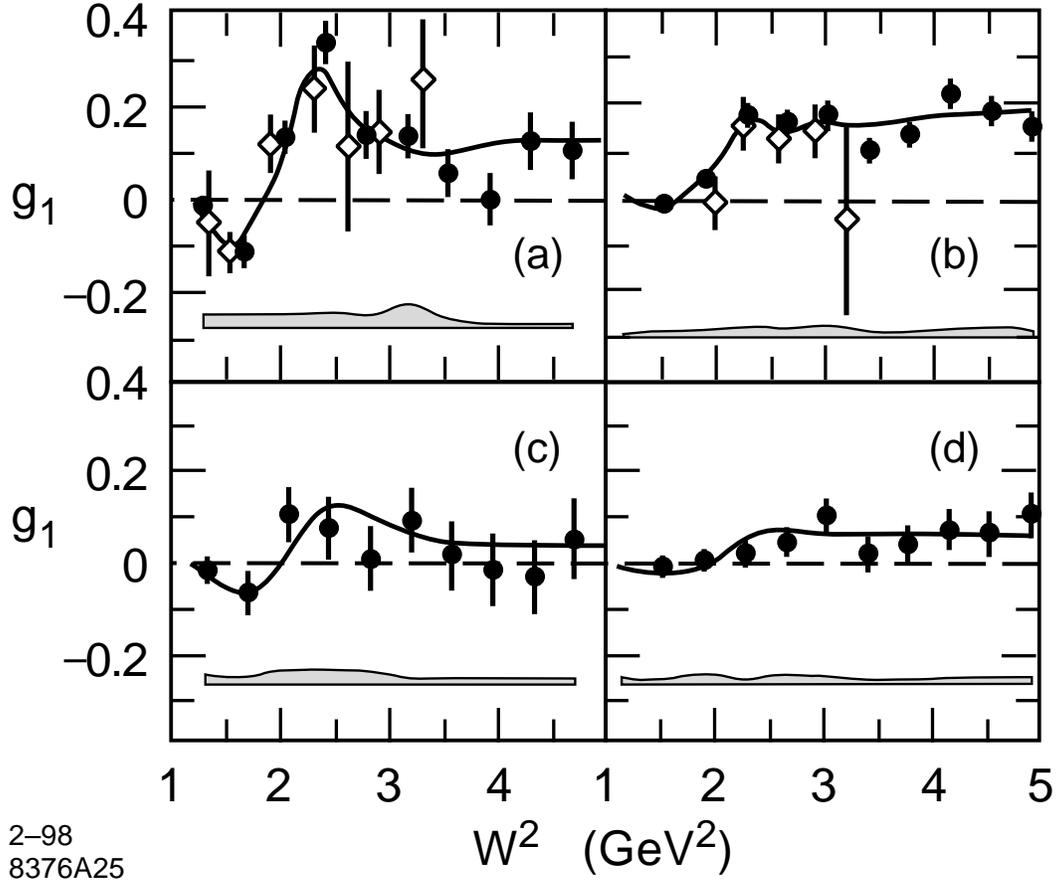,width=5.50 in}}
\vskip 0.7cm
\caption{
Measurements of $g_1(x,Q^2)$ in the resonance region as a function of $W^2$ 
for the proton at  (a) $4.5^\circ$
and (b) $7^\circ$; and for the deuteron at (c) $4.5^\circ$
and (d) $7^\circ$.  The present data (circles) are 
plotted together with the data of
Baum {\it et al.} \protect\cite{baum} (diamonds) and our Monte Carlo 
simulation (solid line).
The errors are indicated as in Fig.~\protect\ref{fig:res4}.}
\label{fig:res5}
\end{figure}

In addition to $A_1+\eta A_2$, Table~\ref{tbl:A1} also shows the results for 
$g_1$ for the resonance region. Figure~\ref{fig:res5} shows $g_1$ for proton and deuteron (per nucleon) measured with the two spectrometers 
as a function of $W^2$. The data of Baum {\it et al.}\cite{baum} are taken at similar kinematics and converted to $g_1$ for 
comparison by assuming $A_2=0$. Within errors, the two measurements agree well.  
Both data sets show a negative contribution in the region of the $\Delta(1232)$ 
resonance at $W^2$$\approx$1.5 GeV$^2$, and a strongly positive contribution
just above $W^2=2$ GeV$^2$ where the $S_{11}$ and $D_{13}$ resonances are 
important. This peak is less pronounced for the deuteron. The solid lines show 
the Monte Carlo simulation.

\begin{figure}[h]
\centerline{\epsfig{file=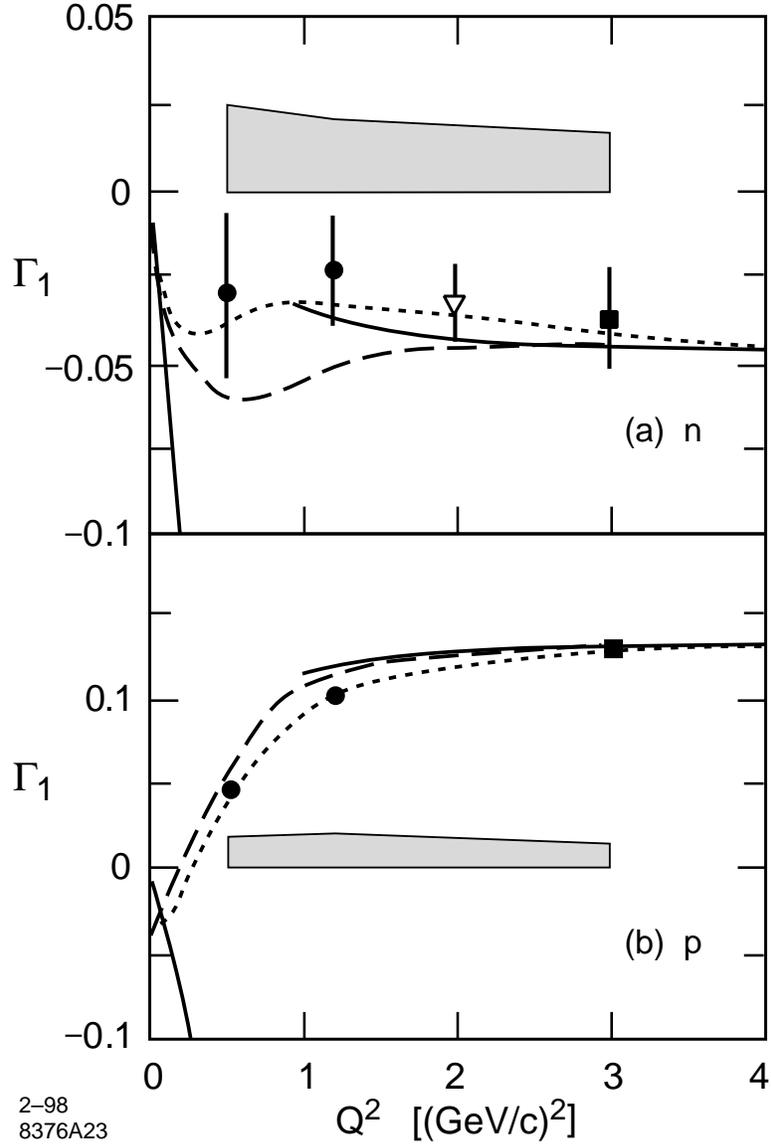,width=4 in}}
\vskip 0.7cm
\caption{
Integrals of $g_1(x,Q^2)$ at several fixed 
values of $Q^2$ for (a) the neutron
and (b) the proton.  The present data (circles) 
are plotted together with data from
E143 deep-inelastic results from this experiment
(squares) and E142\protect\cite{e142} (inverted triangle).
The curves correspond to the
evolution\protect\cite{larin}
of the deep-inelastic results due to changing $\alpha_s$ (solid),
the predictions of Burkert and Ioffe\protect\cite{BuIo} (short dash),
the model of Soffer\protect\cite{soffer} (long dash),
and the GDH approach to $Q^2 = 0$ (solid).
The error bars are statistical, and the shaded
bands correspond to the systematic errors.}
\label{fig:res6}
\end{figure}

Figure~\ref{fig:res6} shows the integrals $\Gamma_1(Q^2)$ for proton and
neutron, evaluated at the average $Q^2$ for the resonance region ($M^2<W^2<4$ 
GeV$^2$). We summed our resonance results directly (where $Q^2$ does not vary 
much) and then added a contribution from smaller $x$ (larger $W^2$) at the same 
fixed $Q^2$ by interpolating the 9.7 and 16 GeV data to the appropriate $Q^2$.
The neutron integrals were derived assuming a 5\% D-state probability for the
deuteron. The statistical errors assigned to the integral over the 
deep-inelastic region ($\Gamma_1^{\rm DIS}$) correspond to the weighted average 
of the statistical errors on the corresponding 9.7 and 16.2 GeV data points used 
in the interpolation. Systematic errors on the total integral $\Gamma_1^{\rm 
tot}$ were calculated using the systematic uncertainties for the measured $g_1$ 
in the resonance region added linearly to the systematic errors for the 
deep-inelastic region, which are highly correlated with each other. For the $ 
x<0.03$ extrapolation we simply took the overall parameterization of the data 
and integrated it from $x=0$ to $x=0.03$. Extrapolation errors for the region 
below the last measured datum at $x=0.03$ were taken to be as large as the
values themselves. Table~\ref{tbl:gamma} lists for each target the numerical 
values for the integrals in the resonance region alone ($\Gamma_1^{\rm res}$), 
in the deep-inelastic region ($\Gamma_1^{\rm DIS}$), for the low-$x$
extrapolation ($\Gamma_1^{\rm ext}$), and for the combined total 
($\Gamma_1^{\rm tot}$).

Although several models for the $Q^2$ evolution of $\Gamma_1(Q^2)$ exist 
\cite{jiU,BKM,LiLi,BuIo,soffer}, we show here only two representative ones, 
together with the evolution\cite{larin} of the world's deep-inelastic data due 
to the changing coupling constant $\alpha_S$. Although the GDH sum rule is 
strictly valid only at $Q^2=0$ where $\Gamma_1(Q^2)$ vanishes, it can be used 
to predict the slope of $\Gamma_1(Q^2)$ for small $Q^2$.  The solid line at low 
$Q^2$ shows $\Gamma_1 = -\kappa^2Q^2/8M^2$ in which $\kappa$ is the anomalous 
magnetic moment of either the proton or neutron. Burkert and Ioffe\cite{BuIo} 
consider the contributions from the resonances using the code AO, and the 
nonresonant contributions using a simple higher-twist-type form fitted to the 
deep-inelastic data. Their model is constrained to fit both the GDH and the 
deep-inelastic limits, and it describes the data quite well. Soffer and 
Teryaev\cite{soffer} assume that the integral over $g_1 + g_2$ varies smoothly 
from high $Q^2$ where $g_2\approx 0$ down to $Q^2=0$.  Using their simple 
prediction for this integral and subtracting the contribution from $g_2$ using 
the Burkhardt-Cottingham sum rule\cite{buco} gives the dashed curves in 
Fig.~\ref{fig:res6}, which also agree quite well with our data.

The present spin structure function data in the region of the nucleon 
resonances allow us to determine the integrals $\Gamma_1(Q^2)$ for the first 
time at $Q^2$ below 2~(GeV/c)$^2$. In contrast to the nearly flat behavior in 
the deep-inelastic region above $Q^2=2$~(GeV/c)$^2$, $\Gamma_1$ varies rapidly 
below $Q^2=2$~(GeV/c)$^2$. Models that interpolate between the deep-inelastic 
and GDH limits describe the data quite well in this non-perturbative regime.

\section{Conclusion}
 In summary, we have presented final results from SLAC Experiment E143 on the
spin structure functions $g_1$ and $g_2$ for proton and deuteron targets
covering a wide range of kinematics from the deep-inelastic to the resonance 
region. For deep-inelastic data the ratio $g_1/F_1$ is consistent with being 
independent of $Q^2$ for $Q^2 \geq 1$~(GeV/c)$^2$, but also consistent with pQCD NLO fits 
which show a weak  $Q^2$ dependence.  We have evaluated  the first moments of 
$g_1$, using a Regge form for the unmeasured low $x$ region. The 
Ellis-Jaffe sum rules are a function of the SU(3) parameters $F$ and $D$ and
the validity of the sum rules depend critically on the errors assigned to these 
parameters. We find $a_0^{inv} =0.33\pm0.06$, and in the parton model 
interpretation we find the average results: $\Delta u =0.84\pm0.02,$ $\Delta d =-0.42\pm0.02,$ and
$\Delta s =-0.09\pm0.02$ or $\pm0.05$ depending on $\delta F/D.$  Combined 
world  data are consistent with the Bjorken sum rule at the one standard
deviation level of 7\%.  Results for the twist-3 matrix element extracted
from the higher moments of $g_1$ and $g_2$ are consistent with  
calculations within the large errors. The resonance 
region data show the theoretically expected asymmetries at the $\Delta(1232)$ 
peak and larger than expected asymmetries (at least at low $Q^2$) in the region 
of the $S_{11}$ and $D_{13}$ resonances. The first moment of $g_1^p(Q^2)$ 
decreases with decreasing $Q^2$ at low $Q^2$ toward the GDH sum rule limit as 
predicted by several models. The asymmetry of pions is close to zero.

With the current round of experiments we now have good knowledge of the
distribution of quark spins for $x\geq 0.003.$  A complete understanding of the 
spin structure of the nucleon awaits experiments to measure directly the gluon 
spin distribution and to probe the quark spin distribution at lower $x.$

\newpage


 \begin{table}[h] 
 \caption{ Quark helicity predictions from the nonrelativistic 
     quark-parton model (NR QPM) where $\Delta G = 0$ and from the relativistic 
     quark-parton model (R QPM) \protect\cite{beyer,schlumpf} with $\Delta G = 0$ and 
  $\Delta G$($Q^2=1$ (GeV/c)$^2$) = $1.6\pm0.9$ \protect\cite{abfr}.}
 \label{tb:delqmod}
 \begin{tabular}{cccc}
            & NR QPM          & R QPM           &  R QPM + gluons\\ \hline
 $\alpha_s\Delta G/2\pi$ & 0               & 0               & $\phantom{-}0.13 \pm 0.08$ \\
 $\Delta u- {\alpha_s\Delta G/ 2\pi}$ & $\phantom{-}1.33$ & $\phantom{-}1.0$  & $\phantom{-}0.87\pm0.08$ \\
 $\Delta d- {\alpha_s\Delta G/ 2\pi}$ & $-0.33$           &  $-0.25$          & $-0.38\pm0.08$ \\
 $\Delta s- {\alpha_s\Delta G/ 2\pi}$ & 0               & 0               & $-0.13\pm0.08$ \\
 $a_0 = \Delta \Sigma- {3\alpha_s\Delta G/ 2\pi}$ & $\phantom{-}1.0$  & $\phantom{-}0.75$ & $\phantom{-}0.36\pm0.24$  
 \end{tabular}
 \end{table}

\begin{table}[h]
\caption[Beam heating results]{Beam heating correction results
at beam intensity of $4\times  10^9$ electrons/pulse.}
\label{tb:beamheat}
\begin{tabular}{ccccc}
Target        & NH$_3$       & NH$_3$      & ND$_3$      & ND$_3$       \\ \hline
Polarization  & Long.        & Tran.       & Long.       & Tran.        \\ 
$P_{i}$ (\%)  & $75\pm 1.9$    & $75\pm 1.9$   & $30\pm1.2$    & $30\pm 1.2$   \\ 
$P_{m}$ (\%)  & $68.3\pm 1.7$  & $68.5\pm1.7$  & $24.9\pm1.0$  & $24.8\pm 1.0$ \\ 
$z$           &$0.924\pm 0.029$ &$0.903\pm 0.033$ & $0.912\pm0.023$ &$0.931\pm 0.021$ \\ 
$P_T$ (\%)    &$67.7\pm 1.7$   &$67.8\pm 1.7$  &$24.4\pm 1.0$  &$24.4\pm 1.0$ \\ 
$C_{heat}$    &$0.0081\pm 0.0036$ &$0.0103\pm 0.0040$ &$0.0197\pm 0.0064$ &$0.0157\pm 0.0053$  \\ 
\end{tabular}
\end{table}


 \begin{table}[t] 
 \caption{Proton    $A_\parallel$ and $A_\perp$ results with statistical errors for E=29.1  GeV at the measured $Q^2$ in (GeV/c)$^2$.  Also shown are the radiative corrections $A_{rc}^\parallel$ and $A^\perp_{rc}$ which were applied to the data. }
\label{tb:APpar29}
 \begin{tabular}{rrrrrr}
$x$ & $<Q^2>$ &$A_\parallel$ & $A_{rc}^\parallel$ & $A_\perp$ & $A_{rc}^\perp$ \\
 \hline
\multicolumn{6}{c}{$\theta=4.5^\circ$} \\
\hline
 0.028&  1.17& $-0.026\pm$ 0.054& $ 0.014 $& $ 0.031\pm$ 0.063& $ 0.004 $\\
 0.031&  1.27& $ 0.048\pm$ 0.026& $ 0.014 $& $ 0.010\pm$ 0.032& $ 0.004 $\\
 0.035&  1.40& $ 0.091\pm$ 0.019& $ 0.013 $& $ 0.012\pm$ 0.024& $ 0.004 $\\
 0.039&  1.52& $ 0.060\pm$ 0.016& $ 0.012 $& $ 0.007\pm$ 0.020& $ 0.004 $\\
 0.044&  1.65& $ 0.076\pm$ 0.015& $ 0.011 $& $ 0.004\pm$ 0.018& $ 0.004 $\\
 0.049&  1.78& $ 0.083\pm$ 0.014& $ 0.010 $& $ 0.008\pm$ 0.017& $ 0.004 $\\
 0.056&  1.92& $ 0.082\pm$ 0.013& $ 0.009 $& $ 0.003\pm$ 0.016& $ 0.004 $\\
 0.063&  2.07& $ 0.082\pm$ 0.012& $ 0.008 $& $ 0.014\pm$ 0.015& $ 0.004 $\\
 0.071&  2.22& $ 0.086\pm$ 0.011& $ 0.007 $& $ 0.012\pm$ 0.014& $ 0.004 $\\
 0.079&  2.38& $ 0.102\pm$ 0.012& $ 0.006 $& $-0.009\pm$ 0.014& $ 0.004 $\\
 0.090&  2.53& $ 0.081\pm$ 0.012& $ 0.005 $& $-0.006\pm$ 0.014& $ 0.004 $\\
 0.101&  2.69& $ 0.114\pm$ 0.012& $ 0.004 $& $ 0.004\pm$ 0.014& $ 0.004 $\\
 0.113&  2.84& $ 0.108\pm$ 0.013& $ 0.004 $& $-0.012\pm$ 0.015& $ 0.004 $\\
 0.128&  3.00& $ 0.097\pm$ 0.013& $ 0.003 $& $ 0.010\pm$ 0.015& $ 0.004 $\\
 0.144&  3.15& $ 0.086\pm$ 0.013& $ 0.003 $& $-0.031\pm$ 0.015& $ 0.004 $\\
 0.162&  3.30& $ 0.113\pm$ 0.013& $ 0.003 $& $ 0.023\pm$ 0.016& $ 0.004 $\\
 0.182&  3.45& $ 0.110\pm$ 0.014& $ 0.002 $& $-0.021\pm$ 0.016& $ 0.004 $\\
 0.205&  3.59& $ 0.097\pm$ 0.014& $ 0.002 $& $ 0.043\pm$ 0.017& $ 0.004 $\\
 0.230&  3.73& $ 0.118\pm$ 0.015& $ 0.002 $& $-0.005\pm$ 0.018& $ 0.004 $\\
 0.259&  3.85& $ 0.107\pm$ 0.015& $ 0.002 $& $ 0.017\pm$ 0.019& $ 0.004 $\\
 0.292&  3.98& $ 0.096\pm$ 0.016& $ 0.002 $& $-0.055\pm$ 0.020& $ 0.004 $\\
 0.329&  4.09& $ 0.110\pm$ 0.018& $ 0.002 $& $-0.005\pm$ 0.022& $ 0.004 $\\
 0.370&  4.20& $ 0.080\pm$ 0.020& $ 0.002 $& $ 0.012\pm$ 0.024& $ 0.003 $\\
 0.416&  4.30& $ 0.140\pm$ 0.023& $ 0.002 $& $ 0.002\pm$ 0.028& $ 0.003 $\\
 0.468&  4.40& $ 0.140\pm$ 0.026& $ 0.002 $& $-0.048\pm$ 0.032& $ 0.003 $\\
 0.526&  4.47& $ 0.134\pm$ 0.031& $ 0.002 $& $-0.037\pm$ 0.038& $ 0.002 $\\
 0.592&  4.55& $ 0.066\pm$ 0.037& $ 0.003 $& $-0.029\pm$ 0.045& $ 0.002 $\\
 0.666&  4.63& $ 0.075\pm$ 0.045& $ 0.000 $& $-0.013\pm$ 0.055& $ 0.002 $\\
 0.749&  4.70& $ 0.128\pm$ 0.062& $-0.007 $& $-0.114\pm$ 0.074& $ 0.004 $\\
 \hline
\multicolumn{6}{c}{$\theta=7.0^\circ$} \\
\hline
 0.071&  2.91& $ 0.261\pm$ 0.095& $ 0.018 $& $-0.122\pm$ 0.102& $ 0.006 $\\
 0.079&  3.17& $ 0.159\pm$ 0.043& $ 0.015 $& $-0.002\pm$ 0.049& $ 0.006 $\\
 0.090&  3.48& $ 0.115\pm$ 0.029& $ 0.012 $& $ 0.040\pm$ 0.034& $ 0.006 $\\
 0.101&  3.79& $ 0.143\pm$ 0.024& $ 0.010 $& $-0.016\pm$ 0.027& $ 0.006 $\\
 0.113&  4.11& $ 0.158\pm$ 0.022& $ 0.008 $& $ 0.011\pm$ 0.025& $ 0.006 $\\
 0.128&  4.43& $ 0.164\pm$ 0.021& $ 0.006 $& $-0.026\pm$ 0.023& $ 0.006 $\\
 0.144&  4.78& $ 0.159\pm$ 0.020& $ 0.005 $& $ 0.007\pm$ 0.022& $ 0.006 $\\
 0.162&  5.13& $ 0.171\pm$ 0.019& $ 0.004 $& $ 0.018\pm$ 0.021& $ 0.006 $\\
 0.182&  5.49& $ 0.192\pm$ 0.019& $ 0.003 $& $ 0.052\pm$ 0.021& $ 0.006 $\\
 0.205&  5.86& $ 0.215\pm$ 0.019& $ 0.003 $& $ 0.012\pm$ 0.021& $ 0.006 $\\
 0.230&  6.24& $ 0.150\pm$ 0.019& $ 0.002 $& $ 0.007\pm$ 0.022& $ 0.006 $\\
 0.259&  6.60& $ 0.250\pm$ 0.020& $ 0.002 $& $ 0.001\pm$ 0.022& $ 0.006 $\\
 0.292&  6.97& $ 0.197\pm$ 0.021& $ 0.002 $& $-0.023\pm$ 0.023& $ 0.005 $\\
 0.329&  7.34& $ 0.195\pm$ 0.022& $ 0.002 $& $-0.012\pm$ 0.025& $ 0.005 $\\
 0.370&  7.69& $ 0.190\pm$ 0.024& $ 0.002 $& $-0.018\pm$ 0.027& $ 0.005 $\\
 0.416&  8.04& $ 0.244\pm$ 0.026& $ 0.002 $& $-0.009\pm$ 0.030& $ 0.004 $\\
 0.468&  8.37& $ 0.223\pm$ 0.030& $ 0.002 $& $ 0.005\pm$ 0.034& $ 0.004 $\\
 0.526&  8.68& $ 0.233\pm$ 0.034& $ 0.002 $& $-0.015\pm$ 0.039& $ 0.003 $\\
 0.592&  8.99& $ 0.224\pm$ 0.041& $ 0.002 $& $-0.011\pm$ 0.047& $ 0.003 $\\
 0.666&  9.26& $ 0.155\pm$ 0.051& $ 0.002 $& $-0.093\pm$ 0.063& $ 0.002 $\\
 0.749&  9.53& $ 0.223\pm$ 0.069& $ 0.005 $& $-0.097\pm$ 0.094& $ 0.001 $\\
\end{tabular}
 \end{table}
 \begin{table}[t] 
 \caption{Proton    $A_\parallel$ results with statistical errors for E=16.2 GeV  at the measured $Q^2$ in (GeV/c)$^2$.  Also shown are the radiative corrections $A_{rc}^\parallel$ which were applied to the data. }
\label{tb:APpar16}
 \begin{tabular}{rrrr}
$x$ & $<Q^2>$ &$A_\parallel$  & $A_{rc}^\parallel$              \\ 
 \hline
\multicolumn{4}{c}{$\theta=4.5^\circ$} \\
\hline
 0.022&  0.47& $ 0.023\pm$ 0.024& $ 0.005$ \\
 0.024&  0.51& $ 0.066\pm$ 0.019& $ 0.005$ \\
 0.027&  0.55& $ 0.058\pm$ 0.018& $ 0.005$ \\
 0.031&  0.59& $ 0.054\pm$ 0.016& $ 0.005$ \\
 0.035&  0.64& $ 0.036\pm$ 0.015& $ 0.005$ \\
 0.039&  0.68& $ 0.041\pm$ 0.014& $ 0.005$ \\
 0.044&  0.73& $ 0.051\pm$ 0.013& $ 0.005$ \\
 0.049&  0.78& $ 0.074\pm$ 0.012& $ 0.005$ \\
 0.056&  0.83& $ 0.052\pm$ 0.012& $ 0.004$ \\
 0.063&  0.88& $ 0.068\pm$ 0.012& $ 0.004$ \\
 0.071&  0.92& $ 0.043\pm$ 0.012& $ 0.004$ \\
 0.079&  0.97& $ 0.046\pm$ 0.012& $ 0.003$ \\
 0.090&  1.01& $ 0.062\pm$ 0.012& $ 0.003$ \\
 0.101&  1.06& $ 0.071\pm$ 0.012& $ 0.003$ \\
 0.113&  1.10& $ 0.049\pm$ 0.012& $ 0.003$ \\
 0.128&  1.14& $ 0.060\pm$ 0.012& $ 0.002$ \\
 0.144&  1.18& $ 0.056\pm$ 0.012& $ 0.002$ \\
 0.162&  1.22& $ 0.068\pm$ 0.013& $ 0.002$ \\
 0.182&  1.26& $ 0.051\pm$ 0.013& $ 0.002$ \\
 0.205&  1.29& $ 0.060\pm$ 0.013& $ 0.002$ \\
 0.230&  1.32& $ 0.047\pm$ 0.014& $ 0.002$ \\
 0.259&  1.35& $ 0.041\pm$ 0.014& $ 0.002$ \\
 0.292&  1.38& $ 0.055\pm$ 0.015& $ 0.002$ \\
 0.329&  1.40& $ 0.043\pm$ 0.016& $ 0.003$ \\
 0.370&  1.43& $ 0.055\pm$ 0.017& $ 0.003$ \\
 0.416&  1.45& $ 0.079\pm$ 0.019& $ 0.000$ \\
 0.468&  1.46& $ 0.094\pm$ 0.021& $ 0.003$ \\
 0.527&  1.48& $ 0.082\pm$ 0.024& $ 0.010$ \\
 0.593&  1.49& $ 0.089\pm$ 0.028& $ 0.004$ \\
 0.668&  1.51& $ 0.052\pm$ 0.031& $-0.009$ \\
 0.752&  1.52& $ 0.258\pm$ 0.449& $-0.029$ \\
 \hline
\multicolumn{4}{c}{$\theta=7.0^\circ$} \\
\hline
 0.044&  0.98& $ 0.072\pm$ 0.390& $ 0.012$ \\
 0.049&  1.06& $ 0.055\pm$ 0.078& $ 0.012$ \\
 0.056&  1.16& $ 0.091\pm$ 0.043& $ 0.011$ \\
 0.063&  1.26& $ 0.034\pm$ 0.031& $ 0.010$ \\
 0.071&  1.37& $ 0.092\pm$ 0.026& $ 0.009$ \\
 0.079&  1.47& $ 0.082\pm$ 0.023& $ 0.008$ \\
 0.090&  1.58& $ 0.113\pm$ 0.022& $ 0.007$ \\
 0.101&  1.69& $ 0.101\pm$ 0.021& $ 0.006$ \\
 0.113&  1.80& $ 0.108\pm$ 0.019& $ 0.005$ \\
 0.128&  1.91& $ 0.115\pm$ 0.018& $ 0.005$ \\
 0.144&  2.03& $ 0.120\pm$ 0.018& $ 0.004$ \\
 0.162&  2.14& $ 0.103\pm$ 0.017& $ 0.003$ \\
 0.182&  2.26& $ 0.105\pm$ 0.018& $ 0.003$ \\
 0.205&  2.36& $ 0.116\pm$ 0.018& $ 0.003$ \\
 0.230&  2.47& $ 0.136\pm$ 0.018& $ 0.002$ \\
 0.259&  2.57& $ 0.165\pm$ 0.019& $ 0.002$ \\
 0.292&  2.67& $ 0.159\pm$ 0.019& $ 0.002$ \\
 0.329&  2.76& $ 0.127\pm$ 0.020& $ 0.002$ \\
 0.370&  2.85& $ 0.157\pm$ 0.022& $ 0.002$ \\
 0.416&  2.94& $ 0.142\pm$ 0.023& $ 0.003$ \\
 0.468&  3.02& $ 0.150\pm$ 0.025& $ 0.003$ \\
 0.527&  3.08& $ 0.127\pm$ 0.029& $ 0.004$ \\
 0.593&  3.15& $ 0.070\pm$ 0.032& $-0.001$ \\
 0.668&  3.21& $ 0.136\pm$ 0.038& $ 0.003$ \\
 0.752&  3.27& $-0.280\pm$ 0.242& $ 0.009$ \\
\end{tabular}
 \end{table}
 \begin{table}[t] 
 \caption{Proton    $A_\parallel$ results with statistical errors for E= 9.7 GeV  at the measured $Q^2$ in (GeV/c)$^2$.  Also shown are the radiative corrections $A_{rc}^\parallel$ which were applied to the data. }
\label{tb:APpar9}
 \begin{tabular}{rrrr}
$x$ & $<Q^2>$ &$A_\parallel$  & $A_{rc}^\parallel$              \\ 
 \hline
\multicolumn{4}{c}{$\theta=4.5^\circ$} \\
\hline
 0.028&  0.28& $ 0.021\pm$ 0.046& $ 0.003$ \\
 0.031&  0.30& $ 0.021\pm$ 0.022& $ 0.002$ \\
 0.035&  0.31& $ 0.048\pm$ 0.019& $ 0.002$ \\
 0.039&  0.33& $ 0.011\pm$ 0.017& $ 0.002$ \\
 0.044&  0.35& $ 0.046\pm$ 0.016& $ 0.001$ \\
 0.049&  0.36& $ 0.043\pm$ 0.016& $ 0.001$ \\
 0.056&  0.38& $ 0.040\pm$ 0.015& $ 0.001$ \\
 0.063&  0.40& $ 0.026\pm$ 0.015& $ 0.001$ \\
 0.071&  0.41& $ 0.030\pm$ 0.015& $ 0.001$ \\
 0.080&  0.43& $ 0.032\pm$ 0.015& $ 0.001$ \\
 0.090&  0.44& $ 0.018\pm$ 0.014& $ 0.000$ \\
 0.101&  0.45& $ 0.024\pm$ 0.014& $ 0.000$ \\
 0.113&  0.47& $ 0.041\pm$ 0.013& $ 0.000$ \\
 0.128&  0.48& $ 0.002\pm$ 0.013& $ 0.000$ \\
 0.144&  0.49& $ 0.002\pm$ 0.013& $ 0.000$ \\
 0.162&  0.50& $ 0.018\pm$ 0.013& $ 0.000$ \\
 0.182&  0.51& $ 0.024\pm$ 0.013& $-0.002$ \\
 0.205&  0.52& $ 0.015\pm$ 0.013& $-0.008$ \\
 0.231&  0.53& $ 0.054\pm$ 0.014& $-0.004$ \\
 0.259&  0.53& $ 0.047\pm$ 0.014& $ 0.015$ \\
 0.292&  0.54& $ 0.051\pm$ 0.016& $ 0.014$ \\
 0.329&  0.55& $ 0.020\pm$ 0.018& $ 0.001$ \\
 0.370&  0.55& $-0.002\pm$ 0.019& $-0.007$ \\
 0.417&  0.56& $-0.004\pm$ 0.021& $-0.015$ \\
 0.469&  0.56& $-0.034\pm$ 0.021& $-0.020$ \\
 0.527&  0.57& $ 0.007\pm$ 0.029& $-0.034$ \\
 0.594&  0.57& $-0.013\pm$ 0.064& $-0.043$ \\
 0.669&  0.57& $ 0.010\pm$ 0.179& $-0.039$ \\
 0.753&  0.58& $ 0.021\pm$ 0.214& $-0.036$ \\
 0.847&  0.58& $ 0.031\pm$ 0.257& $-0.032$ \\
 \hline
\multicolumn{4}{c}{$\theta=7.0^\circ$} \\
\hline
 0.063&  0.60& $-0.013\pm$ 0.093& $ 0.003$ \\
 0.071&  0.64& $ 0.031\pm$ 0.038& $ 0.003$ \\
 0.080&  0.69& $ 0.076\pm$ 0.025& $ 0.003$ \\
 0.090&  0.74& $ 0.064\pm$ 0.019& $ 0.003$ \\
 0.101&  0.78& $ 0.057\pm$ 0.016& $ 0.003$ \\
 0.113&  0.82& $ 0.073\pm$ 0.015& $ 0.003$ \\
 0.128&  0.86& $ 0.048\pm$ 0.015& $ 0.003$ \\
 0.144&  0.90& $ 0.047\pm$ 0.014& $ 0.002$ \\
 0.162&  0.93& $ 0.065\pm$ 0.013& $ 0.002$ \\
 0.182&  0.97& $ 0.087\pm$ 0.013& $ 0.002$ \\
 0.205&  1.00& $ 0.068\pm$ 0.013& $ 0.002$ \\
 0.231&  1.03& $ 0.084\pm$ 0.013& $ 0.002$ \\
 0.259&  1.06& $ 0.070\pm$ 0.013& $ 0.003$ \\
 0.292&  1.09& $ 0.055\pm$ 0.013& $ 0.004$ \\
 0.329&  1.12& $ 0.088\pm$ 0.014& $ 0.005$ \\
 0.370&  1.14& $ 0.088\pm$ 0.014& $-0.004$ \\
 0.417&  1.16& $ 0.082\pm$ 0.016& $ 0.001$ \\
 0.469&  1.18& $ 0.084\pm$ 0.016& $ 0.018$ \\
 0.527&  1.21& $ 0.086\pm$ 0.018& $ 0.010$ \\
 0.594&  1.22& $-0.001\pm$ 0.018& $-0.004$ \\
 0.669&  1.23& $ 0.022\pm$ 0.018& $-0.026$ \\
 0.753&  1.25& $-0.044\pm$ 0.047& $-0.061$ \\
 0.847&  1.26& $-0.018\pm$ 0.149& $-0.064$ \\
\end{tabular}
 \end{table}
 \begin{table}[t] 
 \caption{Deuteron  $A_\parallel$ and $A_\perp$ results with statistical errors for E=29.1  GeV at the measured $Q^2$ in (GeV/c)$^2$.  Also shown are the radiative corrections $A_{rc}^\parallel$ and $A^\perp_{rc}$ which were applied to the data. }
\label{tb:ADpar29}
 \begin{tabular}{rrrrrr}
$x$ & $<Q^2>$ &$A_\parallel$ & $A_{rc}^\parallel$ & $A_\perp$ & $A_{rc}^\perp$ \\
 \hline
\multicolumn{6}{c}{$\theta=4.5^\circ$} \\
\hline
 0.028&  1.17& $-0.042\pm$ 0.075& $-0.004 $& $-0.138\pm$ 0.175& $ 0.002 $\\
 0.031&  1.27& $ 0.035\pm$ 0.030& $-0.004 $& $-0.114\pm$ 0.146& $ 0.002 $\\
 0.035&  1.40& $ 0.004\pm$ 0.021& $-0.004 $& $ 0.060\pm$ 0.083& $ 0.002 $\\
 0.039&  1.52& $ 0.043\pm$ 0.019& $-0.004 $& $ 0.038\pm$ 0.046& $ 0.002 $\\
 0.044&  1.65& $-0.011\pm$ 0.017& $-0.004 $& $ 0.050\pm$ 0.038& $ 0.002 $\\
 0.049&  1.78& $-0.009\pm$ 0.016& $-0.004 $& $ 0.022\pm$ 0.034& $ 0.002 $\\
 0.056&  1.92& $ 0.012\pm$ 0.015& $-0.003 $& $-0.043\pm$ 0.031& $ 0.002 $\\
 0.063&  2.07& $ 0.010\pm$ 0.014& $-0.003 $& $-0.040\pm$ 0.029& $ 0.002 $\\
 0.071&  2.22& $ 0.014\pm$ 0.013& $-0.003 $& $ 0.009\pm$ 0.028& $ 0.002 $\\
 0.079&  2.38& $ 0.023\pm$ 0.013& $-0.003 $& $-0.009\pm$ 0.029& $ 0.002 $\\
 0.090&  2.53& $ 0.038\pm$ 0.014& $-0.003 $& $ 0.032\pm$ 0.029& $ 0.002 $\\
 0.101&  2.69& $ 0.028\pm$ 0.014& $-0.003 $& $-0.031\pm$ 0.030& $ 0.002 $\\
 0.113&  2.84& $ 0.037\pm$ 0.015& $-0.002 $& $ 0.012\pm$ 0.031& $ 0.002 $\\
 0.128&  3.00& $ 0.079\pm$ 0.015& $-0.002 $& $-0.001\pm$ 0.032& $ 0.002 $\\
 0.144&  3.15& $ 0.053\pm$ 0.016& $-0.002 $& $-0.023\pm$ 0.033& $ 0.002 $\\
 0.162&  3.30& $ 0.046\pm$ 0.016& $-0.001 $& $ 0.043\pm$ 0.035& $ 0.002 $\\
 0.182&  3.45& $ 0.054\pm$ 0.017& $-0.001 $& $-0.042\pm$ 0.036& $ 0.002 $\\
 0.205&  3.59& $ 0.049\pm$ 0.017& $ 0.000 $& $-0.013\pm$ 0.038& $ 0.002 $\\
 0.230&  3.73& $ 0.020\pm$ 0.019& $ 0.000 $& $ 0.008\pm$ 0.040& $ 0.002 $\\
 0.259&  3.85& $ 0.021\pm$ 0.020& $ 0.000 $& $ 0.095\pm$ 0.043& $ 0.002 $\\
 0.292&  3.98& $ 0.054\pm$ 0.021& $ 0.001 $& $-0.016\pm$ 0.047& $ 0.002 $\\
 0.329&  4.09& $ 0.078\pm$ 0.023& $ 0.001 $& $-0.025\pm$ 0.052& $ 0.002 $\\
 0.370&  4.20& $ 0.072\pm$ 0.026& $ 0.001 $& $-0.026\pm$ 0.059& $ 0.002 $\\
 0.416&  4.30& $ 0.063\pm$ 0.030& $ 0.001 $& $-0.050\pm$ 0.068& $ 0.001 $\\
 0.468&  4.40& $ 0.010\pm$ 0.036& $ 0.001 $& $-0.071\pm$ 0.081& $ 0.001 $\\
 0.526&  4.47& $ 0.065\pm$ 0.043& $ 0.001 $& $ 0.077\pm$ 0.098& $ 0.001 $\\
 0.592&  4.55& $ 0.057\pm$ 0.052& $ 0.001 $& $-0.116\pm$ 0.120& $ 0.001 $\\
 0.666&  4.62& $ 0.023\pm$ 0.066& $-0.002 $& $ 0.160\pm$ 0.150& $ 0.002 $\\
 0.749&  4.70& $-0.190\pm$ 0.091& $-0.006 $& $ 0.150\pm$ 0.203& $ 0.004 $\\
 \hline
\multicolumn{6}{c}{$\theta=7.0^\circ$} \\
\hline
 0.071&  2.91& $ 0.044\pm$ 0.108& $-0.004 $& $ 0.088\pm$ 0.237& $ 0.003 $\\
 0.079&  3.17& $-0.020\pm$ 0.049& $-0.004 $& $ 0.081\pm$ 0.094& $ 0.003 $\\
 0.090&  3.48& $ 0.044\pm$ 0.033& $-0.004 $& $-0.024\pm$ 0.062& $ 0.003 $\\
 0.101&  3.79& $ 0.044\pm$ 0.027& $-0.004 $& $ 0.075\pm$ 0.050& $ 0.003 $\\
 0.113&  4.11& $ 0.023\pm$ 0.025& $-0.004 $& $-0.036\pm$ 0.046& $ 0.003 $\\
 0.128&  4.44& $ 0.053\pm$ 0.023& $-0.003 $& $-0.014\pm$ 0.043& $ 0.003 $\\
 0.144&  4.78& $ 0.110\pm$ 0.023& $-0.003 $& $-0.038\pm$ 0.041& $ 0.003 $\\
 0.162&  5.13& $ 0.051\pm$ 0.022& $-0.003 $& $-0.027\pm$ 0.041& $ 0.003 $\\
 0.182&  5.49& $ 0.133\pm$ 0.022& $-0.002 $& $ 0.035\pm$ 0.040& $ 0.003 $\\
 0.205&  5.86& $ 0.067\pm$ 0.022& $-0.002 $& $-0.085\pm$ 0.041& $ 0.003 $\\
 0.230&  6.23& $ 0.088\pm$ 0.023& $-0.001 $& $-0.056\pm$ 0.043& $ 0.003 $\\
 0.259&  6.60& $ 0.051\pm$ 0.024& $ 0.000 $& $ 0.065\pm$ 0.045& $ 0.003 $\\
 0.292&  6.97& $ 0.102\pm$ 0.026& $ 0.000 $& $-0.028\pm$ 0.048& $ 0.003 $\\
 0.329&  7.33& $ 0.108\pm$ 0.028& $ 0.001 $& $-0.034\pm$ 0.052& $ 0.002 $\\
 0.370&  7.69& $ 0.143\pm$ 0.030& $ 0.001 $& $-0.024\pm$ 0.057& $ 0.002 $\\
 0.416&  8.03& $ 0.089\pm$ 0.034& $ 0.001 $& $-0.018\pm$ 0.064& $ 0.002 $\\
 0.468&  8.37& $ 0.125\pm$ 0.039& $ 0.000 $& $-0.003\pm$ 0.073& $ 0.002 $\\
 0.526&  8.67& $ 0.172\pm$ 0.046& $ 0.000 $& $-0.023\pm$ 0.086& $ 0.002 $\\
 0.592&  8.98& $ 0.094\pm$ 0.056& $-0.001 $& $ 0.235\pm$ 0.108& $ 0.002 $\\
 0.666&  9.26& $ 0.086\pm$ 0.070& $-0.001 $& $-0.125\pm$ 0.147& $ 0.002 $\\
 0.749&  9.52& $ 0.193\pm$ 0.096& $ 0.000 $& $-0.068\pm$ 0.212& $ 0.002 $\\
\end{tabular}
 \end{table}
 \begin{table}[t] 
 \caption{Deuteron  $A_\parallel$ results with statistical errors for E=16.2 GeV  at the measured $Q^2$ in (GeV/c)$^2$.  Also shown are the radiative corrections $A_{rc}^\parallel$ which were applied to the data. }
\label{tb:ADpar16}
 \begin{tabular}{rrrr}
$x$ & $<Q^2>$ &$A_\parallel$  & $A_{rc}^\parallel$              \\ 
 \hline
\multicolumn{4}{c}{$\theta=4.5^\circ$} \\
\hline
 0.022&  0.47& $-0.018\pm$ 0.064& $-0.006$ \\
 0.024&  0.51& $-0.035\pm$ 0.049& $-0.006$ \\
 0.027&  0.55& $-0.012\pm$ 0.045& $-0.005$ \\
 0.031&  0.59& $ 0.012\pm$ 0.041& $-0.005$ \\
 0.035&  0.64& $-0.009\pm$ 0.038& $-0.004$ \\
 0.039&  0.68& $ 0.056\pm$ 0.034& $-0.004$ \\
 0.044&  0.73& $-0.021\pm$ 0.028& $-0.004$ \\
 0.049&  0.78& $ 0.046\pm$ 0.026& $-0.003$ \\
 0.056&  0.83& $ 0.003\pm$ 0.025& $-0.003$ \\
 0.063&  0.87& $ 0.021\pm$ 0.025& $-0.003$ \\
 0.071&  0.92& $ 0.026\pm$ 0.025& $-0.003$ \\
 0.079&  0.97& $-0.016\pm$ 0.025& $-0.002$ \\
 0.090&  1.01& $ 0.038\pm$ 0.025& $-0.002$ \\
 0.101&  1.06& $ 0.019\pm$ 0.025& $-0.002$ \\
 0.113&  1.10& $ 0.016\pm$ 0.025& $-0.001$ \\
 0.128&  1.14& $ 0.063\pm$ 0.025& $-0.001$ \\
 0.144&  1.18& $ 0.050\pm$ 0.025& $-0.001$ \\
 0.162&  1.22& $-0.025\pm$ 0.025& $-0.001$ \\
 0.182&  1.25& $ 0.043\pm$ 0.026& $ 0.000$ \\
 0.205&  1.29& $ 0.042\pm$ 0.027& $ 0.000$ \\
 0.230&  1.32& $ 0.016\pm$ 0.027& $ 0.000$ \\
 0.259&  1.35& $ 0.022\pm$ 0.029& $ 0.001$ \\
 0.292&  1.37& $ 0.062\pm$ 0.031& $ 0.001$ \\
 0.329&  1.40& $ 0.023\pm$ 0.033& $ 0.001$ \\
 0.370&  1.42& $-0.031\pm$ 0.036& $ 0.001$ \\
 0.416&  1.44& $ 0.013\pm$ 0.040& $-0.001$ \\
 0.468&  1.46& $ 0.089\pm$ 0.045& $ 0.001$ \\
 0.527&  1.48& $ 0.014\pm$ 0.053& $ 0.003$ \\
 0.593&  1.49& $ 0.071\pm$ 0.063& $-0.002$ \\
 0.668&  1.50& $ 0.037\pm$ 0.076& $-0.013$ \\
 \hline
\multicolumn{4}{c}{$\theta=7.0^\circ$} \\
\hline
 0.049&  1.06& $-0.075\pm$ 0.201& $-0.005$ \\
 0.056&  1.16& $ 0.062\pm$ 0.106& $-0.005$ \\
 0.063&  1.26& $-0.003\pm$ 0.075& $-0.004$ \\
 0.071&  1.36& $ 0.054\pm$ 0.060& $-0.004$ \\
 0.079&  1.47& $-0.073\pm$ 0.054& $-0.004$ \\
 0.090&  1.58& $ 0.121\pm$ 0.049& $-0.004$ \\
 0.101&  1.69& $ 0.054\pm$ 0.046& $-0.003$ \\
 0.113&  1.80& $ 0.027\pm$ 0.041& $-0.003$ \\
 0.128&  1.91& $ 0.018\pm$ 0.038& $-0.002$ \\
 0.144&  2.03& $ 0.039\pm$ 0.036& $-0.002$ \\
 0.162&  2.14& $ 0.060\pm$ 0.036& $-0.002$ \\
 0.182&  2.25& $ 0.097\pm$ 0.036& $-0.001$ \\
 0.205&  2.36& $ 0.045\pm$ 0.036& $-0.001$ \\
 0.230&  2.47& $ 0.041\pm$ 0.038& $ 0.000$ \\
 0.259&  2.57& $ 0.041\pm$ 0.039& $ 0.000$ \\
 0.292&  2.67& $ 0.115\pm$ 0.040& $ 0.001$ \\
 0.329&  2.76& $ 0.160\pm$ 0.042& $ 0.001$ \\
 0.370&  2.85& $ 0.039\pm$ 0.045& $ 0.001$ \\
 0.416&  2.93& $ 0.055\pm$ 0.049& $ 0.001$ \\
 0.468&  3.01& $ 0.168\pm$ 0.054& $ 0.001$ \\
 0.527&  3.08& $ 0.139\pm$ 0.062& $ 0.001$ \\
 0.593&  3.15& $ 0.055\pm$ 0.073& $-0.002$ \\
 0.668&  3.21& $-0.013\pm$ 0.088& $-0.001$ \\
\end{tabular}
 \end{table}
 \begin{table}[t] 
 \caption{Deuteron  $A_\parallel$ results with statistical errors for E= 9.7 GeV  at the measured $Q^2$ in (GeV/c)$^2$.  Also shown are the radiative corrections $A_{rc}^\parallel$ which were applied to the data. }
\label{tb:ADpar9}
 \begin{tabular}{rrrr}
$x$ & $<Q^2>$ &$A_\parallel$  & $A_{rc}^\parallel$              \\ 
 \hline
\multicolumn{4}{c}{$\theta=4.5^\circ$} \\
\hline
 0.028&  0.28& $ 0.053\pm$ 0.069& $-0.005$ \\
 0.031&  0.30& $-0.067\pm$ 0.034& $-0.005$ \\
 0.035&  0.31& $-0.014\pm$ 0.028& $-0.004$ \\
 0.039&  0.33& $ 0.017\pm$ 0.025& $-0.004$ \\
 0.044&  0.35& $ 0.001\pm$ 0.024& $-0.004$ \\
 0.050&  0.36& $ 0.022\pm$ 0.023& $-0.004$ \\
 0.056&  0.38& $-0.017\pm$ 0.022& $-0.003$ \\
 0.063&  0.40& $-0.002\pm$ 0.021& $-0.003$ \\
 0.071&  0.41& $ 0.032\pm$ 0.022& $-0.003$ \\
 0.080&  0.43& $-0.004\pm$ 0.022& $-0.003$ \\
 0.090&  0.44& $ 0.008\pm$ 0.021& $-0.002$ \\
 0.101&  0.45& $ 0.038\pm$ 0.020& $-0.002$ \\
 0.113&  0.47& $-0.002\pm$ 0.019& $-0.002$ \\
 0.128&  0.48& $-0.017\pm$ 0.019& $-0.002$ \\
 0.144&  0.49& $-0.004\pm$ 0.019& $-0.002$ \\
 0.162&  0.50& $ 0.015\pm$ 0.019& $-0.002$ \\
 0.182&  0.51& $ 0.028\pm$ 0.019& $-0.002$ \\
 0.205&  0.52& $ 0.012\pm$ 0.020& $-0.005$ \\
 0.231&  0.53& $ 0.016\pm$ 0.021& $-0.003$ \\
 0.259&  0.53& $ 0.009\pm$ 0.022& $ 0.005$ \\
 0.292&  0.54& $ 0.028\pm$ 0.024& $ 0.005$ \\
 0.329&  0.55& $ 0.031\pm$ 0.026& $-0.004$ \\
 0.370&  0.55& $-0.032\pm$ 0.027& $-0.009$ \\
 0.417&  0.56& $ 0.029\pm$ 0.030& $-0.014$ \\
 0.469&  0.56& $-0.039\pm$ 0.032& $-0.019$ \\
 0.527&  0.57& $-0.005\pm$ 0.045& $-0.029$ \\
 0.594&  0.57& $-0.008\pm$ 0.074& $-0.036$ \\
 0.669&  0.57& $-0.021\pm$ 0.110& $-0.039$ \\
 0.753&  0.58& $ 0.001\pm$ 0.117& $-0.042$ \\
 0.847&  0.58& $ 0.010\pm$ 0.142& $-0.046$ \\
 \hline
\multicolumn{4}{c}{$\theta=7.0^\circ$} \\
\hline
 0.063&  0.60& $ 0.116\pm$ 0.133& $-0.005$ \\
 0.071&  0.64& $-0.036\pm$ 0.054& $-0.005$ \\
 0.080&  0.69& $ 0.013\pm$ 0.037& $-0.004$ \\
 0.090&  0.74& $-0.013\pm$ 0.028& $-0.003$ \\
 0.101&  0.78& $ 0.016\pm$ 0.024& $-0.003$ \\
 0.113&  0.82& $ 0.024\pm$ 0.022& $-0.003$ \\
 0.128&  0.86& $ 0.052\pm$ 0.021& $-0.002$ \\
 0.144&  0.90& $ 0.012\pm$ 0.020& $-0.002$ \\
 0.162&  0.93& $ 0.036\pm$ 0.020& $-0.001$ \\
 0.182&  0.97& $ 0.027\pm$ 0.019& $-0.001$ \\
 0.205&  1.00& $ 0.005\pm$ 0.019& $ 0.000$ \\
 0.231&  1.03& $ 0.044\pm$ 0.019& $ 0.000$ \\
 0.259&  1.06& $ 0.019\pm$ 0.020& $ 0.001$ \\
 0.292&  1.09& $ 0.021\pm$ 0.020& $ 0.002$ \\
 0.329&  1.12& $ 0.034\pm$ 0.021& $ 0.002$ \\
 0.370&  1.14& $ 0.036\pm$ 0.022& $-0.003$ \\
 0.417&  1.16& $ 0.068\pm$ 0.023& $-0.001$ \\
 0.469&  1.19& $ 0.020\pm$ 0.025& $ 0.007$ \\
 0.527&  1.21& $ 0.024\pm$ 0.029& $ 0.001$ \\
 0.594&  1.22& $ 0.022\pm$ 0.033& $-0.011$ \\
 0.669&  1.24& $ 0.028\pm$ 0.039& $-0.029$ \\
 0.753&  1.25& $ 0.000\pm$ 0.068& $-0.054$ \\
 0.847&  1.26& $-0.018\pm$ 0.118& $-0.066$ \\
\end{tabular}
 \end{table}


 \begin{table}[t] 
 \caption{Results for $g_1/F_1$ in the DIS region ($W^2\geq 4 {\rm ~GeV^2}$). There is an additional normalization uncertainty due to beam and target polarization shown in Table~ \protect\ref{tb:norm_err}.  }
 \label{tb:g1f1}
 \begin{tabular}{rrrrrr}
$x$ & $<Q^2>$ & E(GeV) & $g_1^p/F_1^p\pm {\rm stat} \pm {\rm syst}$ & $g_1^d/F_1^d\pm {\rm stat} \pm {\rm syst}$ & $g_1^n/F_1^n\pm {\rm stat} \pm {\rm syst}$  \\
 \hline
 0.024 &  0.51 &  16.2 & $ 0.092\pm 0.027\pm 0.014$ & $-0.048\pm 0.068\pm 0.009$ & $-0.205\pm 0.153\pm 0.023$ \\
 0.027 &  0.55 &  16.2 & $ 0.086\pm 0.026\pm 0.013$ & $-0.018\pm 0.065\pm 0.009$ & $-0.131\pm 0.146\pm 0.022$ \\
 0.027 &  1.17 &  29.1 & $-0.032\pm 0.068\pm 0.010$ & $-0.058\pm 0.092\pm 0.009$ & $-0.099\pm 0.225\pm 0.020$ \\
 0.031 &  0.59 &  16.2 & $ 0.084\pm 0.026\pm 0.012$ & $ 0.019\pm 0.065\pm 0.008$ & $-0.048\pm 0.144\pm 0.020$ \\
 0.031 &  1.27 &  29.1 & $ 0.064\pm 0.034\pm 0.009$ & $ 0.041\pm 0.040\pm 0.008$ & $ 0.021\pm 0.095\pm 0.019$ \\
 0.035 &  0.31 &   9.7 & $ 0.096\pm 0.037\pm 0.018$ & $-0.027\pm 0.055\pm 0.008$ & $-0.161\pm 0.125\pm 0.024$ \\
 0.035 &  0.64 &  16.2 & $ 0.059\pm 0.024\pm 0.011$ & $-0.015\pm 0.062\pm 0.007$ & $-0.097\pm 0.139\pm 0.018$ \\
 0.035 &  1.40 &  29.1 & $ 0.123\pm 0.025\pm 0.008$ & $ 0.008\pm 0.029\pm 0.007$ & $-0.115\pm 0.069\pm 0.017$ \\
 0.039 &  0.33 &   9.7 & $ 0.024\pm 0.037\pm 0.018$ & $ 0.035\pm 0.053\pm 0.008$ & $ 0.050\pm 0.122\pm 0.024$ \\
 0.039 &  0.68 &  16.2 & $ 0.072\pm 0.024\pm 0.011$ & $ 0.099\pm 0.060\pm 0.007$ & $ 0.140\pm 0.136\pm 0.017$ \\
 0.039 &  1.52 &  29.1 & $ 0.083\pm 0.023\pm 0.008$ & $ 0.062\pm 0.026\pm 0.007$ & $ 0.049\pm 0.063\pm 0.016$ \\
 0.044 &  0.35 &   9.7 & $ 0.106\pm 0.037\pm 0.019$ & $ 0.003\pm 0.054\pm 0.008$ & $-0.109\pm 0.126\pm 0.024$ \\
 0.044 &  0.73 &  16.2 & $ 0.096\pm 0.023\pm 0.011$ & $-0.038\pm 0.052\pm 0.006$ & $-0.191\pm 0.120\pm 0.017$ \\
 0.044 &  0.98 &  16.2 & $ 0.097\pm 0.520\pm 0.010$ & $-0.943\pm 1.296\pm 0.008$ & $-2.222\pm 2.964\pm 0.019$ \\
 0.044 &  1.65 &  29.1 & $ 0.110\pm 0.021\pm 0.008$ & $-0.013\pm 0.025\pm 0.006$ & $-0.150\pm 0.060\pm 0.014$ \\
 0.049 &  0.36 &   9.7 & $ 0.108\pm 0.039\pm 0.020$ & $ 0.055\pm 0.057\pm 0.008$ & $ 0.003\pm 0.134\pm 0.025$ \\
 0.049 &  0.78 &  16.2 & $ 0.147\pm 0.024\pm 0.011$ & $ 0.092\pm 0.052\pm 0.006$ & $ 0.045\pm 0.121\pm 0.016$ \\
 0.049 &  1.06 &  16.2 & $ 0.078\pm 0.109\pm 0.010$ & $-0.104\pm 0.280\pm 0.008$ & $-0.321\pm 0.646\pm 0.018$ \\
 0.049 &  1.78 &  29.1 & $ 0.125\pm 0.020\pm 0.008$ & $-0.013\pm 0.024\pm 0.005$ & $-0.168\pm 0.058\pm 0.013$ \\
 0.056 &  0.38 &   9.7 & $ 0.108\pm 0.041\pm 0.021$ & $-0.045\pm 0.059\pm 0.008$ & $-0.221\pm 0.141\pm 0.027$ \\
 0.056 &  0.57 &   9.7 & $-0.214\pm 1.891\pm 0.014$ & $ 2.992\pm 2.465\pm 0.007$ & $ 7.028\pm 5.977\pm 0.020$ \\
 0.056 &  0.83 &  16.2 & $ 0.110\pm 0.025\pm 0.011$ & $ 0.007\pm 0.054\pm 0.006$ & $-0.106\pm 0.127\pm 0.016$ \\
 0.056 &  1.16 &  16.2 & $ 0.132\pm 0.063\pm 0.010$ & $ 0.091\pm 0.153\pm 0.007$ & $ 0.060\pm 0.357\pm 0.017$ \\
 0.056 &  1.92 &  29.1 & $ 0.130\pm 0.020\pm 0.008$ & $ 0.016\pm 0.023\pm 0.005$ & $-0.109\pm 0.058\pm 0.012$ \\
 0.063 &  0.40 &   9.7 & $ 0.078\pm 0.043\pm 0.022$ & $-0.006\pm 0.062\pm 0.009$ & $-0.101\pm 0.149\pm 0.029$ \\
 0.063 &  0.60 &   9.7 & $-0.024\pm 0.177\pm 0.014$ & $ 0.221\pm 0.252\pm 0.007$ & $ 0.531\pm 0.609\pm 0.019$ \\
 0.063 &  0.87 &  16.2 & $ 0.157\pm 0.027\pm 0.012$ & $ 0.049\pm 0.057\pm 0.005$ & $-0.064\pm 0.135\pm 0.016$ \\
 0.063 &  1.26 &  16.2 & $ 0.052\pm 0.046\pm 0.009$ & $-0.004\pm 0.112\pm 0.006$ & $-0.068\pm 0.262\pm 0.015$ \\
 0.063 &  2.07 &  29.1 & $ 0.138\pm 0.020\pm 0.008$ & $ 0.014\pm 0.023\pm 0.004$ & $-0.123\pm 0.057\pm 0.011$ \\
 0.063 &  2.69 &  29.1 & $ 1.138\pm 0.833\pm 0.009$ & $-0.446\pm 2.060\pm 0.006$ & $-2.318\pm 4.857\pm 0.016$ \\
 0.071 &  0.41 &   9.7 & $ 0.096\pm 0.049\pm 0.023$ & $ 0.104\pm 0.070\pm 0.011$ & $ 0.129\pm 0.171\pm 0.032$ \\
 0.071 &  0.64 &   9.7 & $ 0.064\pm 0.077\pm 0.014$ & $-0.072\pm 0.109\pm 0.006$ & $-0.240\pm 0.267\pm 0.019$ \\
 0.071 &  0.92 &  16.2 & $ 0.107\pm 0.029\pm 0.012$ & $ 0.064\pm 0.061\pm 0.005$ & $ 0.026\pm 0.145\pm 0.016$ \\
 0.071 &  1.36 &  16.2 & $ 0.144\pm 0.041\pm 0.009$ & $ 0.085\pm 0.095\pm 0.005$ & $ 0.033\pm 0.225\pm 0.014$ \\
 0.071 &  2.22 &  29.1 & $ 0.150\pm 0.020\pm 0.008$ & $ 0.025\pm 0.023\pm 0.004$ & $-0.114\pm 0.059\pm 0.011$ \\
 0.071 &  2.91 &  29.1 & $ 0.324\pm 0.121\pm 0.010$ & $ 0.063\pm 0.139\pm 0.006$ & $-0.223\pm 0.353\pm 0.015$ \\
 0.079 &  0.43 &   9.7 & $ 0.113\pm 0.053\pm 0.025$ & $-0.014\pm 0.075\pm 0.014$ & $-0.161\pm 0.185\pm 0.038$ \\
 0.079 &  0.69 &   9.7 & $ 0.164\pm 0.053\pm 0.014$ & $ 0.029\pm 0.078\pm 0.006$ & $-0.121\pm 0.194\pm 0.019$ \\
 0.079 &  0.97 &  16.2 & $ 0.122\pm 0.031\pm 0.013$ & $-0.043\pm 0.066\pm 0.005$ & $-0.242\pm 0.159\pm 0.016$ \\
 0.079 &  1.47 &  16.2 & $ 0.135\pm 0.039\pm 0.009$ & $-0.119\pm 0.088\pm 0.005$ & $-0.436\pm 0.212\pm 0.013$ \\
 0.079 &  2.38 &  29.1 & $ 0.188\pm 0.021\pm 0.009$ & $ 0.041\pm 0.025\pm 0.004$ & $-0.121\pm 0.063\pm 0.012$ \\
 0.079 &  3.17 &  29.1 & $ 0.209\pm 0.057\pm 0.010$ & $-0.020\pm 0.065\pm 0.005$ & $-0.289\pm 0.167\pm 0.014$ \\
 0.090 &  0.44 &   9.7 & $ 0.069\pm 0.055\pm 0.027$ & $ 0.031\pm 0.078\pm 0.018$ & $-0.008\pm 0.193\pm 0.045$ \\
 0.090 &  0.74 &   9.7 & $ 0.145\pm 0.043\pm 0.014$ & $-0.028\pm 0.062\pm 0.007$ & $-0.236\pm 0.156\pm 0.019$ \\
 0.090 &  1.01 &  16.2 & $ 0.178\pm 0.034\pm 0.013$ & $ 0.110\pm 0.071\pm 0.006$ & $ 0.054\pm 0.173\pm 0.017$ \\
 0.090 &  1.58 &  16.2 & $ 0.196\pm 0.037\pm 0.009$ & $ 0.209\pm 0.085\pm 0.004$ & $ 0.268\pm 0.207\pm 0.013$ \\
 0.090 &  2.53 &  29.1 & $ 0.159\pm 0.023\pm 0.010$ & $ 0.077\pm 0.027\pm 0.004$ & $-0.003\pm 0.070\pm 0.013$ \\
 0.090 &  3.48 &  29.1 & $ 0.157\pm 0.040\pm 0.010$ & $ 0.057\pm 0.045\pm 0.004$ & $-0.049\pm 0.117\pm 0.013$ \\
 0.101 &  0.45 &   9.7 & $ 0.102\pm 0.057\pm 0.030$ & $ 0.158\pm 0.081\pm 0.023$ & $ 0.252\pm 0.203\pm 0.054$ \\
 0.101 &  0.78 &   9.7 & $ 0.138\pm 0.040\pm 0.014$ & $ 0.039\pm 0.058\pm 0.007$ & $-0.070\pm 0.147\pm 0.021$ \\
 0.101 &  1.06 &  16.2 & $ 0.221\pm 0.037\pm 0.012$ & $ 0.058\pm 0.077\pm 0.006$ & $-0.121\pm 0.189\pm 0.017$ \\
 0.101 &  1.69 &  16.2 & $ 0.184\pm 0.037\pm 0.009$ & $ 0.098\pm 0.083\pm 0.004$ & $ 0.016\pm 0.205\pm 0.012$ \\
 0.101 &  2.69 &  29.1 & $ 0.237\pm 0.025\pm 0.010$ & $ 0.056\pm 0.030\pm 0.004$ & $-0.148\pm 0.078\pm 0.013$ \\
 0.101 &  3.79 &  29.1 & $ 0.196\pm 0.033\pm 0.010$ & $ 0.067\pm 0.038\pm 0.004$ & $-0.075\pm 0.099\pm 0.013$ \\
 0.113 &  0.47 &   9.7 & $ 0.185\pm 0.059\pm 0.033$ & $-0.008\pm 0.086\pm 0.027$ & $-0.237\pm 0.215\pm 0.064$ \\
 0.113 &  0.82 &   9.7 & $ 0.191\pm 0.040\pm 0.015$ & $ 0.063\pm 0.057\pm 0.009$ & $-0.076\pm 0.147\pm 0.023$ \\
 0.113 &  1.10 &  16.2 & $ 0.168\pm 0.041\pm 0.012$ & $ 0.054\pm 0.083\pm 0.007$ & $-0.071\pm 0.208\pm 0.018$ \\
 0.113 &  1.80 &  16.2 & $ 0.208\pm 0.037\pm 0.009$ & $ 0.052\pm 0.078\pm 0.004$ & $-0.126\pm 0.196\pm 0.012$ \\
 0.113 &  2.84 &  29.1 & $ 0.239\pm 0.028\pm 0.012$ & $ 0.082\pm 0.033\pm 0.005$ & $-0.090\pm 0.087\pm 0.015$ \\
 0.113 &  4.11 &  29.1 & $ 0.225\pm 0.031\pm 0.010$ & $ 0.029\pm 0.036\pm 0.004$ & $-0.203\pm 0.094\pm 0.013$ \\
 0.128 &  0.48 &   9.7 & $ 0.011\pm 0.064\pm 0.036$ & $-0.082\pm 0.092\pm 0.031$ & $-0.209\pm 0.233\pm 0.072$ \\
 0.128 &  0.86 &   9.7 & $ 0.135\pm 0.041\pm 0.015$ & $ 0.146\pm 0.059\pm 0.010$ & $ 0.191\pm 0.152\pm 0.026$ \\
 0.128 &  1.14 &  16.2 & $ 0.221\pm 0.045\pm 0.013$ & $ 0.230\pm 0.090\pm 0.008$ & $ 0.293\pm 0.227\pm 0.020$ \\
 0.128 &  1.91 &  16.2 & $ 0.234\pm 0.037\pm 0.010$ & $ 0.037\pm 0.077\pm 0.005$ & $-0.197\pm 0.194\pm 0.013$ \\
 0.128 &  3.00 &  29.1 & $ 0.230\pm 0.030\pm 0.013$ & $ 0.186\pm 0.036\pm 0.005$ & $ 0.171\pm 0.096\pm 0.017$ \\
 0.128 &  4.44 &  29.1 & $ 0.237\pm 0.030\pm 0.010$ & $ 0.077\pm 0.035\pm 0.004$ & $-0.106\pm 0.093\pm 0.013$ \\
 0.144 &  0.90 &   9.7 & $ 0.141\pm 0.042\pm 0.016$ & $ 0.036\pm 0.060\pm 0.012$ & $-0.087\pm 0.158\pm 0.029$ \\
 0.144 &  1.18 &  16.2 & $ 0.222\pm 0.049\pm 0.013$ & $ 0.200\pm 0.099\pm 0.009$ & $ 0.217\pm 0.251\pm 0.022$ \\
 0.144 &  2.03 &  16.2 & $ 0.259\pm 0.038\pm 0.011$ & $ 0.085\pm 0.078\pm 0.005$ & $-0.114\pm 0.198\pm 0.015$ \\
 0.144 &  3.15 &  29.1 & $ 0.213\pm 0.033\pm 0.014$ & $ 0.132\pm 0.039\pm 0.006$ & $ 0.057\pm 0.106\pm 0.018$ \\
 0.144 &  4.78 &  29.1 & $ 0.242\pm 0.030\pm 0.010$ & $ 0.163\pm 0.034\pm 0.004$ & $ 0.097\pm 0.093\pm 0.013$ \\
 0.162 &  0.93 &   9.7 & $ 0.212\pm 0.043\pm 0.017$ & $ 0.117\pm 0.063\pm 0.014$ & $ 0.024\pm 0.166\pm 0.032$ \\
 0.162 &  1.22 &  16.2 & $ 0.293\pm 0.054\pm 0.013$ & $-0.107\pm 0.108\pm 0.011$ & $-0.639\pm 0.279\pm 0.025$ \\
 0.162 &  2.14 &  16.2 & $ 0.237\pm 0.040\pm 0.012$ & $ 0.137\pm 0.082\pm 0.007$ & $ 0.039\pm 0.211\pm 0.018$ \\
 0.162 &  3.30 &  29.1 & $ 0.306\pm 0.036\pm 0.014$ & $ 0.128\pm 0.044\pm 0.007$ & $-0.075\pm 0.119\pm 0.020$ \\
 0.162 &  5.13 &  29.1 & $ 0.272\pm 0.030\pm 0.010$ & $ 0.078\pm 0.035\pm 0.005$ & $-0.159\pm 0.096\pm 0.014$ \\
 0.182 &  0.97 &   9.7 & $ 0.303\pm 0.045\pm 0.018$ & $ 0.095\pm 0.067\pm 0.015$ & $-0.151\pm 0.178\pm 0.035$ \\
 0.182 &  1.25 &  16.2 & $ 0.239\pm 0.060\pm 0.014$ & $ 0.201\pm 0.121\pm 0.012$ & $ 0.196\pm 0.315\pm 0.028$ \\
 0.182 &  2.25 &  16.2 & $ 0.257\pm 0.043\pm 0.014$ & $ 0.236\pm 0.087\pm 0.008$ & $ 0.261\pm 0.227\pm 0.021$ \\
 0.182 &  3.45 &  29.1 & $ 0.313\pm 0.040\pm 0.014$ & $ 0.152\pm 0.048\pm 0.008$ & $-0.030\pm 0.134\pm 0.021$ \\
 0.182 &  5.49 &  29.1 & $ 0.320\pm 0.031\pm 0.011$ & $ 0.222\pm 0.036\pm 0.006$ & $ 0.136\pm 0.101\pm 0.016$ \\
 0.205 &  1.00 &   9.7 & $ 0.253\pm 0.048\pm 0.017$ & $ 0.017\pm 0.071\pm 0.016$ & $-0.291\pm 0.192\pm 0.036$ \\
 0.205 &  1.29 &  16.2 & $ 0.304\pm 0.066\pm 0.015$ & $ 0.211\pm 0.135\pm 0.014$ & $ 0.135\pm 0.355\pm 0.031$ \\
 0.205 &  2.36 &  16.2 & $ 0.301\pm 0.046\pm 0.015$ & $ 0.116\pm 0.094\pm 0.009$ & $-0.110\pm 0.250\pm 0.023$ \\
 0.205 &  3.59 &  29.1 & $ 0.303\pm 0.044\pm 0.015$ & $ 0.150\pm 0.054\pm 0.009$ & $-0.026\pm 0.151\pm 0.023$ \\
 0.205 &  5.86 &  29.1 & $ 0.371\pm 0.032\pm 0.012$ & $ 0.107\pm 0.038\pm 0.007$ & $-0.235\pm 0.108\pm 0.018$ \\
 0.230 &  1.03 &   9.7 & $ 0.332\pm 0.050\pm 0.018$ & $ 0.175\pm 0.075\pm 0.017$ & $ 0.001\pm 0.205\pm 0.039$ \\
 0.230 &  1.32 &  16.2 & $ 0.255\pm 0.073\pm 0.017$ & $ 0.086\pm 0.150\pm 0.015$ & $-0.126\pm 0.401\pm 0.034$ \\
 0.230 &  2.47 &  16.2 & $ 0.376\pm 0.050\pm 0.015$ & $ 0.112\pm 0.104\pm 0.010$ & $-0.230\pm 0.281\pm 0.026$ \\
 0.230 &  3.73 &  29.1 & $ 0.389\pm 0.049\pm 0.015$ & $ 0.066\pm 0.062\pm 0.010$ & $-0.374\pm 0.176\pm 0.025$ \\
 0.230 &  6.23 &  29.1 & $ 0.273\pm 0.035\pm 0.012$ & $ 0.153\pm 0.042\pm 0.009$ & $ 0.014\pm 0.120\pm 0.021$ \\
 0.259 &  1.35 &  16.2 & $ 0.236\pm 0.082\pm 0.018$ & $ 0.130\pm 0.169\pm 0.015$ & $ 0.011\pm 0.459\pm 0.036$ \\
 0.259 &  2.57 &  16.2 & $ 0.485\pm 0.055\pm 0.016$ & $ 0.121\pm 0.114\pm 0.012$ & $-0.378\pm 0.315\pm 0.028$ \\
 0.259 &  3.85 &  29.1 & $ 0.384\pm 0.055\pm 0.016$ & $ 0.089\pm 0.071\pm 0.011$ & $-0.321\pm 0.205\pm 0.027$ \\
 0.259 &  6.60 &  29.1 & $ 0.477\pm 0.038\pm 0.014$ & $ 0.105\pm 0.046\pm 0.010$ & $-0.422\pm 0.137\pm 0.024$ \\
 0.292 &  1.37 &  16.2 & $ 0.340\pm 0.091\pm 0.020$ & $ 0.385\pm 0.192\pm 0.015$ & $ 0.538\pm 0.530\pm 0.036$ \\
 0.292 &  2.67 &  16.2 & $ 0.498\pm 0.061\pm 0.017$ & $ 0.359\pm 0.126\pm 0.013$ & $ 0.232\pm 0.355\pm 0.031$ \\
 0.292 &  3.98 &  29.1 & $ 0.361\pm 0.063\pm 0.017$ & $ 0.205\pm 0.082\pm 0.013$ & $ 0.014\pm 0.243\pm 0.031$ \\
 0.292 &  6.97 &  29.1 & $ 0.395\pm 0.043\pm 0.015$ & $ 0.202\pm 0.053\pm 0.012$ & $-0.051\pm 0.158\pm 0.028$ \\
 0.329 &  2.76 &  16.2 & $ 0.419\pm 0.068\pm 0.021$ & $ 0.533\pm 0.142\pm 0.017$ & $ 0.830\pm 0.409\pm 0.040$ \\
 0.329 &  4.09 &  29.1 & $ 0.456\pm 0.075\pm 0.018$ & $ 0.317\pm 0.097\pm 0.014$ & $ 0.168\pm 0.297\pm 0.034$ \\
 0.329 &  7.33 &  29.1 & $ 0.418\pm 0.048\pm 0.020$ & $ 0.227\pm 0.060\pm 0.013$ & $-0.031\pm 0.186\pm 0.033$ \\
 0.370 &  2.85 &  16.2 & $ 0.553\pm 0.077\pm 0.023$ & $ 0.135\pm 0.161\pm 0.014$ & $-0.511\pm 0.477\pm 0.037$ \\
 0.370 &  4.20 &  29.1 & $ 0.362\pm 0.089\pm 0.020$ & $ 0.319\pm 0.118\pm 0.016$ & $ 0.321\pm 0.371\pm 0.037$ \\
 0.370 &  7.69 &  29.1 & $ 0.432\pm 0.055\pm 0.022$ & $ 0.325\pm 0.070\pm 0.015$ & $ 0.217\pm 0.225\pm 0.038$ \\
 0.416 &  2.93 &  16.2 & $ 0.529\pm 0.087\pm 0.020$ & $ 0.204\pm 0.186\pm 0.014$ & $-0.297\pm 0.569\pm 0.035$ \\
 0.416 &  4.30 &  29.1 & $ 0.676\pm 0.108\pm 0.021$ & $ 0.295\pm 0.145\pm 0.017$ & $-0.298\pm 0.474\pm 0.039$ \\
 0.416 &  8.03 &  29.1 & $ 0.597\pm 0.065\pm 0.021$ & $ 0.216\pm 0.083\pm 0.017$ & $-0.413\pm 0.278\pm 0.040$ \\
 0.468 &  3.01 &  16.2 & $ 0.590\pm 0.101\pm 0.027$ & $ 0.669\pm 0.216\pm 0.013$ & $ 0.990\pm 0.689\pm 0.038$ \\
 0.468 &  4.40 &  29.1 & $ 0.713\pm 0.136\pm 0.022$ & $ 0.037\pm 0.185\pm 0.017$ & $-1.205\pm 0.628\pm 0.040$ \\
 0.468 &  8.37 &  29.1 & $ 0.584\pm 0.078\pm 0.023$ & $ 0.328\pm 0.102\pm 0.019$ & $-0.077\pm 0.356\pm 0.044$ \\
 0.526 &  4.47 &  29.1 & $ 0.731\pm 0.169\pm 0.022$ & $ 0.375\pm 0.237\pm 0.015$ & $-0.224\pm 0.838\pm 0.038$ \\
 0.526 &  8.67 &  29.1 & $ 0.652\pm 0.096\pm 0.024$ & $ 0.478\pm 0.129\pm 0.020$ & $ 0.257\pm 0.471\pm 0.047$ \\
 0.592 &  4.55 &  29.1 & $ 0.382\pm 0.217\pm 0.021$ & $ 0.310\pm 0.310\pm 0.012$ & $ 0.243\pm 1.129\pm 0.032$ \\
 0.592 &  8.98 &  29.1 & $ 0.670\pm 0.123\pm 0.025$ & $ 0.325\pm 0.168\pm 0.021$ & $-0.331\pm 0.640\pm 0.048$ \\
 0.666 &  9.26 &  29.1 & $ 0.478\pm 0.165\pm 0.026$ & $ 0.251\pm 0.226\pm 0.019$ & $-0.192\pm 0.876\pm 0.046$ \\
 0.749 &  9.52 &  29.1 & $ 0.744\pm 0.237\pm 0.031$ & $ 0.646\pm 0.331\pm 0.033$ & $ 0.569\pm 1.233\pm 0.073$ \\
 \end{tabular}
 \end{table}


 \begin{table}[t] 
 \caption{Results for  $A_1$ in the DIS region ($W^2\geq 4 {\rm ~GeV^2}$). There is an additional normalization uncertainty due to beam and target polarization shown in Table~ \protect\ref{tb:norm_err}.  }
 \label{tb:a1}
 \begin{tabular}{rrrrrr}
$x$ & $<Q^2>$ & E(GeV) & $A_1^p\pm {\rm stat} \pm {\rm syst}$ & $A_1^d\pm {\rm stat} \pm {\rm syst}$ & $A_1^n\pm {\rm stat} \pm {\rm syst}$ \\
 \hline
 0.024 &  0.51 &  16.2 & $ 0.091\pm 0.027\pm 0.014$ & $-0.049\pm 0.068\pm 0.009$ & $-0.205\pm 0.153\pm 0.023$ \\
 0.027 &  0.55 &  16.2 & $ 0.085\pm 0.026\pm 0.013$ & $-0.018\pm 0.065\pm 0.009$ & $-0.131\pm 0.146\pm 0.022$ \\
 0.027 &  1.17 &  29.1 & $-0.034\pm 0.068\pm 0.010$ & $-0.048\pm 0.092\pm 0.009$ & $-0.073\pm 0.225\pm 0.020$ \\
 0.031 &  0.59 &  16.2 & $ 0.084\pm 0.026\pm 0.012$ & $ 0.019\pm 0.065\pm 0.008$ & $-0.048\pm 0.144\pm 0.020$ \\
 0.031 &  1.27 &  29.1 & $ 0.063\pm 0.034\pm 0.009$ & $ 0.050\pm 0.040\pm 0.008$ & $ 0.044\pm 0.095\pm 0.019$ \\
 0.035 &  0.31 &   9.7 & $ 0.093\pm 0.037\pm 0.018$ & $-0.027\pm 0.055\pm 0.008$ & $-0.159\pm 0.125\pm 0.024$ \\
 0.035 &  0.64 &  16.2 & $ 0.058\pm 0.024\pm 0.011$ & $-0.015\pm 0.062\pm 0.007$ & $-0.097\pm 0.139\pm 0.018$ \\
 0.035 &  1.40 &  29.1 & $ 0.122\pm 0.025\pm 0.008$ & $ 0.002\pm 0.028\pm 0.007$ & $-0.126\pm 0.069\pm 0.017$ \\
 0.039 &  0.33 &   9.7 & $ 0.022\pm 0.037\pm 0.018$ & $ 0.035\pm 0.053\pm 0.008$ & $ 0.053\pm 0.122\pm 0.024$ \\
 0.039 &  0.68 &  16.2 & $ 0.072\pm 0.024\pm 0.011$ & $ 0.098\pm 0.060\pm 0.007$ & $ 0.140\pm 0.136\pm 0.017$ \\
 0.039 &  1.52 &  29.1 & $ 0.083\pm 0.023\pm 0.008$ & $ 0.059\pm 0.026\pm 0.007$ & $ 0.041\pm 0.063\pm 0.016$ \\
 0.044 &  0.35 &   9.7 & $ 0.103\pm 0.037\pm 0.019$ & $ 0.003\pm 0.054\pm 0.008$ & $-0.106\pm 0.126\pm 0.025$ \\
 0.044 &  0.73 &  16.2 & $ 0.095\pm 0.023\pm 0.011$ & $-0.039\pm 0.052\pm 0.006$ & $-0.191\pm 0.120\pm 0.017$ \\
 0.044 &  0.98 &  16.2 & $ 0.096\pm 0.520\pm 0.010$ & $-0.944\pm 1.296\pm 0.008$ & $-2.222\pm 2.964\pm 0.019$ \\
 0.044 &  1.65 &  29.1 & $ 0.110\pm 0.021\pm 0.008$ & $-0.019\pm 0.025\pm 0.006$ & $-0.162\pm 0.060\pm 0.014$ \\
 0.049 &  0.36 &   9.7 & $ 0.103\pm 0.039\pm 0.020$ & $ 0.054\pm 0.057\pm 0.008$ & $ 0.007\pm 0.135\pm 0.026$ \\
 0.049 &  0.78 &  16.2 & $ 0.146\pm 0.024\pm 0.011$ & $ 0.091\pm 0.052\pm 0.006$ & $ 0.045\pm 0.121\pm 0.016$ \\
 0.049 &  1.06 &  16.2 & $ 0.077\pm 0.109\pm 0.010$ & $-0.104\pm 0.280\pm 0.008$ & $-0.321\pm 0.646\pm 0.018$ \\
 0.049 &  1.78 &  29.1 & $ 0.125\pm 0.020\pm 0.008$ & $-0.016\pm 0.024\pm 0.005$ & $-0.173\pm 0.058\pm 0.013$ \\
 0.056 &  0.38 &   9.7 & $ 0.103\pm 0.041\pm 0.021$ & $-0.046\pm 0.059\pm 0.009$ & $-0.217\pm 0.141\pm 0.027$ \\
 0.056 &  0.57 &   9.7 & $-0.217\pm 1.891\pm 0.014$ & $ 2.991\pm 2.465\pm 0.007$ & $ 7.031\pm 5.978\pm 0.020$ \\
 0.056 &  0.83 &  16.2 & $ 0.109\pm 0.025\pm 0.011$ & $ 0.007\pm 0.054\pm 0.006$ & $-0.106\pm 0.127\pm 0.016$ \\
 0.056 &  1.16 &  16.2 & $ 0.131\pm 0.063\pm 0.010$ & $ 0.090\pm 0.153\pm 0.007$ & $ 0.061\pm 0.358\pm 0.017$ \\
 0.056 &  1.92 &  29.1 & $ 0.129\pm 0.020\pm 0.008$ & $ 0.021\pm 0.023\pm 0.005$ & $-0.095\pm 0.058\pm 0.012$ \\
 0.063 &  0.40 &   9.7 & $ 0.071\pm 0.043\pm 0.022$ & $-0.007\pm 0.062\pm 0.010$ & $-0.096\pm 0.149\pm 0.029$ \\
 0.063 &  0.60 &   9.7 & $-0.027\pm 0.177\pm 0.014$ & $ 0.220\pm 0.252\pm 0.007$ & $ 0.532\pm 0.609\pm 0.020$ \\
 0.063 &  0.87 &  16.2 & $ 0.155\pm 0.027\pm 0.012$ & $ 0.048\pm 0.057\pm 0.005$ & $-0.064\pm 0.135\pm 0.016$ \\
 0.063 &  1.26 &  16.2 & $ 0.050\pm 0.046\pm 0.010$ & $-0.005\pm 0.112\pm 0.006$ & $-0.068\pm 0.262\pm 0.016$ \\
 0.063 &  2.07 &  29.1 & $ 0.136\pm 0.020\pm 0.008$ & $ 0.020\pm 0.023\pm 0.004$ & $-0.107\pm 0.058\pm 0.012$ \\
 0.063 &  2.69 &  29.1 & $ 1.146\pm 0.834\pm 0.009$ & $-0.309\pm 2.063\pm 0.006$ & $-2.009\pm 4.864\pm 0.016$ \\
 0.071 &  0.41 &   9.7 & $ 0.088\pm 0.049\pm 0.023$ & $ 0.102\pm 0.070\pm 0.012$ & $ 0.135\pm 0.171\pm 0.033$ \\
 0.071 &  0.64 &   9.7 & $ 0.060\pm 0.077\pm 0.014$ & $-0.074\pm 0.109\pm 0.007$ & $-0.239\pm 0.267\pm 0.019$ \\
 0.071 &  0.92 &  16.2 & $ 0.105\pm 0.029\pm 0.012$ & $ 0.063\pm 0.061\pm 0.006$ & $ 0.026\pm 0.145\pm 0.017$ \\
 0.071 &  1.36 &  16.2 & $ 0.143\pm 0.041\pm 0.009$ & $ 0.084\pm 0.095\pm 0.005$ & $ 0.033\pm 0.225\pm 0.014$ \\
 0.071 &  2.22 &  29.1 & $ 0.149\pm 0.020\pm 0.008$ & $ 0.023\pm 0.023\pm 0.004$ & $-0.115\pm 0.059\pm 0.011$ \\
 0.071 &  2.91 &  29.1 & $ 0.340\pm 0.122\pm 0.010$ & $ 0.052\pm 0.139\pm 0.006$ & $-0.267\pm 0.353\pm 0.015$ \\
 0.079 &  0.43 &   9.7 & $ 0.102\pm 0.053\pm 0.025$ & $-0.016\pm 0.075\pm 0.015$ & $-0.154\pm 0.186\pm 0.039$ \\
 0.079 &  0.69 &   9.7 & $ 0.160\pm 0.053\pm 0.014$ & $ 0.027\pm 0.078\pm 0.007$ & $-0.120\pm 0.194\pm 0.019$ \\
 0.079 &  0.97 &  16.2 & $ 0.120\pm 0.031\pm 0.013$ & $-0.044\pm 0.066\pm 0.006$ & $-0.241\pm 0.159\pm 0.017$ \\
 0.079 &  1.47 &  16.2 & $ 0.134\pm 0.039\pm 0.009$ & $-0.120\pm 0.088\pm 0.005$ & $-0.437\pm 0.212\pm 0.014$ \\
 0.079 &  2.38 &  29.1 & $ 0.191\pm 0.021\pm 0.009$ & $ 0.043\pm 0.025\pm 0.004$ & $-0.119\pm 0.064\pm 0.012$ \\
 0.079 &  3.17 &  29.1 & $ 0.210\pm 0.057\pm 0.010$ & $-0.031\pm 0.065\pm 0.005$ & $-0.316\pm 0.167\pm 0.014$ \\
 0.090 &  0.44 &   9.7 & $ 0.057\pm 0.055\pm 0.027$ & $ 0.028\pm 0.078\pm 0.018$ & $ 0.000\pm 0.193\pm 0.046$ \\
 0.090 &  0.74 &   9.7 & $ 0.140\pm 0.043\pm 0.014$ & $-0.030\pm 0.062\pm 0.007$ & $-0.235\pm 0.156\pm 0.020$ \\
 0.090 &  1.01 &  16.2 & $ 0.175\pm 0.034\pm 0.013$ & $ 0.109\pm 0.071\pm 0.006$ & $ 0.054\pm 0.173\pm 0.018$ \\
 0.090 &  1.58 &  16.2 & $ 0.194\pm 0.037\pm 0.010$ & $ 0.208\pm 0.085\pm 0.005$ & $ 0.268\pm 0.207\pm 0.013$ \\
 0.090 &  2.53 &  29.1 & $ 0.160\pm 0.023\pm 0.010$ & $ 0.070\pm 0.027\pm 0.004$ & $-0.022\pm 0.071\pm 0.013$ \\
 0.090 &  3.48 &  29.1 & $ 0.152\pm 0.040\pm 0.010$ & $ 0.061\pm 0.045\pm 0.004$ & $-0.033\pm 0.117\pm 0.014$ \\
 0.101 &  0.45 &   9.7 & $ 0.087\pm 0.057\pm 0.030$ & $ 0.154\pm 0.081\pm 0.023$ & $ 0.262\pm 0.203\pm 0.055$ \\
 0.101 &  0.78 &   9.7 & $ 0.132\pm 0.040\pm 0.014$ & $ 0.037\pm 0.058\pm 0.008$ & $-0.069\pm 0.147\pm 0.021$ \\
 0.101 &  1.06 &  16.2 & $ 0.217\pm 0.037\pm 0.012$ & $ 0.057\pm 0.077\pm 0.006$ & $-0.121\pm 0.189\pm 0.018$ \\
 0.101 &  1.69 &  16.2 & $ 0.183\pm 0.037\pm 0.009$ & $ 0.097\pm 0.083\pm 0.004$ & $ 0.016\pm 0.205\pm 0.012$ \\
 0.101 &  2.69 &  29.1 & $ 0.237\pm 0.026\pm 0.011$ & $ 0.064\pm 0.030\pm 0.004$ & $-0.129\pm 0.079\pm 0.014$ \\
 0.101 &  3.79 &  29.1 & $ 0.199\pm 0.033\pm 0.010$ & $ 0.055\pm 0.038\pm 0.004$ & $-0.106\pm 0.099\pm 0.013$ \\
 0.113 &  0.47 &   9.7 & $ 0.167\pm 0.059\pm 0.033$ & $-0.012\pm 0.086\pm 0.028$ & $-0.226\pm 0.215\pm 0.064$ \\
 0.113 &  0.82 &   9.7 & $ 0.185\pm 0.040\pm 0.015$ & $ 0.061\pm 0.057\pm 0.009$ & $-0.075\pm 0.147\pm 0.023$ \\
 0.113 &  1.10 &  16.2 & $ 0.164\pm 0.041\pm 0.013$ & $ 0.052\pm 0.083\pm 0.007$ & $-0.071\pm 0.208\pm 0.019$ \\
 0.113 &  1.80 &  16.2 & $ 0.206\pm 0.037\pm 0.009$ & $ 0.051\pm 0.078\pm 0.004$ & $-0.126\pm 0.196\pm 0.012$ \\
 0.113 &  2.84 &  29.1 & $ 0.243\pm 0.028\pm 0.012$ & $ 0.079\pm 0.033\pm 0.005$ & $-0.103\pm 0.088\pm 0.016$ \\
 0.113 &  4.11 &  29.1 & $ 0.224\pm 0.031\pm 0.010$ & $ 0.035\pm 0.036\pm 0.004$ & $-0.187\pm 0.094\pm 0.013$ \\
 0.128 &  0.48 &   9.7 & $-0.010\pm 0.064\pm 0.036$ & $-0.087\pm 0.092\pm 0.031$ & $-0.196\pm 0.233\pm 0.072$ \\
 0.128 &  0.86 &   9.7 & $ 0.129\pm 0.041\pm 0.015$ & $ 0.143\pm 0.059\pm 0.011$ & $ 0.193\pm 0.152\pm 0.026$ \\
 0.128 &  1.14 &  16.2 & $ 0.217\pm 0.045\pm 0.013$ & $ 0.229\pm 0.090\pm 0.008$ & $ 0.294\pm 0.227\pm 0.021$ \\
 0.128 &  1.91 &  16.2 & $ 0.233\pm 0.037\pm 0.010$ & $ 0.037\pm 0.077\pm 0.005$ & $-0.197\pm 0.194\pm 0.014$ \\
 0.128 &  3.00 &  29.1 & $ 0.228\pm 0.031\pm 0.013$ & $ 0.187\pm 0.037\pm 0.006$ & $ 0.176\pm 0.098\pm 0.018$ \\
 0.128 &  4.44 &  29.1 & $ 0.243\pm 0.031\pm 0.010$ & $ 0.080\pm 0.035\pm 0.005$ & $-0.106\pm 0.093\pm 0.014$ \\
 0.144 &  0.90 &   9.7 & $ 0.134\pm 0.042\pm 0.016$ & $ 0.033\pm 0.060\pm 0.012$ & $-0.085\pm 0.158\pm 0.029$ \\
 0.144 &  1.18 &  16.2 & $ 0.218\pm 0.049\pm 0.013$ & $ 0.199\pm 0.099\pm 0.009$ & $ 0.218\pm 0.251\pm 0.023$ \\
 0.144 &  2.03 &  16.2 & $ 0.258\pm 0.038\pm 0.011$ & $ 0.085\pm 0.078\pm 0.006$ & $-0.114\pm 0.198\pm 0.016$ \\
 0.144 &  3.15 &  29.1 & $ 0.227\pm 0.033\pm 0.014$ & $ 0.141\pm 0.041\pm 0.007$ & $ 0.064\pm 0.109\pm 0.020$ \\
 0.144 &  4.78 &  29.1 & $ 0.242\pm 0.030\pm 0.010$ & $ 0.171\pm 0.035\pm 0.005$ & $ 0.118\pm 0.094\pm 0.014$ \\
 0.162 &  0.93 &   9.7 & $ 0.205\pm 0.043\pm 0.017$ & $ 0.115\pm 0.063\pm 0.014$ & $ 0.027\pm 0.166\pm 0.033$ \\
 0.162 &  1.22 &  16.2 & $ 0.290\pm 0.054\pm 0.014$ & $-0.108\pm 0.108\pm 0.011$ & $-0.638\pm 0.279\pm 0.026$ \\
 0.162 &  2.14 &  16.2 & $ 0.237\pm 0.040\pm 0.013$ & $ 0.137\pm 0.082\pm 0.007$ & $ 0.039\pm 0.211\pm 0.019$ \\
 0.162 &  3.30 &  29.1 & $ 0.297\pm 0.037\pm 0.015$ & $ 0.109\pm 0.045\pm 0.008$ & $-0.112\pm 0.124\pm 0.021$ \\
 0.162 &  5.13 &  29.1 & $ 0.269\pm 0.030\pm 0.011$ & $ 0.084\pm 0.035\pm 0.005$ & $-0.138\pm 0.097\pm 0.015$ \\
 0.182 &  0.97 &   9.7 & $ 0.298\pm 0.045\pm 0.018$ & $ 0.094\pm 0.067\pm 0.016$ & $-0.148\pm 0.178\pm 0.036$ \\
 0.182 &  1.25 &  16.2 & $ 0.237\pm 0.060\pm 0.014$ & $ 0.200\pm 0.121\pm 0.012$ & $ 0.198\pm 0.315\pm 0.029$ \\
 0.182 &  2.25 &  16.2 & $ 0.258\pm 0.043\pm 0.014$ & $ 0.237\pm 0.087\pm 0.008$ & $ 0.261\pm 0.227\pm 0.022$ \\
 0.182 &  3.45 &  29.1 & $ 0.326\pm 0.041\pm 0.015$ & $ 0.175\pm 0.051\pm 0.009$ & $ 0.013\pm 0.140\pm 0.023$ \\
 0.182 &  5.49 &  29.1 & $ 0.309\pm 0.031\pm 0.011$ & $ 0.214\pm 0.037\pm 0.006$ & $ 0.132\pm 0.102\pm 0.017$ \\
 0.205 &  1.00 &   9.7 & $ 0.251\pm 0.048\pm 0.017$ & $ 0.017\pm 0.071\pm 0.016$ & $-0.285\pm 0.190\pm 0.037$ \\
 0.205 &  1.29 &  16.2 & $ 0.305\pm 0.066\pm 0.016$ & $ 0.213\pm 0.135\pm 0.014$ & $ 0.137\pm 0.355\pm 0.032$ \\
 0.205 &  2.36 &  16.2 & $ 0.304\pm 0.046\pm 0.015$ & $ 0.117\pm 0.094\pm 0.010$ & $-0.109\pm 0.250\pm 0.024$ \\
 0.205 &  3.59 &  29.1 & $ 0.278\pm 0.045\pm 0.015$ & $ 0.159\pm 0.057\pm 0.010$ & $ 0.032\pm 0.160\pm 0.025$ \\
 0.205 &  5.86 &  29.1 & $ 0.370\pm 0.033\pm 0.012$ & $ 0.132\pm 0.039\pm 0.008$ & $-0.167\pm 0.110\pm 0.019$ \\
 0.230 &  1.03 &   9.7 & $ 0.335\pm 0.050\pm 0.018$ & $ 0.178\pm 0.075\pm 0.017$ & $ 0.005\pm 0.204\pm 0.039$ \\
 0.230 &  1.32 &  16.2 & $ 0.260\pm 0.073\pm 0.017$ & $ 0.090\pm 0.150\pm 0.015$ & $-0.123\pm 0.401\pm 0.035$ \\
 0.230 &  2.47 &  16.2 & $ 0.381\pm 0.050\pm 0.016$ & $ 0.116\pm 0.104\pm 0.011$ & $-0.228\pm 0.281\pm 0.027$ \\
 0.230 &  3.73 &  29.1 & $ 0.396\pm 0.051\pm 0.016$ & $ 0.061\pm 0.067\pm 0.011$ & $-0.398\pm 0.189\pm 0.027$ \\
 0.230 &  6.23 &  29.1 & $ 0.272\pm 0.036\pm 0.013$ & $ 0.173\pm 0.043\pm 0.009$ & $ 0.067\pm 0.123\pm 0.022$ \\
 0.259 &  1.35 &  16.2 & $ 0.249\pm 0.082\pm 0.019$ & $ 0.138\pm 0.169\pm 0.016$ & $ 0.014\pm 0.457\pm 0.037$ \\
 0.259 &  2.57 &  16.2 & $ 0.496\pm 0.055\pm 0.017$ & $ 0.127\pm 0.114\pm 0.012$ & $-0.373\pm 0.312\pm 0.029$ \\
 0.259 &  3.85 &  29.1 & $ 0.372\pm 0.057\pm 0.017$ & $ 0.005\pm 0.078\pm 0.012$ & $-0.529\pm 0.225\pm 0.029$ \\
 0.259 &  6.60 &  29.1 & $ 0.480\pm 0.039\pm 0.014$ & $ 0.081\pm 0.048\pm 0.011$ & $-0.491\pm 0.141\pm 0.025$ \\
 0.292 &  1.37 &  16.2 & $ 0.362\pm 0.091\pm 0.020$ & $ 0.398\pm 0.192\pm 0.015$ & $ 0.540\pm 0.529\pm 0.037$ \\
 0.292 &  2.67 &  16.2 & $ 0.515\pm 0.061\pm 0.017$ & $ 0.369\pm 0.126\pm 0.013$ & $ 0.234\pm 0.355\pm 0.031$ \\
 0.292 &  3.98 &  29.1 & $ 0.423\pm 0.067\pm 0.018$ & $ 0.225\pm 0.093\pm 0.014$ & $-0.026\pm 0.273\pm 0.033$ \\
 0.292 &  6.97 &  29.1 & $ 0.409\pm 0.043\pm 0.016$ & $ 0.216\pm 0.055\pm 0.012$ & $-0.033\pm 0.164\pm 0.029$ \\
 0.329 &  2.76 &  16.2 & $ 0.446\pm 0.068\pm 0.021$ & $ 0.548\pm 0.142\pm 0.018$ & $ 0.829\pm 0.408\pm 0.041$ \\
 0.329 &  4.09 &  29.1 & $ 0.468\pm 0.079\pm 0.019$ & $ 0.352\pm 0.114\pm 0.015$ & $ 0.248\pm 0.343\pm 0.036$ \\
 0.329 &  7.33 &  29.1 & $ 0.428\pm 0.049\pm 0.021$ & $ 0.247\pm 0.063\pm 0.014$ & $ 0.009\pm 0.196\pm 0.035$ \\
 0.370 &  2.85 &  16.2 & $ 0.594\pm 0.077\pm 0.024$ & $ 0.157\pm 0.161\pm 0.015$ & $-0.515\pm 0.476\pm 0.038$ \\
 0.370 &  4.20 &  29.1 & $ 0.349\pm 0.096\pm 0.021$ & $ 0.362\pm 0.143\pm 0.017$ & $ 0.469\pm 0.443\pm 0.040$ \\
 0.370 &  7.69 &  29.1 & $ 0.448\pm 0.057\pm 0.023$ & $ 0.343\pm 0.075\pm 0.016$ & $ 0.243\pm 0.239\pm 0.040$ \\
 0.416 &  2.93 &  16.2 & $ 0.589\pm 0.087\pm 0.021$ & $ 0.235\pm 0.186\pm 0.015$ & $-0.310\pm 0.569\pm 0.037$ \\
 0.416 &  4.30 &  29.1 & $ 0.681\pm 0.118\pm 0.023$ & $ 0.390\pm 0.184\pm 0.018$ & $-0.020\pm 0.588\pm 0.043$ \\
 0.416 &  8.03 &  29.1 & $ 0.611\pm 0.067\pm 0.022$ & $ 0.232\pm 0.091\pm 0.018$ & $-0.392\pm 0.300\pm 0.043$ \\
 0.468 &  3.01 &  16.2 & $ 0.677\pm 0.101\pm 0.028$ & $ 0.710\pm 0.216\pm 0.015$ & $ 0.962\pm 0.689\pm 0.040$ \\
 0.468 &  4.40 &  29.1 & $ 0.828\pm 0.151\pm 0.024$ & $ 0.190\pm 0.246\pm 0.019$ & $-0.940\pm 0.814\pm 0.044$ \\
 0.468 &  8.37 &  29.1 & $ 0.589\pm 0.082\pm 0.025$ & $ 0.335\pm 0.113\pm 0.020$ & $-0.064\pm 0.392\pm 0.048$ \\
 0.526 &  4.47 &  29.1 & $ 0.841\pm 0.194\pm 0.025$ & $ 0.182\pm 0.335\pm 0.018$ & $-1.063\pm 1.150\pm 0.044$ \\
 0.526 &  8.67 &  29.1 & $ 0.678\pm 0.102\pm 0.028$ & $ 0.508\pm 0.147\pm 0.022$ & $ 0.304\pm 0.530\pm 0.053$ \\
 0.592 &  4.55 &  29.1 & $ 0.479\pm 0.255\pm 0.026$ & $ 0.672\pm 0.464\pm 0.016$ & $ 1.249\pm 1.639\pm 0.042$ \\
 0.592 &  8.98 &  29.1 & $ 0.695\pm 0.133\pm 0.031$ & $ 0.065\pm 0.198\pm 0.024$ & $-1.283\pm 0.744\pm 0.057$ \\
 0.666 &  9.26 &  29.1 & $ 0.612\pm 0.182\pm 0.033$ & $ 0.423\pm 0.284\pm 0.024$ & $ 0.112\pm 1.078\pm 0.059$ \\
 0.749 &  9.52 &  29.1 & $ 0.914\pm 0.273\pm 0.041$ & $ 0.769\pm 0.443\pm 0.039$ & $ 0.613\pm 1.605\pm 0.088$ \\
 \end{tabular}
 \end{table}

 \begin{table}[t] 
 \caption{
 Coefficients for fits to all available data with
  $Q^2\geq Q^2_{min}$ of the form
 $ax^{\alpha}(1+bx+cx^2)[1+Cf(Q^2)]$, along with the
 $\chi^2$ for the indicated number of degrees of freedom,
 calculated with the statistical errors only.  Fits I to IV
 are to $g_1/F_1$, while fit V is to $A_1.$}
 \label{tb:g1f1_fit}
 \begin{tabular}{rccccccccc}
fit to   & $Q^2_{min}$ & f($Q^2$) & $\alpha$ & $a$ & $b$ & $ c $ & $C$ & $\chi^2$  & df\\
 \hline
 I. $g_1^p/F_1^p$ & 0.3 & none & $\phantom{+}0.62$ & $\phantom{+}0.641$ & $\phantom{+}2.231$ & $-2.666$ & $\phantom{+}0.000$ & 145 & 117 \\
 II. $g_1^p/F_1^p$ & 1.0 & none & $\phantom{+}0.64 $ & $\phantom{+}0.749 $ & $ \phantom{+}1.466 $ & $ -1.982 $ & $ \phantom{+}0.000$ & 112 & 106 \\
 III. $g_1^p/F_1^p$ & 0.3 & $1/Q^2 $ & $  \phantom{+}0.62 $ & $ \phantom{+}0.762 $ & $ \phantom{+}1.434 $ & $ -1.917 $ & $ -0.160$ & 116 & 116 \\
 IV. $g_1^p/F_1^p$ & 0.3 & $-\ln (Q^2) $ & $  \phantom{+}0.66 $ & $ \phantom{+}0.728 $ & $ \phantom{+}0.850 $ & $ -1.384 $ & $ -0.100$ & 120 & 116 \\
 V. $\phantom{g_1^p/}A_1^p$ & 0.3 & $1/Q^2 $ & $ \phantom{+}0.66 $ & $ \phantom{+}0.898 $ & $ \phantom{+}0.595 $ & $ -0.371 $ & $ -0.180$ & 118 & 116 \\
 \hline
 I. $g_1^d/F_1^d$ & 0.3 & none & $\phantom{+}1.52 $ & $\phantom{+}2.439 $ & $-1.718 $ & $\phantom{+}0.867 $ & $\phantom{+}0.000$ & 122 & 111 \\
 II. $g_1^d/F_1^d$ & 1.0 & none & $\phantom{+}1.46 $ & $\phantom{+}2.222 $ & $-1.666 $ & $\phantom{+}0.829 $ & $\phantom{+}0.000$ & 115 & 100 \\
 III. $g_1^d/F_1^d$ & 0.3 & $1/Q^2 $ & $\phantom{+}1.44 $ & $\phantom{+}2.342 $ & $-1.724 $ & $\phantom{+}0.902 $ & $-0.260$ & 119 & 110 \\
 IV. $g_1^d/F_1^d$ & 0.3 & $-\ln (Q^2) $ & $\phantom{+}1.48 $ & $\phantom{+}2.030 $ & $-1.812 $ & $\phantom{+}0.979 $ & $-0.100$ & 120 & 110 \\
 V. $\phantom{g_1^p/}A_1^d$ & 0.3 & $1/Q^2 $ & $\phantom{+}1.46 $ & $\phantom{+}2.493 $ & $-1.915 $ & $\phantom{+}1.376 $ & $-0.260$ & 119 & 110 \\
 \end{tabular}
 \end{table}


 \begin{table}[t] 
 \caption{Results for averaged  $g_1/F_1$ for $Q^2\geq 1~{\rm (GeV/c)^2}$.}
 \label{tb:g1f1av}
 \begin{tabular}{rrrrr}
$x$ & $<Q^2>$ & $ g_1^p/F_1^p\pm {\rm stat} \pm {\rm syst}$ & $ g_1^d/F_1^d\pm {\rm stat} \pm {\rm syst}$  &  $ g_1^n/F_1^n\pm {\rm stat} \pm {\rm syst}$  \\
 \hline
 0.031 &  1.27 & $ 0.064\pm 0.034\pm 0.009$ & $ 0.041\pm 0.040\pm 0.008$ & $ 0.021\pm 0.095\pm 0.019$ \\
 0.035 &  1.40 & $ 0.123\pm 0.025\pm 0.008$ & $ 0.008\pm 0.029\pm 0.007$ & $-0.115\pm 0.069\pm 0.017$ \\
 0.039 &  1.52 & $ 0.083\pm 0.023\pm 0.008$ & $ 0.062\pm 0.026\pm 0.007$ & $ 0.049\pm 0.063\pm 0.016$ \\
 0.044 &  1.65 & $ 0.110\pm 0.021\pm 0.008$ & $-0.013\pm 0.025\pm 0.006$ & $-0.150\pm 0.060\pm 0.014$ \\
 0.049 &  1.78 & $ 0.124\pm 0.020\pm 0.008$ & $-0.014\pm 0.024\pm 0.005$ & $-0.169\pm 0.058\pm 0.013$ \\
 0.056 &  1.91 & $ 0.130\pm 0.019\pm 0.008$ & $ 0.017\pm 0.023\pm 0.005$ & $-0.105\pm 0.057\pm 0.012$ \\
 0.063 &  2.04 & $ 0.125\pm 0.018\pm 0.008$ & $ 0.013\pm 0.022\pm 0.004$ & $-0.121\pm 0.056\pm 0.012$ \\
 0.071 &  2.19 & $ 0.153\pm 0.018\pm 0.008$ & $ 0.029\pm 0.022\pm 0.004$ & $-0.108\pm 0.056\pm 0.011$ \\
 0.079 &  2.41 & $ 0.179\pm 0.018\pm 0.009$ & $ 0.024\pm 0.022\pm 0.004$ & $-0.163\pm 0.057\pm 0.012$ \\
 0.090 &  2.55 & $ 0.169\pm 0.016\pm 0.011$ & $ 0.084\pm 0.021\pm 0.004$ & $ 0.012\pm 0.055\pm 0.013$ \\
 0.101 &  2.85 & $ 0.215\pm 0.016\pm 0.010$ & $ 0.063\pm 0.022\pm 0.004$ & $-0.110\pm 0.056\pm 0.013$ \\
 0.113 &  3.13 & $ 0.217\pm 0.016\pm 0.011$ & $ 0.057\pm 0.022\pm 0.005$ & $-0.135\pm 0.058\pm 0.014$ \\
 0.128 &  3.41 & $ 0.232\pm 0.017\pm 0.011$ & $ 0.128\pm 0.023\pm 0.005$ & $ 0.025\pm 0.061\pm 0.015$ \\
 0.144 &  3.71 & $ 0.235\pm 0.018\pm 0.012$ & $ 0.146\pm 0.024\pm 0.005$ & $ 0.068\pm 0.064\pm 0.016$ \\
 0.162 &  4.03 & $ 0.276\pm 0.019\pm 0.012$ & $ 0.090\pm 0.025\pm 0.006$ & $-0.139\pm 0.068\pm 0.017$ \\
 0.182 &  4.34 & $ 0.296\pm 0.020\pm 0.013$ & $ 0.200\pm 0.027\pm 0.007$ & $ 0.102\pm 0.074\pm 0.019$ \\
 0.205 &  4.15 & $ 0.319\pm 0.020\pm 0.014$ & $ 0.109\pm 0.027\pm 0.009$ & $-0.165\pm 0.075\pm 0.023$ \\
 0.230 &  4.37 & $ 0.322\pm 0.021\pm 0.015$ & $ 0.130\pm 0.029\pm 0.011$ & $-0.103\pm 0.083\pm 0.026$ \\
 0.259 &  5.26 & $ 0.434\pm 0.026\pm 0.015$ & $ 0.103\pm 0.036\pm 0.011$ & $-0.369\pm 0.104\pm 0.026$ \\
 0.292 &  5.53 & $ 0.405\pm 0.029\pm 0.016$ & $ 0.227\pm 0.041\pm 0.012$ & $ 0.029\pm 0.121\pm 0.029$ \\
 0.329 &  6.01 & $ 0.427\pm 0.035\pm 0.020$ & $ 0.284\pm 0.048\pm 0.014$ & $ 0.130\pm 0.147\pm 0.034$ \\
 0.370 &  6.29 & $ 0.451\pm 0.040\pm 0.022$ & $ 0.300\pm 0.056\pm 0.015$ & $ 0.139\pm 0.178\pm 0.038$ \\
 0.416 &  6.56 & $ 0.592\pm 0.047\pm 0.021$ & $ 0.232\pm 0.067\pm 0.017$ & $-0.371\pm 0.221\pm 0.039$ \\
 0.468 &  6.79 & $ 0.608\pm 0.056\pm 0.024$ & $ 0.319\pm 0.083\pm 0.018$ & $-0.126\pm 0.282\pm 0.042$ \\
 0.526 &  7.72 & $ 0.671\pm 0.083\pm 0.024$ & $ 0.455\pm 0.113\pm 0.019$ & $ 0.141\pm 0.411\pm 0.045$ \\
 0.592 &  7.97 & $ 0.600\pm 0.107\pm 0.025$ & $ 0.322\pm 0.148\pm 0.019$ & $-0.191\pm 0.557\pm 0.045$ \\
 0.666 &  9.26 & $ 0.478\pm 0.165\pm 0.026$ & $ 0.251\pm 0.226\pm 0.019$ & $-0.192\pm 0.876\pm 0.046$ \\
 0.749 &  9.52 & $ 0.744\pm 0.237\pm 0.031$ & $ 0.646\pm 0.331\pm 0.033$ & $ 0.569\pm 1.233\pm 0.073$ \\
 \end{tabular}
 \end{table}


 \begin{table}[t] 
 \caption{Results for averaged  $A_1$ for $Q^2\geq 1~{\rm (GeV/c)^2}$.}
 \label{tb:a1av}
 \begin{tabular}{rrrrr}
$x$ & $<Q^2>$ & $ A_1^p\pm {\rm stat} \pm {\rm syst}$ & $ A_1^d\pm {\rm stat} \pm {\rm syst}$  &  $ A_1^n\pm {\rm stat} \pm {\rm syst}$  \\
 \hline
 0.031 &  1.27 & $ 0.063\pm 0.034\pm 0.009$ & $ 0.050\pm 0.040\pm 0.008$ & $ 0.044\pm 0.095\pm 0.019$ \\
 0.035 &  1.40 & $ 0.122\pm 0.025\pm 0.008$ & $ 0.002\pm 0.028\pm 0.007$ & $-0.126\pm 0.069\pm 0.017$ \\
 0.039 &  1.52 & $ 0.083\pm 0.023\pm 0.008$ & $ 0.059\pm 0.026\pm 0.007$ & $ 0.041\pm 0.063\pm 0.016$ \\
 0.044 &  1.65 & $ 0.110\pm 0.021\pm 0.008$ & $-0.019\pm 0.025\pm 0.006$ & $-0.162\pm 0.060\pm 0.014$ \\
 0.049 &  1.78 & $ 0.123\pm 0.020\pm 0.008$ & $-0.016\pm 0.024\pm 0.005$ & $-0.175\pm 0.058\pm 0.013$ \\
 0.056 &  1.91 & $ 0.130\pm 0.019\pm 0.008$ & $ 0.023\pm 0.023\pm 0.005$ & $-0.091\pm 0.057\pm 0.012$ \\
 0.063 &  2.04 & $ 0.123\pm 0.018\pm 0.008$ & $ 0.019\pm 0.022\pm 0.004$ & $-0.105\pm 0.056\pm 0.012$ \\
 0.071 &  2.19 & $ 0.152\pm 0.018\pm 0.009$ & $ 0.028\pm 0.022\pm 0.004$ & $-0.110\pm 0.056\pm 0.012$ \\
 0.079 &  2.41 & $ 0.180\pm 0.018\pm 0.009$ & $ 0.024\pm 0.022\pm 0.004$ & $-0.165\pm 0.057\pm 0.012$ \\
 0.090 &  2.55 & $ 0.168\pm 0.016\pm 0.011$ & $ 0.080\pm 0.022\pm 0.004$ & $ 0.004\pm 0.055\pm 0.014$ \\
 0.101 &  2.85 & $ 0.214\pm 0.016\pm 0.011$ & $ 0.063\pm 0.022\pm 0.004$ & $-0.110\pm 0.057\pm 0.014$ \\
 0.113 &  3.13 & $ 0.217\pm 0.017\pm 0.011$ & $ 0.058\pm 0.022\pm 0.005$ & $-0.135\pm 0.059\pm 0.015$ \\
 0.128 &  3.41 & $ 0.232\pm 0.017\pm 0.011$ & $ 0.128\pm 0.023\pm 0.005$ & $ 0.025\pm 0.061\pm 0.016$ \\
 0.144 &  3.71 & $ 0.238\pm 0.018\pm 0.012$ & $ 0.154\pm 0.024\pm 0.006$ & $ 0.081\pm 0.065\pm 0.017$ \\
 0.162 &  4.03 & $ 0.272\pm 0.019\pm 0.012$ & $ 0.086\pm 0.026\pm 0.007$ & $-0.142\pm 0.069\pm 0.018$ \\
 0.182 &  4.34 & $ 0.294\pm 0.020\pm 0.013$ & $ 0.204\pm 0.027\pm 0.008$ & $ 0.115\pm 0.075\pm 0.020$ \\
 0.205 &  4.15 & $ 0.314\pm 0.020\pm 0.014$ & $ 0.123\pm 0.027\pm 0.010$ & $-0.122\pm 0.076\pm 0.024$ \\
 0.230 &  4.37 & $ 0.325\pm 0.022\pm 0.015$ & $ 0.142\pm 0.030\pm 0.011$ & $-0.075\pm 0.085\pm 0.027$ \\
 0.259 &  5.26 & $ 0.437\pm 0.026\pm 0.016$ & $ 0.071\pm 0.038\pm 0.011$ & $-0.457\pm 0.108\pm 0.027$ \\
 0.292 &  5.53 & $ 0.432\pm 0.030\pm 0.017$ & $ 0.245\pm 0.043\pm 0.013$ & $ 0.036\pm 0.127\pm 0.031$ \\
 0.329 &  6.01 & $ 0.441\pm 0.036\pm 0.021$ & $ 0.308\pm 0.052\pm 0.015$ & $ 0.180\pm 0.157\pm 0.036$ \\
 0.370 &  6.29 & $ 0.471\pm 0.041\pm 0.023$ & $ 0.319\pm 0.061\pm 0.016$ & $ 0.162\pm 0.192\pm 0.040$ \\
 0.416 &  6.56 & $ 0.616\pm 0.049\pm 0.022$ & $ 0.258\pm 0.074\pm 0.018$ & $-0.314\pm 0.242\pm 0.042$ \\
 0.468 &  6.79 & $ 0.654\pm 0.059\pm 0.026$ & $ 0.383\pm 0.093\pm 0.019$ & $ 0.019\pm 0.314\pm 0.046$ \\
 0.526 &  7.72 & $ 0.713\pm 0.090\pm 0.027$ & $ 0.456\pm 0.135\pm 0.022$ & $ 0.064\pm 0.481\pm 0.051$ \\
 0.592 &  7.97 & $ 0.649\pm 0.118\pm 0.030$ & $ 0.159\pm 0.182\pm 0.023$ & $-0.851\pm 0.677\pm 0.054$ \\
 0.666 &  9.26 & $ 0.612\pm 0.182\pm 0.033$ & $ 0.423\pm 0.284\pm 0.024$ & $ 0.112\pm 1.078\pm 0.059$ \\
 0.749 &  9.52 & $ 0.914\pm 0.273\pm 0.041$ & $ 0.769\pm 0.443\pm 0.039$ & $ 0.613\pm 1.605\pm 0.088$ \\
 \end{tabular}
 \end{table}


 \begin{table}[t] 
 \caption{ Results for $g_1$ obtained from average $g_1/F_1$ for   $Q^2\geq 1~{\rm(GeV/c)^2)}.$
 Not included in the systematic errors listed are 
  normalization uncertaintird shown in 
Table~ \protect\ref{tb:norm_err}.  }
 \label{tb:g1}
 \begin{tabular}{rrrrrr}
$<x>$ & $x$ range & $<Q^2>$ &  $g_1^p\pm {\rm stat} \pm {\rm syst}$ & $g_1^d\pm {\rm stat} \pm {\rm syst}$ & $g_1^n\pm {\rm stat} \pm {\rm syst}$ \\
 \hline
 0.031 & 0.029-0.033 &  1.27 & $ 0.248\pm 0.132\pm 0.034$ & $ 0.150\pm 0.147\pm 0.030$ & $ 0.077\pm 0.345\pm 0.074$ \\
 0.035 & 0.033-0.037 &  1.40 & $ 0.436\pm 0.089\pm 0.032$ & $ 0.027\pm 0.097\pm 0.025$ & $-0.378\pm 0.227\pm 0.063$ \\
 0.039 & 0.037-0.042 &  1.52 & $ 0.269\pm 0.073\pm 0.025$ & $ 0.191\pm 0.080\pm 0.021$ & $ 0.145\pm 0.188\pm 0.054$ \\
 0.044 & 0.042-0.047 &  1.65 & $ 0.322\pm 0.062\pm 0.022$ & $-0.036\pm 0.068\pm 0.017$ & $-0.401\pm 0.160\pm 0.043$ \\
 0.049 & 0.047-0.053 &  1.78 & $ 0.327\pm 0.053\pm 0.019$ & $-0.034\pm 0.059\pm 0.013$ & $-0.405\pm 0.139\pm 0.036$ \\
 0.056 & 0.053-0.059 &  1.91 & $ 0.309\pm 0.045\pm 0.017$ & $ 0.039\pm 0.052\pm 0.010$ & $-0.224\pm 0.121\pm 0.029$ \\
 0.063 & 0.059-0.067 &  2.04 & $ 0.266\pm 0.039\pm 0.015$ & $ 0.027\pm 0.045\pm 0.008$ & $-0.230\pm 0.106\pm 0.023$ \\
 0.071 & 0.067-0.075 &  2.19 & $ 0.294\pm 0.034\pm 0.014$ & $ 0.053\pm 0.041\pm 0.006$ & $-0.182\pm 0.095\pm 0.021$ \\
 0.079 & 0.075-0.084 &  2.41 & $ 0.310\pm 0.031\pm 0.014$ & $ 0.039\pm 0.037\pm 0.005$ & $-0.245\pm 0.086\pm 0.019$ \\
 0.090 & 0.084-0.095 &  2.55 & $ 0.260\pm 0.024\pm 0.012$ & $ 0.123\pm 0.031\pm 0.005$ & $ 0.015\pm 0.073\pm 0.019$ \\
 0.101 & 0.095-0.107 &  2.85 & $ 0.299\pm 0.022\pm 0.012$ & $ 0.082\pm 0.029\pm 0.004$ & $-0.130\pm 0.067\pm 0.017$ \\
 0.113 & 0.107-0.120 &  3.13 & $ 0.272\pm 0.021\pm 0.011$ & $ 0.067\pm 0.026\pm 0.003$ & $-0.142\pm 0.061\pm 0.015$ \\
 0.128 & 0.120-0.136 &  3.41 & $ 0.262\pm 0.019\pm 0.010$ & $ 0.135\pm 0.024\pm 0.004$ & $ 0.023\pm 0.057\pm 0.017$ \\
 0.144 & 0.136-0.153 &  3.71 & $ 0.239\pm 0.018\pm 0.010$ & $ 0.138\pm 0.022\pm 0.004$ & $ 0.056\pm 0.052\pm 0.017$ \\
 0.162 & 0.153-0.172 &  4.03 & $ 0.253\pm 0.017\pm 0.010$ & $ 0.075\pm 0.021\pm 0.003$ & $-0.100\pm 0.049\pm 0.014$ \\
 0.182 & 0.172-0.193 &  4.34 & $ 0.243\pm 0.016\pm 0.009$ & $ 0.147\pm 0.020\pm 0.005$ & $ 0.064\pm 0.046\pm 0.018$ \\
 0.205 & 0.193-0.218 &  4.15 & $ 0.231\pm 0.014\pm 0.009$ & $ 0.070\pm 0.017\pm 0.005$ & $-0.088\pm 0.040\pm 0.015$ \\
 0.230 & 0.218-0.245 &  4.37 & $ 0.206\pm 0.014\pm 0.008$ & $ 0.072\pm 0.016\pm 0.005$ & $-0.047\pm 0.038\pm 0.014$ \\
 0.259 & 0.245-0.276 &  5.26 & $ 0.242\pm 0.014\pm 0.008$ & $ 0.049\pm 0.017\pm 0.004$ & $-0.140\pm 0.040\pm 0.013$ \\
 0.292 & 0.276-0.310 &  5.53 & $ 0.192\pm 0.014\pm 0.007$ & $ 0.089\pm 0.016\pm 0.004$ & $ 0.009\pm 0.037\pm 0.013$ \\
 0.329 & 0.310-0.349 &  6.01 & $ 0.168\pm 0.014\pm 0.006$ & $ 0.091\pm 0.015\pm 0.004$ & $ 0.032\pm 0.036\pm 0.013$ \\
 0.370 & 0.349-0.393 &  6.29 & $ 0.144\pm 0.013\pm 0.005$ & $ 0.076\pm 0.014\pm 0.004$ & $ 0.026\pm 0.034\pm 0.011$ \\
 0.416 & 0.393-0.442 &  6.56 & $ 0.148\pm 0.012\pm 0.005$ & $ 0.045\pm 0.013\pm 0.003$ & $-0.051\pm 0.030\pm 0.009$ \\
 0.468 & 0.442-0.497 &  6.79 & $ 0.114\pm 0.011\pm 0.004$ & $ 0.045\pm 0.012\pm 0.002$ & $-0.012\pm 0.027\pm 0.007$ \\
 0.526 & 0.497-0.559 &  7.72 & $ 0.084\pm 0.010\pm 0.003$ & $ 0.043\pm 0.011\pm 0.002$ & $ 0.009\pm 0.025\pm 0.006$ \\
 0.592 & 0.559-0.629 &  7.97 & $ 0.048\pm 0.009\pm 0.002$ & $ 0.019\pm 0.009\pm 0.001$ & $-0.007\pm 0.021\pm 0.003$ \\
 0.666 & 0.629-0.708 &  9.26 & $ 0.021\pm 0.007\pm 0.001$ & $ 0.008\pm 0.007\pm 0.001$ & $-0.004\pm 0.017\pm 0.001$ \\
 0.749 & 0.708-0.791 &  9.52 & $ 0.015\pm 0.005\pm 0.001$ & $ 0.009\pm 0.005\pm 0.001$ & $ 0.005\pm 0.011\pm 0.001$ \\
 \end{tabular}
 \end{table}

\begin{table}
\caption{Normalization systematic uncertainties (\%).}
\label{tb:norm_err}
\begin{tabular}{lcc}
Parameter           & Proton  & Deuteron   \\ \hline
Beam Polarization   & 2.4   & 2.4\\
Target Polarization & 2.5   & 4.0 \\
Dilution Factor     & 1.2   & 1.4 \\
Nitrogen Correction & 0.4   &   -  \\
\hline
Total               &3.7    & 4.9 \\
\end{tabular}
\end{table}

 \begin{table}[t] 
 \caption{Significant systematic errors for 
$g_1^d$ for  
 E=29 GeV. The beam and target 
polarization errors are each completely correlated and 
the $F_2$ error includes a correlated normalization error of
 $\sim$ 1\%.  }
 \label{tb:syserg10}
 \begin{tabular}{ccccccccc}
$x$ & $Q^2$ & $f$ & $F_2$ & R & $P_B$ & $P_T$ & $A_{rc}$ & TOTAL \\
 \hline
\multicolumn{9}{c}{$\theta=4.5^\circ$} \\
\hline
 0.031 &  1.30 & 0.0010 & 0.0009 & 0.0016 & 0.0018 & 0.0030 & 0.0298 & 0.0301\\
 0.035 &  1.40 & 0.0010 & 0.0009 & 0.0014 & 0.0018 & 0.0030 & 0.0249 & 0.0252\\
 0.050 &  1.80 & 0.0011 & 0.0012 & 0.0011 & 0.0020 & 0.0033 & 0.0130 & 0.0137\\
 0.080 &  2.30 & 0.0013 & 0.0015 & 0.0007 & 0.0023 & 0.0038 & 0.0044 & 0.0066\\
 0.125 &  2.90 & 0.0014 & 0.0017 & 0.0005 & 0.0025 & 0.0041 & 0.0024 & 0.0058\\
 0.175 &  3.40 & 0.0014 & 0.0018 & 0.0003 & 0.0025 & 0.0042 & 0.0031 & 0.0062\\
 0.250 &  3.80 & 0.0013 & 0.0017 & 0.0001 & 0.0023 & 0.0038 & 0.0039 & 0.0063\\
 0.350 &  4.10 & 0.0009 & 0.0014 & 0.0001 & 0.0017 & 0.0029 & 0.0037 & 0.0053\\
 0.500 &  4.40 & 0.0005 & 0.0008 & 0.0000 & 0.0009 & 0.0015 & 0.0018 & 0.0027\\
 \hline
\multicolumn{9}{c}{$\theta=7.0^\circ$} \\
\hline
 0.080 &  3.40 & 0.0013 & 0.0016 & 0.0029 & 0.0024 & 0.0040 & 0.0067 & 0.0089\\
 0.125 &  4.40 & 0.0014 & 0.0017 & 0.0016 & 0.0026 & 0.0044 & 0.0027 & 0.0064\\
 0.175 &  5.30 & 0.0014 & 0.0018 & 0.0009 & 0.0026 & 0.0044 & 0.0028 & 0.0063\\
 0.250 &  6.40 & 0.0013 & 0.0017 & 0.0005 & 0.0023 & 0.0039 & 0.0035 & 0.0061\\
 0.350 &  7.50 & 0.0009 & 0.0013 & 0.0003 & 0.0017 & 0.0028 & 0.0032 & 0.0048\\
 0.500 &  8.60 & 0.0004 & 0.0006 & 0.0001 & 0.0007 & 0.0012 & 0.0018 & 0.0024\\
 0.700 &  9.30 & 0.0001 & 0.0001 & 0.0000 & 0.0002 & 0.0003 & 0.0003 & 0.0005\\
 \end{tabular}
 \end{table}
 \begin{table}[t] 
 \caption{Significant systematic errors for 
$g_1^d/F_1^d$ for 
E=29 GeV. The beam and target 
polarization errors are each completely correlated.}
 \label{tb:sysergf13}
 \begin{tabular}{cccccccc}
$x$ & $Q^2$ & $f$ & R & $P_B$ & $P_T$ & $A_{rc}$ & TOTAL  \\
 \hline
\multicolumn{8}{c}{$\theta=4.5^\circ$} \\
\hline
 0.031 &  1.30 & 0.0003 & 0.0005 & 0.0005 & 0.0008 & 0.0080 & 0.0081\\
 0.035 &  1.40 & 0.0003 & 0.0006 & 0.0005 & 0.0009 & 0.0074 & 0.0075\\
 0.050 &  1.80 & 0.0004 & 0.0009 & 0.0008 & 0.0013 & 0.0052 & 0.0055\\
 0.080 &  2.30 & 0.0008 & 0.0020 & 0.0014 & 0.0023 & 0.0027 & 0.0044\\
 0.125 &  2.90 & 0.0013 & 0.0040 & 0.0023 & 0.0039 & 0.0023 & 0.0066\\
 0.175 &  3.40 & 0.0018 & 0.0053 & 0.0033 & 0.0055 & 0.0041 & 0.0095\\
 0.250 &  3.80 & 0.0025 & 0.0063 & 0.0046 & 0.0077 & 0.0080 & 0.0138\\
 0.350 &  4.10 & 0.0032 & 0.0077 & 0.0059 & 0.0099 & 0.0124 & 0.0189\\
 0.500 &  4.40 & 0.0037 & 0.0084 & 0.0068 & 0.0114 & 0.0139 & 0.0213\\
 \hline
\multicolumn{8}{c}{$\theta=7.0^\circ$} \\
\hline
 0.080 &  3.40 & 0.0008 & 0.0017 & 0.0014 & 0.0023 & 0.0039 & 0.0051\\
 0.125 &  4.40 & 0.0013 & 0.0030 & 0.0024 & 0.0039 & 0.0024 & 0.0061\\
 0.175 &  5.30 & 0.0019 & 0.0041 & 0.0034 & 0.0056 & 0.0035 & 0.0087\\
 0.250 &  6.40 & 0.0026 & 0.0060 & 0.0047 & 0.0078 & 0.0071 & 0.0132\\
 0.350 &  7.50 & 0.0032 & 0.0082 & 0.0060 & 0.0099 & 0.0115 & 0.0185\\
 0.500 &  8.60 & 0.0037 & 0.0099 & 0.0069 & 0.0114 & 0.0166 & 0.0238\\
 0.700 &  9.30 & 0.0037 & 0.0101 & 0.0069 & 0.0115 & 0.0132 & 0.0217\\
 \end{tabular}
 \end{table}
 \begin{table}[t] 
 \caption{Significant systematic errors for 
$g_1^p$ for  
 E=29 GeV. The beam and target 
polarization errors are each completely correlated and 
the $F_2$ error includes a correlated normalization error of
 $\sim$ 1\%.  }
 \label{tb:syserg23}
 \begin{tabular}{ccccccccc}
$x$ & $Q^2$ & $f$ & $F_2$ & R & $P_B$ & $P_T$ & $A_{rc}$ & TOTAL \\
 \hline
\multicolumn{9}{c}{$\theta=4.5^\circ$} \\
\hline
 0.031 &  1.30 & 0.0052 & 0.0062 & 0.0104 & 0.0071 & 0.0074 & 0.0320 & 0.0361\\
 0.035 &  1.40 & 0.0053 & 0.0061 & 0.0085 & 0.0070 & 0.0073 & 0.0269 & 0.0310\\
 0.050 &  1.80 & 0.0051 & 0.0059 & 0.0049 & 0.0068 & 0.0071 & 0.0156 & 0.0206\\
 0.080 &  2.30 & 0.0047 & 0.0056 & 0.0023 & 0.0064 & 0.0067 & 0.0094 & 0.0153\\
 0.125 &  2.90 & 0.0043 & 0.0055 & 0.0014 & 0.0060 & 0.0063 & 0.0060 & 0.0128\\
 0.175 &  3.40 & 0.0041 & 0.0054 & 0.0008 & 0.0057 & 0.0060 & 0.0047 & 0.0117\\
 0.250 &  3.80 & 0.0036 & 0.0050 & 0.0003 & 0.0052 & 0.0054 & 0.0038 & 0.0104\\
 0.350 &  4.10 & 0.0028 & 0.0039 & 0.0001 & 0.0041 & 0.0043 & 0.0033 & 0.0083\\
 0.500 &  4.40 & 0.0016 & 0.0022 & 0.0000 & 0.0024 & 0.0025 & 0.0017 & 0.0047\\
 \hline
\multicolumn{9}{c}{$\theta=7.0^\circ$} \\
\hline
 0.080 &  3.40 & 0.0046 & 0.0059 & 0.0095 & 0.0066 & 0.0069 & 0.0130 & 0.0202\\
 0.125 &  4.40 & 0.0046 & 0.0057 & 0.0042 & 0.0064 & 0.0067 & 0.0070 & 0.0143\\
 0.175 &  5.30 & 0.0043 & 0.0056 & 0.0022 & 0.0060 & 0.0063 & 0.0043 & 0.0122\\
 0.250 &  6.40 & 0.0037 & 0.0050 & 0.0012 & 0.0053 & 0.0055 & 0.0027 & 0.0102\\
 0.350 &  7.50 & 0.0027 & 0.0037 & 0.0006 & 0.0040 & 0.0041 & 0.0020 & 0.0076\\
 0.500 &  8.60 & 0.0014 & 0.0018 & 0.0002 & 0.0020 & 0.0021 & 0.0012 & 0.0039\\
 0.700 &  9.30 & 0.0003 & 0.0005 & 0.0000 & 0.0005 & 0.0005 & 0.0004 & 0.0010\\
 \end{tabular}
 \end{table}
 \begin{table}[t] 
 \caption{Significant systematic errors for 
$g_1^p/F_1^p$ for 
E=29 GeV. The beam and target 
polarization errors are each completely correlated.}
 \label{tb:sysergf23}
 \begin{tabular}{cccccccc}
$x$ & $Q^2$ & $f$ & R & $P_B$ & $P_T$ & $A_{rc}$ & TOTAL  \\
 \hline
\multicolumn{8}{c}{$\theta=4.5^\circ$} \\
\hline
 0.031 &  1.30 & 0.0013 & 0.0030 & 0.0018 & 0.0019 & 0.0082 & 0.0092\\
 0.035 &  1.40 & 0.0015 & 0.0034 & 0.0020 & 0.0021 & 0.0076 & 0.0089\\
 0.050 &  1.80 & 0.0019 & 0.0040 & 0.0026 & 0.0027 & 0.0059 & 0.0083\\
 0.080 &  2.30 & 0.0027 & 0.0062 & 0.0037 & 0.0039 & 0.0055 & 0.0102\\
 0.125 &  2.90 & 0.0037 & 0.0095 & 0.0052 & 0.0055 & 0.0052 & 0.0138\\
 0.175 &  3.40 & 0.0048 & 0.0111 & 0.0068 & 0.0071 & 0.0055 & 0.0165\\
 0.250 &  3.80 & 0.0062 & 0.0122 & 0.0089 & 0.0093 & 0.0066 & 0.0199\\
 0.350 &  4.10 & 0.0079 & 0.0146 & 0.0114 & 0.0119 & 0.0091 & 0.0251\\
 0.500 &  4.40 & 0.0096 & 0.0175 & 0.0140 & 0.0146 & 0.0100 & 0.0301\\
 \hline
\multicolumn{8}{c}{$\theta=7.0^\circ$} \\
\hline
 0.080 &  3.40 & 0.0025 & 0.0051 & 0.0036 & 0.0038 & 0.0072 & 0.0106\\
 0.125 &  4.40 & 0.0038 & 0.0073 & 0.0053 & 0.0055 & 0.0057 & 0.0126\\
 0.175 &  5.30 & 0.0048 & 0.0086 & 0.0069 & 0.0071 & 0.0049 & 0.0148\\
 0.250 &  6.40 & 0.0063 & 0.0115 & 0.0090 & 0.0094 & 0.0045 & 0.0190\\
 0.350 &  7.50 & 0.0079 & 0.0156 & 0.0115 & 0.0120 & 0.0057 & 0.0248\\
 0.500 &  8.60 & 0.0096 & 0.0206 & 0.0142 & 0.0148 & 0.0081 & 0.0317\\
 0.700 &  9.30 & 0.0100 & 0.0229 & 0.0149 & 0.0156 & 0.0129 & 0.0355\\
 \end{tabular}
 \end{table}


 \begin{table}[t] 
 \caption{Deuteron  results for $g_1$ at fixed $Q^2$ of
 2, 3, and 5 (GeV/c)$^2$
 evaluated assuming $g_1/F_1$ is independent of $Q^2.$
 In addition to the systematic errors shown, there are 
 normalization uncertainties shown in 
Table~ \protect\ref{tb:norm_err}.  }
 \label{tb:g1q0_1}
 \begin{tabular}{rrrr}
$x$ &   $g_1^d(Q^2=2)\pm {\rm stat} \pm {\rm syst} $ & $g_1^d(Q^2=3)\pm {\rm stat} \pm {\rm syst} $ & $g_1^d(Q^2=5)\pm {\rm stat} \pm {\rm syst} $ \\
 \hline
 0.031  & $ 0.166\pm 0.162\pm 0.033$ & $ 0.180\pm 0.176\pm 0.036$ & $ 0.198\pm 0.194\pm 0.040$ \\
 0.035  & $ 0.029\pm 0.104\pm 0.027$ & $ 0.031\pm 0.112\pm 0.029$ & $ 0.034\pm 0.124\pm 0.032$ \\
 0.039  & $ 0.202\pm 0.084\pm 0.022$ & $ 0.218\pm 0.091\pm 0.024$ & $ 0.239\pm 0.100\pm 0.027$ \\
 0.044  & $-0.038\pm 0.071\pm 0.017$ & $-0.040\pm 0.076\pm 0.019$ & $-0.044\pm 0.083\pm 0.020$ \\
 0.049  & $-0.035\pm 0.060\pm 0.014$ & $-0.037\pm 0.064\pm 0.014$ & $-0.040\pm 0.070\pm 0.016$ \\
 0.056  & $ 0.040\pm 0.052\pm 0.011$ & $ 0.042\pm 0.056\pm 0.011$ & $ 0.046\pm 0.061\pm 0.012$ \\
 0.063  & $ 0.027\pm 0.045\pm 0.008$ & $ 0.028\pm 0.048\pm 0.008$ & $ 0.031\pm 0.052\pm 0.009$ \\
 0.071  & $ 0.053\pm 0.040\pm 0.006$ & $ 0.056\pm 0.042\pm 0.007$ & $ 0.060\pm 0.046\pm 0.007$ \\
 0.079  & $ 0.038\pm 0.036\pm 0.005$ & $ 0.040\pm 0.038\pm 0.005$ & $ 0.044\pm 0.041\pm 0.006$ \\
 0.090  & $ 0.120\pm 0.030\pm 0.005$ & $ 0.125\pm 0.032\pm 0.005$ & $ 0.135\pm 0.034\pm 0.005$ \\
 0.101  & $ 0.079\pm 0.028\pm 0.004$ & $ 0.083\pm 0.029\pm 0.004$ & $ 0.089\pm 0.031\pm 0.004$ \\
 0.113  & $ 0.064\pm 0.025\pm 0.003$ & $ 0.067\pm 0.026\pm 0.004$ & $ 0.072\pm 0.028\pm 0.004$ \\
 0.128  & $ 0.128\pm 0.023\pm 0.004$ & $ 0.133\pm 0.024\pm 0.004$ & $ 0.141\pm 0.025\pm 0.004$ \\
 0.144  & $ 0.130\pm 0.021\pm 0.004$ & $ 0.135\pm 0.022\pm 0.004$ & $ 0.142\pm 0.023\pm 0.004$ \\
 0.162  & $ 0.071\pm 0.020\pm 0.003$ & $ 0.073\pm 0.020\pm 0.004$ & $ 0.076\pm 0.021\pm 0.004$ \\
 0.182  & $ 0.141\pm 0.019\pm 0.004$ & $ 0.143\pm 0.019\pm 0.004$ & $ 0.148\pm 0.020\pm 0.005$ \\
 0.205  & $ 0.068\pm 0.017\pm 0.005$ & $ 0.069\pm 0.017\pm 0.005$ & $ 0.070\pm 0.017\pm 0.005$ \\
 0.230  & $ 0.071\pm 0.016\pm 0.005$ & $ 0.071\pm 0.016\pm 0.005$ & $ 0.072\pm 0.016\pm 0.005$ \\
 0.259  & $ 0.049\pm 0.017\pm 0.004$ & $ 0.049\pm 0.017\pm 0.004$ & $ 0.049\pm 0.017\pm 0.004$ \\
 0.292  & $ 0.094\pm 0.017\pm 0.004$ & $ 0.092\pm 0.016\pm 0.004$ & $ 0.090\pm 0.016\pm 0.004$ \\
 0.329  & $ 0.100\pm 0.017\pm 0.004$ & $ 0.096\pm 0.016\pm 0.004$ & $ 0.092\pm 0.016\pm 0.004$ \\
 0.370  & $ 0.089\pm 0.017\pm 0.004$ & $ 0.083\pm 0.016\pm 0.004$ & $ 0.078\pm 0.015\pm 0.004$ \\
 0.416  & $ 0.057\pm 0.016\pm 0.004$ & $ 0.051\pm 0.015\pm 0.003$ & $ 0.047\pm 0.014\pm 0.003$ \\
 0.468  & $ 0.064\pm 0.017\pm 0.004$ & $ 0.055\pm 0.014\pm 0.003$ & $ 0.048\pm 0.012\pm 0.002$ \\
 0.526  & $ 0.074\pm 0.019\pm 0.004$ & $ 0.060\pm 0.015\pm 0.003$ & $ 0.049\pm 0.012\pm 0.002$ \\
 0.592  & $ 0.043\pm 0.020\pm 0.002$ & $ 0.031\pm 0.014\pm 0.002$ & $ 0.023\pm 0.011\pm 0.001$ \\
 0.666  & $ 0.026\pm 0.024\pm 0.002$ & $ 0.017\pm 0.016\pm 0.001$ & $ 0.011\pm 0.010\pm 0.001$ \\
 0.749  & $ 0.048\pm 0.025\pm 0.002$ & $ 0.028\pm 0.014\pm 0.001$ & $ 0.016\pm 0.008\pm 0.001$ \\
 \end{tabular}
 \end{table}
 \begin{table}[t] 
 \caption{Proton    results for $g_1$ at fixed $Q^2$ of
 2, 3, and 5 (GeV/c)$^2$
 evaluated assuming $g_1/F_1$ is independent of $Q^2.$
 In addition to the systematic errors shown, there are 
 normalization uncertainties shown in 
Table~ \protect\ref{tb:norm_err}.  }
 \label{tb:g1q0_2}
 \begin{tabular}{rrrr}
$x$ &   $g_1^p(Q^2=2)\pm {\rm stat} \pm {\rm syst} $ & $g_1^p(Q^2=3)\pm {\rm stat} \pm {\rm syst} $ & $g_1^p(Q^2=5)\pm {\rm stat} \pm {\rm syst} $ \\
 \hline
 0.031  & $ 0.272\pm 0.145\pm 0.037$ & $ 0.294\pm 0.157\pm 0.040$ & $ 0.322\pm 0.172\pm 0.044$ \\
 0.035  & $ 0.468\pm 0.096\pm 0.032$ & $ 0.504\pm 0.103\pm 0.035$ & $ 0.552\pm 0.113\pm 0.041$ \\
 0.039  & $ 0.283\pm 0.077\pm 0.026$ & $ 0.304\pm 0.083\pm 0.028$ & $ 0.333\pm 0.091\pm 0.031$ \\
 0.044  & $ 0.334\pm 0.065\pm 0.022$ & $ 0.357\pm 0.069\pm 0.024$ & $ 0.390\pm 0.075\pm 0.027$ \\
 0.049  & $ 0.334\pm 0.054\pm 0.019$ & $ 0.357\pm 0.058\pm 0.020$ & $ 0.389\pm 0.063\pm 0.023$ \\
 0.056  & $ 0.312\pm 0.046\pm 0.017$ & $ 0.332\pm 0.049\pm 0.018$ & $ 0.362\pm 0.053\pm 0.020$ \\
 0.063  & $ 0.267\pm 0.039\pm 0.014$ & $ 0.283\pm 0.041\pm 0.016$ & $ 0.308\pm 0.045\pm 0.017$ \\
 0.071  & $ 0.292\pm 0.034\pm 0.013$ & $ 0.309\pm 0.036\pm 0.014$ & $ 0.336\pm 0.039\pm 0.016$ \\
 0.079  & $ 0.305\pm 0.030\pm 0.013$ & $ 0.322\pm 0.032\pm 0.014$ & $ 0.349\pm 0.034\pm 0.015$ \\
 0.090  & $ 0.257\pm 0.024\pm 0.011$ & $ 0.270\pm 0.025\pm 0.012$ & $ 0.292\pm 0.027\pm 0.013$ \\
 0.101  & $ 0.292\pm 0.022\pm 0.011$ & $ 0.306\pm 0.023\pm 0.012$ & $ 0.330\pm 0.025\pm 0.013$ \\
 0.113  & $ 0.264\pm 0.020\pm 0.010$ & $ 0.276\pm 0.021\pm 0.011$ & $ 0.296\pm 0.022\pm 0.011$ \\
 0.128  & $ 0.253\pm 0.019\pm 0.010$ & $ 0.263\pm 0.019\pm 0.010$ & $ 0.281\pm 0.021\pm 0.010$ \\
 0.144  & $ 0.229\pm 0.017\pm 0.009$ & $ 0.238\pm 0.018\pm 0.009$ & $ 0.252\pm 0.019\pm 0.009$ \\
 0.162  & $ 0.242\pm 0.016\pm 0.009$ & $ 0.250\pm 0.017\pm 0.009$ & $ 0.263\pm 0.018\pm 0.009$ \\
 0.182  & $ 0.232\pm 0.016\pm 0.008$ & $ 0.238\pm 0.016\pm 0.008$ & $ 0.248\pm 0.017\pm 0.009$ \\
 0.205  & $ 0.225\pm 0.014\pm 0.008$ & $ 0.229\pm 0.014\pm 0.008$ & $ 0.236\pm 0.014\pm 0.009$ \\
 0.230  & $ 0.202\pm 0.013\pm 0.008$ & $ 0.204\pm 0.013\pm 0.008$ & $ 0.208\pm 0.014\pm 0.008$ \\
 0.259  & $ 0.241\pm 0.014\pm 0.008$ & $ 0.241\pm 0.014\pm 0.008$ & $ 0.242\pm 0.014\pm 0.008$ \\
 0.292  & $ 0.197\pm 0.014\pm 0.007$ & $ 0.195\pm 0.014\pm 0.007$ & $ 0.192\pm 0.014\pm 0.007$ \\
 0.329  & $ 0.179\pm 0.015\pm 0.006$ & $ 0.174\pm 0.014\pm 0.006$ & $ 0.169\pm 0.014\pm 0.006$ \\
 0.370  & $ 0.161\pm 0.014\pm 0.005$ & $ 0.153\pm 0.014\pm 0.005$ & $ 0.146\pm 0.013\pm 0.005$ \\
 0.416  & $ 0.177\pm 0.014\pm 0.005$ & $ 0.164\pm 0.013\pm 0.005$ & $ 0.151\pm 0.012\pm 0.005$ \\
 0.468  & $ 0.151\pm 0.014\pm 0.005$ & $ 0.134\pm 0.012\pm 0.004$ & $ 0.118\pm 0.011\pm 0.004$ \\
 0.526  & $ 0.139\pm 0.017\pm 0.004$ & $ 0.115\pm 0.014\pm 0.003$ & $ 0.095\pm 0.012\pm 0.003$ \\
 0.592  & $ 0.105\pm 0.019\pm 0.003$ & $ 0.078\pm 0.014\pm 0.002$ & $ 0.059\pm 0.011\pm 0.002$ \\
 0.666  & $ 0.070\pm 0.024\pm 0.002$ & $ 0.046\pm 0.016\pm 0.001$ & $ 0.030\pm 0.010\pm 0.001$ \\
 0.749  & $ 0.080\pm 0.025\pm 0.002$ & $ 0.046\pm 0.014\pm 0.001$ & $ 0.026\pm 0.008\pm 0.001$ \\
 \end{tabular}
 \end{table}
 \begin{table}[t] 
 \caption{Neutron   results for $g_1$ at fixed $Q^2$ of 
 2, 3, and 5 (GeV/c)$^2$
 evaluated from  $g_1^p$ and $g_1^d$ assuming $g_1/F_1$ is 
 independent of $Q^2.$
 In addition there is
 a normalization uncertainty common to all data of 
2.4\% due to beam 
polarization.}
 \label{tb:g1q0_3}
 \begin{tabular}{rrrr}
$x$ &   $g_1^n(Q^2=2)\pm {\rm stat} \pm {\rm syst} $ & $g_1^n(Q^2=3)\pm {\rm stat} \pm {\rm syst} $ & $g_1^n(Q^2=5)\pm {\rm stat} \pm {\rm syst} $ \\
 \hline
 0.031  & $ 0.085\pm 0.378\pm 0.081$ & $ 0.091\pm 0.408\pm 0.088$ & $ 0.100\pm 0.448\pm 0.097$ \\
 0.035  & $-0.406\pm 0.244\pm 0.067$ & $-0.437\pm 0.262\pm 0.072$ & $-0.479\pm 0.287\pm 0.080$ \\
 0.039  & $ 0.153\pm 0.197\pm 0.056$ & $ 0.164\pm 0.212\pm 0.060$ & $ 0.180\pm 0.232\pm 0.066$ \\
 0.044  & $-0.415\pm 0.166\pm 0.044$ & $-0.444\pm 0.177\pm 0.048$ & $-0.485\pm 0.194\pm 0.053$ \\
 0.049  & $-0.413\pm 0.141\pm 0.036$ & $-0.440\pm 0.151\pm 0.039$ & $-0.480\pm 0.164\pm 0.043$ \\
 0.056  & $-0.225\pm 0.122\pm 0.029$ & $-0.240\pm 0.130\pm 0.030$ & $-0.261\pm 0.142\pm 0.033$ \\
 0.063  & $-0.229\pm 0.106\pm 0.023$ & $-0.243\pm 0.112\pm 0.024$ & $-0.264\pm 0.122\pm 0.027$ \\
 0.071  & $-0.180\pm 0.094\pm 0.019$ & $-0.190\pm 0.099\pm 0.021$ & $-0.206\pm 0.107\pm 0.022$ \\
 0.079  & $-0.240\pm 0.084\pm 0.017$ & $-0.252\pm 0.088\pm 0.018$ & $-0.273\pm 0.095\pm 0.020$ \\
 0.090  & $ 0.015\pm 0.071\pm 0.017$ & $ 0.016\pm 0.074\pm 0.018$ & $ 0.017\pm 0.080\pm 0.019$ \\
 0.101  & $-0.125\pm 0.064\pm 0.015$ & $-0.131\pm 0.067\pm 0.016$ & $-0.141\pm 0.072\pm 0.017$ \\
 0.113  & $-0.136\pm 0.058\pm 0.013$ & $-0.141\pm 0.061\pm 0.014$ & $-0.151\pm 0.065\pm 0.015$ \\
 0.128  & $ 0.022\pm 0.054\pm 0.015$ & $ 0.023\pm 0.056\pm 0.016$ & $ 0.024\pm 0.059\pm 0.017$ \\
 0.144  & $ 0.053\pm 0.050\pm 0.015$ & $ 0.055\pm 0.051\pm 0.016$ & $ 0.058\pm 0.054\pm 0.016$ \\
 0.162  & $-0.095\pm 0.047\pm 0.012$ & $-0.098\pm 0.048\pm 0.012$ & $-0.102\pm 0.050\pm 0.013$ \\
 0.182  & $ 0.061\pm 0.044\pm 0.016$ & $ 0.062\pm 0.045\pm 0.016$ & $ 0.064\pm 0.046\pm 0.017$ \\
 0.205  & $-0.086\pm 0.039\pm 0.014$ & $-0.087\pm 0.039\pm 0.014$ & $-0.089\pm 0.040\pm 0.014$ \\
 0.230  & $-0.047\pm 0.038\pm 0.014$ & $-0.047\pm 0.038\pm 0.014$ & $-0.047\pm 0.038\pm 0.014$ \\
 0.259  & $-0.144\pm 0.041\pm 0.012$ & $-0.141\pm 0.040\pm 0.012$ & $-0.140\pm 0.040\pm 0.012$ \\
 0.292  & $ 0.009\pm 0.040\pm 0.013$ & $ 0.009\pm 0.039\pm 0.013$ & $ 0.009\pm 0.038\pm 0.013$ \\
 0.329  & $ 0.036\pm 0.041\pm 0.013$ & $ 0.034\pm 0.039\pm 0.013$ & $ 0.032\pm 0.037\pm 0.012$ \\
 0.370  & $ 0.031\pm 0.040\pm 0.012$ & $ 0.029\pm 0.037\pm 0.011$ & $ 0.027\pm 0.035\pm 0.011$ \\
 0.416  & $-0.067\pm 0.040\pm 0.010$ & $-0.060\pm 0.036\pm 0.009$ & $-0.054\pm 0.032\pm 0.009$ \\
 0.468  & $-0.018\pm 0.040\pm 0.009$ & $-0.015\pm 0.034\pm 0.008$ & $-0.013\pm 0.029\pm 0.007$ \\
 0.526  & $ 0.016\pm 0.047\pm 0.009$ & $ 0.013\pm 0.037\pm 0.008$ & $ 0.010\pm 0.029\pm 0.006$ \\
 0.592  & $-0.018\pm 0.051\pm 0.007$ & $-0.012\pm 0.036\pm 0.005$ & $-0.009\pm 0.026\pm 0.004$ \\
 0.666  & $-0.015\pm 0.067\pm 0.005$ & $-0.009\pm 0.041\pm 0.003$ & $-0.006\pm 0.026\pm 0.002$ \\
 0.749  & $ 0.034\pm 0.073\pm 0.007$ & $ 0.018\pm 0.039\pm 0.004$ & $ 0.009\pm 0.021\pm 0.002$ \\
 \end{tabular}
 \end{table}


 \begin{table}[t] 
 \caption{  $\int_{.03}^{.8}{g_1(x)dx}$ at
 different $Q^2$ by different methods. The first error is 
 statistical and the second is systematic.  There are
 additional  normalization uncertainties shown in 
Table~ \protect\ref{tb:norm_err}.  }
 \label{tb:gam1m}
 \begin{tabular}{rrrrr}
 & method & $Q^2= 2{\rm ~(GeV/c)^2}$ & $Q^2= 3{\rm ~(GeV/c)^2}$ & $Q^2= 5{\rm ~(GeV/c)^2}$  \\
 \hline
 Deuteron &$g_1/F_1$& $ 0.050\pm 0.004\pm 0.003$ & $ 0.046\pm 0.003\pm 0.003$ & $ 0.043\pm 0.003\pm 0.002$ \\
 Deuteron &$A_1$    & $ 0.047\pm 0.005\pm 0.003$ & $ 0.044\pm 0.004\pm 0.003$ & $ 0.043\pm 0.003\pm 0.002$ \\
 \hline
 Proton   &$g_1/F_1$& $ 0.129\pm 0.004\pm 0.006$ & $ 0.121\pm 0.003\pm 0.006$ & $ 0.117\pm 0.003\pm 0.006$ \\
 Proton   &$A_1$    & $ 0.120\pm 0.004\pm 0.006$ & $ 0.116\pm 0.003\pm 0.006$ & $ 0.116\pm 0.003\pm 0.006$ \\
 \hline
 Neutron  &$g_1/F_1$& $-0.022\pm 0.011\pm 0.006$ & $-0.023\pm 0.008\pm 0.006$ & $-0.025\pm 0.007\pm 0.006$ \\
 Neutron  &$A_1$    & $-0.019\pm 0.013\pm 0.005$ & $-0.021\pm 0.009\pm 0.005$ & $-0.023\pm 0.007\pm 0.005$ \\
 \hline
 \end{tabular}
 \end{table}


 \begin{table}[t] 
 \caption{ Systematic errors on the measured
 integral at $Q^2=$ 3~(GeV/c)$^2$.}
 \label{tb:gam1_se}
 \begin{tabular}{rrrrr}
 SOURCE & Deuteron & Proton & Neutron & p-n \\
 \hline
$F_2$    &  0.001 &  0.003 &  0.001 &  0.003   \\
$R$      &  0.000 &  0.001 &  0.000 &  0.001   \\
$A_{rc}$ &  0.001 &  0.002 &  0.003 &  0.005   \\
f        &  0.001 &  0.002 &  0.001 &  0.003   \\
$P_B$    &  0.001 &  0.003 &  0.001 &  0.003   \\
$P_T$    &  0.002 &  0.003 &  0.005 &  0.007   \\
TOTAL    &  0.003 &  0.006 &  0.006 &  0.010   \\

 \end{tabular}
 \end{table}


 \begin{table}[t] 
 \caption{ Estimates of 
 $\int_0^{0.03}g_1(x,Q^2)dx$ at $Q^2=3
 $~(GeV/c)$^2$ using various hypotheses.
 Columns 3-5 have
 the Regge form $g_1 = \beta x^\alpha$ fitted
 to $g_1(x,Q^2)$ at the $Q^2$ shown in the second row,
 with $\alpha$ shown in the first row,
 in the  range $0.03 \leq x \leq x_{cut}$ 
  shown in column 2.
 Column 6 has a fit of the form $ln(1/x)$.
 Column 7 has results of global fit II of
 Table \protect\ref{tb:g1f1_fit}.
 The last column is the integral of the SMC data with
 flat Regge extrapolation ($\alpha=0$) below $x=0.003.$}
 \label{tb:lowx}
 \begin{tabular}{dccccccc}
  & $x_{cut} $ & $ \alpha=0$ & $\alpha=0$ &
 $\alpha=0.5$ &    ln(1/x) & global II  & SMC  \\
   &    & $Q^2$=3 &  $Q^2$=1 &  $Q^2$=3 &    $Q^2$=3 &  $Q^2$=3 &  $Q^2$=3    \\
\hline
Deuteron & $0.10$ & $ \phantom{+}0.002$ & $ \phantom{+}0.002$ & $ \phantom{+}0.001$ & $ \phantom{+}0.003$ & $ \phantom{+}0.001$ & $ -0.005\pm$0.003\\
Proton   & $0.10$ & $ \phantom{+}0.009$ & $ \phantom{+}0.010$ & $ \phantom{+}0.004$ & $ \phantom{+}0.015$ & $ \phantom{+}0.018$ & $ \phantom{+}0.014\pm$0.003\\
Neutron  & $0.10$ & $ -0.005$ & $ -0.006$ & $ -0.002$ & $ -0.009$ & $ -0.016$ & $ -0.025\pm$0.007\\
p-n      & $0.10$ & $ \phantom{+}0.014$ & $ \phantom{+}0.016$ & $ \phantom{+}0.006$ & $ \phantom{+}0.024$ & $ \phantom{+}0.034$ & $ \phantom{+}0.039\pm$0.009\\
\hline
Deuteron & $0.06$  & $ \phantom{+}0.001$  & $ \phantom{+}0.001$  & $ \phantom{+}0.000$ & $ \phantom{+}0.002$ & & \\Proton   & $\phantom{+}.06$  & $ \phantom{+}0.010$  & $ \phantom{+}0.011$  & $ \phantom{+}0.005$ & $ \phantom{+}0.015$ & & \\Neutron  & $\phantom{+}.06$  & $ -0.008$  & $ -0.009$  & $ -0.004$ & $ -0.012$ & & \\p-n      & $\phantom{+}.06$  & $ \phantom{+}0.018$  & $ \phantom{+}0.020$  & $ \phantom{+}0.010$ & $ \phantom{+}0.026$ & & \\
 \end{tabular}
 \end{table}


 \begin{table}[t] 
\caption{Integral of $g_1$ in the measured region
 as well as extrapolations to high and low $x$ as described
 in the text. 
 Slight differences between the measured
 targets and derived targets, n and p-n, are due to
 correlations among systematic errors. The structure 
  function $g_1$ was calculated
 at fixed $Q^2$ assuming $g_1/F_1$ independent of $Q^2.$}
 \label{tb:integral}
 \begin{tabular}{cccccccc}
      & $<Q^2>$ &  Measured &  high $x$ & low $x$ &  Total \\
 & ${\rm (GeV/c)^2}$ &  $\int_{.03}^{.8}g_1$ & $ \int_{.8}^{1}g_1 $ & $ \int_{0}^{.03}g_1$ & $ \int_{0}^{1}g_1$   \\ 
\hline
 Deuteron & 2& $\phantom{+}0.050\pm 0.004\pm 0.003$ & $\phantom{+}0.000\pm 0.001$ & $\phantom{+}0.001\pm 0.006$ & $\phantom{+}0.051\pm 0.004\pm0.006$   \\
 Proton   & 2& $\phantom{+}0.129\pm 0.004\pm 0.006$ & $\phantom{+}0.001\pm 0.001$ & $\phantom{+}0.011\pm 0.007$ & $\phantom{+}0.140\pm 0.004\pm0.010$   \\
 Neutron  & 2& $-0.022\pm 0.011\pm 0.006$ & $\phantom{+}0.001\pm 0.001$ & $-0.009\pm 0.016$ & $-0.030\pm 0.011\pm0.017$   \\
 p-n      & 2& $\phantom{+}0.149\pm 0.012\pm 0.011$ & $\phantom{+}0.001\pm 0.001$ & $\phantom{+}0.020\pm 0.019$ & $\phantom{+}0.169\pm 0.012\pm0.022$   \\
\hline
 Deuteron & 3& $\phantom{+}0.046\pm 0.003\pm 0.003$ & $\phantom{+}0.000\pm 0.001$ & $\phantom{+}0.001\pm 0.006$ & $\phantom{+}0.047\pm 0.003\pm0.006$   \\
 Proton   & 3& $\phantom{+}0.121\pm 0.003\pm 0.006$ & $\phantom{+}0.001\pm 0.001$ & $\phantom{+}0.011\pm 0.007$ & $\phantom{+}0.133\pm 0.003\pm0.009$   \\
 Neutron  & 3& $-0.023\pm 0.008\pm 0.006$ & $\phantom{+}0.001\pm 0.001$ & $-0.010\pm 0.015$ & $-0.032\pm 0.008\pm0.016$   \\
 p-n      & 3& $\phantom{+}0.143\pm 0.009\pm 0.010$ & $\phantom{+}0.001\pm 0.001$ & $\phantom{+}0.021\pm 0.018$ & $\phantom{+}0.164\pm 0.009\pm0.021$   \\
\hline
 Deuteron & 5& $\phantom{+}0.043\pm 0.003\pm 0.002$ & $\phantom{+}0.000\pm 0.001$ & $\phantom{+}0.001\pm 0.006$ & $\phantom{+}0.044\pm 0.003\pm0.006$   \\
 Proton   & 5& $\phantom{+}0.117\pm 0.003\pm 0.006$ & $\phantom{+}0.001\pm 0.001$ & $\phantom{+}0.012\pm 0.008$ & $\phantom{+}0.129\pm 0.003\pm0.010$   \\
 Neutron  & 5& $-0.025\pm 0.007\pm 0.006$ & $\phantom{+}0.001\pm 0.001$ & $-0.010\pm 0.015$ & $-0.034\pm 0.007\pm0.016$   \\
 p-n      & 5& $\phantom{+}0.141\pm 0.008\pm 0.010$ & $\phantom{+}0.001\pm 0.001$ & $\phantom{+}0.022\pm 0.017$ & $\phantom{+}0.164\pm 0.008\pm0.020$   \\
 \end{tabular}
 \end{table}


 \begin{table}[t] 
\caption{Comparison of integrals from this
 experiment and E142\protect\cite{e142}
 and SMC\protect\cite{smcp,smcd}.
 Note that the SMC and E143 results used $g_1/F_1$
 independent of $Q^2$ and E142 used $A_1$ independent
 of $Q^2$  to evaluate $g_1$
 at fixed $Q^2$ from measurements at different $Q^2$,
  and that
 the different experiments had different mean $Q^2$.}
 \label{tb:int_comp}
 \begin{tabular}{ccccccc}
  & $Q^2$  & method & $x$ range & this experiment &  \multicolumn{2}{c}{$\leftarrow$ other experiments $\rightarrow$}  \\
 \hline
Proton &  5& $g_1/F_1$ & $0.03\leq x \leq 0.7$ & $ \phantom{+}0.115\pm0.006$ & SMC  & $ \phantom{+}0.128\pm0.006$ \\
 &  &  & $0 \leq x \leq 1$ & $ \phantom{+}0.129\pm0.010$ &   & $ \phantom{+}0.140\pm0.011$ \\
Deuteron &  5& $g_1/F_1$ & $0.03\leq x \leq 0.7$ & $ \phantom{+}0.041\pm0.004$ & SMC  & $ \phantom{+}0.043\pm0.007$ \\
 &  &  & $0 \leq x \leq 1$ & $ \phantom{+}0.044\pm0.007$ &   & $ \phantom{+}0.039\pm0.008$ \\
Neutron &  2& $A_1$ & $0.03\leq x \leq 0.6 $ & $ -0.021\pm0.009$ & E142  & $ -0.028\pm0.008$ \\
 &  &  & $0 \leq x \leq 1 $  & $ -0.030 \pm0.020$ &    & $ -0.031\pm0.011$ \\
 \end{tabular}
 \end{table}


 \begin{table}[t] 
\caption{Experimental value of $\Gamma_1$ 
 compared to the Ellis-Jaffe sum rule and Bjorken sum
 rule (p$-$n).  For the theoretical input we take
 $\alpha_s(M_Z)=0.118\pm0.003$ 
 and  $3F-D$ =0.58 with uncertainties  of either 
0.032 (small) or 
0.120 (large).
 The Ellis-Jaffe sum is evaluated with both
 the invariant and $Q^2$-dependent pQCD singlet corrections.
 The Bjorken sum rule depends only on the non-singlet
  correction. }
 \label{tb:ej}
 \begin{tabular}{ccccccc}
 & $<Q^2>$ & $\Gamma_1^{exp}$ & $ \Gamma_1^{theory} $ & $ \Gamma_1^{theory} $ & error &  error   \\ 
 &  (GeV/c)$^2$ &  & invariant & $Q^2$-dependent  & (small) &  (large) \\
 \hline
 Deuteron & 2& $\phantom{+}0.051\pm 0.008$ & $\phantom{+}0.070$  & $\phantom{+}0.065$  & $ \pm 0.004$ & $ \pm 0.014$  \\
 Proton   & 2& $\phantom{+}0.140\pm 0.010$ & $\phantom{+}0.161$  & $\phantom{+}0.156$  & $ \pm 0.005$ & $ \pm 0.016$  \\
 Neutron  & 2& $-0.030\pm 0.020$ & $-0.010$  & $-0.015$  & $ \pm 0.005$ & $ \pm 0.016$  \\
 p-n      & 2& $\phantom{+}0.169\pm 0.025$ & $\phantom{+}0.171$  & $\phantom{+}0.171$  & $ \pm 0.006$ & $ \pm 0.006$  \\
 \hline
 Deuteron & 3& $\phantom{+}0.047\pm 0.007$ & $\phantom{+}0.071$  & $\phantom{+}0.066$  & $ \pm 0.004$ & $ \pm 0.014$  \\
 Proton   & 3& $\phantom{+}0.133\pm 0.010$ & $\phantom{+}0.165$  & $\phantom{+}0.160$  & $ \pm 0.005$ & $ \pm 0.016$  \\
 Neutron  & 3& $-0.032\pm 0.018$ & $-0.012$  & $-0.017$  & $ \pm 0.004$ & $ \pm 0.016$  \\
 p-n      & 3& $\phantom{+}0.164\pm 0.023$ & $\phantom{+}0.177$  & $\phantom{+}0.177$  & $ \pm 0.004$ & $ \pm 0.004$  \\
 \hline
 Deuteron & 5& $\phantom{+}0.044\pm 0.007$ & $\phantom{+}0.072$  & $\phantom{+}0.068$  & $ \pm 0.004$ & $ \pm 0.015$  \\
 Proton   & 5& $\phantom{+}0.129\pm 0.010$ & $\phantom{+}0.169$  & $\phantom{+}0.164$  & $ \pm 0.005$ & $ \pm 0.016$  \\
 Neutron  & 5& $-0.034\pm 0.017$ & $-0.014$  & $-0.018$  & $ \pm 0.004$ & $ \pm 0.016$  \\
 p-n      & 5& $\phantom{+}0.164\pm 0.021$ & $\phantom{+}0.182$  & $\phantom{+}0.182$  & $ \pm 0.003$ & $ \pm 0.003$  \\
 \end{tabular}
 \end{table}


 \begin{table}[t] 
 \caption{ The evaluated quark spins 
  using both the ``invariant"
 and $Q^2$-dependent pQCD singlet coefficients,
 with an assumed error on $F/D$ of 0.016. 
  If the more conservative
 estimate of $\delta(3F-D)=0.12$ is used, the only
  change is to the error on $\Delta s$ which is
  shown in the last column.}
 \label{tb:delta}
 \begin{tabular}{ccccccc}
  & method & $\Sigma=a_o$ & $\Delta u$ &  $\Delta d$ & $\Delta s$ & Conservative $\delta\Delta s$ \\
 \hline
Deuterium & invariant            & $ 0.35\pm$0.07 & $ 0.84\pm$0.02 & $-0.42\pm$0.02 & $-0.08\pm$0.03 & $\pm$0.05 \\
Proton    & invariant            & $ 0.29\pm$0.09 & $ 0.83\pm$0.03 & $-0.43\pm$0.03 & $-0.10\pm$0.03 & $\pm$0.06 \\
Deuterium & $Q^2=$3~(GeV/c)$^2$  & $ 0.37\pm$0.08 & $ 0.85\pm$0.03 & $-0.41\pm$0.03 & $-0.07\pm$0.03 & $\pm$0.06 \\
Proton    & $Q^2=$3~(GeV/c)$^2$  & $ 0.32\pm$0.10 & $ 0.83\pm$0.03 & $-0.43\pm$0.03 & $-0.09\pm$0.04 & $\pm$0.06 \\
 \end{tabular}
 \end{table}


\begin{table}[t]
\caption{ Results for $A_2$, $g_2$ and $\overline{g_2}$ for the proton 
measured in the $4.5^\circ$ and  $7.0^\circ$ spectrometers at the 
indicated average values of 
$x$ and $Q^2$ and beam energy of 29.1 GeV. The highest $x$ bin shown is in 
the resonance region defined by missing mass $W^2<4$ GeV$^2$. }
\label{tb:g2A2p}
\begin{tabular}{cccccc} 
$x$ interval & $<x>$ & $<Q^2>$ & $A_2^p$ & $g_2^p$ & $\overline{g_2^p}$ \\
  & &  (GeV/c)$^2$ &$\pm$stat $\pm$syst  &$\pm$stat $\pm$syst  & $\pm$stat $\pm$syst    \\
\hline

$  0.029-0.047$ & 0.038& 1.49& $  0.016\pm 0.018\pm 0.006$& $\phantom{+} 0.489\pm 0.980\pm 0.332$& $\phantom{+} 0.223\pm 0.983\pm 0.332$\\
$  0.047-0.075$ & 0.060& 2.01& $  0.025\pm 0.014\pm 0.005$& $\phantom{+} 0.397\pm 0.374\pm 0.138$& $\phantom{+} 0.223\pm 0.375\pm 0.138$\\
$  0.075-0.120$ & 0.095& 2.60& $  0.004\pm 0.015\pm 0.006$& $      -0.236\pm 0.203\pm 0.074$&      $ -0.295\pm 0.204\pm 0.074$\\
$  0.120-0.193$ & 0.152& 3.21& $  0.021\pm 0.021\pm 0.008$& $      -0.136\pm 0.125\pm 0.049$&      $ -0.127\pm 0.127\pm 0.049$\\
$  0.193-0.310$ & 0.241& 3.77& $  0.091\pm 0.032\pm 0.011$& $      -0.046\pm 0.079\pm 0.026$&$\phantom{+}  0.027\pm 0.080\pm 0.026$\\
$  0.310-0.498$ & 0.379& 4.22& $  0.135\pm 0.060\pm 0.014$& $      -0.050\pm 0.048\pm 0.009$&$\phantom{+}  0.051\pm 0.050\pm 0.009$\\
$  0.498-0.799$ & 0.595& 4.55& $  0.061\pm 0.154\pm 0.028$& $      -0.037\pm 0.020\pm 0.004$&      $ -0.005\pm 0.022\pm 0.004$\\
\hline
$  0.075-0.120$ & 0.101& 3.76& $  0.025\pm 0.025\pm 0.007$& $\phantom{+} 0.060\pm 0.366\pm 0.116$& $\phantom{+} 0.000\pm 0.368\pm 0.116$\\
$  0.120-0.193$ & 0.155& 4.97& $  0.048\pm 0.019\pm 0.007$& $\phantom{+} 0.171\pm 0.141\pm 0.049$& $\phantom{+} 0.198\pm 0.142\pm 0.049$\\
$  0.193-0.310$ & 0.243& 6.37& $  0.053\pm 0.022\pm 0.007$& $      -0.068\pm 0.070\pm 0.021$& $\phantom{+} 0.023\pm 0.071\pm 0.021$\\
$  0.310-0.498$ & 0.382& 7.76& $  0.077\pm 0.035\pm 0.008$& $      -0.039\pm 0.034\pm 0.007$& $\phantom{+} 0.043\pm 0.035\pm 0.007$\\
$  0.498-0.799$ & 0.584& 8.85& $  0.106\pm 0.083\pm 0.016$& $      -0.022\pm 0.011\pm 0.002$& $\phantom{+} 0.001\pm 0.012\pm 0.002$\\
\end{tabular}
\end{table}

\begin{table}[t]
\caption{ Results for $A_2$, $g_2$ and $\overline{g_2}$ for the deuteron 
measured in the $4.5^\circ$ and $7^\circ$ spectrometers at the indicated 
average values of $x$ and $Q^2$ and beam energy of 29.1 GeV. 
The highest $x$ bin shown is in the resonance region defined 
by missing mass $W^2<4$ GeV$^2$. } 
\label{tb:g2A2d}
\begin{tabular}{cccccc}
 $x$ interval & $<x>$ & $<Q^2>$ & $A_2^d$ & $g_2^d$ & $\overline{g_2^d}$ \\
 &  &  (GeV/c)$^2$ & $\pm$stat $\pm$syst & $\pm$stat $\pm$syst &    $\pm$stat $\pm$syst \\
\hline
 $0.029-0.047$ & 0.038&  1.49& $\phantom{+} 0.070\pm 0.045\pm 0.010$& $       3.426\pm 2.157\pm 0.575$& $        3.275\pm  2.161\pm 0.575$\\
 $0.047-0.075$ & 0.060&  2.01& $      -0.025\pm 0.028\pm 0.006$& $      -0.655\pm 0.707\pm 0.157$& $       -0.799\pm  0.709\pm 0.157$\\
 $0.075-0.120$ & 0.095&  2.60& $\phantom{+} 0.008\pm 0.032\pm 0.010$& $\phantom{+} 0.008\pm 0.390\pm 0.118$& $       -0.048\pm  0.392\pm 0.118$\\
 $0.120-0.193$ & 0.152&  3.21& $\phantom{+} 0.005\pm 0.045\pm 0.016$& $      -0.118\pm 0.243\pm 0.080$& $       -0.095\pm  0.245\pm 0.080$\\
 $0.193-0.310$ & 0.241&  3.77& $\phantom{+} 0.078\pm 0.072\pm 0.020$& $\phantom{+} 0.127\pm 0.154\pm 0.041$& $\phantom{+}  0.134\pm  0.156\pm 0.041$\\
 $0.310-0.498$ & 0.378&  4.22& $      -0.079\pm 0.144\pm 0.017$& $      -0.127\pm 0.094\pm 0.010$& $       -0.095\pm  0.096\pm 0.010$\\
 $0.498-0.799$ & 0.595&  4.56& $\phantom{+} 0.327\pm 0.390\pm 0.044$& $\phantom{+} 0.037\pm 0.039\pm 0.004$& $\phantom{+}  0.027\pm  0.041\pm 0.004$\\
\hline
 $0.075-0.120$ & 0.101&  3.77& $\phantom{+} 0.024\pm 0.046\pm 0.009$& $\phantom{+} 0.172\pm 0.621\pm 0.133$& $\phantom{+}  0.084\pm  0.624\pm 0.133$\\
 $0.120-0.193$ & 0.154&  4.97& $      -0.007\pm 0.036\pm 0.011$& $      -0.109\pm 0.235\pm 0.076$& $       -0.086\pm  0.237\pm 0.076$\\
 $0.193-0.310$ & 0.242&  6.37& $      -0.043\pm 0.043\pm 0.015$& $      -0.133\pm 0.116\pm 0.041$& $       -0.117\pm  0.118\pm 0.041$\\
 $0.310-0.498$ & 0.381&  7.76& $\phantom{+} 0.000\pm 0.073\pm 0.014$& $      -0.042\pm 0.056\pm 0.011$& $       -0.006\pm  0.057\pm 0.011$\\
 $0.498-0.799$ & 0.584&  8.86& $\phantom{+} 0.235\pm 0.183\pm 0.030$& $\phantom{+} 0.000\pm 0.018\pm 0.003$& $\phantom{+}  0.011\pm  0.019\pm 0.003$\\
\end{tabular}
\end{table}


\begin{table}[t]
\caption{ Results for $A_2$, $g_2$ and $\overline{g_2}$ for the neutron 
measured in the $4.5^\circ$ and $7^\circ$ spectrometers at the indicated 
average values of $x$ and $Q^2$ and beam energy of 29.1 GeV. 
The highest $x$ bin shown is in the resonance region defined 
by missing mass $W^2<4$ GeV$^2$. } 
\label{tb:g2A2n}
\begin{tabular}{cccccc}
 $x$ interval & $<x>$ & $<Q^2>$ & $A_2^n$ & $g_2^n$ & $\overline{g_2^n}$ \\
 &  &  (GeV/c)$^2$ & $\pm$stat $\pm$syst & $\pm$stat $\pm$syst &    $\pm$stat $\pm$syst \\
\hline
 $0.029-0.047$ & 0.038&  1.49& $\phantom{+} 0.143\pm 0.105\pm 0.023$& $\phantom{+} 7.024\pm 4.777\pm 1.288$& $\phantom{+}  6.963\pm  4.787\pm 1.288$\\
 $0.047-0.075$ & 0.060&  2.01& $      -0.085\pm 0.065\pm 0.015$& $     -1.811\pm  1.574\pm 0.367$& $       -1.948\pm  1.579\pm 0.367$\\
 $0.075-0.120$ & 0.095&  2.60& $\phantom{+} 0.013\pm 0.076\pm 0.025$& $\phantom{+} 0.254\pm 0.868\pm 0.266$& $\phantom{+}  0.192\pm  0.872\pm 0.266$\\
 $0.120-0.193$ & 0.152&  3.21& $      -0.013\pm 0.114\pm 0.040$& $      -0.118\pm 0.541\pm 0.180$& $       -0.079\pm  0.545\pm 0.180$\\
 $0.193-0.310$ & 0.241&  3.77& $\phantom{+} 0.067\pm 0.192\pm 0.056$& $\phantom{+} 0.320\pm 0.342\pm 0.093$& $\phantom{+}  0.263\pm  0.347\pm 0.093$\\
 $0.310-0.498$ & 0.378&  4.22& $      -0.433\pm 0.419\pm 0.058$& $      -0.225\pm 0.209\pm 0.024$& $       -0.256\pm  0.214\pm 0.024$\\
 $0.498-0.799$ & 0.595&  4.56& $\phantom{+} 0.926\pm 1.302\pm 0.200$& $\phantom{+} 0.116\pm 0.086\pm 0.010$& $\phantom{+}  0.064\pm  0.091\pm 0.010$\\
\hline
 $0.075-0.120$ & 0.101&  3.77& $\phantom{+} 0.028\pm 0.113\pm 0.024$& $\phantom{+} 0.315\pm 1.392\pm 0.312$& $\phantom{+}  0.184\pm  1.399\pm 0.312$\\
 $0.120-0.193$ & 0.154&  4.97& $      -0.078\pm 0.091\pm 0.029$& $      -0.407\pm 0.528\pm 0.172$& $       -0.384\pm  0.533\pm 0.172$\\
 $0.193-0.310$ & 0.242&  6.37& $      -0.195\pm 0.118\pm 0.041$& $      -0.219\pm 0.261\pm 0.091$& $       -0.276\pm  0.265\pm 0.091$\\
 $0.310-0.498$ & 0.381&  7.76& $      -0.131\pm 0.218\pm 0.045$& $      -0.052\pm 0.125\pm 0.025$& $       -0.057\pm  0.128\pm 0.025$\\
 $0.498-0.799$ & 0.584&  8.86& $\phantom{+} 0.544\pm 0.627\pm 0.120$& $\phantom{+} 0.021\pm 0.041\pm 0.006$& $\phantom{+}  0.023\pm  0.042\pm 0.006$\\
\end{tabular}
\end{table}

\begin{table}
\caption{ Results for the moments 
$\Gamma_1^{(2)}$ and $\Gamma_2^{(2)}$ evaluated at $Q^2=5$ (GeV/c)$^2$, 
and the extracted twist-3 matrix elements $d_2$ for proton (p),
deuteron (d), and neutron(n). The errors include statistical (which dominate) and 
systematic contributions. }  
\label{tb:moments}
\begin{tabular}{cccc}

& $a_2/2=\Gamma_1^{(2)}\times 10^3$ & $\Gamma_2^{(2)}\times 10^3$ &
 $d_2\times 10^3$\\
\hline
p &$12.4\pm1.0$ & $-6.3\pm1.8$  & $5.8\pm5.0$\\
d &$4.6\pm0.8$ & $-1.4\pm3.0$  &  $5.1\pm9.2$\\
n &$-2.4\pm1.6$ & $3.3\pm6.5$  &  $5.0\pm21.0$\\
\end{tabular}
\end{table}

\begin{table}
\caption{ Theoretical predictions for the 
twist-3 matrix elements $d_2^p$ and $f_2^p$  for proton and $d_2^d$ and  $f_2^d$ for deuteron. Also shown is $\mu$, the 
higher twist correction to $\Gamma_1$ described in the text. }
\label{tb:moments-th}
\begin{tabular}{cccccccc}
& \multicolumn{4}{c}{Bag models} &  \multicolumn{3}{c}{QCD sum rules } \\
                  & Ref.~\cite{song} & Ref.~\cite{strat} & Ref.~\cite{jiU}  & Ref.~\cite{Ji} & Ref.~\cite{schafer} & Ref.~\cite{balitsky}  & Ref.~\cite{RossRoberts}\\   \hline
$Q^2$ (GeV/c)$^2$ &  5               &      5            &      1           &        1       &     1               &      1                &       -              \\ 
$d_2^p\times 10^3$ & $17.6$      &       $6.0$       &       21         &     10         &  $-6\pm 3$          &        $-3\pm3$       &       -              \\       
$f_2^p\times 10^3$ &   -         &         -         &       35         &     28         &  $-37\pm6$          &      $-50\pm34$       &     $-69\pm 5$        \\
$\mu_2^p\times 10^3$&  -         &         -         &       27         &     15         &  $-15\pm7$          &      $-20\pm13$       &     $-27\pm 2$        \\   
$d_2^d\times 10^3$ & $6.6$       &      $2.9$        &       11         &      5         &  $-17\pm 5$         &      $-13\pm5$        &        -              \\ 
$f_2^d\times 10^3$ &  -          &         -         &       17         &     14         &  $-25\pm 4$         &      $-34\pm20$       &    $-38\pm 5$         \\
$\mu_2^d\times 10^3$&  -         &         -         &       13         &      7         &  $-10\pm 3$         &      $-13\pm 8$       &    $-15\pm 2$         \\ 
\end{tabular}
\end{table}

 \begin{table}[t] 
 \caption{ Pion asymmetries versus momentum $E'$ for proton and deuteron targets at E=29.1 GeV.}
 \label{tb:ppion}
 \begin{tabular}{ccccc}
 $E'$(GeV) & $A_\parallel^p(\pi^-)$ & $A_\parallel^p(\pi^+)$  &  $A_\parallel^d(\pi^-)$  & $A_\parallel^d(\pi^+)$\\
 \hline
\multicolumn{5}{c}{$\theta=4.5^\circ$} \\
 \hline
  8.04&  \phantom{+}0.028$\pm$  0.010 & \phantom{+}0.016$\pm$  0.019  &\phantom{+}0.001$\pm$  0.011  &  \phantom{+}0.013$\pm$  0.021    \\
 10.65&  \phantom{+}0.019$\pm$  0.012 & \phantom{+}0.043$\pm$  0.022  &$-0.017$ $\pm$  0.013         &  \phantom{+}0.005$\pm$  0.025   \\
 13.57&  \phantom{+}0.019$\pm$  0.016 & \phantom{+}0.019$\pm$  0.031  &$-0.015$ $\pm$  0.018         &  \phantom{+}0.016$\pm$  0.036  \\
 16.58&  \phantom{+}0.001$\pm$  0.023 & \phantom{+}0.065$\pm$  0.043  &$-0.017$$\pm$  0.025          &  \phantom{+}0.020$\pm$  0.052  \\
 19.41&  \phantom{+}0.002$\pm$  0.036 & \phantom{+}0.041$\pm$  0.069  &\phantom{+}0.108$\pm$  0.041  &  \phantom{+}0.031$\pm$  0.087  \\
 21.88&  $-0.044$        $\pm$  0.056 & \phantom{+}0.028$\pm$  0.111  &$-0.108$$\pm$  0.068          &  \phantom{+}0.031$\pm$  0.146   \\
 23.90&  \phantom{+}0.018$\pm$  0.063 & \phantom{+}0.273$\pm$  0.127  &$-0.084$$\pm$  0.079          &  $-0.172$ $\pm$  0.171   \\
 25.44&  \phantom{+}0.058$\pm$  0.082 & \phantom{+}0.057$\pm$  0.168  &\phantom{+}0.057$\pm$  0.111  &   \phantom{+}0.094$\pm$  0.242  \\
 \hline   				                               	  
 \multicolumn{5}{c}{$\theta=7.0^\circ$}             \\		  
 \hline								  
  5.61&  \phantom{+}0.123$\pm$  0.069 & $-0.010$        $\pm$  0.135   & \phantom{+}0.022$\pm$  0.072    &  $-0.272$        $\pm$  0.195 \\
  7.72&  \phantom{+}0.018$\pm$  0.010 &  \phantom{+}0.027$\pm$  0.019   & $-0.021$        $\pm$  0.010   &  $-0.038$        $\pm$  0.031 \\
 10.29&  \phantom{+}0.001$\pm$  0.012 &  \phantom{+}0.081$\pm$  0.023  & $-0.019$        $\pm$  0.013    &  \phantom{+}0.000$\pm$  0.037  \\
 13.19&  \phantom{+}0.016$\pm$  0.022 &  \phantom{+}0.057$\pm$  0.043  &  \phantom{+}0.063$\pm$  0.024   &  \phantom{+}0.034$\pm$  0.071 \\
 16.19&  \phantom{+}0.085$\pm$  0.036 &  \phantom{+}0.071$\pm$  0.075  & $-0.021$         $\pm$  0.042   &  \phantom{+}0.142$\pm$  0.129 \\
 19.06&  \phantom{+}0.096$\pm$  0.052 &  $-0.035$        $\pm$  0.107  &  \phantom{+}0.070$\pm$  0.063   &  \phantom{+}0.231$\pm$  0.196 \\
 21.59&  $-0.034$        $\pm$  0.064 &  \phantom{+}0.106$\pm$  0.136  &  \phantom{+}0.023$\pm$  0.085   &  \phantom{+}0.060$\pm$  0.263\\
 \end{tabular}
 \end{table}

\begin{table}
\begin{center}
\caption{
The measured  virtual photon-nucleon asymmetry $A_1 + \eta A_2$ and the spin 
structure function $g_1$ for the resonance region. The values of $W^2$ and $Q^2$ are given at bin centers. The dilution factor 
$f$ and applied correction term $A_{rc}$, which for these data also includes a 
resolution correction, are from Eqs.~\protect\ref{eq:dilution} and \protect\ref{eq:arc}. The 
value of $g_1$ in the last column is calculated from $A_\parallel$ under the 
assumption that $A_2=0$.}
\label{tbl:A1}
\begin{tabular}{rrrrrr}
\multicolumn{1}{c}{$W^2$} & \multicolumn{1}{c}{$Q^2$} &  
\multicolumn{1}{c}{$f$} &
\multicolumn{1}{c}{${A_{rc}}$} & \multicolumn{1}{c}{$A_1+\eta A_2$} & 
\multicolumn{1}{c}{$g_1$} \\
\multicolumn{1}{c}{GeV$^2$}& \multicolumn{1}{c}{(GeV/c)$^2$} &&&
\multicolumn{1}{c}{$\pm$ stat.$\pm$ syst.} & 
\multicolumn{1}{c}{$\pm$ stat.$\pm$ syst.} \\
\hline
\multicolumn{6}{c}{Proton \hskip 1cm $\theta=4.5^\circ$} \\
\hline
1.31 & 0.55 & 0.180 & $-$0.0444 & $-$0.086$\pm$0.126$\pm$0.225 & $-$0.011$\pm$0.017$\pm$0.030\\
1.69 & 0.54 & 0.164 & $-$0.0284 & $-$0.453$\pm$0.125$\pm$0.127 & $-$0.112$\pm$0.031$\pm$0.032\\
2.06 & 0.53 & 0.155 & 0.0101 & 0.461$\pm$0.106$\pm$0.104 & 0.139$\pm$0.032$\pm$0.031\\
2.44 & 0.52 & 0.153 & 0.0204 & 0.694$\pm$0.091$\pm$0.086 & 0.344$\pm$0.045$\pm$0.035\\
2.81 & 0.50 & 0.154 & $-$0.0051 & 0.222$\pm$0.078$\pm$0.046 & 0.142$\pm$0.050$\pm$0.027\\
3.19 & 0.49 & 0.144 & $-$0.0049 & 0.242$\pm$0.079$\pm$0.092 & 0.142$\pm$0.046$\pm$0.050\\
3.56 & 0.48 & 0.144 & $-$0.0047 & 0.090$\pm$0.072$\pm$0.030 & 0.061$\pm$0.049$\pm$0.018\\
3.94 & 0.47 & 0.147 & $-$0.0032 & 0.002$\pm$0.064$\pm$0.011 & 0.002$\pm$0.054$\pm$0.010\\
4.31 & 0.46 & 0.144 & $-$0.0013 & 0.134$\pm$0.059$\pm$0.017 & 0.129$\pm$0.057$\pm$0.008\\
4.69 & 0.45 & 0.143 & $-$0.0013 & 0.105$\pm$0.056$\pm$0.014 & 0.110$\pm$0.058$\pm$0.009\\
\hline
\multicolumn{6}{c}{Deuteron \hskip 1cm $\theta=4.5^\circ$} \\
\hline
1.31 & 0.55 & 0.247 & $-$0.0242 & $-$0.173$\pm$0.278$\pm$0.236 & $-$0.015$\pm$0.024$\pm$0.021\\
1.69 & 0.54 & 0.235 & $-$0.0221 & $-$0.305$\pm$0.231$\pm$0.078 & $-$0.061$\pm$0.047$\pm$0.015\\
2.06 & 0.53 & 0.232 & $-$0.0009 & 0.290$\pm$0.157$\pm$0.095 & 0.107$\pm$0.058$\pm$0.034\\
2.44 & 0.52 & 0.227 & 0.0115 & 0.184$\pm$0.152$\pm$0.072 & 0.078$\pm$0.064$\pm$0.030\\
2.81 & 0.50 & 0.232 & 0.0030 & 0.021$\pm$0.128$\pm$0.055 & 0.011$\pm$0.068$\pm$0.029\\
3.19 & 0.49 & 0.230 & $-$0.0033 & 0.147$\pm$0.113$\pm$0.021 & 0.093$\pm$0.072$\pm$0.011\\
3.56 & 0.48 & 0.230 & $-$0.0040 & 0.023$\pm$0.102$\pm$0.011 & 0.017$\pm$0.074$\pm$0.008\\
3.94 & 0.47 & 0.230 & $-$0.0037 & $-$0.017$\pm$0.096$\pm$0.010 & $-$0.014$\pm$0.078$\pm$0.008\\
4.31 & 0.46 & 0.233 & $-$0.0031 & $-$0.034$\pm$0.091$\pm$0.009 & $-$0.030$\pm$0.080$\pm$0.008\\
4.69 & 0.45 & 0.229 & $-$0.0028 & 0.055$\pm$0.086$\pm$0.010 & 0.053$\pm$0.084$\pm$0.008\\
\hline
\multicolumn{6}{c}{Proton \hskip 1cm $\theta=7.0^\circ$} \\
\hline
1.56 & 1.26 & 0.159 & $-$0.0524 & $-$0.143$\pm$0.128$\pm$0.113 & $-$0.015$\pm$0.013$\pm$0.012\\
1.94 & 1.23 & 0.154 & $-$0.0012 & 0.349$\pm$0.110$\pm$0.105 & 0.038$\pm$0.012$\pm$0.012\\
2.31 & 1.20 & 0.156 & 0.0251 & 0.795$\pm$0.087$\pm$0.088 & 0.177$\pm$0.019$\pm$0.020\\
2.69 & 1.18 & 0.158 & 0.0036 & 0.593$\pm$0.077$\pm$0.058 & 0.163$\pm$0.021$\pm$0.014\\
3.06 & 1.15 & 0.154 & 0.0046 & 0.507$\pm$0.069$\pm$0.078 & 0.177$\pm$0.024$\pm$0.025\\
3.44 & 1.12 & 0.150 & 0.0029 & 0.262$\pm$0.066$\pm$0.035 & 0.101$\pm$0.025$\pm$0.010\\
3.81 & 1.10 & 0.148 & 0.0027 & 0.299$\pm$0.063$\pm$0.039 & 0.136$\pm$0.028$\pm$0.008\\
4.19 & 1.07 & 0.148 & 0.0038 & 0.433$\pm$0.059$\pm$0.062 & 0.221$\pm$0.030$\pm$0.012\\
4.56 & 1.04 & 0.148 & 0.0035 & 0.324$\pm$0.056$\pm$0.054 & 0.185$\pm$0.032$\pm$0.017\\
4.94 & 1.02 & 0.147 & 0.0033 & 0.237$\pm$0.053$\pm$0.037 & 0.150$\pm$0.034$\pm$0.010\\
\hline
\multicolumn{6}{c}{Deuteron \hskip 1cm $\theta=7.0^\circ$} \\
\hline
1.56 & 1.26 & 0.232 & $-$0.0312 & 0.043$\pm$0.243$\pm$0.103 & 0.003$\pm$0.017$\pm$0.007\\
1.94 & 1.23 & 0.232 & $-$0.0131 & 0.163$\pm$0.179$\pm$0.128 & 0.018$\pm$0.019$\pm$0.014\\
2.31 & 1.20 & 0.230 & 0.0108 & 0.183$\pm$0.154$\pm$0.033 & 0.030$\pm$0.025$\pm$0.005\\
2.69 & 1.18 & 0.231 & 0.0110 & 0.237$\pm$0.134$\pm$0.072 & 0.054$\pm$0.030$\pm$0.016\\
3.06 & 1.15 & 0.234 & 0.0019 & 0.389$\pm$0.114$\pm$0.046 & 0.112$\pm$0.033$\pm$0.012\\
3.44 & 1.12 & 0.228 & 0.0000 & 0.081$\pm$0.105$\pm$0.017 & 0.028$\pm$0.037$\pm$0.006\\
3.81 & 1.10 & 0.228 & $-$0.0005 & 0.121$\pm$0.099$\pm$0.019 & 0.049$\pm$0.040$\pm$0.006\\
4.19 & 1.07 & 0.230 & $-$0.0002 & 0.178$\pm$0.094$\pm$0.030 & 0.080$\pm$0.042$\pm$0.009\\
4.56 & 1.04 & 0.230 & $-$0.0006 & 0.145$\pm$0.091$\pm$0.026 & 0.073$\pm$0.046$\pm$0.008\\
4.94 & 1.02 & 0.230 & $-$0.0006 & 0.201$\pm$0.085$\pm$0.036 & 0.113$\pm$0.048$\pm$0.012\\
\end{tabular}
\end{center}
\end{table}

\begin{table}
\begin{center}
\caption{
 Systematic errors (absolute) on $A_1 + \eta A_2$ 
 by category for the resonance region (See text for details).}
\label{tbl:A1err}
\begin{tabular}{rrrrrrrrrr} \hline
\multicolumn{1}{c}{$W^2$} & \multicolumn{1}{c}{$A_{rc}$} & 
\multicolumn{1}{c}{model} & \multicolumn{1}{c}{$\theta$} &
\multicolumn{1}{c}{$E^\prime$} & \multicolumn{1}{c}{Resol} &
\multicolumn{1}{c}{$P_b P_t$} & \multicolumn{1}{c}{$f$} &
\multicolumn{1}{c}{$R$} & \multicolumn{1}{c}{$g_2$} \\
\hline
\multicolumn{10}{c}{Proton \hskip 1cm $\theta=4.5^\circ$} \\
\hline
1.31 & 0.0083 & 0.1842 & 0.0033 & 0.1156 & 0.0328 & 0.0354 & 0.0260 & 0.0188 & 0.0057\\ 
1.69 & 0.0006 & 0.1245 & 0.0051 & 0.0163 & 0.0052 & 0.0019 & 0.0014 & 0.0169 & 0.0046\\
2.06 & 0.0837 & 0.0082 & 0.0077 & 0.0541 & 0.0019 & 0.0145 & 0.0106 & 0.0224 & 0.0040\\
2.44 & 0.0548 & 0.0614 & 0.0070 & 0.0114 & 0.0098 & 0.0211 & 0.0155 & 0.0520 & 0.0088\\
2.81 & 0.0339 & 0.0019 & 0.0014 & 0.0201 & 0.0040 & 0.0106 & 0.0078 & 0.0191 & 0.0042\\
3.19 & 0.0544 & 0.0832 & 0.0008 & 0.0118 & 0.0027 & 0.0106 & 0.0078 & 0.0349 & 0.0040\\
3.56 & 0.0146 & 0.0256 & 0.0004 & 0.0003 & 0.0008 & 0.0045 & 0.0033 & 0.0150 & 0.0036\\
3.94 & 0.0102 & 0.0015 & 0.0000 & 0.0022 & 0.0020 & 0.0009 & 0.0006 & 0.0003 & 0.0032\\
4.31 & 0.0011 & 0.0028 & 0.0006 & 0.0012 & 0.0010 & 0.0055 & 0.0040 & 0.0156 & 0.0031\\
4.69 & 0.0062 & 0.0017 & 0.0004 & 0.0008 & 0.0001 & 0.0043 & 0.0032 & 0.0115 & 0.0029\\
\hline
\multicolumn{10}{c}{Deuteron \hskip 1cm $\theta=4.5^\circ$} \\
\hline
1.31 & 0.1965 & 0.1531 & 0.0242 & 0.1203 & 0.0375 & 0.0236 & 0.0132 & 0.0027 & 0.0033\\ 
1.69 & 0.0638 & 0.0613 & 0.0129 & 0.0331 & 0.0048 & 0.0015 & 0.0008 & 0.0259 & 0.0049\\
2.06 & 0.0220 & 0.0861 & 0.0180 & 0.0197 & 0.0044 & 0.0161 & 0.0090 & 0.0234 & 0.0058\\
2.44 & 0.0107 & 0.0702 & 0.0059 & 0.0073 & 0.0010 & 0.0046 & 0.0026 & 0.0126 & 0.0040\\
2.81 & 0.0074 & 0.0542 & 0.0017 & 0.0047 & 0.0013 & 0.0004 & 0.0002 & 0.0016 & 0.0022\\
3.19 & 0.0110 & 0.0047 & 0.0067 & 0.0046 & 0.0016 & 0.0099 & 0.0056 & 0.0108 & 0.0034\\
3.56 & 0.0037 & 0.0099 & 0.0014 & 0.0003 & 0.0002 & 0.0027 & 0.0015 & 0.0017 & 0.0022\\
3.94 & 0.0029 & 0.0094 & 0.0003 & 0.0001 & 0.0002 & 0.0003 & 0.0002 & 0.0016 & 0.0019\\
4.31 & 0.0021 & 0.0084 & 0.0011 & 0.0008 & 0.0002 & 0.0009 & 0.0005 & 0.0037 & 0.0016\\
4.69 & 0.0024 & 0.0056 & 0.0027 & 0.0007 & 0.0001 & 0.0039 & 0.0022 & 0.0060 & 0.0016\\
\hline
\multicolumn{10}{c}{Proton \hskip 1cm $\theta=7.0^\circ$} \\
\hline
1.56 & 0.0186 & 0.1033 & 0.0010 & 0.0126 & 0.0386 & 0.0163 & 0.0090 & 0.0030 & 0.0010\\
1.94 & 0.0165 & 0.0186 & 0.0086 & 0.1006 & 0.0010 & 0.0163 & 0.0089 & 0.0137 & 0.0019\\
2.31 & 0.0742 & 0.0028 & 0.0150 & 0.0171 & 0.0141 & 0.0258 & 0.0141 & 0.0251 & 0.0081\\
2.69 & 0.0103 & 0.0287 & 0.0088 & 0.0196 & 0.0012 & 0.0237 & 0.0130 & 0.0370 & 0.0064\\
3.06 & 0.0342 & 0.0630 & 0.0057 & 0.0105 & 0.0043 & 0.0200 & 0.0110 & 0.0385 & 0.0064\\
3.44 & 0.0106 & 0.0196 & 0.0029 & 0.0039 & 0.0024 & 0.0100 & 0.0055 & 0.0262 & 0.0033\\
3.81 & 0.0075 & 0.0030 & 0.0030 & 0.0021 & 0.0013 & 0.0115 & 0.0063 & 0.0353 & 0.0040\\
4.19 & 0.0054 & 0.0056 & 0.0040 & 0.0004 & 0.0004 & 0.0164 & 0.0090 & 0.0591 & 0.0061\\
4.56 & 0.0071 & 0.0235 & 0.0028 & 0.0005 & 0.0005 & 0.0127 & 0.0070 & 0.0463 & 0.0046\\
4.94 & 0.0009 & 0.0099 & 0.0018 & 0.0004 & 0.0003 & 0.0089 & 0.0049 & 0.0341 & 0.0039\\
\hline
\multicolumn{10}{c}{Deuteron \hskip 1cm $\theta=7.0^\circ$} \\
\hline
1.56 & 0.0887 & 0.0335 & 0.0008 & 0.0355 & 0.0288 & 0.0216 & 0.0091 & 0.0030 & 0.0018\\
1.94 & 0.0144 & 0.1146 & 0.0172 & 0.0523 & 0.0004 & 0.0158 & 0.0066 & 0.0016 & 0.0019\\
2.31 & 0.0206 & 0.0020 & 0.0111 & 0.0185 & 0.0051 & 0.0064 & 0.0027 & 0.0113 & 0.0028\\
2.69 & 0.0072 & 0.0686 & 0.0095 & 0.0040 & 0.0004 & 0.0098 & 0.0041 & 0.0163 & 0.0043\\
3.06 & 0.0085 & 0.0136 & 0.0209 & 0.0052 & 0.0110 & 0.0229 & 0.0096 & 0.0269 & 0.0083\\
3.44 & 0.0045 & 0.0129 & 0.0051 & 0.0033 & 0.0032 & 0.0049 & 0.0020 & 0.0056 & 0.0031\\
3.81 & 0.0045 & 0.0101 & 0.0068 & 0.0003 & 0.0030 & 0.0073 & 0.0030 & 0.0115 & 0.0031\\
4.19 & 0.0046 & 0.0095 & 0.0101 & 0.0062 & 0.0004 & 0.0113 & 0.0047 & 0.0232 & 0.0045\\
4.56 & 0.0015 & 0.0075 & 0.0079 & 0.0032 & 0.0004 & 0.0091 & 0.0038 & 0.0208 & 0.0037\\
4.94 & 0.0063 & 0.0059 & 0.0109 & 0.0082 & 0.0011 & 0.0130 & 0.0054 & 0.0289 & 0.0054\\
\end{tabular}
\end{center}
\end{table}

\begin{table}[t]
\caption{Integrals $\Gamma_1(Q^2)$ of the structure functions
$g_1$ for the proton (p), deuteron and neutron (n) at low $Q^2$.  Listed are the
measured sums $\Gamma_1^{\rm res}$ for the  resonance region ($W^2<4$ GeV$^2$)
and $\Gamma_1^{\rm DIS}$ for the deep-inelastic region ($W^2>4$ GeV$^2$ using
data from 9.7 and 16 GeV beam energies), the low-$x$ extrapolation $\Gamma_1^{\rm ext}$
for $x<0.03$, and the combined total $\Gamma_1^{\rm tot}$.}
\label{tbl:gamma}
\begin{tabular}{ccrrrr}
$Q^2$&&\multicolumn{1}{c}{$\Gamma_1^{\rm res}$}&
\multicolumn{1}{c}{$\Gamma_1^{\rm DIS}$} &
\multicolumn{1}{c}{$\Gamma_1^{\rm ext}$} &
\multicolumn{1}{c}{$\Gamma_1^{\rm tot}$}\\
(GeV/c)$^2$ &&\multicolumn{1}{c}{$\pm$stat.$\pm$syst. }&
\multicolumn{1}{c}{$\pm$stat.$\pm$syst.}&&
\multicolumn{1}{c}{$\pm$stat.$\pm$syst.}\\
\hline
0.5&Proton&$0.022\pm0.007\pm0.008$&$0.017\pm0.002\pm0.003$&$0.009$&$0.047\pm0.007\pm0.015$\\
0.5&Deuteron&$0.004\pm0.010\pm0.008$&$0.004\pm0.003\pm0.002$&$0.000$&$0.008\pm0.011\pm0.009$\\
0.5&Neutron&$-0.013\pm0.023\pm0.018$&&&$-0.030\pm0.024\pm0.025$\\
0.5&p-n&&&&$0.077\pm0.027\pm0.036$\\
\hline
1.2&Proton&$0.039\pm0.003\pm0.003$&$0.051\pm0.003\pm0.003$&$0.014$&$0.104\pm0.005\pm0.016$\\
1.2&Deuteron&$0.018\pm0.005\pm0.003$&$0.019\pm0.006\pm0.002$&$0.001$&$0.037\pm0.007\pm0.006$\\
1.2&Neutron&$-0.001\pm0.010\pm0.008$&&&$-0.023\pm0.016\pm0.020$\\
1.2&p-n&&&&$0.127\pm0.018\pm0.034$\\
\end{tabular}
\end{table}

\vfill

\vfill

\vfill

\vfill

\vfill

\vfill

\vfill

\vfill

\vfill

\vfill

\vfill

\vfill

\vfill

\vfill

\vfill

\vfill

\vfill

\vfill

\vfill

\vfill

\vfill

\vfill

\vfill

\vfill

\vfill

\end{document}